%% file: main.tex
\documentclass[12pt,twoside,draft]{report}
\usepackage{latexsym,amsbsy,graphicx,amsmath}
\def\bce{\begin{center}}
\def\ece{\end{center}} 
\def\beq{\begin{equation}} 
\def\eeq{\end{equation}}
\def\VEV#1{\left\langle #1 \right\rangle} 
\def\der#1#2{\frac{d#1}{d#2}}

\def\s{\sigma}

\def\E{{\cal E}}                            
\def\N{{\cal N}}                            
                           
\def\T{{\cal T}}                            
\def\V{{\cal V}}                            
\def\Z{{\cal Z}}                            

\def\a{\alpha}

\def\b{\beta}

\def\d{\delta}
\def\g{\gamma}

\def\L{\Lambda}

\def\l{\lambda}

\def\s{\sigma}
\def\om{\omega}
\def\Om{\Omega}
\def\t{\theta}

\def\bds#1{\boldsymbol{#1}}
\def\lowmp{\lower.11em\hbox{${\scriptstyle\mp}$}}
\def\intf{\int_{-\infty}^{+\infty}}
\def\Im{{\rm Im\,}}
\def\Re{{\rm Re\,}}

\def\abs#1{\left| #1\right|}                
\def\VEV#1{\left\langle #1 \right\rangle}   
\def\frac#1#2{{\textstyle{ #1 \over #2 }}}  

\def\bra#1{\left\langle #1\right|}               
\def\ket#1{\left| #1\right\rangle}               
\def\der#1#2{{d#1\over d#2}}
\def\pdif#1#2{\dfrac{\partial #1}{\partial #2}}
\def\braket#1#2{\left\langle #1 \right| \left. #2 \right\rangle}
\def\VEV#1{\left\langle #1 \right\rangle}
\def\vev#1{\langle #1 \rangle}

\def\mbare{m^2_{\rm b}}
\def\lbare{\lambda_{\rm b}}

\def\intL{\int_{-\Lambda}^\Lambda\dfrac{dk}{2\pi}}

\begin{document}
\begin{titlepage}
\vskip -2in
\title{\bf Aspects of Non-Equilibrium Dynamics in Quantum Field Theory }
\author{by \\ \\ Emanuele Manfredini\\ \\
Dipartimento di Fisica ``G. Occhialini''\\
Universit\`a di Milano-Bicocca\\
P.zza della Scienza, 3, I-20126, Milano, ITALIA.\\
emanuele.manfredini@mib.infn.it} 
\date{\vskip 2in 
Dottorato di Ricerca in Fisica - XII ciclo \\
Universit\`a degli Studi di Milano\\
a.a. 1999/2000\\
Bicocca-FT-01-03\\
\vskip 1cm
\begin{flushleft}
Tutore: Prof. Claudio Destri
\end{flushleft}
}
\maketitle
\end{titlepage}
\begin{abstract}
\input{abs.tex}

\end{abstract}
\subsection*{Preface}
\input{pre.tex}
\tableofcontents
\chapter{Introduction}\label{introduction}
\input{ch1.tex}

\chapter{Out of Equilibrium Dynamics in $\Phi^4$ Quantum Field Theory}\label{sqft}
\input{ch2.tex}

\chapter{Out of Equilibrium Dynamics for the Non linear $\sigma$-model in $2D$}\label{smodel}
\input{ch3.tex}

\chapter{First steps in gauge theories}\label{gauge}
\input{ch4.tex}
\chapter{Conclusions}\label{conclusion}
\input{ch5.tex}

\addcontentsline{toc}{chapter}{Appendix} 
\chapter*{Appendix}
\input{ch6.tex}

\addcontentsline{toc}{chapter}{Bibliography}
\bibliography{biblio}
\bibliographystyle{prsty}

\newpage
\subsection*{Ringraziamenti}
\input{dedica.tex}

\newpage
\subsection*{Parentesi Musicale}
\input{canzone.tex}

\newpage
\subsection*{Adie\`u}
\input{pub-articolo.tex}

\end{document}

%% file: abs.tex

This work is devoted to the study of relaxation--dissipation processes
in systems described by Quantum Field Theory. After a brief
introduction to the main stream of applications and to the general CTP
formalism, a preparatory study in quantum mechanics is presented.

I then introduce the scalar quantum field theory $\l \phi ^4$ in
finite volume, which is studied in the infinite $N$ limit, both at
equilibrium and out of equilibrium. The dynamical equations are
derived and solved numerically. I find that the zero-mode quantum
fluctuations cannot grow macroscopically large starting from
microscopic initial conditions. which leads to the conclusion that
there is no evidence for a dynamical Bose-Einstein condensation, in
the usual sense. On the other hand, out of equilibrium the
long-wavelength fluctuations do scale with the linear size of the
system, signalling dynamical infrared properties quite different from
the equilibrium ones characteristic of the same approximation scheme.

With the aim of going beyond the gaussian approximation intrinsic in
the large $N$ limit, I introduce a non-gaussian Hartree-Fock
approximation (tdHF). I derive the mean-field coupled time-dependent
Schroedinger equations for the modes of the scalar field and I
renormalize them properly. The dynamical evolution in a further
controlled gaussian approximation of our tdHF approach, for $N=1$, is
studied from non-equilibrium initial conditions characterized by a
uniform condensate. I find that, during the slow rolling down, the
long-wavelength quantum fluctuations do not grow to a macroscopic size
but do scale with the linear size of the system, as happens in the
large $N$ approximation of the O(N) model. This behavior shows an
internal inconsistency of this approximation. I also study the
dynamics of the system in infinite volume with particular attention to
the asymptotic evolution in the broken symmetry phase. We are able to
show that the fixed points of the evolution cover at most the
classically metastable part of the static effective potential.

Finally, the dynamical evolution in the $O(N+1)$ nonlinear sigma model
in $1+1$ dimensions is investigated in the large N limit. I first of
all verify that the large coupling limit of the $O(N+1) \, \phi^4$
model, which renders the model non linear, commutes with the large $N$
limit, so that the $O(\infty)$ nonlinear sigma model is uniquely
defined. I then numerically study the evolution of several
observables, with a particular attention to the spectrum of produced
particles during the relaxation of an initial condensate and find no
evidence for parametric resonance, a result that is consistent with
the presence of the nonlinear contraint. Only a weak nonlinear
resonance at late times is observed.

I conclude with some remarks on the ``state of art'' in gauge theories
and some comments about the open issues in the subject.

%% file: pre.tex

This work is devoted to some aspects of the dynamical evolution in
{\em Quantum Field Theory} (QFT). Before describing the specific attitude I
will take and the applications I will be considering in the following
chapters, I would like first to briefly introduce the general setting
which the subject of this thesis belongs to and, at the same time,
give at least one motivation to keep studying QFT. 

If we go back in time and want to talk about the history of
Theoretical Physics, few simple words might be enough: {\em
Unification of Concepts and Descriptions}. If we look at the evolution
of Theoretical Physics, starting from Newton's theory of gravitation,
up to the Standard Model of elementary interactions, passing through
Maxwell's electromagnetism, we realize that the dream of reducing the
complexity of phenomena to a unique fundamental principle (or to the
lowest number of them), has been one of the powerful and successful
ideas, which have been leading theoretical physicists not only to
describe in a simple and beautiful fashion the Nature as was known,
but even to make important predictions and discoveries, later
confirmed by the experiments
\cite{QED,fabri}. We might cite three great historical examples: the
prediction of the existence of a new planet, Neptune, in the solar
system (discovered later in 1846), according to the theory of
gravitation and to the observational data on the orbits of the already
known planets, the prediction of the Hertzian waves (experimentally
observed in 1888), according to Maxwell's theory of electromagnetism
and the prediction of the existence of the electron's antiparticle,
the positron, according to the Dirac's relativistic theory of the
electron (discovered in 1932).

From this point of view, during the 20$^{\rm th}$ century, Theoretical
Physics went very far on the path of Unification. The formulation of
the Standard Model of the Electroweak Interactions, by S. Glashow,
S. Weinberg and A. Salam
\cite{Weinberg:1979pi,Salam:1980jd,Glashow:1979pj} which worth the
Nobel prize to its inventors, represents one of the brightest result
of Theoretical Physics. Three of the four fundamental interactions of
Nature, namely the electromagnetic, the weak and the strong forces,
and all the phenomena which are related to them (almost the entire
world as we know it), can be described in a single conceptual
framework, using a unique ``language'' (as a side-product, again the
$W$ and $Z$ vector bosons were predicted by the Standard Model before
their discovery in 1983). This was possible thanks to the merging of
two of the most important achievements of Theoretical Physics in the
first half of XX$^{\rm th}$ century: Special Relativity
\cite{Einstein:1905ve} and Quantum Mechanics \cite{QM}. Since the
first attempts to reconcile the two theories, it became evident that
internal consistency asked for a Quantum Theory of Relativistic Fields
to be formulated \cite{Wein}. 

Then, in spite of the initial mistrust theoreticians devoted to QFT as
the framework for a fundamental theory, the experiments has been
showing at a deeper and deeper level the capability of such a
``language'' to describe with some simple words almost all the
phenomena happening in our world (for completeness' sake, I should say
that this reductionist point of view may be applicable and justified
when limited, for instance, to particle Physics, but its extension to
the whole Physics, or the whole Science, has been deeply criticized
\cite{science0}; for recent reviews and criticism of QFT, see also
\cite{Weinberg:1996kw,Jackiw:1997tg,Wilczek:1998ma}) ...

All but one. In fact, the gravitational interaction is still described
by the Einstein's General theory of Relativity (GR), which dates back
to 1915 and is a classical (non-quantum) field theory. It is not the
subject of this work to talk about the efforts made to include
gravitation in a QFT description of Nature. Thus it will suffice to
say that, even though gravity still remains out of this scheme, QFT
represents a sort of partial Unification, in the sense specified
above, and in any case, it provides a broad framework within which
problems in different branches of physics can be formulated and
studied.

One of the greatest success of QFT, when it is applied to particle
physics, consists of the ability to predict the scattering cross
sections and decay widths of elementary particles as measured in
collision experiments, like those performed in the accelerators at CERN,
the European Laboratory for Particle Physics near Geneva, and at
Fermilab, the Fermi National Accelerator Laboratory near Chicago. The
mathematical formalism is based on the formal theory of scattering,
where the S matrix elements are computed using a covariant
perturbation theory, based on the expansion on powers of a small
parameter, which is usually a coupling constant of the theory. The
coefficients of the series expansion are obtained by using the Feynman
diagram technique. The crucial point in such an approach, is the
computation of transition amplitudes from an asymptotic state in the
remote past (at $t=-\infty$) to a different asymptotic state in the
remote future (at $t=+\infty$). To this end, one needs to compute the
matrix element between free particle states of the time--evolution
operator $U(t=-\infty,t=+\infty)$.

Although the (perturbative) scattering theory has been very useful, it
is able to address a very limited subset of problems one might want to
solve in QFT. For example, the coupling may not be weak enough to
justify a perturbative expansion, there may not exist the free
asymptotic states of the $0^{\rm th}$ order of perturbation theory or
we may really need something more than just the scattering
probabilities. 

Moreover, the area of applicability of QFT is not limited to particle
physics. In fact, it is now clear that QFT provides a convenient and
unifying formulation also for condensed matter and statistical
mechanics and it represents a valid description and a powerful tool of
computation for different phenomena like the behavior of a metal or an
alloy in the superconducting phase \cite{Bardeen:1957mv} or a
statistical system near the critical temperature of a phase transition
\cite{Wilson:1974jj}.

Thus, while remaining inside such a fruitful conceptual scheme, the
goal of this work is to study in detail some aspects of and put some
light on the out of equilibrium, finite time evolution for systems
described by a QFT, from a point of view which is more appropriate, as
we will see in detail in the following, in situations where
relaxation/dissipation and decoherence effects are important and the
formal theory of scattering is not able to give a complete information
on the process under consideration.

%% file: ch1.tex

\section{Motivations}

There are many interesting physical situations in which the system
under consideration evolves through a series of highly excited
states (i.e., states of finite energy density).

As an example consider any model of cosmological inflation, where the
inflaton drives the universe exponential evolution by staying for a
certain period in states far away from the vacuum
\cite{Boyanovsky:1997mq,Kofman:1994rk,Riotto:1998bt}.

On the side of particle physics, the ultra-relativistic heavy-ion
collisions, scheduled in the forthcoming years at the Relativistic
Heavy Ion Collider of Brookhaven National Laboratory (BNL-RHIC) and at
the Large Hadron Collider of the European Laboratory for Particle
Physics (CERN-LHC), are supposed to produce hadron matter at very high
densities and temperatures and out of thermal equilibrium; in such a
regime, an approach based on particle scattering at low density cannot
be considered a good interpretative tool. To extract sensible
information from the theory new computational schemes are necessary.

We need, first of all, to go beyond the simple Feynman diagram
description at finite order. The use of resummation schemes, like the
Hartree--Fock
\cite{Kerman:1976yn,cactus} approximation (HF) and the large $N$ limit
\cite{largen_exp}, or the Hard Thermal Loop resummation (HTL) for systems at
finite temperature
\cite{Pisarski:1989cs,Pisarski:1989vd}, can be considered a first step
in this direction. They, in fact, enforce a sum over an infinite
subset of Feynman diagrams that, in the case of HTL, are dominant in a
given region of the parameter space, where the simple truncation of
the usual perturbative series at finite order is not justified. In the
case of HF, instead, the approximation resums the diagrams which
become dominant when the number of spatial dimensions are high. In any
case, all of these schemes allow for a non--perturbative treatment of
the dynamics under consideration and resummation techniques are
already applied to the computation of scattering cross sections and
decay width.

This is not enough, however, when one wants to describe systems under
extreme conditions of density and temperature during their approach to
thermal equilibrium. In this case, the suitable approach to QFT
consists in setting up an initial value problem. To this end, the
standard formulation based on time--ordered Green's functions is not
viable, while a Schroedinger picture approach \cite{Eboli:1988fm} may
help in clarifying some aspects, by shifting the interest upon the
evolution of local field expectation values rather than on the
computation of in--out matrix elements. Thus, one specifies an initial
state (generally described by a density matrix functional) and then
follows its evolution under the time--evolution operator generated by
the quantum Hamiltonian. 

A similarity with a classical mechanics problem may be useful to
clarify the difference between this approach and the in--out
formalism. Studying out of equilibrium evolution in QFT is like
solving a classical dynamics problem, where one follows the time
history of the mechanical system, once the initial conditions (for
position and velocity) have been specified. On the other hand, one
could also fix the mechanical configurations at an initial time
$t_{\rm in}$ and a final time $t_{\rm out}$, and then look for the
specific trajectory, if any, which connects the two configurations at
the two different times; this is exactly what one does to compute
cross sections using S matrix elements, where $t_{\rm in} \to
-\infty$, $t_{\rm out} \to +\infty$ and the initial and final state
describe free particles.

As a consequence, the standard functional formalism used in S matrix
calculations, is not appropriate to perform the task of finding the
solution of an initial value problem in QFT, because the argument of
the standard effective action $\Gamma$ is the matrix element of the
quantum field operator between an {\em in} and an {\em out} state. In
fact, if one tries to derive field equations from such an effective
action, complex and non causal quantum corrections appear. On the
contrary, the correct order parameter is obtained as the average value
of the field operator at a fixed time. Thus, while the effective
potential (or the Landau--Ginzburg--Helmholtz free energy) has been
very useful in clarifying matters in static configurations, the non
equilibrium evolution of the expectation value of quantum fields
should be studied from {\em first principles}, resorting to
non--equilibrium formalism and addressing issues like dissipative
dynamics, decoherence and phase transitions out of equilibrium
\cite{Calzetta:1988bw1,Calzetta:1988bw2,Paz:1990jg,Paz:1990sd,Gleiser:1994ea}.

Of course, the Cauchy problem in QFT is far too difficult to be solved
exactly (which would correspond to solve exactly either the time
dependent Schroedinger equation for the wave--functional or the
quantum Heisenberg equations of motion for the time-dependent field
operators). Actually, as we will see in great detail in the following
chapters, some sort of approximation is always applied, usually in the
form of a dynamics reduced to a subspace of the Hilbert space of the
system.

To summarize, by {\em Quantum Field Theory Out of Equilibrium} I mean
the study of the dynamics of non-equilibrium processes from a
fundamental field-theoretical description, by deriving and solving the
dynamical equations of motion for the expectation values and
correlation functions of physically relevant observables, in the
underlying $(3+1)-$dimensional quantum field theory. Typical examples are:
phase transitions and particle production out of equilibrium, symmetry
breaking and dissipative processes, relaxation and transport processes
in extreme environment.

The study of real time evolution in Quantum Field Theory dates back to
the early 60's and can be rigorously formulated by means of the
so--called Closed Time Path (CTP) generating functional
\cite{CTP1,CTP2,CTP3,Chou:1985es,Landsman:1987uw} (cfr. section
\ref{CTPsec}), or related formalisms like the Feynman--Vernon
influence functional technique \cite{Feynman:1963fq} and the Zwanzig's
projection method
\cite{Anastopoulos:1997qp}. The problem, even if well defined, is so
complex that little can be done analytically even in the simplest
situation. Thus, the method was used quite rarely until the beginning
of the 90's. A decade ago, a large research program was started by
different research groups
\cite{QFTooE1,Boyanovsky:1993pf,QFTooE3,Baacke:1997se}, who obtained useful
results in realistic physical systems, by heavily exploiting the power
of the modern computing facilities. In fact, it should be clear that
one will have to resort to numerical techniques in order to solve the
partial non linear differential equations of any realistic theory;
thus, a formulation of the problem practical for numerical methods is
crucial. However, it must be also said that the real time dynamics of
phase transitions can be studied in a concrete way with presently
available computers.

I finally want to emphasize from the beginning the interdisciplinary
character of such an approach, which can be applied with considerable
impact in the fields of cosmology, astrophysics, particle physics,
quantum optics and statistical mechanics out of equilibrium. Relevant
results have been obtained in the study of pion condensates,
supercooled phase transitions (strongly out of equilibrium),
inflationary cosmology and early universe dynamics, strong field
electrodynamics in astrophysical plasmas, the hadronization stage of
the quark gluon plasma, particle production in heavy ion collisions,
dynamical symmetry breaking and dissipative processes, decoherence
processes in the transition from quantum to classical regime.

For the reader's benefit, I will summarize here the main topics
addressed and some remarkable results obtained in the framework of
Quantum Field Theory Out of Equilibrium.

\subsection*{Cosmological issues}
The method has been used to study the non--equilibrium aspects of
cosmological phase transitions, in an inflationary
scenario. The Inflationary paradigm \cite{Guth:1981zm} (for recent
developments see also ref. \cite{Kolb:1999ar}) is one of the greatest
application of QFT; ``rolling'' equations in the new inflationary model 
and the corrections introduced by thermal and quantum fluctuations 
\cite{Boyanovsky:1997mq,Boyanovsky:1994pa,Boyanovsky:1996sv,
Cormier:1998nt,Boyanovsky:1999jg} are studied using the methods
of non equilibrium quantum statistical mechanics. 

This formalism is able to give the quantum evolution equations for the
inflaton field coupled to Einstein's equation for the metric, allowing
for a semiclassical treatment of the coupled gravity-matter problem
and showing the existence of inflationary solutions also in this case
\cite{Boyanovsky:1997mq}. The subject is relevant to the study of {\em
Reheating} of the Universe at the end of inflationary epoch, which can
be treated in a full dynamical fashion, without the use of static
quantities like the effective potential. Reheating proceeds by means
of either parametric (for chaotic models) or spinodal (for new
inflation models) instabilities, with profuse particle
production. Both are non--perturbative out of equilibrium mechanisms
and require a self--consistent non--perturbative approximation scheme
to be used, in order to be correctly described in a quantitative
manner. The first stage leads to a non thermal momentum spectrum for
the produced particles, and is called {\em Preheating} for this
reason. After and maybe together with that, thermalization occurs via
standard scattering processes among particles \cite{Felder:2000hr}.
Moreover, it is widely accepted that our Universe undertook several
phase transitions during its cooling down from the initial Big
Bang. The last transitions, namely the deconfining and the chiral
phase transitions, may be experimentally proved at the modern
accelerators, as the BNL-RHIC, which started its activity quite
recently, and the CERN-LHC. Even the baryon asymmetry (the
predominance of matter over antimatter) may be explained by an
electroweak phase transition in non--equilibrium conditions
\cite{Sakharov:1967dj}. The efficiency of such an electroweak baryogenesis may 
be quantitatively measured by the use of the non equilibrium formalism. In 
fact, while the equilibrium and static properties of phase transitions are by 
now well understood, we still need to clarify many points about the dynamics 
of the processes involved as a phase transition proceeds in time and out of 
thermal equilibrium.

Early attempts to analyze the temperature-induced symmetry--changing
phase transitions, that are the crucial ingredients in inflationary
cosmology, were based on the use of the static, temperature dependent,
effective potential, with the temperature changing in time according
to some plausible rule. But the methods of thermal field theory are of
no help, unless we are able to show that the evolution proceeds so
slowly that local thermal equilibrium is maintained all along the
phase transition. If this is not the case, it should be noticed that
the effective potential is useless, being a quantity defined strictly
at equilibrium. As such, it can give information about static
properties like possible stable or metastable equilibrium states and
critical temperature, but it loses all information about real time
processes, like the approach to equilibrium.

An other important goal is to study the process of domain formation
and growth in an expanding cosmology, which is very relevant for our
understanding of the formation of scalar density perturbations during
the phase transition, may have left an imprint in the fluctuations of
the cosmic microwave background and should have allowed the formation
of large scale structures in the Universe.

Such theoretical efforts are justified by an intensive experimental
work that has transformed Cosmology in a truly observational
discipline. Indeed, the precise determination of the power spectrum of
the anisotropies in the Cosmic Microwave Background Radiation
temperature, yielded first by COBE \cite{cobe}, very recently by
BOOMERANG \cite{boomerang} (much more detailed measurements are
expected from MAP \cite{map} and PLANCK \cite{planck} missions),
allows to test the quantitative predictions of inflationary models
based on particle physics scenarios and in the near future will
certainly validate or rule out specific proposal.

\subsection*{Quantum Decoherence}
An other interesting subject is the study of the decoherence process,
which means the transition from quantum to classical behavior. Since
quantum superpositions of different mean field states are so difficult
to observe in nature, the dynamics should be such that the
interference between macroscopically distinguishable states is
dynamically suppressed
\cite{Cooper:1994hr,Habib:1996ee,Anastopoulos:1998wh}. The subject may clarify
the dynamics of phase separation in strongly supercooled phase transitions
both in Minkowsky
\cite{Boyanovsky:1994pa,Cooper:1997ii} (like the formation and evolution of 
defects in $^4$He after a sudden quench) and in cosmological backgrounds
\cite{Boyanovsky:1999jg} and also some aspects of charged particles dynamics 
\cite{Anastopoulos:1998wh}; 
with this regard, a formulation based on the evolution of a density matrix is 
very useful because reveals the emergence of a semiclassical stochastic 
description. In mean field approximations, the decoherence process seems to 
proceed through a dephasing of the different modes and causes a growth of a 
suitably defined effective entropy \cite{Cooper:1997ii}.

\subsection*{Bose--Einstein Condensation}
From the condensed--matter side, the recent experiments on the
Bose--Einstein condensation of dilute atomic gases in laser traps at
ultracold temperatures have raised a wide interest \cite{science1,
science2,science3}. I think that a true understanding of this
phenomenon requires the control of non equilibrium properties, since
the characteristic time scales and the temperatures are not those
typical of thermalization processes.

\subsection*{Disoriented Chiral Condensates}
One of the phenomena which may occur in hadron physics at high
energies, is the possible formation of Disoriented Chiral Condensates
(DCC's), which are regions of misaligned vacuum in the isospin
internal space of the pions, analogous to misaligned domains in a
ferromagnet. Such regions might act as pion lasers, in the sense that
they may relax to the standard ground state by coherent emission of
pions, with highly non gaussian charge distribution. This may provide
an explanation for the Centauro (overwhelmingly charged) and
Anti-Centauro (overwhelmingly neutral) cosmic-ray events and no other
processes besides a QCD phase transition out of equilibrium have been
proposed, which could produce such a signature.

The influence of quantum and thermal fluctuations on the dynamics of
the formation of Disoriented Chiral Condensate has been studied in the
framework of Quantum Field Theory Out of Equilibrium; this subject may
be experimentally relevant at present accelerator energies
\cite{Boyanovsky:1994pa,Boyanovsky:1995yk}.

\section{Layout presentation}

Of course, such a broad field of applicability cannot be covered even
partially by the issues addressed in a Ph.D. thesis. Many of the topics 
illustrated in the previous section, 
however, have been already clarified by means of a thorough analysis
of various dynamical aspects in the scalar $\phi^4$ theory in
$(3+1)$D. Many physical systems, in fact, at different levels of description,
may be modeled using such a field theory. Nevertheless, a complete
understanding of the topic is still lacking and a relatively little
work has been done to clarify the matter with respect to other field
theories.

Thus, the specific subject of this thesis is the study of the real
time dynamics in two specific models with opposite renormalization
properties in the ultraviolet, namely the ``trivial'' $\lambda \phi^4$
model in $(3+1)$D and the asymptotically free non linear $\sigma$
model in $(1+1)$D, mainly considering the evolution of translation
(and rotation) invariant states.

Thus, in chapter \ref{sqft}, after a brief introduction to the CTP
formalism, an amusing exercise is performed in Quantum Mechanics,
considering a harmonic oscillator plus a quartic perturbation. Then,
the $\lambda \phi^4$ model in $(3+1)$D is studied in a finite volume
and in the infinite $N$ limit both at equilibrium and out of
equilibrium, with particular attention to certain fundamental features
of the broken symmetry phase. The numerical solution of the dynamical
evolution equations shows that the zero--mode quantum fluctuations
cannot grow macroscopically large starting from microscopic initial
conditions. I conclude that a Bose--Einstein condensation of a
``novel'' form is implied by the non--equilibrium dynamics. On the
other hand, out of equilibrium the long--wavelength fluctuations do
scale with the linear size of the system, signalling dynamical
infrared properties quite different from the equilibrium ones
characteristic of the same approximation scheme.

Now, it is to be noticed that the large $N$ limit of $O(N)$ $\phi^4$ model is 
a completely {\em gaussian} theory of {\em transverse} modes. I am  
interested in going beyond both this approximations and in finding a scheme 
which allows an interaction between the longitudinal and the transverse modes 
and is able to retain some non gaussian feature of the complete theory.

Consequently, I consider also an other non--perturbative scheme,
namely a non--gaussian time dependent Hartree--Fock approximation
(tdHF) both at equilibrium and out of equilibrium. I concentrate
myself to the simplest case $N=1$, with particular attention to the
structure of the ground state and of certain dynamical features in the
broken symmetry phase. I derive the mean--field coupled
time--dependent Schroedinger equations for the modes of the scalar
field and I outline a suitable procedure to renormalize them. A
further controlled gaussian approximation of this new tdHF approach is used
in order to study the dynamical evolution of the system from
non--equilibrium initial conditions characterized by an uniform
condensate. I find that, during the slow rolling down, the
long--wavelength quantum fluctuations do not grow to a macroscopic
size but do scale with the linear size of the system, in accordance
with similar results valid for the large $N$ limit of the $O(N)$
model. This behavior is at the basis of an argument proving that the
gaussian approximation within this tdHF approach is inconsistent. It
would be interesting to numerically solve the quantum Schroedinger
equations (\ref{H_k}) and (\ref{H_k1}), in order to have a better 
comprehension of this inconsistencies.

In addition, I study the dynamics of the system in infinite volume
with particular attention to the asymptotic evolution in the broken
symmetry phase.  I am able to show that the fixed points of the
evolution cover at most the classically metastable part of the static
effective potential.

As a first step in future directions, I derive the fundamental equations
for the case $N>1$ and I outline some problems in completing the
renormalization procedure safely.

I also relax the hypothesis of spatial uniformity, deriving the
evolution equations for a rotationally invariant scalar condensate in
the large $N$ limit. A suitable algorithm to solve the partial
derivatives equations is presented and a possible implementation on a
PC's cluster of the {\em Beowulf} class is outlined.

The $\phi^4$ model in $(3+1)$D dimension is a {\em trivial} theory
\cite{Wilson:1974jj}: it suffers from complete screening and would
have a vanishing coupling constant in the absence of a cutoff. Thus,
it must be always considered as an effective theory, valid below a
certain scale of energy. It is well known that there exist theories
with a completely different ultraviolet behavior, like the non linear
$\sigma$ model in $(1+1)$D and QCD. They are asymptotically free
theories in the ultraviolet, which means that the running coupling
decreases with the increase of the energy scale.

With the aim of starting an analysis of the dynamical properties of
such theories, the out--of--equilibrium dynamics of the $O(N+1)$
nonlinear $\sigma$--model in $1+1$ dimensions is investigated in chapter 
\ref{smodel}, performing the large $N$ limit. Regarding the nonlinearity as 
the effect of a suitable large coupling limit of the $O(N+1) \, \phi^4$ model, 
I first of all verify that the two limits commute, so that the $O(\infty)$
nonlinear $\sigma$ model is uniquely defined. Thanks to asymptotic
freedom, such model can be completely renormalized also in the
out--of--equilibrium context. I numerically study the spectrum of
produced particles during the relaxation of an initial condensate and
find no evidence for parametric resonance, a result that is consistent
with the presence of the nonlinear constraint. Only a weak nonlinear
resonance at late times is observed.

I present in chapter \ref{gauge} the results obtained so far by other research 
groups, studying the dynamical evolution in abelian gauge theories like scalar 
and fermionic QED. The subject is relevant to the comprehension of the 
formation and evolution out of equilibrium of a plasma of electrically charged 
particles, as it may happen in certain astrophysical configurations. I also 
outline the main difficulties which one has to face, when the same approach is 
applied to the study of non abelian gauge theories. 

Finally, in chapter \ref{conclusion} I summarize the results and I comment 
on some open issues in the subject.

I close with an appendix containing some details on technical issues 
about the analytical and numerical computations performed in this work.

In conclusion, understanding out of equilibrium processes in Quantum
Field Theory requires the use of non perturbative approximation
schemes. On the other hand, the results derived by the use of such
schemes should be carefully checked with respect to the approximations
used, in order to understand to what extent those conclusions can
be extended to the full theory and to the phenomena that theory is
modeling. With this respect, the use of more and more powerful
computers will be of huge help in the near future, in order to
investigate quantitatively the phenomenology of QFT out of equilibrium
at a deeper and deeper level, especially in the still almost
unexplored arena of non abelian gauge theories.

\vskip 0.5cm
Let me conclude this introductory chapter with the words of Frank Wilczek 
\cite{Wilczek:1999fr}: {\em As physicists we should not, however, be satisfied 
with hoarding up formal, abstract knowledge.  There are concrete experimental
situations and astrophysical objects we must speak to.  Hopefully,
having mastered some of the basic vocabulary and grammar, we will soon
be in a better position to participate in a two-way dialogue with
Nature.}

%% file: ch2.tex

\section{Introduction}
As I said in the previous chapter, Quantum Field Theory Out of
Equilibrium can be defined as the study of real time dynamics of
quantum fields and addresses the fundamental issue of obtaining the
evolution equations for time dependent order parameters or field
condensates including the effect of quantum fluctuations.

Regarding the field theoretical models, whose non equilibrium
properties one wants to understand, one of the most studied is the
$\phi^4$ model in $(3+1)$D. In fact, there are many physical systems
which can be modeled by a self-interacting scalar field. To give an
example, spontaneous symmetry breaking by a scalar order parameter
occurs in systems as diverse as $^4$He at temperatures of 2 $K$ to the
Standard Model at temperatures of $10^{15}$ $K$.

\subsection{Dissipation and Decoherence}
From the numerical computations in such a scalar model, we can
conclude that an effectively dissipative dynamics is observed, where
the large energy density, present only in few modes at early times, is
continuously transferred to the quantum fluctuations resulting a
spectacular production of particles for Bose fields (Pauli blocking
prevents the same phenomenon from occurring for fermions).  In
particular, two questions have found a satisfactory answer through the
use of Quantum Field Theory Out of Equilibrium: (i) how can a
microscopic dissipative behavior be derived from a time reversal
invariant dynamics ? (ii) can a phenomenological friction term of the
form $\Gamma \dot\phi$ be derived from the underlying equations ? The
first question implies the study of dissipative effects in quantum
mechanical systems and its answer is well-known and lies
in some kind of ``coarse graining'' of the degrees of freedom of the
field, which consists in the distinction between {\em system} modes,
which are out of equilibrium, and {\em environment} modes (the thermal
bath), which drive the system into equilibrium. This makes a closed
system to be divided into an open one coupled to a heat bath. The
classical prototypes are Boltzmann's equation and the Brownian motion,
where an averaging over the characteristics of the {\em environment}
is performed in the first place (the theories of quantum Brownian
motion acted as an intellectual gym to clarify the main issues
regarding dissipation and decoherence). Of course, the complete
dynamics is unitary and energy conserving, but the dissipation and the
decoherence may be introduced in the game by a necessary
approximation, for instance in the form of a truncation of the
infinite tower of coupled Schwinger--Dyson equations of QFT.

However, a dissipative behavior can be extracted without any sort of
averaging procedure from the beginning. Usually, in situations
described by QFT, the long wavelength modes (often the condensate
alone) act as the ``system'' under consideration, while the faster
short wavelength modes provide the thermal bath. The infinite set of
coupled ordinary differential equations account for the continuous
energy transfer from the macroscopic condensate to the quantum
modes. Thus, an effectively irreversible energy flow from coherent
mean field to quantum fluctuations is observed, resulting in a
spectacular creation of elementary excitations (massless Goldstone
bosons in case of Spontaneous Symmetry Breaking)
\cite{Cooper:1997ii}. 

If one addresses the problem in such a way, it becomes possible to
answer the second question rigorously. In fact, one gets
integro-differential equation generally non local in the time
variable. It turns out that the term linear in the field velocity,
has a kernel which does not become local in any limit
\cite{Gleiser:1994ea,Boyanovsky:1995me}, meaning that the interaction
with the faster modes acts effectively as a noise term with memory.

In addition, recent works have shown that the familiar picture of
``rolling'' is dramatically modified, when quantum fluctuations are
taken into account
\cite{Cormier:1998nt,Boyanovsky:1994xf,Boyanovsky:2000hs}. 
They are indeed extremely important and enhanced by the instabilities
that are the hallmark of the phase transition.

\subsection{Inflationary Cosmology}
One other field in which the application of non--equilibrium techniques is
mandatory is inflationary cosmology. In some inflationary models, the
inflaton potential is taken to be quartic and $O(N)$
symmetric. Essentially one studies a linear $\sigma$ model coupled to
a cosmological background. The inflaton must be treated as a
quantum field out of equilibrium, because the constraint on the quartic
self-coupling ($\lambda \leq 10^{-12}$) does not allow the complete
thermalization of all modes. In fact, the long wavelength modes interact too
slowly compared with the universe expansion rate, in order to reach
the thermal equilibrium. A great effort was made to understand the
dynamics of quantum fields evolving in a cosmological background, from
the point of view of Quantum Field Theory Out of Equilibrium (for a
review of the results, see for instance
\cite{Boyanovsky:1997mq} and references therein). One can divide the
approach to this problem in three levels of increasing complexity: a)
the non linear dynamics of quantum fields is studied in Minkowsky
space time, with experimental application to high energy particle
collisions
\cite{Boyanovsky:1997mq}; b) the influences of a fixed
cosmological background on the dynamics of the quantum fields are
analyzed, studying the evolution in matter and radiation dominated
Friedman-Robertson-Walker (FRW) and de Sitter universes
\cite{Boyanovsky:1997mq,Boyanovsky:1994xf}; c) a self--consistent
treatment of quantum fields semiclassically coupled to a cosmological
background is used and the consequences of the out--equilibrium
evolution of the inflaton field on the scale factor and vice versa are
elucidated \cite{Boyanovsky:1997mq}: in the new inflation model, the
spinodal instabilities drive the growth of non--perturbatively large
quantum fluctuations, which eventually shut off the growth of the
scale factor, turning the exponential expansion into a power law
one. In addition, the method accounts for a mechanism of generation of
density perturbations and is able also to give precise prediction on
the power spectrum of the anisotropies in the cosmic microwave
background radiation.

\subsection{Approximation schemes}

In any case, it is not possible to limit oneself to the use of
perturbation theory, as the validity of the perturbative expansion
holds at early times only. The presence of parametric or spinodal
amplification is indeed responsible for the exponential growth of
quantum fluctuation. Thus, dissipation and out of equilibrium
evolution in general, can only be understood beyond perturbation
theory. Basically, two non perturbative approximation schemes have
been used so far: the large $N$ expansion (cfr. section \ref{ln}), which might
in principle be improved systematically, and the time dependent
Hartree-Fock approximation (cfr. section \ref{impHF}), which allows a lower
level of control, but has a somewhat larger applicability.


Both schemes have these three good properties: they are renormalizable
schemes, maintain all of the conservation laws and lend themselves to a
detailed analytical and numerical study. In addition, large $N$ has
the following advantage: it can be consistently improved, in
principle, considering next to leading powers in $1/N$, although this
appears extremely awkward from the numerical point of view.

As far as the renormalization is concerned, as we will see in the
following, the main point is to obtain finite evolution equations in
terms of suitably defined renormalized parameters. I will show later
how this can be done by the use of an ultraviolet cutoff $\L$ (see
also
\cite{Boyanovsky:1994xf}). Another renormalization scheme has been 
considered in \cite{Baacke:1997rs}, which is fully covariant and
independent of initial conditions. It has been applied to the
evolution of a scalar field in a conformally flat FRW universe,
including quantum back-reaction in one--loop approximation, using
dimensional regularization and the $\overline{\rm MS}$
renormalization scheme. 

In any case, the dynamics which follows from these approximation
schemes, shows the presence of two different time scales: the early
time evolution is driven by linear instabilities, the back-reaction of
the quantum fluctuations on themselves can be neglected and analytical
results are available; in fact, the fluctuations satisfy a Lam\`e
equation (a Schroedinger like equation with a two-zone potential); a
second time scale, during which the quantum back-reaction becomes
comparable to the tree level terms and the dynamics turns completely
non linear and non--perturbative; during this stage, the instabilities
in the evolution of the quantum fluctuations are shut off by the
back-reaction itself. To be precise, it should be noticed that a third
asymptotic time scale, which reveals the emergence of a scaling regime
with a dynamical correlation length, is observed at least for the
evolution in radiation and matter dominated FRW backgrounds
\cite{Boyanovsky:1999jg}.

\section{A phenomenological application}

\subsection*{What can Quantum Field Theory Out of Equilibrium say about 
Heavy Ion Collision ?}

The RHIC and LHC experiments will hopefully be able to probe the
Quark--Gluon plasma and the Chiral Phase Transition. In fact, the
current typical estimates of energy densities and temperatures near
the central rapidity region are $\varepsilon \approx 1 - 10 {\rm
Gev/fm^3}$ and $T_0\approx 300 - 900 {\rm MeV}$ (see
ref. \cite{Boyanovsky:1998ba} and references therein).  Thus,
according to the lattice estimates, the central rapidity region will
be well above the transition temperature.

The initial state after the collision will be strongly out of
equilibrium and the evolution will proceed towards thermalization and
hadronization, through perturbative and nonperturbative processes. The
perturbative aspects are studied by means of parton cascade models,
which keep track of the constituent evolution by following the parton
distribution functions, as determined by the perturbative parton
parton dynamics. After thermalization, one basically assume that a
boost invariant hydrodynamic description is suitable. This means that
the system is assumed in local thermodynamical equilibrium with a
local energy--momentum tensor and a local equation of state and such a
picture should emerge from the underlying fundamental physics when the
energy is large compared to the rest mass of the hadrons involved. The
complete justification of this, as well as the description of the
hadronization process and of particle production out of equilibrium,
will certainly require a non perturbative treatment. Whether the system
will reach or not thermal and chemical equilibrium is not clear at all
and a detailed investigation, using rate equations and transport
models, or, better the formalism of non--equilibrium QFT, is necessary.

Among the spectacular phenomena which may be studied in heavy ion collision 
experiments, one of the most fascinating is the possible formation of
Disoriented Chiral Condensates (DCC's), which are regions of misaligned
vacuum in the isospin internal space, analogous to misaligned domains
in a ferromagnet. As the ``baked alaska'' scenario proposed by
Bjorken, Kowalski and Taylor implies, such regions would act as ``pion
lasers'', in the sense that they may relax to the standard ground state
by coherent emission of pions, with highly non gaussian charge
distribution. After a real heavy ion collision, the central rapidity
region cools down by a rapid expansion, which lowers the energy
density until when the individual pions fly toward the detector.

A first attempt to model the dynamics of the
chiral order parameter in a far from equilibrium phase transition was
made in \cite{Rajagopal:1993ah}, where a sudden quench below the
critical temperature was considered for the $O(4)$ linear $\sigma$
model. The Gell-Mann--Levy lagrangian \cite{Gell-Mann:1960np} is
believed to correctly describe the low energy interaction of pions and
represents a concrete way to describe a far from thermal equilibrium
plasma after a heavy ion collision. It can be also obtained as a
Landau--Ginzburg effective action from a Nambu--Jona--Lasinio model
which is often used to describe the phenomenology of chiral symmetry
at the quark level. In any case, since the effective mass squared
is negative, sufficiently long wavelength modes are amplified
exponentially at early times. Even though this is a completely
classical treatment (without any attempt to include quantum
corrections) and relies on a series of idealizations and approximations,
it elucidates a mechanism which could occur in real heavy ion
collisions.

An other interesting analysis of a phenomenologically relevant 
scenario, using out of equilibrium techniques, starts considering the 
spherically symmetric state of the so--called {\em Tsunami} configuration
\cite{Boyanovsky:1998ba,Cao:1999tc}, in a theory which admits Spontaneous 
Symmetry Breaking at
zero density: the dynamics of a dense relativistic quantum fluid out
of equilibrium is studied starting from an initial state described by
a Gaussian wave functional with a large number of particles around
$|\vec{k}_0|$. This is relevant to the Physics of in-medium effects at
high energy density, which may be dominant in scenarios like a heavy ion 
collision or the interior of dense stars. The subsequent dynamical evolution 
shows the existence of a critical density beyond which the symmetry is
restored at the beginning and is dynamically broken at late times by
the presence of spinodal instabilities. Also a dynamical restoration
at late times, of a symmetry broken at the beginning can happen, if the energy 
density is large enough. Due to the strong non
linearities in the dynamics, a deep rearrangement in the particle
distribution takes place, leading to profuse production of soft
particles. The equation of state of the asymptotic gas is ultra-relativistic 
(even if the distribution is not thermal) \cite{Boyanovsky:1998ba}.
In ref. \cite{Cao:1999tc}, analytic solutions for narrow
particle distributions and early times are provided, as well as an
interesting study of the two point correlation function at equal times,
which displays the contribution of two terms: one is due to the
initial particle distribution, while the other is entirely due to the
Goldstone bosons created by the spinodal instabilities. The
asymptotic equation of state is of radiation type if the symmetry is 
asymptotically broken.

Pion production modeled by the same theory is studied also with
non--homogeneous condensates with cylindrical symmetry
\cite{Boyanovsky:1998ka} (chosen to take advantage of experience
achieved in the study of theories in $1+1$ dimensions).

Extension of such analysis for gauge theories would be very useful (cfr. 
chapter \ref{gauge}).

\section{Introduction to the CTP formalism}\label{CTPsec}

There exist an elegant method to obtain the evolution equation for
mean values, which is known as Closed Time Path (CTP) functional formalism
and will be introduced in this section, even though I will not
have to resort to such general technique for the applications
considered in this work.

For a theory defined by a Lagrangian density $L$, in the presence of
an arbitrary source $J(x)$, the generating functional suitable for S
matrix computation is defined as the vacuum persistence amplitude
$$
Z[J] = \langle {\rm out} , +\infty | {\rm in} -\infty \rangle .
$$
By varying with respect to the external source, it is possible to
compute matrix elements of the Heisenberg field operators between
$\ket{\rm in}$ and $\bra{\rm out}$ states. These off-diagonal matrix
elements are in general not real and their equations of motion are not
causal. Thus, they are not appropriate to describe the out of
equilibrium time evolution of the system, in the sense clarified above.

For that purpose, we need to consider diagonal (equal time) matrix
elements of field operators and the natural tool for do that is the CTP
functional.

The CTP formalism can be introduced considering first a diagonal
matrix element of the system at time $t$ and insert a complete set of
states into this matrix elements at a later time $t'$. Now, the
diagonal matrix element can be expressed as a functional integral of
products of transition matrix elements from $t$ to $t'$ and their time
reversed. Thus, the diagonal matrix elements may be expressed using
the standard path integral representation. If the forward time
evolution takes place in the presence of an external source $J^+$ but
the backward evolution takes place in the presence of a different
external source $J^-$, then we get a generating functional depending
on both the external sources $J^+$ and $J^-$, which produces diagonal
matrix elements under derivation with respect to the $J$s:
$$
Z_{\rm in}[J^+,J^-] = \exp \left(i W_{\rm in} [J^+,J^-]\right) = \int [{\cal D}\Psi] \braket{\rm in}{\Psi}_{J^-} 
\braket{\Psi}{\rm in}_{J^+} =
$$
\begin{equation}\label{CTPgf}
\int [{\cal D}\Psi] \bra{\rm in}{\cal T}^* \exp\left[-i \int_0^{t'} d^4x 
J^-(x)\Phi(x)\right] \ket{\Psi,t'} \bra{\Psi,t'}{\cal T} \exp\left[i
\int_0^{t'} d^4x J^+(x)\Phi(x)\right] \ket{\rm in}
\end{equation}
and
\begin{equation}
\dfrac{\d W_{\rm in} [J^+,J^-]}{\d J^+} | _{J^+=J^-=0} = - \dfrac{\d
W_{\rm in} [J^+,J^-]}{\d J^-} | _{J^+=J^-=0}  = \bra{\rm in} \Phi (x)
\ket{\rm in}
\end{equation}

Since the first transition matrix element has a backward time
ordering, from $t'$ to $0$, while the first one has a forward time
ordering, from $0$ to $t'$, the generating functional (\ref{CTPgf}) is
given the name of Closed Time Path generating functional.

This expression can be generalized to arbitrary initial density
matrix. Introducing a path integral representation for each transition
matrix element results in the doubling of fields and leads to the
expression
$$
Z_{\rm in}[J^+,J^-] = \int [{\cal D} \varphi] [{\cal D} \varphi ']
\bra{\varphi} \rho \ket{\varphi '} \int [{\cal D \psi}] 
\int _{\varphi} ^{\psi} [{\cal D \phi^+}] \int _{\varphi '} ^{\psi}
[{\cal D} \phi^-] 
$$
$$
\exp\left[i \int_0^{\infty} d^4x (L[\phi^+] - L[\phi^-]
+ J^+\phi^+ - J^-\phi^-)\right]
$$
Now, if we want to consider thermal equilibrium at the beginning, it
will be enough to consider the suitable thermal density matrix,
proportional to $\exp[-\b H(t_0)]$. 

From this generating functional, it is possible to compute $W$, the
generating functional for the connected Green's function, and by means
of a Legendre transform, we get the generating functional of the 1PI
graphs, $\Gamma$, which gives the equations of motion, under derivation
with respect to its argument.  Needless to say, the equations are now
real and causal, as they should be.

An example of application of this formalism can be found in
\cite{Cooper:1994hr}, where the CTP functional computation has been
married with the large$-N$ expansion (see below, section \ref{ln}) in
order to derive the time evolution of a closed system consisting of a
mean field interacting with its quantum fluctuations. Two specific
cases have been considered: the $O(N)$ $\l\phi^4$ model and QED with
$N$ fermion fields. The first model can give interesting information
on the real time dynamics of phase transitions with a scalar order
parameter. The second one accounts for pair creation processes in
strong electric fields and may clarify the scattering and transport
properties of e$^+$ e$^-$ plasmas.

For completeness' sake, I should say that the problem of dynamical
evolution in QFT may be addressed also from a different point of
view. It is possible to write a quantum Liouville equation for the
effective action which contains in a compact form the time evolution
of all equal time correlators
\cite{Nachbagauer:1997ub}.

\section{Warm up in Quantum Mechanics}\label{warmup}
\subsection{Generalities}\label{gen}

The elegant and formal methods introduced in the previous section
have the nice property of being very general. However, when one
considers the problem of finding dynamical evolution equations for
specific cases, it is usually not necessary at all to use them in all
their generality; it is often enough, sometimes better, to derive
equations of motion by more direct procedures, which in turn allow for
a better control of the physics characterizing the case under
consideration.

Thus, let us forget for a while the general formalism and consider the
simplest case one can imagine: one simple harmonic oscillator with a
quartic self-interaction in quantum mechanics. This system has only
one degree of freedom and allows for a clear derivation of the
equations of motion, without the use of CTP formalism.

In this section, I will introduce the main qualitative concepts and the
non perturbative approximation schemes, which will be very useful in
the analysis of the dynamical evolution in Quantum Field Theory, as
will be discussed in the following chapters. Its goal is to make the
reader familiar with some of the ideas and techniques I will be using
during this work. Many of them will be applied with little changes to
QFT, which reduces to a system with finitely many degrees of freedom,
once ultraviolet and infrared cutoffs have been introduced (see
section \ref{cft}).

First of all, I show here how to split the dynamical variables in the
classical and quantum fluctuation parts and how to describe their mutual
interaction. In the specific example of a harmonic oscillator with a
quartic perturbation, I start from the exact Heisenberg equations and
I consider three different approximation schemes: (i) the loop
expansion in powers of $\hbar$ (I get the evolution equations to
order $\hbar$), (ii) the Hartree-Fock approximation and (iii) the
large $N$ expansion (at leading order), both reducing the quartic
potential to a quadratic one with a self-consistent time-dependent
frequency. The main feature of this description lies in its
Hamiltonian nature: it is possible to describe the quantum evolution
by suitable {\em classical} systems (i.e. couples of canonical
variables, Poisson brackets, a suitable Hamiltonian, ...) that however
retain some dependence on $\hbar$. Thus, the dynamical equation we
derive directly from the quantum Heisenberg equations, may be
interpreted as Hamilton equations referring to a particular classical
system. As we will see later on, this property may be maintained and
efficiently used also in the less simple Field Theoretical models we
will be considering in the following (cfr. section \ref{cft} and
chapter \ref{smodel}).

Here, I also start to clarify the meaning of dissipative processes,
as energy transfer from a part of the system to the other. Of course,
in Quantum Mechanics we can not talk about a real dissipation, the
{\em dissipative} channels being finitely many.

\subsubsection{Evolution equations}
\label{eveq}
Consider a non relativistic quantum system, described by an Hamiltonian 
operator of the generic form:
\begin{equation}\label{qmham}
\hat{H} (\hat{q}, \hat{p}) = \dfrac{\hat{p} ^{2}}{2 m} + V \left( \hat{q} 
\right)
\end{equation}
In  the Heisenberg representation, the operators are time-dependent
and they evolve according to the Heisenberg equations:
\begin{eqnarray}
- i \, \hbar \, \dot{\hat{q}} \, = \left[ \hat{H} ,\hat{q} 
\right] & = \dfrac{\hat{p}}{m}  \\
- i \, \hbar \, \dot{\hat{p}} \, = \left[ \hat{H} , \hat{p} 
\right] & = - V' \left( \hat{q} \right)
\end{eqnarray}
Let $x$ be the mean value of the position operator $\hat{q}$ on a state 
described by the vector $\left| \Psi \right\rangle$; if we split $\hat{q} = x 
+ \sqrt{\hbar} \hat{\xi}$, we can expand the equation of motion for $\hat{q}$
\begin{equation}
m \, \stackrel{\cdot \cdot}{\hat{q}} \, = - V' \left( \hat{q} \right)
\end{equation}
as a power series in $\hbar$
\begin{equation}
\stackrel{\cdot \cdot}{x} + \sqrt{\hbar} \stackrel{\cdot
\cdot}{\hat{\xi}} \, = - \dfrac{1}{m} V' \left( x \right) -
\dfrac{1}{m} V '' \left( x \right) 
\sqrt{\hbar}\hat{\xi} - \dfrac{1}{m} \dfrac{V ''' \left( x \right)}{2} \hbar 
\hat{\xi} ^{2} + \cdots
\label{eqsv}
\end{equation}

\subsubsection{Energy}
I will split the total energy of the system as the sum of two pieces,
one referring to the {\em classical} variable (mean value) and the
other to the quantum fluctuations. Being the system isolated, the
total energy (the sum of the two parts) is a constant of motion; on
the other hand, there is no reason to expect the individual parts to
remain constant. Indeed, it is possible that a transfer from the
classical part to the quantum part (and vice versa) will take place
during the evolution. The conserved energy of the system described by
the vector $\left| \Psi
\right\rangle$ is defined as
\begin{equation}\label{totE}
E \left( t \right) = \left \langle H \right \rangle _{\Psi}
\end{equation}
Of course, I can write an approximate conserved energy in the same
way I get approximate evolution equations.

\subsection{$O \left( \hbar \right)$ Expansion}
\label{ohbar}
\subsubsection{Equations of motion}
The simplest approximation leading to an interaction between the mean value 
and its quantum fluctuations is obtained by retaining only the $O \left( \hbar 
\right)$ term in the equation for $x$ and only the leading term in the 
equation for the fluctuation operator $\hat{\xi}$. This approximation
will be valid as long as the neglected terms remain ``small'' compared
to the one I have kept in the equations. As we are considering a time
evolution problem, we might envisage a situation in which the
approximation is justified during certain periods of time, while is
not in other periods. Anyway, at this level of approximation, I get
the coupled equations (from here on, I remove the hats from the
operators, being clear from the context when I am referring to
operators and when to c-numbers):
\begin{equation}
\stackrel{\cdot \cdot}{x} =  - \dfrac{1}{m} V' \left( x \right) - 
\dfrac{\hbar}{2m} V ''' \left( x \right) \vev{\xi ^{2}} _{\Psi}
\label{vm}
\end{equation}
\begin{equation}
\stackrel{\cdot \cdot}{\xi} = - \dfrac{1}{m} V '' \left( x \right) \xi
\label{fl}
\end{equation}
The equation for the operator $\xi$ being linear, its solution can be
written as a linear combination (with operatorial coefficients) of two
real functions $u \left( t \right)$ and $v \left( t \right)$, that are
a basis for the linear space of solutions:
\begin{equation}
\xi \left( t \right) = \xi _{0} u \left( t \right) + 
\stackrel{\cdot}{\xi} _{0} v \left( t \right)
\label{xit}
\end{equation}
The two real functions may be reassembled together in a unique complex
function $f$, as is made for example in
\cite{Habib:1996ee,Cooper:1997ii}, defining:
\begin{equation}\label{omdef}
u = - \sqrt{2} \,\omega(0) \,\Im f \qquad v = \sqrt2 \,\Re f \qquad V '' 
\left[ x \left( t \right) \right] = m \omega ^2 \left( t \right)
\end{equation}
Now the equation (\ref{vm}) can be written as:
\begin{equation}
\begin{array}{ll}
\ddot{x} \left( t \right) &= - \dfrac{1}{m} V' \left( x 
\right) \\
&- \dfrac{\hbar}{2\,m} V ''' \left( x \right) \left[ \VEV{\xi
^{2} _{0}}_{\Psi} u ^{2} \left( t \right) + \VEV{\stackrel{\cdot}{\xi}
^{2} _{0}}_{\Psi} v ^{2} \left( t \right) + \VEV{\left\{ \xi_{0} ,
\dot{\xi} _{0} \right\}}_{\Psi} u \left( t 
\right) v \left( t \right) \right]
\end{array}
\label{vm1}
\end{equation}
\begin{equation}\label{fl1}
\dfrac{d}{dt} \left(
\begin{array}{c}
u \\ v
\end{array} 
\right) = -  V '' \left[ x \left( t \right) \right] \left(
\begin{array}{c}
u \\ v
\end{array} 
\right)
\end{equation}

The Cauchy's conditions are:
\begin{equation}
u \left( 0 \right) \, = \stackrel{\cdot}{v} \left( 0 \right) = 1 \qquad 
v \left( 0 \right) \, = \stackrel{\cdot}{u} \left( 0 \right) = 0
\end{equation}
\begin{equation}
x \left( 0 \right) \, = \left \langle q _{0} \right \rangle _{\Psi}
\qquad 
\stackrel{\cdot}{x} \left( 0 \right) \, = \dfrac{1}{m} \left \langle p _{0}
\right \rangle _{\Psi}
\label{x0v0}
\end{equation}

\subsubsection*{Energy $O \left( \hbar \right)$}
Expanding the energy (\ref{totE}) to first order in $\hbar$ I obtain:
\begin{equation}
E \left( t \right) = \dfrac{m}{2} \VEV{\dot{q} ^{2}}_{\Psi} + \VEV{V
\left( q \right)}_{\Psi} = E ^{cl} \left( t \right) +
\hbar E ^{fl} \left( t \right)
\end{equation}
where
\begin{equation}
E ^{cl} \left( t \right) = \dfrac{m}{2} \dot{x} ^{2} + V \left( x 
\right) 
\end{equation}
and
$$
E ^{fl} \left( t \right) = \VEV{\xi ^{2} _{0}} \left[ \dfrac{m}{2}
\dot{u} ^{2} \left( t \right) +
\dfrac{V'' \left( x \right)}{2} u ^{2} \left( t \right) \right] + 
$$
$$
+ \VEV{\dot{\xi} ^{2} _{0}}_{\Psi} 
\left[ \dfrac{m}{2} \dot{v} ^{2} \left( t \right) + \dfrac{V'' 
\left( x \right)}{2} v ^{2} \left( t \right) \right] + 
$$
\begin{equation}
+ \VEV{\left\{ \xi _{0} , \dot{\xi} _{0} 
\right\}}_{\Psi} \left[ \dfrac{m}{2} \dot{u} \left( 
t \right) \dot{v} \left( t \right) + \dfrac{V'' \left( x \right)}
{2} u \left( t \right) v \left( t \right) \right]
\end{equation}

%
\subsubsection{Hamiltonian formalism}
Once the quantum fluctuations have been expressed in terms of the two
real functions $u \left( t \right)$ and $v \left( t \right)$, the
original operatorial nature of the system has disappeared from the
game; thus, we may look for a classic dynamical system described by
the set of canonical coordinates $\left\{ x \left( t \right) , u
\left( t \right) , v \left( t \right)
\right\}$. It turns out that the following Lagrangian:
\begin{equation}
L \left( x , {\bf z} , \dot{x} , \dot{{\bf z}} \right)
 = \dfrac{m}{2} \dot{x} ^{2} + \hbar \dfrac{m}{2} \dot{\bf z} ^{T} A 
\dot{\bf z} - V \left( x \right) - \hbar 
\dfrac{V '' \left( x \right)}{2} {\bf z} ^{T} A {\bf z}  
\end{equation}
where
$$
A = \left(
\begin{array}{cc}
 \VEV{\xi ^{2} _{0} }_{\Psi} & \dfrac{1}{2} 
\VEV{\left\{ \xi _{0} , \dot{\xi} _{0} \right\}}_{\Psi} \\
\dfrac{1}{2} \VEV{\left\{ \xi _{0} , \dot{\xi}_{0} \right\}}_{\Psi} & 
\VEV{\dot{\xi}^{2}_{0}}_{\Psi}
\end{array}
\right)
,
{\bf z} =
\left(
\begin{array}{c}
u \\
v
\end{array}
\right)
$$
gives exactly the equations of motion derived before, as its
Euler-Lagrange equations.

We can define the momenta conjugated to each canonical coordinate 
$$
p _{x} = \dfrac{\partial L}{\partial \dot{x}} \qquad {\bf p _{z}} = 
\dfrac{\partial L}{\partial \dot{\bf z}}
$$
and we get the Hamiltonian by performing the Legendre transform
\begin{equation}
H \left( x , {\bf z} , p _{x} , {\bf p _{z}} \right) = \dfrac{1}{2 m} p _{x} 
^{2} + \dfrac{1}{2 \hbar m} {\bf p _{z}} ^{T} A ^{-1} {\bf p _{z}} + V \left(
x \right) + \hbar \dfrac{V '' \left( x \right)}{2} {\bf z} ^{T} A {\bf z}
\end{equation}
This Hamiltonian generates a set of Hamilton equations completely equivalent 
to (\ref{vm1}) and (\ref{fl1}).

\subsection{An example}\label{example}
We are now ready to adapt the equation (\ref{vm1}) to a quite
interesting toy--model. I am going to specify both the quantum system,
by giving the precise form of the potential (harmonic oscillator with
a quartic perturbation), and the initial conditions, by choosing a
gaussian wave packet as a trial state. Consequently, it will be
possible to solve numerically the problem. Its solution allows us to
study in details the energy balance between the classic degrees of
freedom and the quantum ones, even if we do not expect to see any kind
of dissipation for the mean value $x$, because of the quantum
mechanical nature of the model; anyway, this problem should be thought
as a preparatory exercise to the more complex computation that will be
made in the framework of a scalar field with quartic interaction
($\lambda
\phi ^{4}$ model, cfr. sections \ref{cft}ff). 
\subsubsection{Gaussian packet}
Let us see what are the initial conditions and parameters, that I need to 
insert in the equations (\ref{vm1}) and (\ref{x0v0}), when I want to 
start from a gaussian state $\left| \Psi \right\rangle$; in the Schroedinger 
representation, the wave function at the time $t=0$ can be written as
\begin{equation}
\braket{x}{\Psi} \, = \Psi _{X , \sigma} \left( x ; 0 \right)
 = \dfrac{1}{(2 \pi \sigma^2_0) ^{1/4}} \exp \left[ - \dfrac{(x -
 \bar{x}) ^{2}}{4 \sigma^2_0} \right] 
\label{pacchetto}
\end{equation}
So, the initial conditions are:
\begin{equation}
x \left( 0 \right) = \VEV{q _{0}}_{\Psi} \,= \bar{x}
\end{equation}
\begin{equation}
m\,\dot{x} \left( 0 \right) = \VEV{ p _{0} }_{\Psi} \,= 0
\end{equation}
while the parameters are:
\begin{equation}
\VEV{\xi ^{2} _{0}}_{\Psi} =
\dfrac{\sigma^2_0}{\hbar} \qquad \VEV{\dot{\xi}^{2} _{0}}_{\Psi} \,= 
\dfrac{\hbar}{4 \sigma^2_0 m ^{2}}
\qquad \VEV{ \left\{ \xi _{0} , \dot{\xi} _{0} \right\}}_{\Psi} = 0
\end{equation}

%
\subsubsection{Harmonic potential + quartic perturbation}
\label{pot_arm}
The system I will be studying in this section is specified by adding a
quartic perturbation to the standard harmonic potential:
\begin{equation}\label{hp}
V \left( q \right) = \dfrac{1}{2} s m \omega ^{2} q ^{2} + 
\dfrac{\lambda}{4 !} \, q ^{4}
\end{equation}
where the parameter $s$ can assume only two values, either $+1$ or
$-1$. For the time being, I concentrate on the first order in $\hbar$,
postponing the discussion of non perturbative approximation schemes to
the next sections. Thus, I compute the first, second and third
derivative of this potential and I insert them in the equation
(\ref{vm1}), together with the parameters obtained previously for a
gaussian packet. The result is the evolution equation describing this
particular case.

At this point it is worth noticing that in order to study the energy
balance, it is not necessary to know the complete time evolution of
the operator $\xi \left( t \right)$ but it is enough to know the mean
values related to it: $\VEV{ \xi ^{2}}_{\Psi}$
and $\VEV{\dot{\xi}^{2}}_{\Psi}$. Then, I define:
\begin{equation}
\VEV{\xi ^{2}}_{\Psi} = \sigma \left( t \right) ^{2}
\label{csiquad_A}
\end{equation}
and
\begin{equation}
\VEV{\dot{\xi} ^{2}}_{\Psi} = 
\dot{\sigma} \left( t \right) ^{2} + \sigma \left( t \right) B \left( t 
\right)
\label{csipuntquad_A}
\end{equation}
It is easy to show by some algebra that (see also the discussion on a
generalized set of coherent states, with application to time dependent
systems, in ref. \cite{rajagopalandmarshall})
\begin{equation}
B \left( t \right) = \dfrac{1}{4 m ^{2} \sigma \left( t \right) ^{3}}
\end{equation}
It is useful also to scale the variables, in order to solve the
equations numerically: I define a dimensionless time variable $\tau =
\omega t$, a dimensionless quantum coupling constant $g = \hbar
\lambda / m^2 \omega ^3$ and the functions
\begin{equation}
\eta \left( \tau \right) = \sqrt{\dfrac{\lambda}{6 m \omega ^2}} x \left( t 
\right)
\label{def_z}
\end{equation}
\begin{equation}
a \left( \tau \right) = \sqrt{\dfrac{m \omega}{2}} \sigma \left( t \right)
\label{def_a}
\end{equation}
Rescaling the evolution equations, I get the following second order
Cauchy problem:
\begin{equation}
\eta '' + s \eta + \eta ^3 + g a ^2 \eta = 0
\label{eqrinx}
\end{equation}
\begin{equation}\label{qfl_onel}
a '' + s a + 3 \eta^2 a - \dfrac{1}{16 a ^3} = 0
\end{equation}
with the following initial conditions:
\begin{equation}
\begin{array}{ll}
\eta \left( 0 \right) = \sqrt{\dfrac{\lambda}{6 m \omega ^2}} \bar{x}
= \alpha & a \left( 0 \right) = \sigma_0\sqrt{\dfrac{m \omega}{2
\hbar}} = \beta\\
\eta' \left( 0 \right) = 0 & a ' \left( 0 \right) = 0 
\end{array}
\label{ci}
\end{equation}

A little comment on various approximation schemes is in order here:
\begin{enumerate}
\item The term $\eta^2\, a$ in eq. (\ref{qfl_onel}) may be treated as a
perturbation, if one wants to generate a perturbative expansion in
terms of the {\it amplitude} of the mean value. The adimensional
variable $a$ is the modulus of the adimensional counterpart of the
complex function $f$ in (\ref{omdef}). Neglecting the term $\eta^2\,
a$ in (\ref{qfl_onel}) yields the $0^{\rm th}$ order equation, which quite
obviously, has the solution $a^0(\tau) = 1/2$. The phase of $f$
instead varies linearly with time. Now, inserting $a^0$ back in
(\ref{qfl_onel}) allows to derive the equation for the $1{\rm st}$
order term, which is then substituted in (\ref{eqrinx}); thus, to
one--loop level and to cubic order in the mean value amplitude, I get
the following evolution equation:
\begin{equation}
\ddot{\eta}(\tau) + (s + 6 \pi g) \eta(\tau) + \eta ^3 (\tau) + 6
\pi g \eta(\tau)\, \left\{ \int _0 ^{\tau} \eta(\tau ') 
\dot{\eta} (\tau ') \cos [ 2 ( \tau - \tau ')] d\tau ' \right\} = 0 
\end{equation}
Notice that while here we have a {\em finite} redefinition of the
frequency, in the field theoretical case we will have an {\em
infinite} renormalization of the mass.
\item Otherwise, we can absorb the term quadratic in the mean value in the
definition of a time dependent frequency and then solve the evolution
equations to first order in $\hbar$ and all orders in $\eta$ (generating
a {\em loop} expansion), and this is what we are doing in this
paragraph.
\item An other possibility is to consider non perturbative
approximation schemes which enforce an infinite resummation of all
orders in the perturbative expansion in $\hbar$. This case will be
treated in the next section.
\end{enumerate}

I can also define a dimensionless energy variable
\begin{equation}
\varepsilon = \dfrac{\lambda}{3 m ^2 \omega ^4} E = \varepsilon  ^{cl} + 
\varepsilon ^{fl} 
\label{E_A}
\end{equation}
which, once expressed in terms of the dimensionless variables defined 
previously, turns out to be:
\begin{equation}
\varepsilon ^{cl} = \eta ^{\prime 2} + s \eta ^{2} + \dfrac{\eta ^{4}}{2}
\label{enclrin}
\end{equation}
\begin{equation}
\varepsilon ^{fl} = \dfrac{g}{3} \left[a ^{\prime 2} + \left( s + 3 \eta ^{2} 
\right) a ^{2} + \frac{1}{16 a ^2} \right]
\label{enflrin}
\end{equation}

%

\subsection{Non perturbative approximation schemes}
\label{sez_en}
\begin{figure} 
\includegraphics[height=8cm,width=15cm]{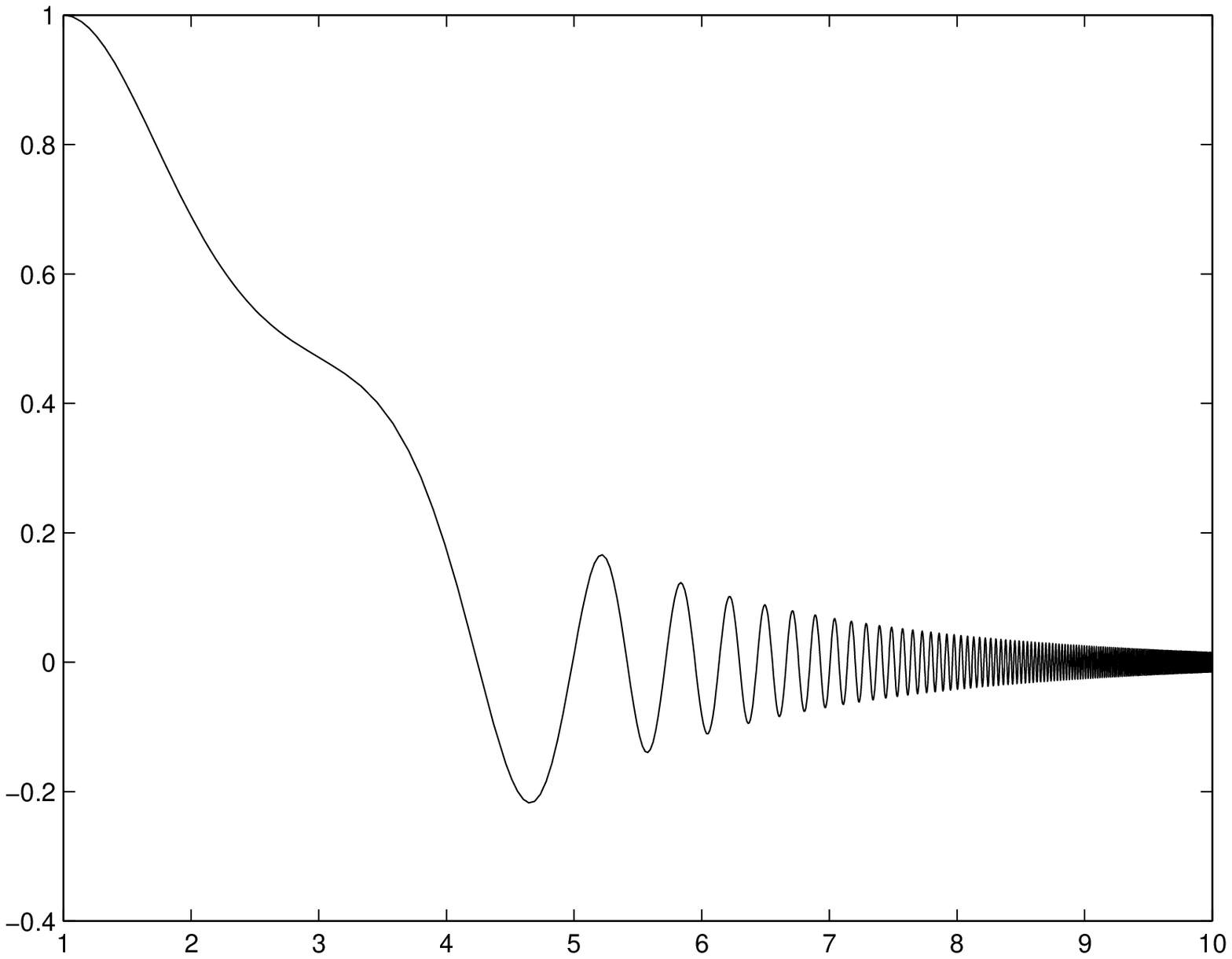}
\caption{\it Evolution of the expectation value $\eta$ according to
eq. (\ref{eqrinx}) and (\ref{qfl_onel}). Here $s=-1$ and $g=1$ }\label{mv1l}
\end{figure}

\begin{figure} 
\includegraphics[height=8cm,width=15cm]{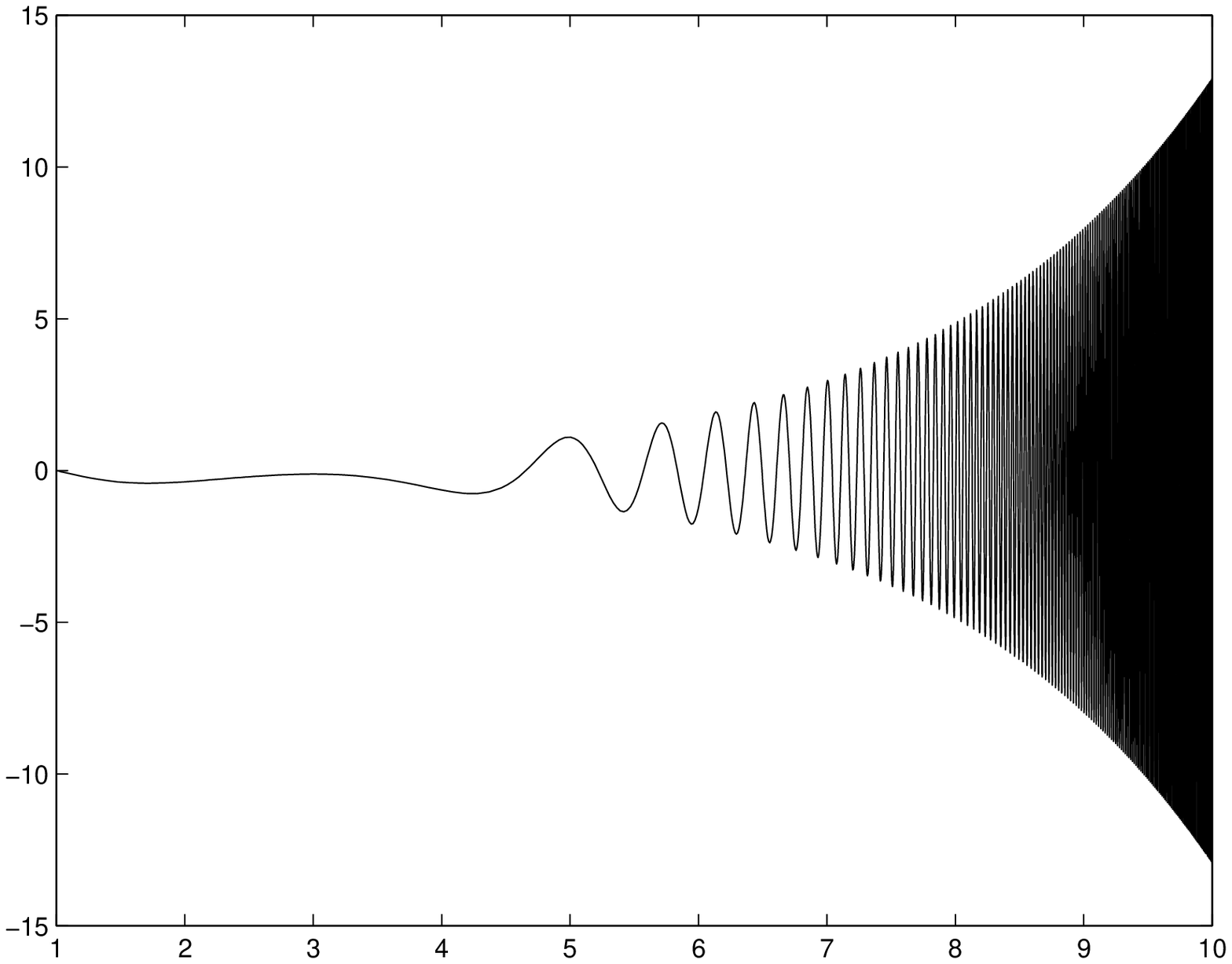}
\caption{\it Evolution of the expectation value speed $\eta'$ according to
eq. (\ref{eqrinx}) and (\ref{qfl_onel}), for the values of the
parameters as in fig \ref{mv1l} }\label{mvs1l}
\end{figure}

\begin{figure} 
\includegraphics[height=8cm,width=15cm]{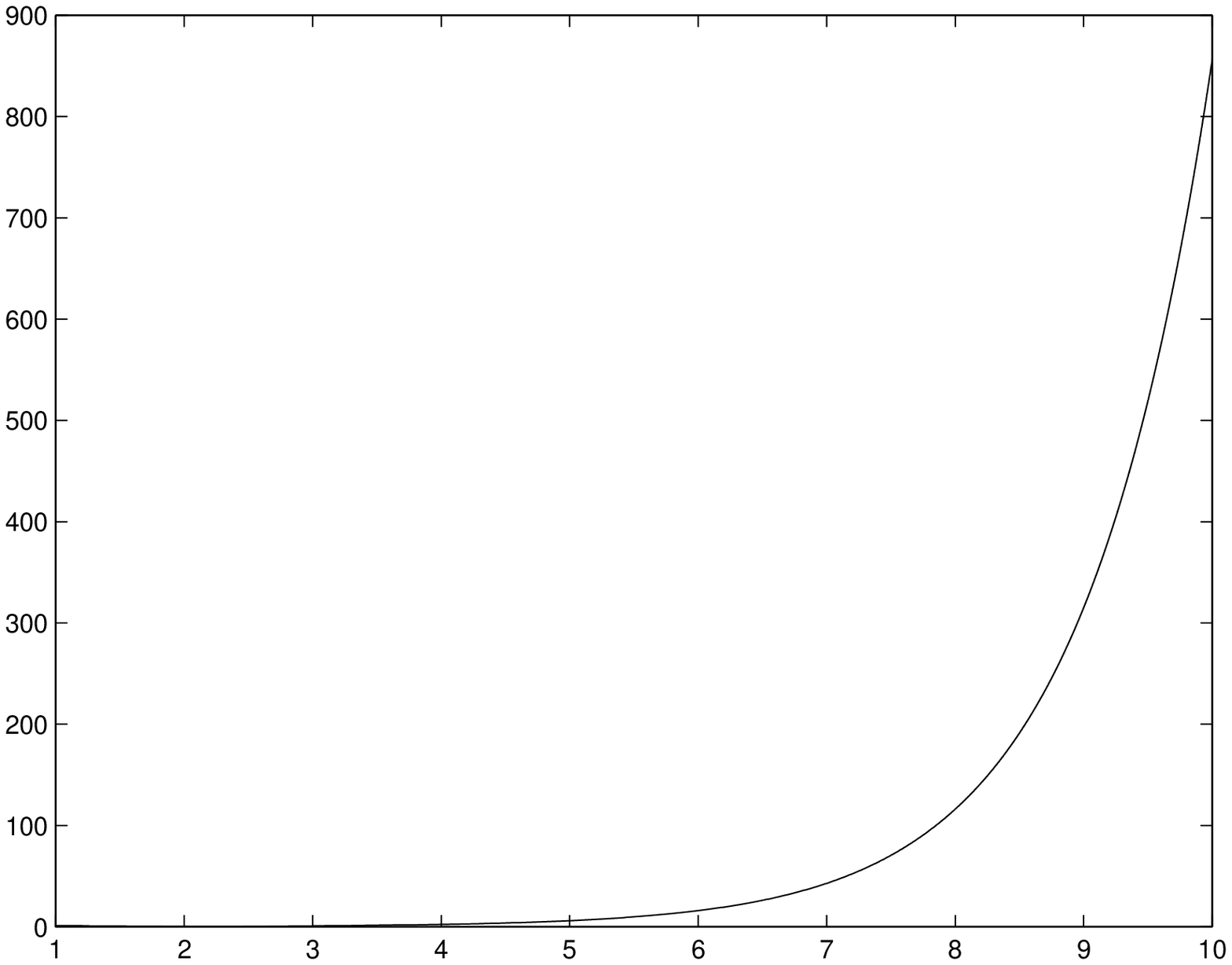}
\caption{\it Evolution of the quantum width $a$, according to
eq. (\ref{eqrinx}) and (\ref{qfl_onel}) }\label{wid1l}
\end{figure}

If the parameter $s$ has the value $-1$, there exist an interval around the 
origin inside which the concavity of the potential, i.e. its second 
derivative, is negative. In this zone, the quantum fluctuations grow up 
exponentially, while the ``classical'' variable starts to oscillate with a 
decreasing amplitude around $0$; the energy balance is granted by the
growing up of the oscillation amplitude for the ``classical'' speed 
(i.e. $\dot{x}$), as figures \ref{mv1l}, \ref{mvs1l} and \ref{wid1l} shows.

This anomalous behavior is due to the fact that the energy of the
fluctuations (\ref{enflrin}) is not bounded from below if $s=-1$. It 
introduces an inconsistency in our computation scheme: the evolution
equations (\ref{vm}), (\ref{fl}) or equally (\ref{eqrinx}) and
(\ref{qfl_onel}) have been obtained supposing the quantum fluctuations
were small (in some sense) with respect to the ``classical''
variable. Instead, the result I get here shows that the fluctuations
grow up exponentially for $t \rightarrow \infty$.

Clearly, this behavior is the proof that a complete dynamical
treatment is only possible in the framework of a non perturbative
approximation scheme, which contains a (at least) partial resummation
of the perturbative series.

\subsubsection{Hartree-Fock}\label{hf_sec}
A first resummation scheme which may be implemented is the
time--dependent Hartree-Fock (HF) approximation, which goes as follows. One
considers the following substitutions:

\begin{eqnarray}
\xi^{2n} &\to& \dfrac{(2n)!}{2^n (n-1)!} \vev{ \xi^2}^{n-1}\,\xi^2 
- \dfrac{(2n)!(n-1)}{2^n n!} \vev{\xi^2}^n \; , \nonumber \\
\xi^{2n+1} &\to& \dfrac{(2n+1)!}{2^n n!} \vev{\xi^2}^n \xi \; .
\label{hfact}
\end{eqnarray}

The coefficients of the terms in the rhs have been fixed by the
requirement that the mean values of $\xi ^n$ is the same both in the
free theory and in the quadratic theory obtained by the substitution
(\ref{hfact}) (cfr. \cite{Boyanovsky:1994xf,Cormier:1998wk}).

Given this factorization, any potential $V(q)$ becomes
\begin{equation}
V(x + \sqrt{\hbar} \xi) = \sum_{n=0}^{\infty} \dfrac{\hbar^n}{n!} \left(
\dfrac{\vev{ \xi^2 }}{2} \right)^n \left\{
V^{(2n)}(x) + \sqrt{\hbar} \xi V^{(2n+1)}(x) +\dfrac\hbar2 \left[ \xi^2 
- \vev{ \xi^2} \right] V^{(2n+2)}(x) \right\} ,
\label{factorF}
\end{equation}
where we use the notation
\begin{equation}
V^{(n)}(q) \equiv \dfrac{\delta^n}{\delta q^n} V(q) \; .
\end{equation}
In the simple case I am considering here [eq. (\ref{hp})], the equations
reduce to
\begin{equation}
\begin{array}{ll}
\stackrel{\cdot \cdot}{x} \; = & - \dfrac{1}{m} V' \left( x \right) -
\dfrac{\hbar}{m} \frac{V ''' \left( x \right)}{2} \VEV{\xi ^{2}}_{\Psi} \\
\stackrel{\cdot \cdot}{\xi} \; = & - \dfrac{1}{m} \left[ V ''
\left( x \right) + \hbar \dfrac{V ^{(IV)} \left( x \right)}{2}
\VEV{\xi ^{2}}_{\Psi} \right] \xi
\end{array}
\label{eqmot}
\end{equation}
The energy becomes
\begin{equation}
E \left( t \right) = E _{cl} ^{(0)} \left( t \right) + \hbar E _{fl} ^{(1)} 
\left( t \right) + \hbar ^{2} E _{fl} ^{(2)} \left( t \right)
\label{en_hf}
\end{equation}
where (omitting the $_{\Psi}$ symbol to the mean value)
\begin{equation}
\begin{array}{ll}
E _{cl} ^{(0)} \left( t \right) \; = & \dfrac{m}{2} \dot{x} ^{2} + V
\left( x \right) \\ E _{fl} ^{(1)} \left( t \right) \; = &
\dfrac{m}{2} \VEV{\dot{\xi} ^{2}} + \frac{V ''
\left( x \right)}{2} \VEV{\xi ^{2}} \\
E _{fl} ^{(2)} \left( t \right) \; = & \dfrac{V ^{(IV)} \left( x
\right)}{8} \VEV{\xi ^{2}}^{2} 
\end{array}
\end{equation}
Even if the evolution equation obtained for the operator $\xi \left( t 
\right)$ contains a cubic term ($\simeq \xi ^3$), still it can be 
solved, writing its general solution as a linear combination (with
operatorial coefficients) of two real functions, in a way completely
similar to that used in the section \ref{eveq}; it is also
labor-saving to use the $\sigma$ variable defined in (\ref{csiquad_A}); the
equations of motion are:
\begin{equation}\label{hfeq}
\begin{split}
\ddot{x} &=  - \dfrac{1}{m} V' \left( x \right) - \dfrac{\hbar}
{m} \dfrac{V ''' \left( x \right)}{2} \sigma ^{2} \\
\ddot{\sigma} &= - \dfrac{1}{m} \left[ V '' \left( x \right) 
+ \hbar \dfrac{V ^{(IV)} \left( x \right)}{2} \sigma ^{2} \right] \sigma +
\dfrac{1}{4 m ^{2} \sigma ^{3}}
\end{split}
\end{equation}
with the initial conditions
\begin{equation}
x \left( 0 \right) = \bar{x} \quad \dot{x} \left( 0 \right) = 0
 \quad \sigma \left( 0 \right) = \dfrac{\sigma_0}{\sqrt{\hbar}}
 \quad \dot{\sigma} \left( 0 \right) = 0
\end{equation}
These equations of motion can be derived by a Lagrangian/Hamiltonian 
principle, starting with a Lagrangian
\begin{equation}
L \left( x , \dot{x} ; \sigma , \dot{\sigma} \right) = \dfrac{m}{2} \dot{x} ^{2}
+ \dfrac{m \hbar}{2} \dot{\sigma} ^{2} -
V \left( x \right) - \dfrac{m \hbar}{2} \dfrac{1}{4 m ^{2} \sigma ^{2}} - 
\dfrac{\hbar}{2} V '' \left( x \right) \sigma ^{2} - \dfrac{\hbar}{8} V ^{(IV)} 
\left( x \right) \sigma ^{4}
\end{equation}
By means of a Legendre transformation I get the Hamiltonian function,
that is the energy (\ref{en_hf}), expressed in terms of the canonical
variables $\left\{ x(t) , \sigma(t) \right\}$ and their conjugated
momenta. Deriving the Hamilton equations for this Hamiltonian yields
exactly the evolution equations (\ref{hfeq}). A more general view on
this subject can be found in the framework of the dissipation and
decoherence in field theory, as explained in
\cite{Cooper:1997ii}.

I come back to the specific case considered in the previous section. By
defining dimensionless dynamical variables as in (\ref{def_z}) and
(\ref{def_a}) I get the following evolution equations
\begin{equation}\label{adim_hfeq}
\begin{split}
&\eta '' + s \eta + \eta ^3 + g a ^2 \eta = 0 \\
&a '' + s a + 3 \eta^2 a - \dfrac{1}{16 a ^3} + g a ^3= 0
\end{split}
\end{equation}
with the initial conditions
\begin{eqnarray}
\eta \left( 0 \right) = \sqrt{\dfrac{\lambda}{6 m \omega ^2}} \bar{x} = \alpha & a
\left( 0 \right) = \sigma_0 \sqrt{\dfrac{m \omega}{2 \hbar}} = \beta\\
\eta' \left( 0 \right) = 0 & a ' \left( 0 \right) = 0 
\label{ci_hf}
\end{eqnarray}
and the following expression for the energy
$$
\varepsilon = \eta ^{\prime 2} + s \eta ^{2} + \dfrac{\eta ^{4}}{2} + 
$$
\begin{equation}
+ \dfrac{g}{3} \left[a ^{\prime 2} + \left( s + 3 \eta ^{2} \right) a ^{2} + 
\dfrac{g}{2} a ^4 + \dfrac{1}{16 a ^2} \right]
\label{ehf}
\end{equation}
From the last expression above we can explicitly see that the energy is now 
bounded from below and the fluctuations cannot grow up indefinitely 
for any value of $\eta$.

\paragraph{Variational Principle}
It is well known that the quantum evolution in time is given by
minimizing the following functional
\begin{equation}\label{qm_varpr}
	\delta \int dt\, \vev{i\partial_t-H}=0 \;,\quad
	\vev{\cdot} \equiv \bra{\Psi(t)}\cdot\ket{\Psi(t)}
\end{equation}
with respect to variations of the wavefunction $\Psi(t)$. Now, let us
restrict the Hilbert space to gaussian wavefunctions
$$
\Psi(x|\a(t),\b(t))=\left( \dfrac{2 \Re \b}{\pi} \right)^{1/4} \exp
\left( - \dfrac{(\Re \a)^2}{4 \Re \b} \right) \exp \left( \a x + \b x^2 \right)
$$
with $\a$ and $\b$ complex parameters, related to mean values and
widths of the position and momentum operators. The expectation value
in eq. (\ref{qm_varpr}) becomes a function of the parameters
specifying the waefunction and of their first derivatives
$$
\int dt\,\vev{i\partial_t-H} = \int dt\, \L
(\a(t),\b(t),\dot{\a}(t),\dot{\b}(t)) = \Sigma
$$
The stationary condition on the ``action'' $\Sigma$, $\d \Sigma = 0$,
yields a set of Euler--Lagrangian equations for the ``Lagrangian''
$\L$, which are first order in $\a$ and $\b$ but are completely
equivalent to (\ref{hfeq}) or (\ref{adim_hfeq}). This argument shows
how the Hartree--Fock approach is based on a gaussian ansatz for the
wavefunction.

\subsubsection{Large $N$}
As we have just seen the HF approximation considers a gaussian state,
which evolves in a self-consistently determined quadratic potential,
as is obtained from the original theory by means of the HF
factorization. The potential felt by the gaussian state and the
consequent evolution are self--consistent, in the sense that they
depend upon the same parameters which specifies the wave function.

I introduce in this section a different non perturbative
approximation scheme, that nevertheless shares similar features with
HF. I generalize our system considering a set of $N$ harmonic
oscillators described by the canonical coordinates ${q _{1}, ... ,q
_{N}}$ and interacting by the $O(N)$ symmetric potential
\begin{equation}
V \left( q _{1}, ... ,q _{N} \right) = \dfrac{1}{2} s m \omega ^{2} q
\cdot q + \dfrac{\lambda}{4 !} \left( q \cdot q \right) ^{2} 
\end{equation}
It is well known that the limit $N \to \infty$ yields a well-defined
theory provided I rescale the coupling constant in such a way that
$\l\,N \sim \,\,$constant. As is shown in \cite{Yaffe:1982vf}, the theory
resulting from the large $N$ limit at leading order can be fully
described by means of a set of generalized coherent states. In other
words, for $N$ large, only Gaussian states are relevant for the
description of the system. Thus, the same factorization as in HF can
be performed in this limit, with the difference that here it is exact
(at $N=\infty$), while previously was only an approximation, basically
out of control.

I now want to get dynamical equations for this system when $N \to
\infty$. To this end, I split the position operator of each
oscillator in this way:
\begin{equation}
q _{i} = \sqrt{N} x _{i} + \sqrt{\hbar} \xi _{i}
\end{equation}
The Heisenberg evolution equations turn out to be:
\begin{equation}
\left( \dfrac{d ^{2}}{dt ^{2}} + s \omega ^{2} \right) q _{i} = - \dfrac{1}
{N} \dfrac{1}{3 !} \dfrac{\Lambda}{m} q ^{2} q _{i}
\end{equation}
where $q ^{2} = \sum _{j=1} ^{N} q _{j} ^{2}$, $i$ going from $1$ to
$N$ and $\Lambda = \lambda N$ (not to be confused with the ultraviolet
cutoff of QFT, which I will introduce later). Again, I adopt an Hartree-Fock 
approximation:
\begin{equation}
\xi _{i} \xi _{j} = \vev{ \xi _{i} \xi _{j}}
\end{equation}
\begin{equation}
\xi ^{2} \xi _{j} = \vev{ \xi ^{2} } \xi _{j} + 2 \vev{\xi _{i} \xi _{j}}
\xi _{i}
\end{equation}
and I obtain the following evolution equations
\begin{equation}
\left( \dfrac{d ^{2}}{dt ^{2}} + s \omega ^{2} \right) x _{i} = - 
\dfrac{1}{3 !} \dfrac{\Lambda}{m} \left( x ^{2} x _{i} + \dfrac{2
\hbar}{N} x _{j} \vev{ \xi _{i} \xi _{j}}+ \dfrac{\hbar}{N} x _{i} 
\vev{ \xi ^{2}} \right)
\end{equation}
\begin{equation}
\left( \dfrac{d ^{2}}{dt ^{2}} + s \omega ^{2} \right) \xi _{j} = - 
\dfrac{1}{3 !} \dfrac{\Lambda}{m} \left[ x ^{2} \xi _{j} + 2 x \cdot 
\xi x _{j} + \dfrac{\hbar}{N} \left( \vev{ \xi ^{2} }
 \xi x _{j} + 2 \vev{\xi _{i} \xi _{j}} 
 \right) \xi x _{i} \right]
\end{equation}
Now, when $N = \infty$ I can consistently assume that $x = \left(0,
\cdots, u\right)$. Neglecting terms $O(1/N)$, the equations have a solution of 
the kind $x _{i} \left( t \right) = 0$ for $i = 1 \cdots N-1$. I split the 
degrees of freedom in transverse ($i = 1 \cdots N-1$) and longitudinal ($i=N$) 
and I get the equations:
\begin{equation}
\left( \dfrac{d ^{2}}{dt ^{2}} + s \omega ^{2} \right) u = - \dfrac{1}{3 !} 
\dfrac{\Lambda}{m} \left( u ^{3} + \dfrac{\hbar}{N}  \vev{\xi _{\perp}
^{2}} u \right)
\end{equation}
\begin{equation}
\left( \dfrac{d ^{2}}{dt ^{2}} + s \omega ^{2} \right) \xi _{j} = - 
\dfrac{1}{3 !} \dfrac{\Lambda}{m} u ^{2} \xi _{j}
\end{equation}
\begin{equation}
\left( \dfrac{d ^{2}}{dt ^{2}} + s \omega ^{2} \right) \xi _{N} = - 
\dfrac{\Lambda}{2 m} u ^{2} \xi _{N} 
\end{equation}
with the obvious definition $\xi _{\perp} ^{2} = \sum _{i=1}^{N-1} \xi_i 
^{2}$. In the large $N$ limit, the difference between $\frac{1}{N} \xi
^{2}$ and $\frac{1}{N} \xi ^{2} _{\perp}$ is $O(1/N)$, thus negligible.

The first two equations (the second equation refers to the
transverse fluctuations only) do not depend in any way upon the
longitudinal fluctuation $\xi _{N}$ and so they form a `closed
system'. I can take advantage of the $O \left( N - 1 \right)$
residual symmetry to write:
\begin{equation}
\vev{ \xi _{i} \xi _{j} } = 
\vev{ \xi _{1} ^{2}} \delta _{ij} 
\end{equation}
\begin{equation}
\vev{ \xi _{i} \xi _{N}} =
\vev{\xi _{1} \xi _{N}}
\end{equation}
Of course, in order to these identities be valid, the initial state
needs to have the same symmetry as the dynamics which I take the
equation from. I get the following evolution equations to leading
order in the $1/N$ expansion:
\begin{equation}
\left( \dfrac{d ^{2}}{dt ^{2}} + s \omega ^{2} \right) u = - \dfrac{1}{3!} 
\dfrac{\Lambda}{m} \left( u ^{3} + \hbar \vev{ \xi _{1} ^{2}} u \right)
\end{equation}
\begin{equation}
\left( \dfrac{d ^{2}}{dt ^{2}} + s \omega ^{2} \right) \xi _{1} = - 
\dfrac{1}{3 !} \dfrac{\Lambda}{m} \left( u ^{2} + \hbar \vev{\xi _{1} 
^{2}} \right) \xi _{1} 
\end{equation}
together with the equation for the longitudinal fluctuation (which
will be neglected):
\begin{equation}
\left( \dfrac{d ^{2}}{dt ^{2}} + \omega ^{2} \right) \xi _{N} = - 
\dfrac{1}{3 !} \dfrac{\Lambda}{m} \left[ \left( 3 u ^{2} + \hbar
\vev{\xi _{1} ^{2}} \right) \xi _{N} + 2 \hbar
\vev{ \xi _{1} \xi _{N}} \sum _{i=1} ^{N-1} \xi _{i} 
 \right] 
\end{equation}
I use similar definitions to (\ref{csiquad_A}), (\ref{csipuntquad_A}) and 
the dimensionless variables already considered previously [cfr. (\ref{def_z}) 
and (\ref{def_a})], obtaining the equations:
\begin{equation}\label{lnadimeq}
\begin{split}
\eta '' + s \eta + \eta ^3 + \dfrac{g}{3} a ^2 \eta = 0 \\
a '' + s a + \eta^2 a - \dfrac{1}{16 a ^3} + \dfrac{g}{3} a ^3= 0
\end{split}
\end{equation}
with the initial conditions
\begin{eqnarray}
\eta \left( 0 \right) = \sqrt{\dfrac{\lambda}{6 m \omega ^2}} \bar{x}
= \alpha & a \left( 0 \right) = \sigma_0 \sqrt{\dfrac{m \omega}{2
\hbar}} = \beta\\ \eta' \left( 0 \right) = 0 & a ' \left( 0 \right) = 0 
\end{eqnarray}
The expression of the energy per oscillator in dimensionful variables
is the following
\begin{equation}
\dfrac{E}{N} = \dfrac{m}{2} \left( \dot{u} ^{2} + \hbar 
\VEV{\dot{\xi} _{1} ^{2}} \right) + \dfrac{s m \omega 
^{2}}{2} \left( u ^{2} + \hbar\vev{ \xi_{1} ^{2}}
\right) + \dfrac{\Lambda}{4 !} \left( u ^2 + \hbar \vev{
\xi_{1} ^{2}} \right) ^{2}
\end{equation}
while I have, in dimensionless variables
$$
\varepsilon = \eta ^{\prime 2} + s \eta ^{2} + \dfrac{\eta ^{4}}{2} + 
$$
\begin{equation}
+ \dfrac{g}{3} \left[a ^{\prime 2} + \left( s + \eta ^{2} \right) a ^{2} + 
\dfrac{g}{6} a ^4 + \dfrac{1}{16 a ^2} \right]
\label{en_N}
\end{equation}
which could have been obtained from eqs. (\ref{qmham}) and (\ref{hp}) by means 
of the substitution $q^2 \to u^2 + \hbar \vev{ \xi_{1} ^{2}}$ and $p^2
\to m^2\,\left( \dot{u} ^{2} + \hbar \vev{\dot{\xi} _{1} ^{2}} \right)$. It is 
interesting to notice the factor of $3$ of difference between this case and 
the HF approximation considered before, which is due to the different coupling 
of the longitudinal mode with respect to the transverse ones. This will have 
important consequences on the renormalizability of the field theoretical model 
considered in the following sections.

The main advantage of the large $N$ limit is the possibility of
obtaining a closed system of equations, considering just the $1-$point
and $2-$point functions, thanks to its Gaussian nature. On the other
hand, it is generally believed that in this way the contribution of
scattering processes is neglected; thus, the resulting theory has
infinitely many conserved quantities, which prevent
thermalization. Considering $O(1/N)$ corrections is supposed to give
an answer to the fundamental question of whether the inclusion of
scattering leads to thermalization. Of course, in this case the equations
become very difficult to study. In fact, the evolution of each single
$n-$point function needs be considered in the treatment, because the
exact system is not closed anymore. This makes the problem impossible
to be treated numerically and, for any practical purpose, some
approximation must be inserted by hand. Bettencourt and Wetterich, for
example, consider also the $4-$point function, but neglect all
contributions from 1PI $6-$point vertices \cite{Bettencourt:1998xb}. As a
result, this truncation method converges for large $N$ and is well
suited to describe an approach to thermal equilibrium; but, isolated
systems do not thermalize even in this further approximation.



%
\subsubsection{Conclusions}
I close this section with some comments on the results obtained. We
have analyzed interaction phenomena between classical degrees of
freedom (mean values) and quantum fluctuations, that produce energy
transfer behavior. Yet, we can not speak about a real dissipation of
the classical energy, or an irreversible flux of energy towards the
quantum degrees of freedom. In fact, 
to a phase in which the energy flows from one side to the
other, it follows immediately an other one, in which the opposite
process takes place.  The scenario will be completely different in the
case of the Quantum Field Theory, where the momentum modes will play
the role of infinitely many dissipative channels, producing an
effectively irreversible transfer of energy from the classical to the
quantum part.

For completeness' sake, it should be noticed that the $N-$dimensional
anisotropic harmonic oscillator in the radial quartic potential is
completely integrable, as it has $N$ integrals of motion, which can be
naturally constructed by means of a Lax type representation
\cite{wojciechowski}. The same procedure has been used in QFT, to get
an infinite hierarchy of sum rules \cite{Salgado:1999td}.

\section{Cutoff field theory}\label{cft}
After this brief excursus in quantum mechanics, let us come to the
main subject of this thesis, the {\em dynamical evolution in quantum
field theory}. I start introducing the basic vocabulary and
instruments I will be using in the following: I consider the
$N-$component scalar field operator $\bds\phi$ in a $D-$dimensional
periodic box of size $L$ and write its Fourier expansion as customary
\begin{equation}
	\bds\phi(x)=L^{-D/2}\sum_k \bds\phi_k\,e^{ik\cdot x} 
	\;,\quad \bds\phi_k^\dag = \bds\phi_{-k}
\end{equation}
with the wavevectors $k$ naturally quantized: $k=(2\pi/L)n$, $n\in\Z^D$.
The canonically conjugated momentum $\bds\pi$ has a similar expansion
\begin{equation}
	\bds\pi(x)=L^{-D/2}\sum_k \bds\pi_k\,e^{ik\cdot x} 
	\;,\quad \bds\pi_k^\dag = \bds\pi_{-k}
\end{equation}
with the commutation rules $[\phi_k^\a\,,\pi_{-k'}^\b]=
i\,\delta^{(D)}_{kk'}\,\delta^{\a\b}$. Of course, when the size $L$
goes to $\infty$, the sums become integrals over a continuum of
momentum modes.

To regularize the ultraviolet behavior, I restrict the sums over
wavevectors to the points lying within the $D-$dimensional sphere of
radius $\Lambda$, that is $k^2\le\Lambda^2$, with $\N=\Lambda L/2\pi$
some large integer. Clearly, as long as both the cutoffs remain
finite, I have reduced the original field--theoretical problem to a
quantum--mechanical framework with finitely many (of order $\N^{D-1}$)
degrees of freedom.

The $\phi^4$ Hamiltonian reads
\begin{eqnarray}\label{lN_ham}
  H &=&\dfrac12\int d^Dx \left[\bds\pi^2 + (\partial\bds\phi)^2 +
  \mbare\,\bds\phi^2 + \dfrac\lbare2 ( \bds\phi^2)^2 \right] \nonumber \\ &=&
  \dfrac12\sum_k\left[\bds\pi_k \!\cdot \bds\pi_{-k} +(k^2+
  \mbare)\,\bds\phi_k \!\cdot \bds\phi_{-k}\right] + \nonumber \\
  &+&\dfrac\lbare{4L^D} \sum_{k_1,k_2,k_3,k_4} (\bds\phi_{k_1}\!\!\cdot
  \bds\phi_{k_2})(\bds\phi_{k_3} \!\!\cdot \bds\phi_{k_4})
  \,\delta^{(D)}_{k_1+k_2+k_3+k_4,0}
\end{eqnarray}
where $\mbare$ and $\lbare$ should depend on the UV cutoff $\Lambda$
in such a way to guarantee a finite limit $\Lambda\to\infty$ for all
observable quantities. As is known
\cite{Wilson:1974jj,Boyanovsky:1995me}, this implies triviality
(that is vanishing of renormalized vertex functions with more than two
external lines) for $D>3$ and very likely also for $D=3$. In the
latter case triviality is manifest in the one--loop approximation and
in large$-N$ limit due to the Landau pole.  For this reason I shall
keep $\Lambda$ finite and regard the $\phi^4$ model as an effective
low--energy theory (here low--energy means practically all energies
below Planck's scale, due to the large value of the Landau pole for
renormalized coupling constants of order one or less).

I shall work in the wavefunction representation where
$\braket{\bds\varphi}\Psi=\Psi(\bds\varphi)$ and
\begin{equation}\label{rep1}
	(\bds\phi_0\Psi)(\bds\varphi)=\bds\varphi_0 \Psi(\bds\varphi)
	  \;,\quad 
	(\bds\pi_0\Psi)(\bds\varphi)= -i\pdif{}{\bds\varphi_0}\Psi(\bds\varphi)
\end{equation}
while for $k>0$ (in lexicographic sense)
\begin{equation}\label{rep2}
	(\bds\phi_{\pm k}\Psi)(\bds\varphi)=\dfrac1{\sqrt2}\left(
	\bds\varphi_k\pm i\,\bds\varphi_{-k}\right)\Psi(\bds\varphi)
	  \;,\quad 
	(\bds\pi_{\pm k}\Psi)(\bds\varphi)= \dfrac1{\sqrt2} \left(-i
	\pdif{}{\bds\varphi_k}\pm \pdif{}{\bds\varphi_{-k}}\right)\Psi(\bds\varphi)
\end{equation}
Notice that by construction the variables $\bds\varphi_k$ are all real.
Of course, when either one of the cutoffs are removed, the wave function
$\Psi(\bds\varphi)$ acquires infinitely many arguments and becomes what is
usually called a {\em wavefunctional}.

In practice, the problem of studying the dynamics of the $\phi^4$
field theory out of equilibrium consists now in trying to solve the
time-dependent Schroedinger equation given an initial wavefunction
$\Psi(\bds\varphi,t=0)$ that describes a state of the field far away
from the vacuum. By this I mean a non--stationary state that, in the
infinite volume limit $L\to\infty$, would lay outside the particle
Fock space constructed upon the vacuum. This approach could be
generalized in a straightforward way to mixtures described by density
matrices, as done, for instance, in
\cite{Boyanovsky:1994pa,Habib:1996ee,Boyanovsky:1994xf}. Here I shall
restrict to pure states, for sake of simplicity and because all
relevant aspects of the problem are already present in this case.

A completely equivalent approach to the time dependent problem in QFT
is based on the Heisenberg representation, where the operators are
time dependent while the states are fixed. In such an approach, the
evolution equations for the field condensate and the correlation
functions may be obtained by a generalization of the tadpole equation
to time dependent situations, starting from the Heisenberg equations
for the operators, as already shown in section \ref{gen}.

\subsection*{A rigorous result: the effective potential is convex}
\label{nogo}

I want to stress that the introduction of both a UV and IR cutoff
allows to easily derive the well--known rigorous result concerning the
flatness of the effective potential. In fact $V_{\rm eff}(\bar\phi)$
is a convex analytic function in a finite neighborhood of
$\bar\phi=0$, as long as the cutoffs are present, due to the
uniqueness of the ground state. This is a well known fact in
statistical mechanics, being directly related to stability
requirements. It would therefore hold also for the field theory in the
Euclidean functional formulation. In our quantum--mechanical context I
may proceed as follows. Suppose the field $\phi$ is coupled to a
uniform external source $J$.  Then the ground state energy $E_0(J)$ is
a concave function of $J$, as can be inferred from the negativity of
the second order term in $\Delta J$ of perturbation around any chosen
value of $J$. Moreover, $E_0(J)$ is analytic in a finite neighborhood
of $J=0$, since $J\phi$ is a perturbation ``small'' compared to the
quadratic and quartic terms of the Hamiltonian. As a consequence, this
effective potential $V_{\rm eff}(\bar\phi)=E_0(J)-J\bar\phi$,
$\bar\phi=E_0'(J)=\vev{\phi}_0$, that is the Legendre transform of
$E_0(J)$, is a convex analytic function in a finite neighborhood of
$\bar\phi=0$.  In the infrared limit $L\to\infty$, $E_0(J)$ might
develop a singularity in $J=0$ and $V_{\rm eff}(\bar\phi)$ might
flatten around $\bar\phi=0$. Of course this possibility would apply in
case of spontaneous symmetry breaking, that is for a double--well
classical potential
\cite{Branchina:1990ja,Ringwald:1990dz}. This is a subtle and
important point that will play a crucial role later on, even if the
effective potential is relevant for the static properties of the model
rather than the dynamical evolution out of equilibrium that interests
us here. In fact such evolution is governed by the CTP effective
action
\cite{CTP1,Chou:1985es} and one might expect that, although non--local
in time, it asymptotically reduces to a multiple of the effective
potential for trajectories of $\bar\phi(t)$ with a fixed point at
infinite time. In such case there should exist a one--to--one
correspondence between fixed points and minima of the effective
potential.

\section{Evolution of a homogeneous background}

The dynamics of uniform strongly out of equilibrium condensates in QFT
has been studied mainly in connection with the phenomenology of heavy
ion collisions and with the evolution of the Early Universe. It has
become clear that phenomena associated with parametric amplification
of quantum fluctuations can play an important role in the process of
reheating and thermalization. It should be emphasized, however, that
the dynamics in cosmological backgrounds differs qualitatively and
quantitatively from the dynamics in Minkovski space. In any case, in
such situations, the quantum state is characterized by a large energy
density, which means a large number of particles per correlation
volume $m^{-3}$.

The simplest case we can start with is the evolution of a translation
invariant state, which has a uniform field mean value. This kind of
simplification is fully justified in cosmological scenarios, where the
exponential expansion make all disuniformities to disappear, while we
need surely something better in order to study the out of equilibirum
phenomena occurring during and after a heavy ion collision. We
consider here the case of an uniform condensate, postponing the
discussion on the evolution of a spherically symmetric state to
section \ref{dis_homo}.

\subsection{Two words on perturbative approaches}
As already pointed out in section \ref{example}, there exist several
approximation schemes to solve a time dependent problem. Enforcing a
double perturbative expansion, both in the number of loops and in the
field amplitude, we get the following equation of motion, for a uniform
expectation value of a quantum scalar field (to one--loop level and to
cubic order in field amplitude) \cite{Boyanovsky:1995me}:
\begin{equation}\label{ampl_exp}
\ddot\phi(t)+m^2\phi(t)+\dfrac{\l}6\phi^3(t)+\dfrac{\l^2}{4}\phi(t)\int_{t_0}^t
dt'\phi(t')\dot\phi(t')\int\dfrac{d^3k}{(2\pi)^3}\dfrac{\cos[2\om_k(t-t')]}
{2\om_k^3}=0
\end{equation}
where $\l$ and $m^2$ are the renormalized coupling constant and mass and 
$\om_k=\sqrt{k^2+m^2}$. de Vega and Salgado \cite{deVega:1997yw} solved 
analytically this non linear and non local equation by RG techniques. The 
exact solution shows that the order parameter oscillates as the classical 
cnoidal solution with slowly time dependent amplitude and frequency. In 
addition, the amplitude reaches an asymptotic value, which is a function of 
the initial amplitude, as $t^{-3/2}$.

I can also solve the one loop equations exactly in the field
amplitude. In this case, I reach the conclusion
\cite{Boyanovsky:1995me} that perturbation theory is not suitable for
the purpose of studying the asymptotic dynamics of a quantum
system. Due to parametric resonances and/or spinodal instabilities
there are modes of the field that grow exponentially in time until
they produce non--perturbative effects for any coupling constant, no
matter how small. For this reason, the perturbative approach can be
considered valid only for the early time evolution. On the other hand,
only few, by now standard, approximate non--perturbative schemes are
available for the $\phi^4$ theory, and to these I have to resort after
all. I shall consider the large $N$ expansion to leading order in
section \ref{ln} (cfr. ref. \cite{Destri:1999hd}), remanding to the
definition of a time-dependent Hartree--Fock (tdHF) approach (a
generalization of the treatment given, for instance, in
\cite{Kerman:1976yn}) to section \ref{impHF}
(cfr. ref. \cite{Destri:1999he}). In fact these two methods are very
closely related, as shown in \cite{Boyanovsky:1998ka}, where several
techniques to derive reasonable dynamical evolution equations for
non--equilibrium $\phi^4$ are compared.

\section{Large $N$ expansion at leading order}\label{ln}
\subsection{Definitions}


In this section I consider a standard non--perturbative approach to
the $\phi^4$ model which is applicable also out of equilibrium, namely
the large $N$ method as presented in \cite{Cooper:1997ii}. However I
shall follow a different derivation which makes the gaussian nature of
the $N\to\infty$ limit more explicit.

It is known that the theory described by the Hamiltonian (\ref{lN_ham})
is well behaved for large $N$, provided that the quartic coupling
constant $\lbare$ is rescaled with $1/N$. For example, it is possible
to define a perturbation theory, based on the small expansion
parameter $1/N$, in the framework of which one can compute any
quantity at any chosen order in $1/N$. From the diagrammatic point of
view, this procedure corresponds to a resummation of the usual
perturbative series that automatically collects all the graphs of a
given order in $1/N$ together \cite{largen_exp}. Moreover, it has been
established since the early 80's that the leading order approximation
(that is the strict limit $N\to\infty$) is actually a classical limit
\cite{Yaffe:1982vf}, in the sense that there exists a classical system (i.e.,
a classical phase space, a Poisson bracket and a classical
Hamiltonian) whose dynamics controls the evolution of all fundamental
quantum observables, such as field correlation functions, in the
$N\to\infty$ limit. For instance, from the absolute minimum of the
classical Hamiltonian one reads the energy of the ground state, while
the spectrum is given by the frequencies of small oscillations about
this minimum, etc. etc..  I am here interested in finding an
efficient and rapid way to compute the quantum evolution equations for
some observables in the $N \to \infty$ limit, and we will see that
this task is easily accomplished just by deriving the canonical
Hamilton equations from the large $N$ classical Hamiltonian.

Following Yaffe \cite{Yaffe:1982vf}, I write the quantum mechanical
hamiltonian as 
\begin{equation}\label{hAC}
	H = N h (A\,,C) 
\end{equation}
in terms of the square matrices $A$, $C$ with operator entries
({\bf $\bds\varpi_k$} is the canonical momentum conjugated to the real mode
{\bf $\bds\varphi_k$})
\begin{equation}
	A_{kk'} = \dfrac1{N} \bds\varphi_k \!\cdot \bds\varphi_{k'}
   \;,\quad C _{kk'} = \dfrac1{N} \bds\varpi_k\!\cdot \bds\varpi_{k'}
\end{equation}
These are example of ``classical'' operators, whose two-point correlation
functions factorize in the $N\to\infty$ limit. This can be shown by
considering the coherent states 
\begin{equation}
	\Psi_{z,q,p}( \bds\varphi) = C(z) \exp\left[ i \sqrt{N} \sum_k
	\bds p_k\cdot \bds\varphi_k - \dfrac12 \sum_{kk'} z_{kk'}
	(\bds\varphi_k-\sqrt{N} \bds q_k) \cdot(\bds\varphi_{k'}-
	\sqrt{N} \bds q_{k'}) \right]
\label{coh_states}
\end{equation}
where the complex symmetric matrix $z$ has a positive definite real
part while $\bds p_k$ and $\bds q_k$ are real and coincide,
respectively, with the coherent state expectation values of $\bds
\varpi_k$ and $\bds \varphi_k$. As these parameters take  all their
possible values, the coherent states form an overcomplete set in the
cutoff Hilbert space of the model. The crucial property which ensures
factorization is that they become all orthogonal in the $N\to\infty$
limit. Moreover one can show \cite{Yaffe:1982vf} that the 
coherent states parameters form a classical phase space with Poisson
brackets 
\begin{equation}
	\left\{q_k^i\,,p_{k'}^j\right\}_{\rm P.B.} = \delta_{kk'}
	\delta^{ij} \;,\quad
	\left\{w_{kk'}\,,v_{qq'}\right\}_{\rm P.B.} = 
	\delta _{kq} \delta _{k'q'} + \delta _{kq'} \delta _{k'q}
\end{equation}
where $w$ and $v$ reparametrize $z$ as $z=\frac12
w^{-1}+i\,v$. It is understood that the dimensionality of the vectors
$\bds q_k$ and $\bds p_k$ is arbitrary but finite [that is, only a finite
number, say $n$, of pairs $(\varphi_k^i\,,\varpi_k^i)$ may take a non vanishing
expectation value as $N\to\infty$].

Once applied to the classical operators $A_{kk'}$ and
$C_{kk'}$ the large $N$ factorization allows to obtain the classical
hamiltonian by simply replacing $A$ and $C$ in eq. (\ref{hAC}) by the
coherent expectation values
\begin{equation}
	\vev{A_{kk'}} = \bds q_k\cdot \bds q_{k'} + w_{kk'} \;,\quad
	\vev{C_{kk'}} = \bds p_k\cdot \bds p_{k'} + 
	(v\,w\,v)_{kk'} + \dfrac14 ( w^{-1})_{kk'}
\end{equation}


In the situation considered, having assumed a uniform background
expectation value for $\bds \phi$, I have $\bds q_k=\bds p_k=0$ for
all $k\neq 0$; moreover, translation invariance implies that $w$ and
$v$ are diagonal matrices, so that I may set
\begin{equation}
	w_{kk'} = \s_k^2\,\delta_{kk'} \;,\quad 
	v_{kk'}=\dfrac{s_k}{\s_k}\,\delta_{kk'}
\end{equation}
in term of the canonical couples $(\s_k,s_k)$ which satisfy 
$\{\s_k\,,s_{k'}\}_{\rm P.B.}=\delta_{kk'}$. Notice that the  $\s_k$ are
just the widths (rescaled by $N^{-1/2}$) of the $O(N)$ symmetric
and translation invariant gaussian coherent states.

Thus I find the classical hamiltonian
\begin{equation}
	h_{\rm cl} =\dfrac12(\bds p_0^2 +\mbare\, \bds q^2_0) +
	\dfrac12\sum_k\left[ s_k^2 +(k^2+\mbare)\s_k^2 + \dfrac1{4\s^2_k}
	\right] + \dfrac\lbare{4L^D} \left(\bds q_0^2+\sum_k\s_k^2\right)^2
\end{equation}
where by Hamilton's equations of motion $\bds p_0=\dot{\bds q}_0$ and
$s_k=\dot\s_k$. The corresponding conserved energy density
$\E=L^{-D}h_{\rm cl}$ may be written
\begin{eqnarray}
	\E &= \T +\V \;,\quad \T = \dfrac12\dot{\bar{\bds\phi}}^2 +
	\dfrac1{2L^D}\sum_k \dot\s_k^2  \\ \V &= \dfrac1{2L^D}\sum_k\left(
	 k^2\,\s_k^2 + \dfrac1{4\s^2_k}\right) +V(\bar{\bds\phi}^2
	+\Sigma) \;,\quad \Sigma=\dfrac1{L^D}\sum_k\s_k^2 \label{ElargeN}
\end{eqnarray}
where $\bar{\bds\phi}=L^{-D/2}\bds q_0$ and $V$ is the
$O(N)-$invariant quartic potential regarded as a function of
$\bds\phi^2$, that is $V(z)=\frac12\mbare z + \frac14\lbare z^2$. It
is worth noticing that eq. (\ref{ElargeN}) would apply as is to
generic $V(z)$ and the potential $\V$, the static part of $h_{\rm
cl}$, is what the authors of \cite{Cooper:1997ii} call the {\em true}
effective, that is the correct potential for studying the evolution of
field configurations far from equilibrium. For state in thermal
equilibrium at a temperature $T$, it would correspond to the internal
energy $U$. The standard effective potential corresponds, instead, to the
free energy $F$. Of course, the information contained in $F$ and in
$U$ is very different.

Basically, the evolution equations are Hamilton's equations for an
effective classical Hamiltonian $h_{\rm cl}$ (which contains $\hbar$
as a parameter) given by the the expectation value of the quantum
Hamiltonian on the gaussian mixed state described by generalized
coherent states. This fact shows that the mean field approximation
does not explicitly introduce any dissipative behavior in the system.


In conclusion, in the limit $N \to \infty$, the mean field and the
two--point correlation function evolve as a self--consistent closed
Hamiltonian system, described by a generalized coherent state (or,
more generally, by a guassian density matrix). This corresponds to a
truncation of the infinite hierarchy of Schwinger--Dyson equations to
$1$ and $2$ point functions.

The connection of the quantum evolution with a classical
hamiltonian formalism in mean field approximation has been studied
also in \cite{Benarous:1997kn,Benarous:1997kp}, generalizing the
time--dependent variational principle of Balian and V\'en\'eroni.

As we will see later on (cfr. sec \ref{impHF}), the $O(N)$ $\l \phi ^4$ at
leading order in $1/N$, is very closely related to the Hartree--Fock
mean field approximation which has been much used in nuclear
many--body, atomic and molecular chemistry applications.

\section{Dynamics in Infinite volume}\label{infV}

The non equilibrium dynamics of the $O(N) \, \Phi^4$ model in the
large $N$ limit, in the broken symmetry phase and for state of large
energy density has been considered in \cite{Boyanovsky:1998yp}. The
new phenomena discovered by the detailed numerical analysis of the
time evolution, are all essentially non perturbative. This character
is very well captured by the large $N$ limit, which is also
consistently renormalizable and can be systematically improved by
considering higher powers in the $1/N$ expansion.

The dynamical analysis in this approximation shows once more that the static 
effective potential is not suitable to study out of equilibrium evolutions. 
In fact, it is well known that, at one loop level, it becomes complex in the 
coexistence region. Since at leading order in $1/N$, the one loop results 
become exact, the problem cannot be ascribed to the inexactness of the one 
loop approximation, but in turn suggests that the static effective potential is 
not able to describe the system when a mixture of phases can occur.

From the dynamical point of view, two different physical
situations \cite{Boyanovsky:1998yp} are possible:
\begin{itemize} 
\item if the (conserved) energy density $\varepsilon$ is smaller than the 
local maximum of the tree level potential $V_0$ (expressed in terms of the 
renormalized parameters), then the presence of spinodal instabilities leads to 
a {\em dynamical Maxwell construction}: all the expectation values between the 
minima are available asymptotically with a vanishing effective mass, 
dynamically flattening the potential in the coexistence region at $t=\infty$; 
the dependence of the asymptotic condensate from the initial condition can be 
well parametrized by the following formula:
\begin{equation}
\vev{\hat\Phi (x,t=\infty } = \sqrt{\dfrac2\l}|m| \left[ 1- 
\dfrac\varepsilon{V_0} \right] ^{0.25}
\end{equation}
\item if, instead, $\varepsilon>V_0$, the evolution is symmetric and the 
condensate samples ergodically both the minima of the tree level potential, 
showing that the symmetry is restored at the dynamical level; the mass squared 
reaches an asymptotic value different from zero while the condensate transfers 
all of its energy to quantum fluctuations and vanishes asymptotically.
\end{itemize}

\subsection{Early time and asymptotic dynamics}
The asymptotic dynamics of dissipation and relaxation in scalar field
theories, starting from large energy densities, must be studied going
beyond the perturbative approach, in an energy conserving and
renormalizable framework, able to include self--consistently the
effects of quantum back-reaction. 

A thorough analytical study and a refined numerical analysis have been
provided in \cite{Boyanovsky:1998zg}, where the relaxation of an
initial state of large density is observed through the copious production
of particles in a collisionless regime, similar to Landau damping in
non relativistic electromagnetic plasmas. As a result, the asymptotic
distribution is not thermal. In case of spontaneous symmetry breaking, the 
effects of the
massless excitations on the evolution of the mean field may be studied
in detail and a linear response analysis shows the presence of a
collective plasma mode \cite{Cooper:1997ii}.

The complete evolution can be divided in two parts. The early time
evolution is dominated by a so--called ``linear regime'', during which
the energy initially stored in one (or few) modes of the field is
transferred to other modes via either parametric or spinodal
instabilities, resulting in a large particle production and a
consequent dissipation for the initial condensate
\cite{Boyanovsky:1995me}. 

It is worth explaining briefly the mechanism of spinodal
decomposition (see also \cite{Boyanovsky:1999wd}). The effective,
coarse-grained description of statistical systems with a spontaneously
broken phase, is based on a phenomenological free energy of the
Landau-Ginzburg form.  This is a quartic functional in the order
parameter, with a temperature dependent coefficient for the second
order term, which becomes negative when the temperature falls down
the critical.  Such a functional admits a spinodal region where the
potential is non-convex, corresponding to thermodynamically unstable
states.  When the system approaches the ordered phase, coming from the
disordered one, the long wavelength modes become {\em critically
slowed down}, in the sense that their relaxation time diverges when
both the reduced temperature and the momentum go to zero. Thus,
they relax to equilibirum on very long time scales. Below the critical
temperature, because of the existence of the spinodal region, there is
a band of unstable wave vectors, for which the frequencies are
positive and the corresponding fluctuations from the mean field grow
exponentially. These instabilities are the hallmark of the process of
phase separation and are the early time indications of the formation
and growth of correlated regions.

The asymptotic evolution at late times is completely non linear and is
associated to power law behavior for the growth of quantum
fluctuations and the relaxation of the condensate; the power laws
contain non--universal and non--perturbative dynamical anomalous
exponents. The two regimes are separated by a time scale $t_{\rm s}$,
which is non perturbative in the coupling and in the initial
amplitude, and is defined as the time when the quantum fluctuations
become of the same order as the tree level term in the equations of
motion and the dynamics turns completely non linear and non
perturbative
\cite{Boyanovsky:1995me,Boyanovsky:1998zg}; the analytical study
of the early time evolution
\cite{Boyanovsky:1998zg} shows that $t_{\rm s} \simeq \log 1/\l$ (where $\l$
is the quartic coupling constant).

The main results may be summarized as follows: (i) thanks to the
hierarchy of time scales, a dynamical renormalization group
resummation can be applied, showing the existence of non linear
resonances which turn the exponential behavior in non--universal power
laws with dynamical anomalous exponents; (ii) the effective squared
mass felt by the modes is time dependent and tends to an asymptotic
value as $O(1/t)$; thus, the quantum modes becomes asymptotically
free; (iii) in the unbroken symmetry phase, the condensate relaxes all
the way to zero, transferring completely its energy to the quantum
fluctuations, in spite of the presence of a perturbative threshold for
particle production; (iv) precise sum rules may be established for the
asymptotic particle distribution and an equation of state
interpolating between the radiation--type and the dust--type is found;
(v) in the broken phase, the mass vanishes asymptotically, providing a
dynamical realization of the Goldstone theorem; (vi) the asymptotic
value of the condensate is a function of the initial amplitude; (vii)
at very large time scale, $t \sim \sqrt{V}$ (where $V$ stands for the
volume of the system), the non--perturbative and non--linear evolution
might eventually produce the onset of a non--equilibrium
Bose--Einstein condensation of the long--wavelength Goldstone bosons
usually present in the broken symmetry phase of the model
\cite{Boyanovsky:1998yp,Boyanovsky:1998zg}; (viii) the asymptotic particle
distribution, obtained as the result of the copious particle
production at the expenses of the ``classical'' energy, is strongly
non-thermal
\cite{Boyanovsky:1995me,Boyanovsky:1998zg}.

Moreover, in this formalism, some aspects of the quantum dynamics of
phase ordering can be studied in some detail. The non perturbative
spinodal time $t_s \simeq \log 1/\l$ divides the evolution in two
regimes: for $t<t_s$ the correlation length $\xi(t)$ grows like
$\sqrt{t}$; for $t>t_s$, $\xi(t) \simeq 2(t-t_s)$, the correlation
function vanishes for $r>2(t-t_s)$ and the zero momentum mode of the
quantum fluctuations grows asymptotically linearly. Thus, the
correlated domains grow at the speed of light and contain inside a
non perturbative condensate of Goldstone bosons, with a correlation
which decreases as $1/r$.

Another very interesting result in \cite{Boyanovsky:1998yp} concerns
the dynamical Maxwell construction, which reproduces the flat region
of the effective potential in case of broken symmetry as asymptotic
fixed points of the background evolution. 

Finally, using a density matrix language, a semiclassical but stochastic 
description emerges: after $t_s$, semiclassical large amplitude field 
configurations are represented in $\rho$ with a finite probability.

\section{Dynamics in Finite volume}

In this section I present a detailed study, {\em in finite volume},
of dynamical evolution out of equilibrium for the $\Phi^4$ scalar
field in the large $N$ limit. More precisely, I determine how such
dynamics scales with the size of the periodic box containing the
system in the case of uniform backgrounds. This is necessary to
address questions like out--of--equilibrium symmetry breaking and
dynamical Bose--Einstein condensation.

The introduction of a finite volume should be regarded as a
regularization of the infrared properties of the model, which allows
to ``count'' the different field modes and is needed especially in the
case of broken symmetry. In fact, all the results I have summarized
in section \ref{infV}, have been obtained simulating the system
directly in infinite volume, where the evolution equations contain
momentum integrals, that must be computed numerically by a proper, but
nonetheless rather arbitrary, discretization in momentum space. Of
course, the final result should be as insensitive as possible to the
particular choice of the integration grid. In such a situation, the
definition of a ``zero'' mode and the interpretation of its late time
behavior might not be rigorous enough, unless, for some reason, it
turns out that a particular mode requires a different treatment
compared to the others. In order to understand this point, it is
necessary to put the system in a finite volume (a box of size $L$);
the periodic boundary conditions let us single out the zero mode in a
rigorous way and thus we can carefully analyze its scaling properties
with $L$ and get some information on the infinite volume limit.
\input{finvN.tex}

\section{Improved Hartree-Fock approximation}\label{impHF}

The main limitation of the large $N$ approximation, as far as the
evolution of the widths $\s_k$ is concerned, is in its intrinsic
gaussian nature. In fact, one might envisage a scenario in which,
while gaussian fluctuations stay microscopic, non--gaussian
fluctuations grow in time to a macroscopic size. In addition, the
$O(\infty$) theory contains only the transverse fluctuations, coupled
by means of a mean field interaction. It would be very interesting to
go beyond both these approximations, for example considering the next
to leading terms, which are of order $1/N$. Of course, I need to write
down equations valid for arbitrary $N$, in such an approximation that
shows the interaction between the longitudinal and transverse modes.

Therefore, in order to clarify these points and go beyond the gaussian
approximation, I am going to consider, in this section (cfr. also
\cite{Destri:1999he}), a time--dependent HF approximation capable in
principle of describing the dynamics of non--gaussian fluctuation of
scalar fields with $\phi^4$ interaction.

Another open question concerns the connection between the minima of
the effective potential and the asymptotic values for the evolution of
the background, within the simplest gaussian approximation. As already
pointed out in \cite{Boyanovsky:1998yp}, a dynamical Maxwell
construction occurs for the $O(N)$ model in infinite volume and at
leading order in $1/N$ in case of broken symmetry, in the sense that
any value of the background within the spinodal region can be obtained
as large time limit of the evolution starting from suitable initial
conditions. It would be very enlightening if we could prove this
``experimental'' result by first principles arguments, based on CTP
formalism. Furthermore, preliminary numerical evidence
\cite{Destri:1999he} suggests that something similar occurs also in
the Hartree approximation for a single field, but a more detailed
analysis is needed.

Moreover, The $O(N)$ symmetric linear $\sigma$ model has been much
studied in the past, not only in the large $N$ limit, but also for
finite values of $N$. The model is very interesting, as laboratory for
the Spontaneous Chiral Symmetry Breaking (SCSB), which manifests
itself in the low energy hadronic world. The same Chiral Symmetry is
well realized even in the underlying (more fundamental) theory, QCD,
due the light $u$ and $d$ quark masses. In fact, QCD with $N_f$
massless quark flavors has a $SU(N_f)_L \times SU(N_f)_R$ symmetry
group, which is isomorphic to $O(4)$ for $N_f=2$. The corresponding
order parameter is $\Phi ^{ij} =
\vev{\bar{q}_L^iq_R^j}$. QCD lattice simulations at finite temperature
suggest that a chiral symmetry restoration may occur at a temperature
of $T=150 MeV$ with possible observable consequences like the
formation of DCC. Such a phase transition occurred even during the
evolution of the early universe and may be reproduced in the Heavy Ion
Collision experiments currently performed at BNL--RHIC and scheduled in
the forthcoming years at CERN--LHC.

The general time--dependent variational principle for the many--body system
associated with the Schroedinger equation was introduced by Dirac in
\cite{dirac} and its classical (Hamiltonian) nature was shown later in
\cite{Kerman:1976yn}. The canonical Hamiltonian formulation was hoped to be 
useful in understanding the reduction of the many body scattering
problem to some sort of fluid dynamics or in identifying suitable
dissipation terms in a reduced description.

This variational approximation scheme lies on the self--consistent
field approach, which has been so useful in describing the ground
state and the collective properties of nuclei. The same approach to
the time dependent wave function gives the so called time dependent
Hartree--Fock approximation (tdHF), where the equations for the time
evolution are determined by a least action principles and produce a
description in terms of canonical variables with a conserved classical
hamiltonian. The variational trial wave function is taken to be a
Slater determinant, when considering many fermion systems, while a 
gaussian wavefunctional is used for a quantum scalar field theory. The
physical assumption behind this approximation is that each particle is
only influenced by the average field of all the others.

From a field theoretical point of view, it can be shown
\cite{jackiwandkerman} that the time--dependent variational principle used 
to derive the tdHF approximation gives a variational definition for
the effective action (that is the generating functional of single
particle irreducible $n-$point functions).

An isoentropic, energy non--conserving, time evolution of a mixed
quantum state was studied in QM and QFT in \cite{Eboli:1988fm}. There,
a variational principle based on a Gaussian ansatz was used in order to
derive a Liouville--Von Neumann equations, which are analogous to
Schroedinger equations and mechanical problems. In particular, the
issue of how a system in thermal equilibrium loses and eventually
regains it, when the Hamiltonian acquires a transient, has been
considered carefully.

Two improvements would be desirable in this approach: we should relax the 
isoentropic requirement, possibly finding a suitable coarse graining 
procedure, and we should go beyond the Gaussian approximation, either using 
more elaborate density matrices or the non--equilibrium effective action.

With the aim of studying the dynamics of the model with the inclusion
of some non--gaussian contributions, I introduce in this section an
improved time--dependent Hartree--Fock approach. Even if it is still
based on a factorized trial wavefunction(al), it has the merit to keep
the quartic interaction diagonal in momentum space, explicitly in the
hamiltonians governing the evolution of each mode of the field. In
this framework, issues like the static spontaneous symmetry breaking
can be better understood, and the further gaussian approximation
needed to study the dynamics can be better controlled.

\input{finvHF.tex}

\section{Evolution of a non-homogeneous background}\label{dis_homo}

\subsection{Motivations and summary}

To get a better comprehension of the phenomena involved in out of
equilibrium phase transitions, like the phase ordering process
evolving through the formation of ordered domains separated by
domain walls, it would be very useful to be able to follow the history
of non--homogeneous condensates, which are large amplitude field
configuration localized in space, in interaction with their quantum
fluctuations. Defect formation during phase transitions may reveal
unifying understanding of phenomena belonging to a wide range of
energy scale. In fact, such objects are relevant in condensed matter
systems, where the kinks (or solitons) represent charge density waves
and conducting polymers, whose transport properties need to be studied
in detail. Spatially dependent semiclassical configurations are also
important in cosmology and particle physics (electroweak domain walls,
called sphaleron, may play a crucial role in understanding
electroweak baryogenesis). Also the bubble and droplet nucleation
during supercooled phase transitions may be addressed with this
formalism. (For a static study of topological defect formation in QFT,
with a discussion of finite temperature and volume effects, see also
\cite{Vitiello:1999cj}).

Interesting results have been obtained from the analysis of domain
walls and kinks out of equilibrium, in model theories like $\l \phi^4$
and sine--Gordon in $(1+1)-$D dimensions in the dilute regime (which
means at a temperature much smaller than the kink's mass)
\cite{Alamoudi:1998cc}. The collective coordinate quantization
technique at one loop order is used to trace out the meson degrees of
freedom and compute the influence functional, which in turn
consistently gives the proper Langevin equation. In this scheme, the
relaxation is given by the interaction between the domain wall and the
meson fluctuations. As a result, the noise is gaussian and additive,
although colored and it is related to the damping kernel by a proper
fluctuation--dissipation relation. The equations describing the
evolution are studied both analytically and numerically and show the
presence of a dynamical friction coefficient. It is shown that a
Markovian approximation fails to describe the dynamics at large
temperatures, the long time relaxation is dominated by classical
Landau damping.

With a slightly different approach, a Langevin--like effective
equation of motion for the linear $\sigma$ model, was derived in 
\cite{Gleiser:1994ea}, which is
valid up to two loops and to $O(\l^2)$. The results for a quartic self 
interaction are compared to those produced by a quadratic interaction
with other fields. The equation describes the evolution towards
equilibrium for a non uniform time dependent background field
configuration, starting not too far from equilibrium, and is obtained
integrating out the faster short wavelength fluctuations. This
equation displays noise and dissipation terms which obey a suitable
dissipation--fluctuation relation. The noise is in general colored and
multiplicative for each finite value of the temperature (the
dissipation coefficient depends quadratically on the field amplitude
and vanishes in the infinite temperature limit), that is different from
the phenomenological Langevin equation usually used with white and
additive noise; this may reduce the relaxation time scales and
accelerate the approach to equilibrium.

The $O(4)$ linear $\sigma$ model is studied also in connection with
the non equilibrium relaxation of an inhomogeneous initial
configuration due to quantum and thermal fluctuations. The subject is
relevant for the physics of heavy ion collisions, because while the
inflaton condensate can be regarded as quasi--uniform due to the
exponential expansion, this is not surely the case for the disoriented
chiral condensates which may form in present high energy
experiments. Indeed, non homogeneous semi--classical configurations
will be produced, which will relax through emission of pions in the
medium. This will reduce the spatial gradient of the condensates,
making the energy decrease. The asymptotic space--time evolution of
such configurations is studied in \cite{Boyanovsky:1996zy}, by means
of the CTP methods combined with a small amplitude expansion which
linearizes the evolution equations in leading order. The relaxation of
an initial gaussian inhomogeneous configuration is studied for $\t \gg
\sqrt{t^2-r^2}$ in terms of the spreading of the packet and of the
decay in spherical waves. Different physical situations are
considered, and it turns out that at one--loop approximation, the
evolution can be described in terms of temperature and a decay rate
which depend on rapidity.  Moreover, the time scales involved are
longer for larger rapidities. At $T \neq 0$ new relaxational processes
are found, due to thermal cuts, which do not have counterpart in the
homogeneous case.

Spatially dependent configuration are considered also in
\cite{Braghin:1998sj,Braghin:1998yw}, in the framework of a time dependent
variational approach which goes beyond the Gaussian ansatz approximation.

In order to study analytically and numerically the relaxation of a
(non perturbative) strongly out of equilibrium inhomogeneous field
configuration and the emergence of a hydrodynamical description from
{\em ab initio} calculation, a self--consistent variational framework
which incorporates the quantum back reaction effects and particle
production is used in \cite{Boyanovsky:1998ka}. It is important to get
update equations which are renormalizable and local. The
renormalization issue is very important, because it guarantees the
independence on the size of the spatial greed, while the locality
limits the memory requirements on the numerical algorithm. The initial
Cauchy data for the inhomogeneous expectation value and for the
corresponding Green's functions is given as a solution of a
self--consistent problem, which, for $\phi^4$ and sine--Gordon models in
$1+1$, is mapped in a Schrodinger like problem by suitable ansatze.

\subsection{Definitions and preliminaries}
\input{non_hom.tex}

%% file: finvN.tex
I have defined the model in finite volume, giving all the relevant
notations and definitions in section \ref{cft}. I derived the large
$N$ approximation of the $O(N)-$invariant version of $\lambda
(\bds\phi ^2)^2$ model, according to the general rules of
ref. \cite{Yaffe:1982vf}. In this derivation it appears evident the
essential property of the $N\to\infty$ limit of being a particular
type of {\em classical} limit, so that it leads to a classical phase
space, a classical hamiltonian with associated Hamilton's equations of
motion [see eqs. (\ref{sk2}), (\ref{Nunbgap}) and (\ref{Nbgap})].

I then minimize the hamiltonian function(al) and determine the
conditions for massless Goldstone bosons (i.e. transverse fluctuations
of the field) to form a Bose--Einstein condensate, delocalizing the
vacuum field expectation value (cfr. also
ref. \cite{Cooper:1997ii}). This necessarily requires that the width
of the zero--mode fluctuations becomes macroscopically large, that is
of the order of the volume. Only when the background takes one of the
extremal values proper of symmetry breaking the width of the
zero--mode fluctuations is of order $L^{1/2}$, as typical of a free
massless spectrum.

The study of the lowest energy states of the model is needed for
comparison with the results of the numerical simulations, which show
that the zero--mode width $\s_0$ stays microscopic (that is such that
$\s_0/$volume$\to 0$ when the volume diverges) whenever it starts from
initial conditions in which it is microscopic. The results, in fact,
show clearly the presence of a time scale $\tau_L$, proportional to
the linear size $L$ of the system, at which finite volume effects
start to manifest. I shall give a very simple physical interpretation
of this time scale in section \ref{numerical}. The important point is
that after $\tau_L$ the zero mode amplitude starts decreasing, then
enters an erratic evolution, but never grows macroscopically
large. This result is at odd with the interpretation of the linear
late--time growth of the zero--mode width as a full dynamical
Bose--Einstein condensation of Goldstone bosons, but is compatible
with the ``novel'' form of BEC reported in
\cite{Boyanovsky:1998ba,Boyanovsky:1998yp,Boyanovsky:1998zg}.
In fact I do find that the size of the low--lying widths at time
$\tau_L$ is of order $L$, to be compared to the equilibrium situation
where they would be of order $L^0$ in the massive case or of order
$L^{1/2}$ in the massless case. Perhaps the denomination
``microscopic'' should be reserved to this two
possibilities. Therefore, since the initial condition are indeed
microscopic in this restricted sense, I do observe in the
out--of--equilibrium evolution a rapid transition to a different
regime intermediate between the microscopic one and the macroscopic
one characteristic of Bose--Einstein condensation. As I shall discuss
more in detail later on, this fully agrees with the result found in
\cite{Boyanovsky:1998yp}, that the time--dependent field correlations vanish at
large separations more slowly than for equilibrium free massless
fields (as $r^{-1}$ rather than $r^{-2}$), but definitely faster than
the equilibrium broken symmetry phase characterized by constant
correlations at large distances. At any rate, when one considers
microscopic initial conditions for the choice of bare mass which
corresponds to broken symmetry, the role itself of symmetry breaking
is not very clear in the large $N$ description of the
out--of--equilibrium dynamics, making equally obscure the issues
concerning the so--called quantum phase ordering
\cite{Boyanovsky:1998yp}. This is because the limit $N\to\infty$ is completely
saturated by gaussian states, which might signal the onset of symmetry
breaking only developing macroscopically large fluctuations. Since
such fluctuations do not appear to be there, the meaning itself of
symmetry breaking, as something happening as times goes on and
accompanied by some kind of phase ordering, is quite unclear. I
postpone to section \ref{impHF} (cfr. also \cite{Destri:1999he}) the
discussion about the possibility of using more comprehensive
approximation schemes, that include some non--gaussian features of the
complete theory. As far as the large $N$ approximation is concerned,
I underline that an important limitation of our approach, as well as
of those of the references mentioned above, is in any case the
assumption of a uniform background. Nonetheless, phenomena like the
asymptotic vanishing of the effective mass and the dynamical Maxwell
construction, taking place in this contest of a uniform background and
large $N$ expansion, are certainly very significant manifestations of
symmetry breaking and in particular of the Goldstone theorem which
applies when a continuous symmetry is broken.

\subsection{Static properties}\label{statprop}
Let us consider first the static aspects embodied in the effective
potential $V_{\rm eff}(\bar{\bds\phi})$, that is the minimum of the
potential energy $\V$ at fixed $\bar{\bds\phi}$. I first define in a
precise way the unbroken symmetry phase, in this large $N$ context, as
the case when $V_{\rm eff}(\bar{\bds\phi})$ has a unique minimum at
$\bar{\bds\phi}=0$ in the limit of infinite volume. Minimizing $\V$
w.r.t. $\s_k$ yields
\begin{eqnarray}\label{massive}
	\s^2_k=\dfrac1{2\sqrt{k^2+M^2}} \;,\quad
	&M^2 = \mbare + 2\,V'(\bar{\bds\phi}^2+\Sigma) \\
	&= \mbare + \lbare\bar{\bds\phi}^2+ \dfrac\lbare{L^D}
	\sum_k\dfrac1{2\sqrt{k^2+M^2}} \nonumber
\end{eqnarray}
that is the widths characteristic of a free theory with
self--consistent mass $M$ fixed by the gap equation. By the assumption
of unbroken symmetry, when $\bar{\bds\phi}=0$ and at infinite volume
$M$ coincides with the equilibrium mass $m$ of the theory, that may be
regarded as an independent scale parameter. Since in the limit
$L\to\infty$ sums are replaced by integrals
\begin{equation}
	\Sigma \to \int_{k^2\le\Lambda^2} \dfrac{d^Dk}{(2\pi)^D} \s_k^2
\end{equation}
I obtain the standard bare mass parameterization
\begin{equation}\label{Nm2ren}
	\mbare= m^2 - \lbare I_D(m^2,\Lambda) \;,\quad 
	I_D(z,\Lambda) \equiv \int_{k^2\le\Lambda^2} 
	\dfrac{d^Dk}{(2\pi)^D} \dfrac1{2\sqrt{k^2+z}}
\end{equation}
and the renormalized gap equation
\begin{equation}\label{rengapN}
	M^2 = m^2+ \l\,\bar\phi^2+\l \left[ I_D(M^2,\Lambda)-
	I_D(m^2,\Lambda) \right]_{\rm finite} 
\end{equation}
which implies, when $D=3$,
\begin{equation}\label{lrenN}
	\lbare =\l \left(1-\dfrac{\l}{8\pi^2}
	    \log\dfrac{2\Lambda}{m\sqrt{e}} \right)^{\!\!-1} 
\end{equation}
with a suitable choice of the finite part. No coupling constant
renormalization occurs instead when $D=1$. The renormalized gap
equation (\ref{rengapN}) may also be written quite concisely
\begin{equation}\label{nice}
	\dfrac{M^2}{\hat\l(M)} = \dfrac{m^2}{\hat\l(m)} + \,\bar{\bds\phi}^2
\end{equation}
in terms of the one--loop running couplings constant
\begin{equation}
	\hat\l(\mu) = \l \left[ 1 - \dfrac{\l}{8\pi^2} 
	\log\dfrac{\mu}m \right]^{-1} \;,\quad \hat\l(m)=\l
	\;,\quad \hat\l(2\Lambda\,e^{-1/2}) =\lbare
\end{equation}
It is the Landau pole in $\hat\l(2\Lambda\,e^{-1/2})$ that actually
forbids the limit $\Lambda\to\infty$. Hence I must keep the cutoff
finite and smaller than $\Lambda_{\rm pole}=\frac12
m\exp(1/2+8\pi^2/\l)$, so that the theory does retain a slight
inverse--power dependence on it. At any rate, there exists a very wide
window where this dependence is indeed very weak for couplings of
order one or less, since $\Lambda_{\rm pole} \gg m$.  Moreover, I see from
eq. (\ref{nice}) that for $\sqrt\l|\bar\phi|$ much smaller than the
Landau pole there are two solutions for $M$, one ``physical'', always
larger than $m$ and of the same order of $m+\sqrt\l|\bar\phi|$, and
one ``unphysical'', close to the Landau pole.

One can now easily verify that the effective potential has indeed a
unique minimum in $\bar{\bds\phi}=0$, as required. In fact, if we
assign arbitrary $\bar\phi-$dependent values to the widths $\s_k$,
(minus) the effective force reads
\begin{equation}\label{Nefforce}
	\der{}{\bar\phi^i}\V(\bar{\bds\phi},\{\s_k(\bar{\bds\phi})\})
	= M^2\,\bar\phi^i + \sum_k \pdif\V{\s_k}
	\der{\s_k}{\bar\phi^i}
\end{equation} 
and reduces to $M^2\,\bar\phi^i$ when the widths are extremal as
in eq. (\ref{massive}); but $M^2$ is positive for unbroken symmetry
and so $\bar{\bds\phi}=0$ is the unique minimum. 

I define the symmetry as broken whenever the infinite volume
$V_{\rm eff}$ has more than one minimum.  Of course, as long as $L$
is finite, $V_{\rm eff}$ has a unique minimum in
$\bar{\bds\phi}=0$, because of the uniqueness of the ground state in
Quantum Mechanics, as already discussed in section \ref{cft}. Let us
therefore proceed more formally and take the limit $L\to\infty$
directly on the potential energy $\V$. It reads
\begin{equation}
	\V = \dfrac12 \int_{k^2\le\Lambda^2} \dfrac{d^Dk}{(2\pi)^D}
	 \left(k^2\,\s_k^2 + \dfrac1{4\s^2_k}\right) 
	+V(\bar{\bds\phi}^2 +\Sigma) \;,\quad  
	\Sigma =\int_{k^2\le\Lambda^2} \dfrac{d^Dk}{(2\pi)^D}\,\s_k^2
\end{equation}
where I write for convenience the tree--level potential $V$ in the
positive definite form $V(z)=\frac14\lbare(z+\mbare/\lbare)^2$.
$\V$ is now the sum of two positive definite terms.
Suppose there exists a configuration such that 
$V(\bar{\bds\phi}^2 +\Sigma)=0$ and the first term in $\V$ is at
its minimum. Then this is certainly the absolute minimum of $\V$.
This configuration indeed exists at infinite volume when $D=3$:
\begin{equation}\label{minimum}
	\s^2_k=\dfrac1{2|k|} \;,\quad  \bar{\bds\phi}^2 =v^2
	\;,\quad \mbare = -\lbare\left[ v^2+ I_3(0,\Lambda) \right]
\end{equation}
where the nonnegative $v$ should be regarded as an independent
parameter fixing the scale of the symmetry breaking. It replaces the
mass parameter $m$ of the unbroken symmetry case: now the theory is
massless in accordance with Goldstone theorem.  On the contrary, if
$D=1$ this configuration is not allowed due to the infrared
divergences caused by the massless nature of the width spectrum. This
is just the standard manifestation of Mermin--Wagner--Coleman theorem
that forbids continuous symmetry breaking in a two--dimensional
space--time \cite{Mermin:1966fe,Coleman:1973ci}.

At finite volumes I cannot minimize the first term in $\V$ since this
requires $\s_0$ to diverge, making it impossible to keep
$V(\bar{\bds\phi}^2 +\Sigma)=0$. In fact we know that the
uniqueness of the ground state with finitely many degrees of freedom
implies the minimization equations (\ref{massive}) to hold always
true with a $M^2$ strictly positive.  Therefore, broken symmetry
should manifest itself as the situation in which the equilibrium value
of $M^2$ is a positive definite function of $L$ which vanishes in the
$L\to\infty$ limit.

I can confirm this qualitative conclusion as follows. I assume that
the bare mass has the form given in eq. (\ref{minimum}) and that 
$\bar{\bds\phi}^2 =v^2$ too. Minimizing the potential energy
leads always to the massive spectrum, eq. (\ref{massive}), with the
gap equation
\begin{equation}\label{gapagain}
	\dfrac{M^2}\lbare = \dfrac1{2L^3M} + \dfrac1{2L^3}
	\sum_{k\neq0}\dfrac1{\sqrt{k^2+M^2}} - \dfrac{\Lambda^2}{8\pi^2}
\end{equation}
If $M^2>0$ does not vanish too fast for large volumes, or stays even
finite, then the sum on the modes has a behavior similar to the
corresponding infinite volume integral: there is a quadratic
divergence that cancels the infinite volume contribution, and a
logarithmic one that renormalizes the bare coupling. The direct
computation of the integral would produce a term containing the
$M^2\log(\Lambda/M)$. This can be split into
$M^2[\log(\Lambda/v)-\log(M/v)]$ by using $v$ as mass scale. The first
term renormalizes the coupling correctly, while the second one
vanishes if $M^2$ vanishes in the infinite volume limit.

When $L\to\infty$, the asymptotic solution of (\ref{gapagain}) reads
\begin{equation}
	M = \left( \dfrac{\l}{2}\right) ^{1/3} L ^{-1} + \rm{h.o.t.}
\end{equation}
that indeed vanishes in the limit. Note also that the exponent is
consistent with the assumption made above that $M$ vanishes slowly
enough to approximate the sum over $k\neq 0$ with an integral with the
same $M$.

Let us now consider a state whose field expectation value
$\bar{\bds\phi}^2$ is different from $v^2$. If $\bar{\bds\phi}^2 >
v^2$, the minimization equations (\ref{massive}) leads to a positive
squared mass spectrum for the fluctuations, with $M^2$ given
self--consistently by the gap equation. On the contrary, as soon as
$\bar{\bds\phi}^2 < v^2$, it is clear that a positive $M^2$ cannot
solve the gap equation
\begin{equation}
	M^2 = \lbare \left( \bar{\bds\phi}^2 -v^2 + \dfrac{\s_0^2}{L^3} +
	\dfrac1{2L^3} \sum_{k\neq0}\dfrac1{\sqrt{k^2+M^2}} - 
	\dfrac{\Lambda^2}{8 \pi ^2} \right)
\end{equation}
if I insist on the requirement that $\s_0$ not be macroscopic. In fact,
the r.h.s. of the previous equation is negative, no matter which
positive value for the effective mass I choose, at least for $L$
large enough. But nothing prevent us to consider a static
configuration for which the amplitude of the zero mode is
macroscopically large (i.e. it rescales with the volume $L^3$).
Actually, if I choose
\begin{equation}
	\dfrac{\s _0 ^2}{L^3} =  v ^2 - \bar{\bds\phi}^2 + \dfrac1{2L^3M}
\end{equation}
I obtain the same equation as I did before and the same value for
the potential, that is the minimum, in the limit $L \to \infty$. Note
that at this level the effective mass $M$ needs not to have the same
behavior in the $L \to \infty$ limit, but it is free of rescaling
with a different power of $L$. I can be even more precise: I isolate
the part of the potential that refers to the zero mode width $\s_0$
($\Sigma'$ does not contain the $\s_0$ contribution)
\begin{equation}
\dfrac12 \left[ \mbare + \lbare \left( \bar{\bds\phi}^2 +\Sigma' \right)
\right] \dfrac{\s _0 ^2}{L ^3} + \dfrac{\lbare}4 \dfrac{\s _0 ^4}{L ^6} +
\dfrac{1}{8 L ^3 \sigma _0 ^2}
\end{equation}
and I minimize it, keeping $\bar{\bds\phi}^2$ fixed. The minimum is
attained at $t=\s_0^2/L^3$ solution of the cubic equation
\begin{equation}
\lbare t^3 + \a \lbare t^2 - \dfrac14 L^{-6} = 0
\end{equation}
where $\a = \bar{\bds\phi}^2 - v ^2 + \Sigma ' - I _3 \left( 0 ,
\Lambda \right)$. Note that $\lbare \alpha$ depends on $L$ and it has
a finite limit in infinite volume: $\l (\bar{\bds\phi}^2 - v ^2)$. The
solution of the cubic equation is
\begin{equation}
	\lbare t = \lbare ( v ^2 - \bar{\bds\phi}^2 ) + 
	\dfrac14 [L^3(v^2 -\bar{\bds\phi}^2)]^{-2} + \rm{h.o.t.}
\end{equation}
from which the effective mass can be identified as proportional
to $L^{-3}$. The stability equations for all the other modes can now
be solved by a massive spectrum, in a much similar way as before. 

Since $\s_0$ is now macroscopically large, the infinite volume limit
of the $\s_k$ distribution (that gives a measure of the {\em transverse}
fluctuations in the $O(N)$ model) develop a $\delta-$like singularity, 
signalling  a Bose condensation of the Goldstone bosons: 
\begin{equation}\label{BE}
	\s_k^2 = ( v ^2 - \bar{\bds\phi}^2 )\,\delta^{(D)}(k)+
	\dfrac1{2k}
\end{equation}
At the same time it is evident that the minimal potential energy is
the same as when $\bar{\bds\phi}^2=v^2$, that is  the effective
potential flattens, in accord with the Maxwell construction.

Eq. (\ref{BE}) corresponds in configuration space to the $2-$point
correlation function
\begin{equation}\label{2pN}
	\lim_{N \to \infty} \dfrac{\vev{\bds{\phi}(x)
	\cdot\bds{\phi}(y)}}{N} = \bar{\bds{\phi}}^2+ \int\dfrac{d^Dk}
	{(2\pi)^D}\,\s_k^2\, e^{ik\cdot(x-y)}= C(\bar{\bds{\phi}}^2)
	+\Delta_D(x-y)
\end{equation}
where $\Delta_D(x-y)$ is the massless free--field equal--time
correlator, while  
\begin{equation}\label{broken}
	C(\bar{\bds{\phi}}^2)= v^2\,\Theta(v^2-\bar{\bds{\phi}}^2) +
	\bar{\bds{\phi}}^2\, \Theta(\bar{\bds{\phi}}^2-v^2) =
	{\rm max}(v^2,\bar{\bds{\phi}}^2)
\end{equation}
This expression can be extended to unbroken symmetry by
setting in that case $C(\bar{\bds{\phi}}^2)=\bar{\bds{\phi}}^2$.

Quite evidently, when eq. (\ref{broken}) holds, symmetry breaking can
be inferred from the limit $|x-y|\to\infty$, if clustering is assumed
\cite{zj,allthat}, since $\Delta_D(x-y)$ vanishes for large
separations. Of course this contradicts the infinite volume limit of
the finite--volume definition, $\bar{\bds\phi}=\lim_{N\to\infty}
N^{-1/2}\vev{\bds\phi(x)}$, except at the extremal points
$\bar{\bds{\phi}}^2=v^2$.

In fact the $L\to\infty$ limit of the finite volume states with
$\bar{\bds\phi}^2<v^2$ does violate clustering, because they are linear
superpositions of vectors belonging to superselected sectors and
therefore they are indistinguishable from statistical mixtures. I can
give the following intuitive picture for large $N$. Consider any one
of the superselected sectors based on a physical vacuum with
$\bar{\bds\phi}^2=v^2$. By condensing a macroscopic number of
transverse Goldstone bosons at zero--momentum, one can build coherent
states with rotated $\bar{\bds\phi}$. By incoherently averaging over
such rotated states one obtains new states with field expectation
values shorter than $v$ by any prefixed amount.  In the large $N$
approximation this averaging is necessarily uniform and is forced upon
us by the residual $O(N-1)$ symmetry.

\subsection{Out--of--equilibrium dynamics}\label{ooedN}

I now turn to the dynamics out of equilibrium in this large $N$
context. It is governed by the equations of motion derived from the
total energy density $\E$ in eq. (\ref{ElargeN}), that is
\begin{equation}\label{sk2}
	\der{^2\bar{\bds\phi}}{t^2} = -M^2\,\bar{\bds\phi}
	\;,\quad \der{^2 \s_k}{t^2}= -(k^2+M^2)\,\s_k + \dfrac1{4 \s_k^3}
\end{equation}
where the generally time--dependent effective squared mass $M^2$ is
given by
\begin{equation}\label{Nunbgap}
	M^2 = m^2 + \lbare
	\left[\bar{\bds\phi}^2+\Sigma-I_D(m^2,\Lambda)\right]
\end{equation}
in case of unbroken symmetry and
\begin{equation}\label{Nbgap}
	M^2 = \lbare \left[\bar{\bds\phi}^2- v^2 +\Sigma-I_3(0,\Lambda)\right]
\end{equation}
for broken symmetry in $D=3$.

At time zero, the specific choice of initial conditions for $\s_k$ that
give the smallest energy contribution, that is
\begin{equation}\label{inis}
	\dot \s_k=0 \;,\quad \s_k^2=\dfrac1{2\sqrt{k^2+M^2}}
\end{equation}
turns eq. (\ref{Nunbgap}) into the usual gap equation
(\ref{massive}). For any value of $\bar\phi$ this equation has one
solution smoothly connected to the value $M=m$ at $\bar\phi=0$.  Of
course other initial conditions are possible. The only requirement is
that the corresponding energy must differ from that of the ground
state by an ultraviolet finite amount, as it occurs for the choice
(\ref{inis}). In fact this is guaranteed by the gap equation
itself, as evident from eq. (\ref{Nefforce}): when the widths $\s_k$ 
are extremal the effective force is finite, and therefore so are all
potential energy differences.

This simple argument needs a refinement in two respects. 

Firstly, in case of symmetry breaking the formal energy minimization
w.r.t. $\s_k$ leads always to eqs. (\ref{inis}), but these are
acceptable initial conditions only if the gap equation that follows
from eq. (\ref{Nbgap}) in the $L\to\infty$ limit, namely
\begin{equation}\label{Nbgap2}
	M^2 = \lbare \left[\bar{\bds\phi}^2-v^2 +
	I_D(M^2,\Lambda)-I_D(0,\Lambda) \right] 
\end{equation}
admits a nonnegative, physical solution for $M^2$. 

Secondly, ultraviolet finiteness only requires that the sum over $k$
in eq. (\ref{Nefforce}) be finite and this follows if eq. (\ref{inis})
holds at least for $k$ large enough, solving the issue raised in the
first point: negative $M^2$ are allowed by imposing a new form of gap
equation
\begin{equation}\label{newgap}
	M^2 = \lbare \left[\bar{\bds\phi}^2-v^2 + \dfrac1{L^D} 
	\sum_{k^2<|M^2|}\s_k^2 + \dfrac1{L^D}\sum_{k^2>|M^2|}
	\dfrac1{2\sqrt{k^2-|M^2|}} -I_D(0,\Lambda) \right] 
\end{equation}
where all $\s_k$ with $k^2<|M^2|$ are kept free (but all by hypothesis
microscopic) initial conditions. Of course there is no energy
minimization in this case. To determine when this new form is
required, I observe that, neglecting the inverse--power corrections
in the UV cutoff I may write eq. (\ref{Nbgap2}) in the following
form 
\begin{equation}\label{NiceN}
	\dfrac{M^2}{\hat\l(M)} = \bar{\bds\phi}^2 -v^2
\end{equation}
There exists a positive solution $M^2$ smoothly connected to the
ground state, $\bar{\bds\phi}^2=v^2$ and $M^2=0$, only provided
$\bar{\bds\phi}^2\ge v^2$. So, in the large $N$ limit, as soon as I start
with $\bar{\bds\phi}^2\le v^2$, I cannot satisfy the gap equation with a
positive value of $M^2$.

Once a definite choice of initial conditions is made, the system of
differential equations (\ref{sk2}) can be solved numerically with
standard integration algorithms. This has been already done by several
authors
\cite{Boyanovsky:1995me,Boyanovsky:1998yp,Boyanovsky:1998zg}, working
directly in infinite volume, with the following general results. In
the case of unbroken symmetry it has been established that the $\s_k$
corresponding to wavevectors $k$ in the so--called forbidden bands
with parametric resonances grow exponentially in time until their
growth is shut off by the back--reaction. For broken symmetry it is
the region in $k-$space with the spinodal instabilities caused by an
initially negative $M^2$, whose widths grow exponentially before the
back--reaction shutoff.  After the shutoff time the effective mass
tends to a positive constant for unbroken symmetry and to zero for
broken symmetry (in D=3), so that the only width with a chance to keep
growing indefinitely is $\s_0$ for broken symmetry.

Of course, in all these approaches the integration over modes in the
back--reaction $\Sigma$ cannot be done exactly and is always replaced
by a discrete sum of a certain type, depending on the details of the
algorithms. Hence there exists always an effective infrared cutoff,
albeit too small to be detectable in the numerical outputs.  A
possible troublesome aspect of this is the proper identification of
the zero--mode width $\s_0$. Even if a (rather arbitrary) choice of
discretization is made where a $\s_0$ appears, it is not really
possible to determine whether during the exponential growth or after
such width becomes of the order of the volume. The aim is just to
answer this question and therefore I will perform my numerical
evolution in finite volumes of several growing sizes.

\subsection{Numerical results}\label{numerical}
After a careful study in $D=3$ of the scaling behavior of the
dynamics with respect to different values of $L$, the linear size of
the system, I can reach the following conclusion: there exist a
$L-$dependent time, that I denote by $\tau_L$, that splits
the evolution in two parts; for $t \leq \tau_L$, the
behavior of the system does not differ appreciably from its
counterpart at infinite volume, while finite volume effects abruptly
alter the evolution as soon as $t$ exceeds $\tau_L$; moreover
\begin{itemize}
\item $\tau_L$ is proportional to the linear size of the
box $L$ and so it rescales as the cubic root of the volume.
\item $\tau_L$ does not depend on the value of the quartic
coupling constant $\lambda$, at least in a first approximation.
\end{itemize}

\begin{figure} 
\includegraphics[height=8cm,width=15cm]{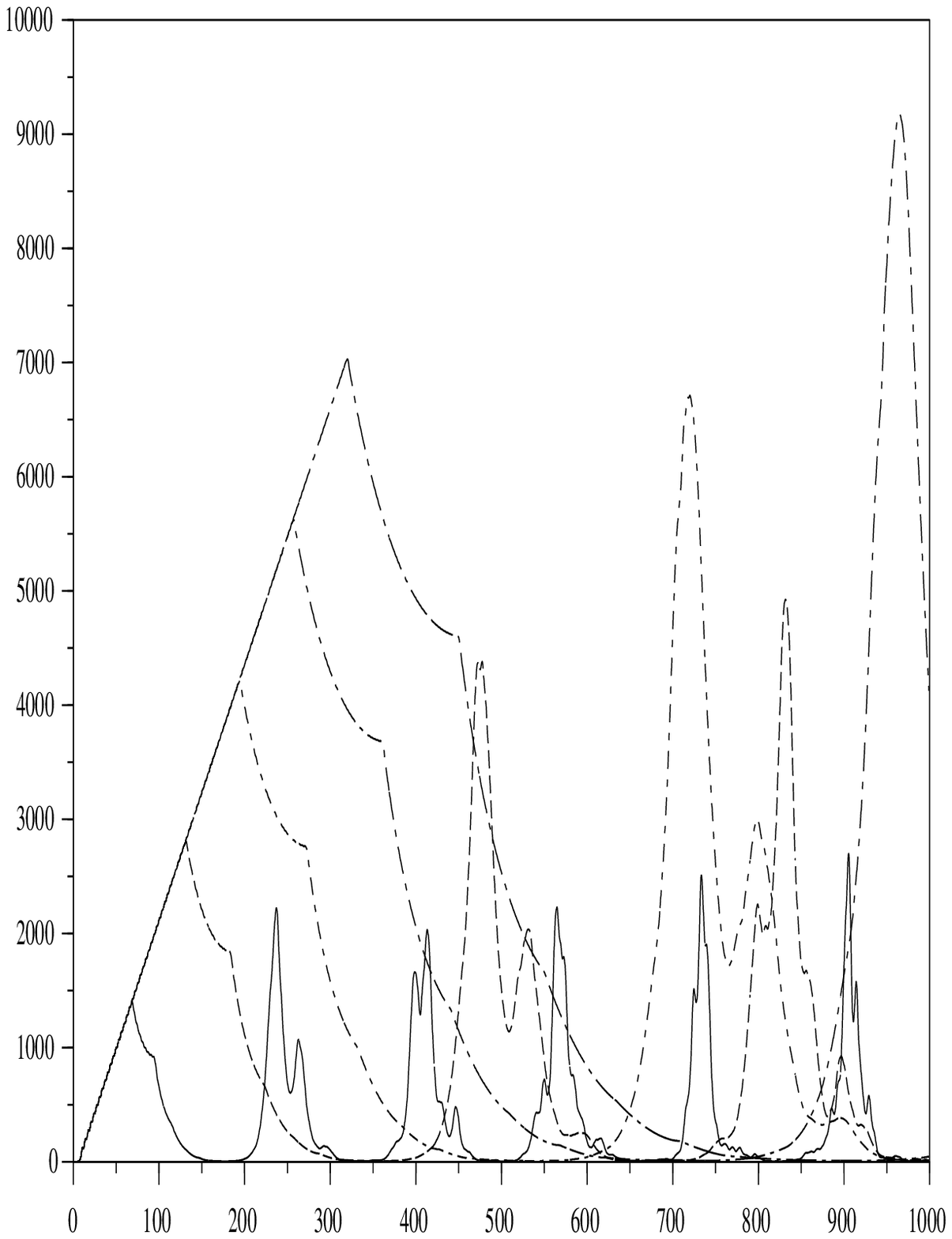}
\caption{\it Zero--mode amplitude evolution for 
different values of the size
$L/2\pi=20,40,60,80,100$, for $\lambda = 0.1$ and broken symmetry,
with $\bar\phi=0$. }\label{fig:m0}
\end{figure}
\vskip 0.5cm
\begin{figure} 
\includegraphics[height=8cm,width=15cm]{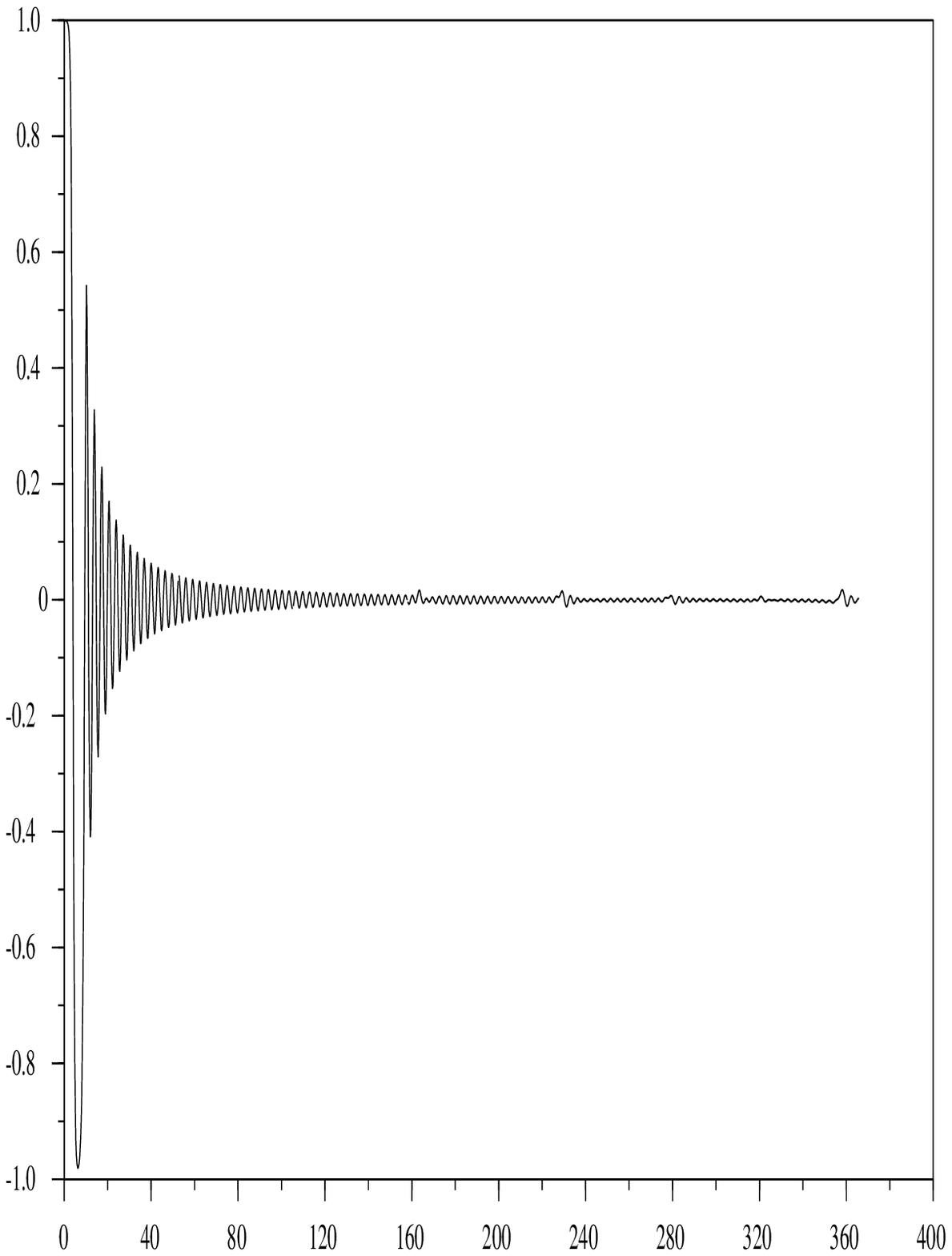}
\caption{\it Time evolution of the squared effective mass $M^2$ in broken
symmetry, for $L/2\pi=50$ and $\lambda=0.1$. }\label{fig:mass2}
\end{figure}

\begin{figure} 
\includegraphics[height=9cm,width=15cm]{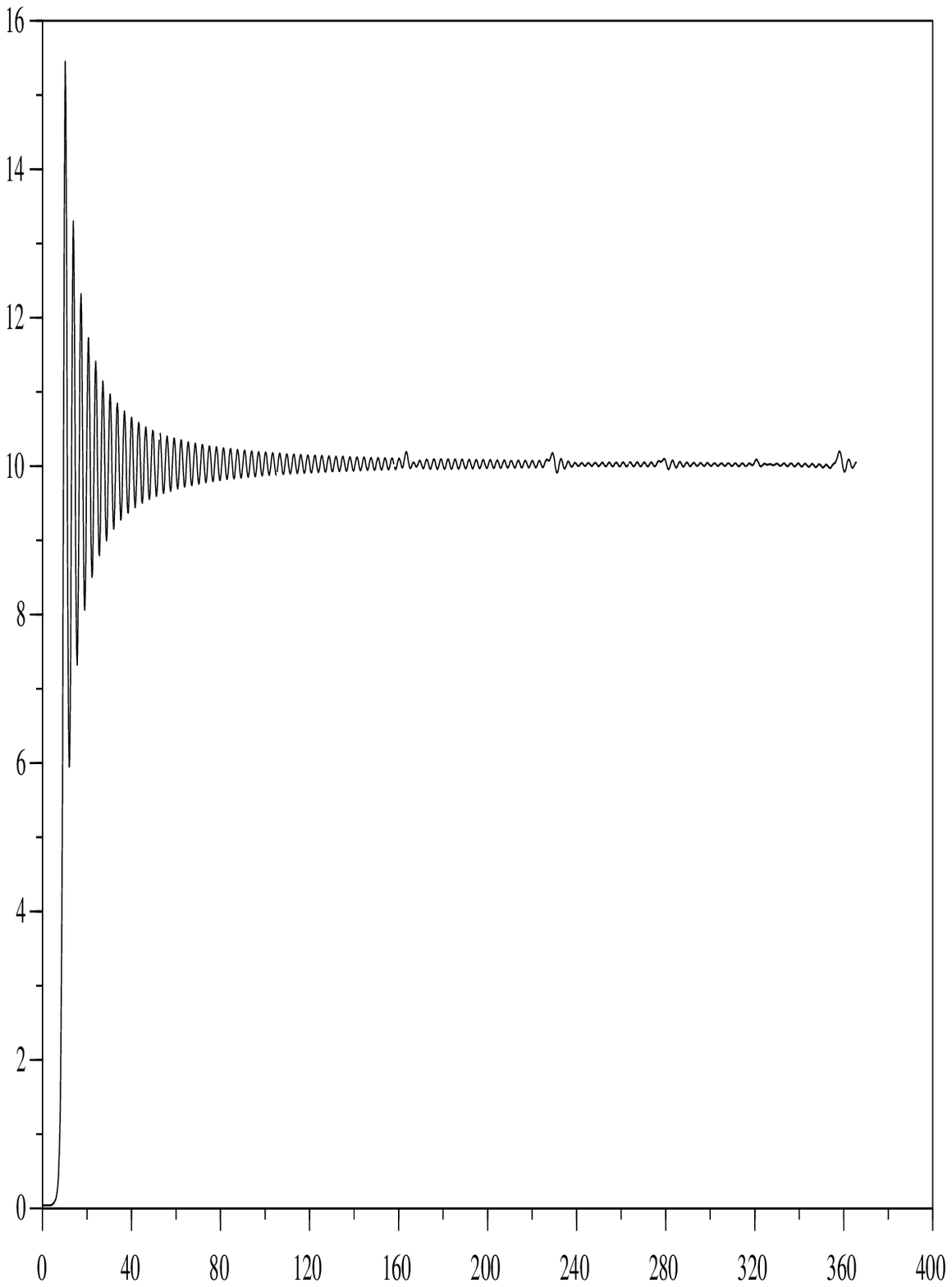}
\caption{\it The quantum back--reaction $\Sigma$, with the parameters as
in Fig. \ref{fig:mass2} }\label{fig:sigma}
\end{figure}

\begin{figure} 
\includegraphics[height=9cm,width=15cm]{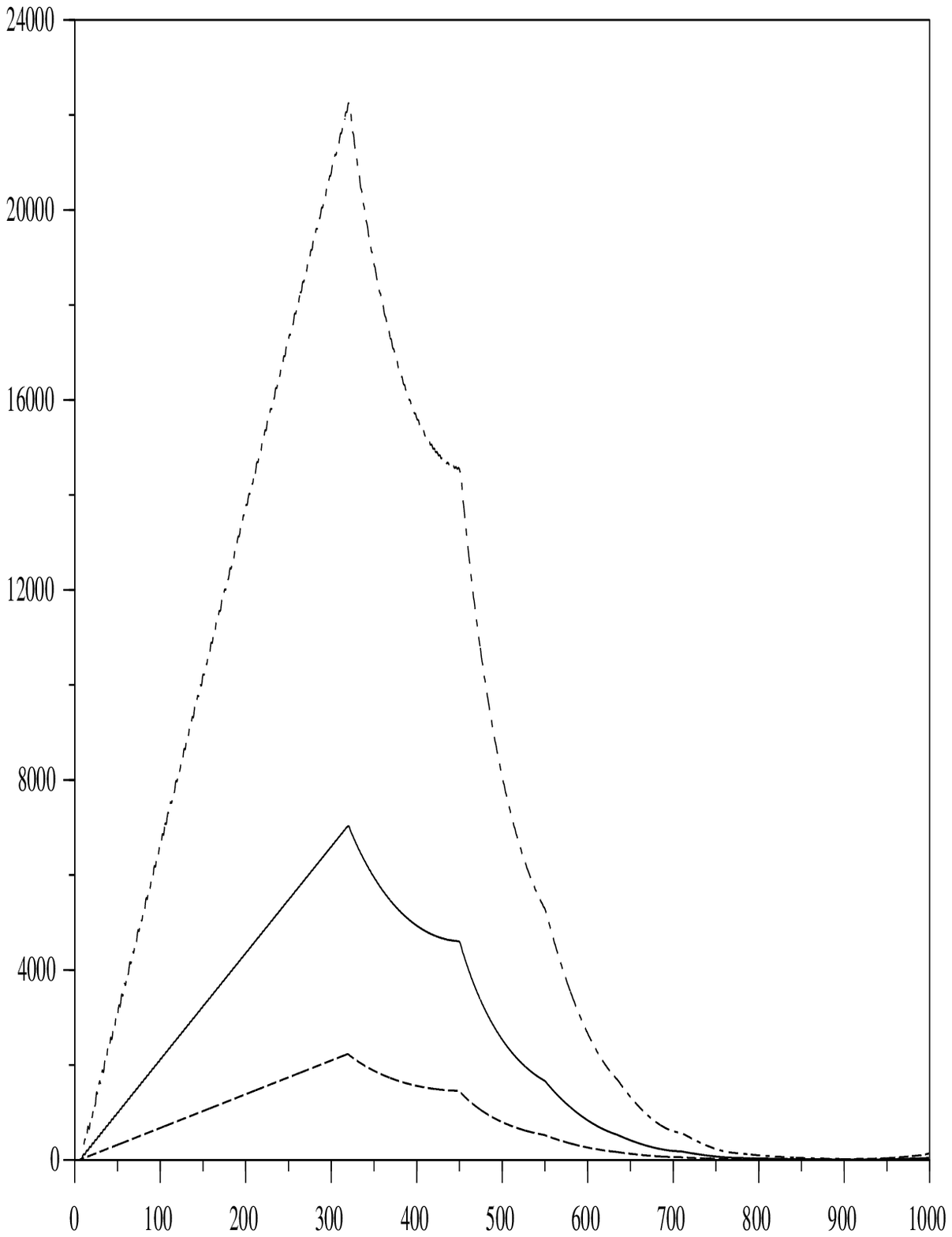}
\caption{\it Zero--mode amplitude evolution for different values of the
renormalized coupling constant $\l=0.01,0.1,1$, for $L/2\pi=100$ and
broken symmetry, with $\bar\phi=0$. }\label{fig:m0_l}
\end{figure}

\begin{figure} 
\includegraphics[height=8cm,width=15cm]{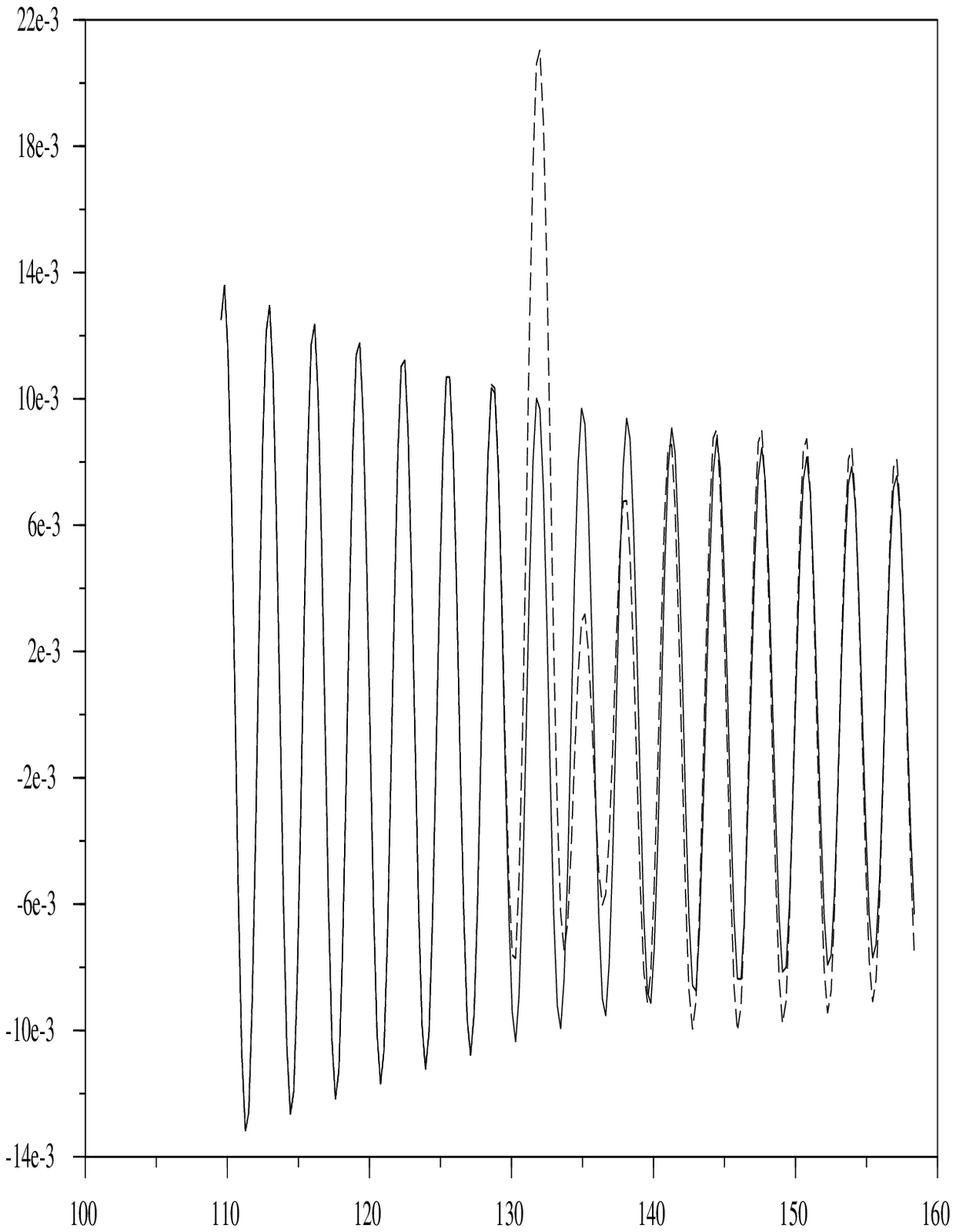}
\caption{\it Detail of $M^2$ near $t=\tau_L$ for $L/2\pi=40$ (dotted
line). The case $L/2\pi=80$ is plotted for comparison (solid line).}\label{fig:usc_mass}
\end{figure}

\begin{figure} 
\includegraphics[height=8cm,width=15cm]{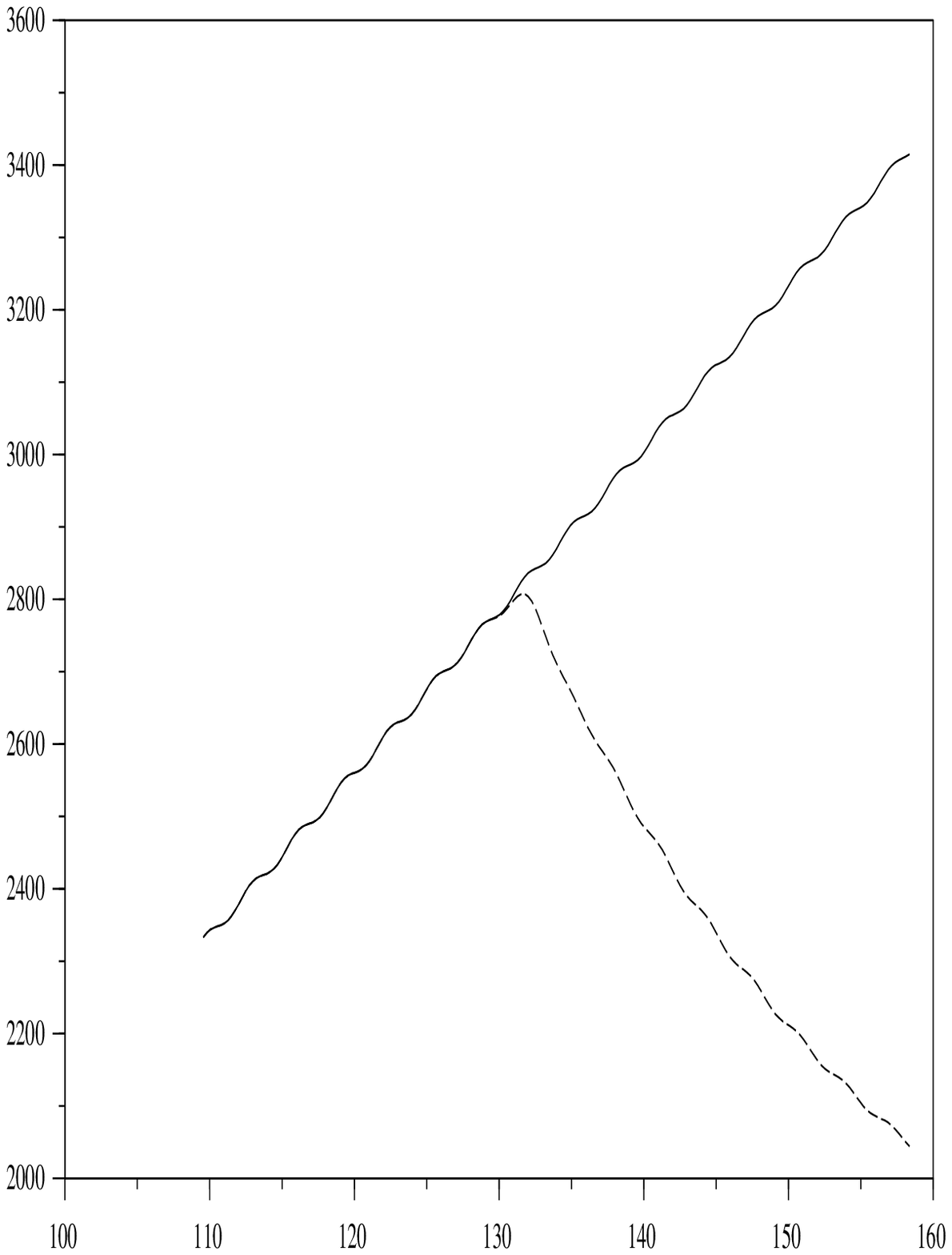}
\caption{\it Detail of $\s_0$ near $t=\tau_L$ for $\tfrac{L}{2\pi}=40$ (dotted
line). The case $L/2\pi=80$ is plotted for comparison (solid
line).}\label{fig:usc_zm}
\end{figure}

The figures show the behavior of the width of the zero mode $\s_0$
(see Fig. \ref{fig:m0}), of the squared effective mass $M^2$ (see Fig.
\ref{fig:mass2}) and of the back--reaction $\Sigma$ (see
Fig. \ref{fig:sigma}), in the more interesting case of broken
symmetry. The initial conditions are chosen in several different ways
(see appendix \ref{num} for details), but correspond to a negative
$M^2$ at early times with the initial widths all microscopic, that is
at most of order $L^{1/2}$. This is particularly relevant for the
zero--mode width $\s_0$, which is instead macroscopic in the lowest
energy state when $\bar{\bds\phi}^2<v^2$, as discussed above. As for
the background, the figures are relative to the simplest case
$\bar{\bds\phi}= 0 =\dot{\bar{\bds\phi}}$, but I have considered also
initial conditions with $\bar{\bds\phi}> 0$, reproducing the
``dynamical Maxwell construction'' observed in
ref. \cite{Boyanovsky:1998yp}.  At any rate, for the purposes of this
work, above all it is important to observe that, due to the quantum
back--reaction, $M^2$ rapidly becomes positive, within the so--called
{\em spinodal time}
\cite{Boyanovsky:1995me,Boyanovsky:1998yp,Boyanovsky:1998zg}, and
then, for times before $\tau_L$, the {\em weakly dissipative} regime
takes place where $M^2$ oscillates around zero with amplitude
decreasing as $t^{-1}$ and a frequency fixed by the largest spinodal
wavevector, in complete agreement with the infinite--volume results
\cite{Boyanovsky:1998yp}. Correspondingly, after the exponential
growth until the spinodal time, the width of the zero--mode grows on
average linearly with time, reaching a maximum for $t
\simeq\tau_L$. Precisely, $\s_0$ performs small amplitude oscillations
with the same frequency of $M^2$ around a linear function of the form
$A + B t$, where $A, B \approx \lambda ^{-1/2}$ (see
Fig. \ref{fig:m0_l}), confirming what already found in
refs. \cite{Boyanovsky:1998yp,Boyanovsky:1998zg}; then quite suddenly
it turns down and enters long irregular Poincar\'e--like cycles. Since
the spinodal oscillation frequency does not depend appreciably on $L$,
the curves of $\s_0$ at different values of $L$ are practically
identical for $t<\tau_L$. After a certain number of complete
oscillations, a number that scales linearly with $L$, a small change
in the behavior of $M^2$ (see Fig. \ref{fig:usc_mass}) determines an
inversion in $\s_0$ (see Fig. \ref{fig:usc_zm}), evidently because of
a phase crossover between the two oscillation patterns. Shortly after
$\tau_L$ dissipation practically stops as the oscillations of $M^2$
stop decreasing in amplitude and become more and more irregular,
reflecting the same irregularity in the evolution of the widths.

I can give a straightforward physical interpretation for the presence
of the time scale $\tau_L$. As shown in \cite{Boyanovsky:1998yp}, long
after the spinodal time $t_{\rm s}$, the effective mass oscillates
around zero with a decreasing amplitude and affects the quantum
fluctuations in such a way that the equal--time two--point correlation
function contains a time--dependent non--perturbative disturbance
growing at twice the speed of light. This is interpreted in terms of
large numbers of Goldstone bosons equally produced at any point in
space (due to translation invariance) and radially propagating at the
speed of light.  This picture applies also at finite volumes, in the
bulk, for volumes large enough. Hence, due to our periodic boundary
conditions, after a time exactly equal to $L/2$ the forward wave front
meets the backward wave front at the opposite point with respect to
the source, and the propagating wave starts interfering with itself
and heavily changes the dynamics with respect to that in infinite
volume. This argument leads us to give the value of $\pi$ for the
proportionality coefficient between $\tau_L$ and $L/2\pi$, prevision
very well verified by the numerical results, as can be inferred by a
look at the figures.

The main consequence of this scenario is that the linear
growth of the zero--mode width at infinite volume should not be
interpreted as a standard form of Bose--Einstein Condensation (BEC),
occurring with time, but should be consistently considered as ``novel''
form of dynamical BEC, as found by the authors of \cite{Boyanovsky:1998yp}. In
fact, if a macroscopic condensation were really there, the zero mode
would develop a $\delta$ function in infinite volume, that would be
announced by a width of the zero mode growing to values $O(L^{3/2})$
at any given size $L$. Now, while it is surely true that when I push
$L$ to infinity, also the time $\tau_L$ tends to infinity, allowing
the zero mode to grow indefinitely, it is also true that, at any fixed
though arbitrarily large volume, the zero mode never reaches a width
$O(L^{3/2})$, just because $\tau_L \propto L$. In other words, if we
start from initial conditions where $\s_0$ is microscopic, then it
never becomes macroscopic later on.
\begin{figure} 
\includegraphics[height=8cm,width=15cm]{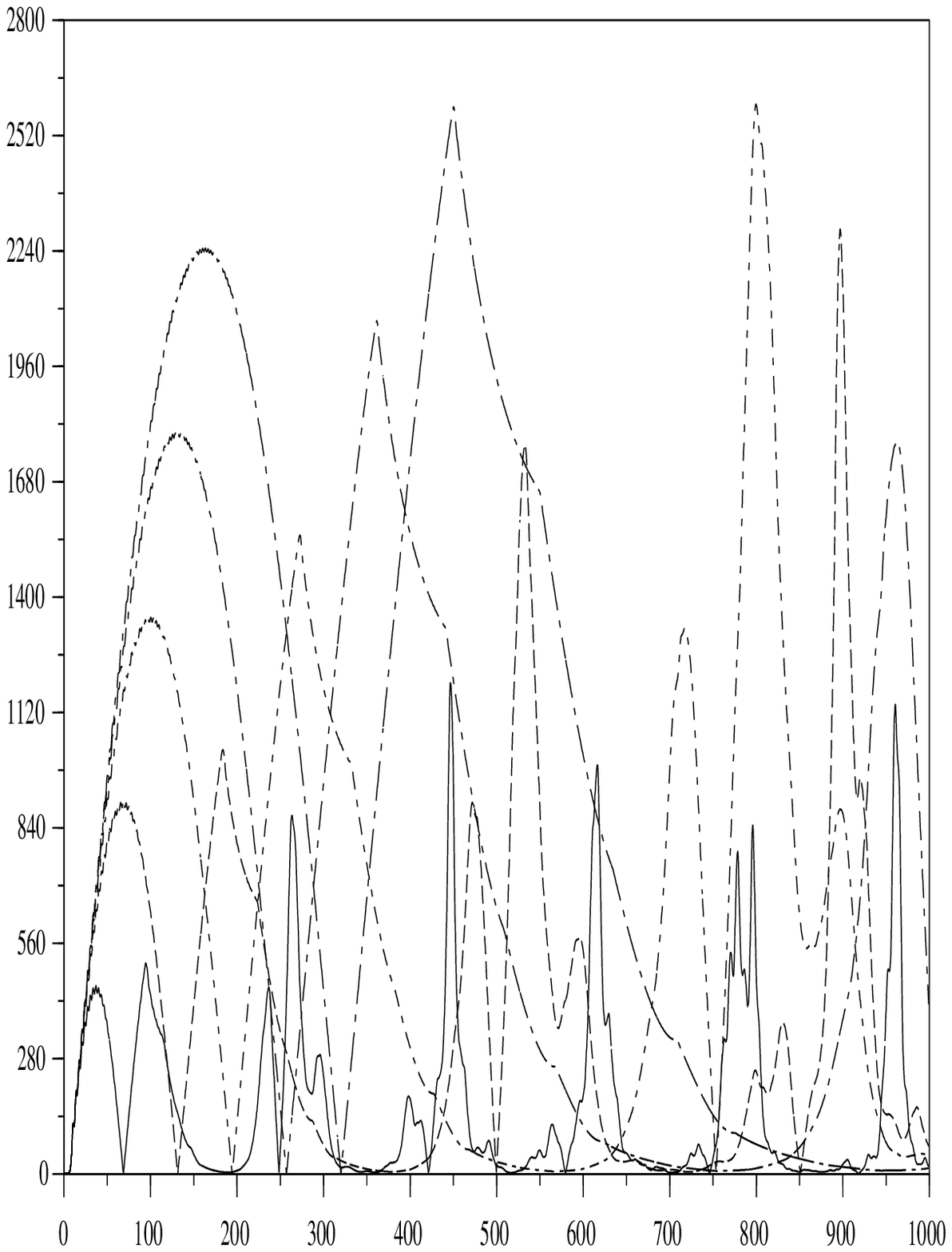}
\caption{\it Next--to--zero mode ($k=2\pi/L$) amplitude evolution for
different values of the size $L/2\pi=20,40,60,80,100$, for $\lambda =
0.1$ and broken symmetry, with $\bar\phi=0$.}\label{fig:m1}
\end{figure}
\vskip 0.5cm
\begin{figure} 
\includegraphics[height=8cm,width=15cm]{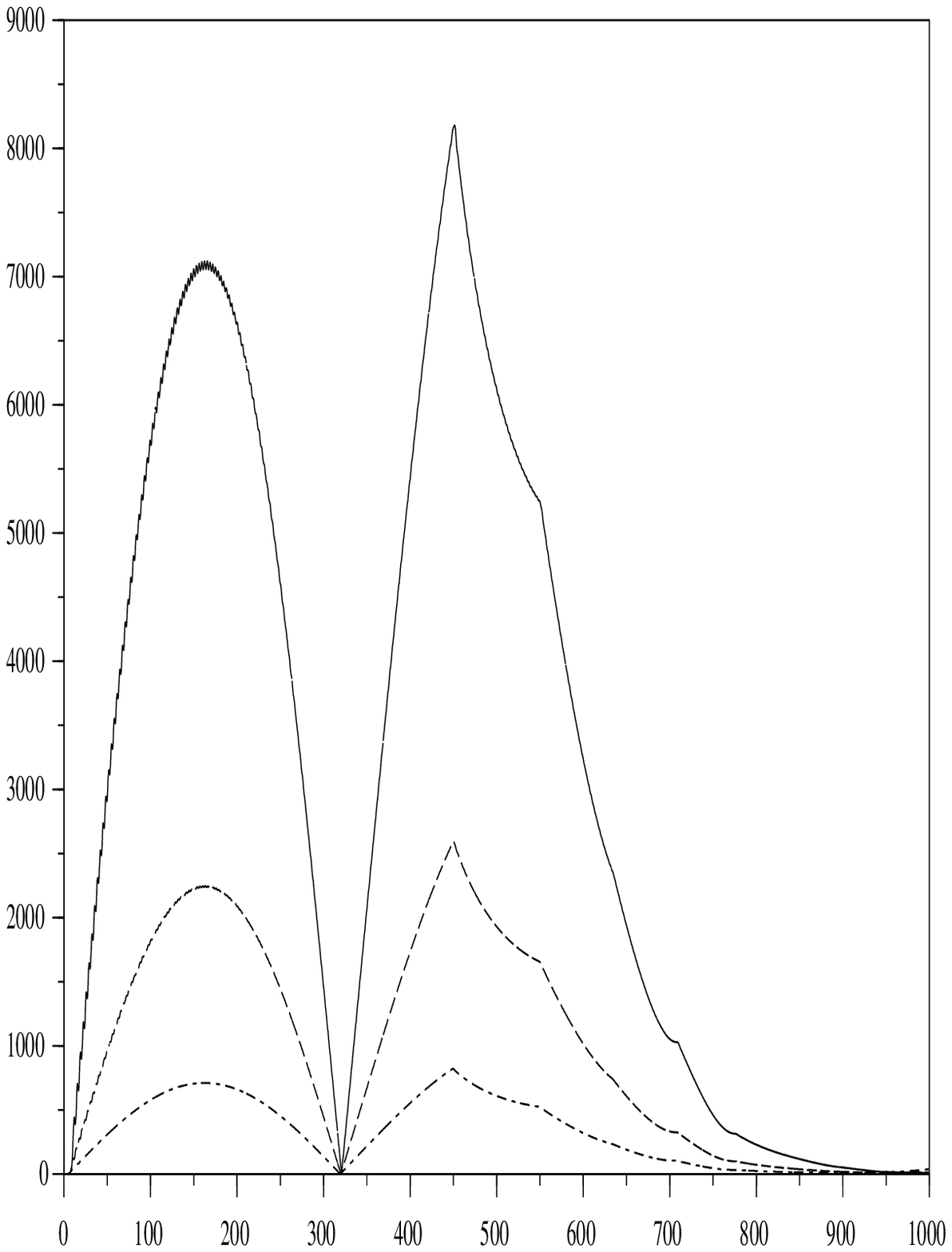}
\caption{\it Next--to--zero mode ($k=2\pi/L$) amplitude evolution for
different values of the renormalized coupling constant $\l=0.01,0.1,1$,
for $L/2\pi=100$ and broken symmetry, with $\bar\phi=0$.}
\label{fig:m1_l}
\end{figure}
On the other hand, looking at the behavior of the mode functions of
momenta $k=(2\pi/L)n$ for $n$ fixed but for different values of $L$,
one realizes that they obey a scaling similar to that observed for the
zero--mode: they oscillate in time with an amplitude and a period that
are $O(L)$ (see fig. \ref{fig:m1} and \ref{fig:m1_l}). Thus, each mode shows a
behavior that is exactly half a way between a macroscopic amplitude
[i.e. $O(L^{3/2})$] and a usual microscopic one [i.e. at most
$O(L^{1/2})$]. This means that the spectrum of the quantum
fluctuations at times of the order of the diverging volume can be
interpreted as a {\em massless} spectrum of {\em interacting}
Goldstone modes, because their power spectrum develops in the limit a
$1/k^2$ singularity, rather than the $1/k$ pole typical of free
massless modes. As a consequence the equal--time field correlation
function [see eq. (\ref{2pN})] will fall off as $|\bds x-\bds y|^{-1}$
for large separations smaller only than the diverging elapsed time.
This is in accord with what found in \cite{Boyanovsky:1998yp}, where the same
conclusion were reached after a study of the correlation function for
the scalar field in infinite volume.

The fact that each mode never becomes macroscopic, if it started
microscopic, might be regarded as a manifestation of unitarity in the
large $N$ approximation: an initial gaussian state with only
microscopic widths satisfies clustering and clustering cannot be
spoiled by a unitary time evolution. As a consequence, in the
infinite--volume late--time dynamics, the zero--mode width $\s_0$ does
not play any special role and only the behavior of $\s_k$ as $k\to 0$
is relevant. As already stated above, it turns out from our numerics
as well as from
refs. \cite{Boyanovsky:1998ba,Boyanovsky:1998yp,Boyanovsky:1998zg}
that this behavior is of a novel type characteristic both of the
out--of--equilibrium dynamics and of the equilibrium
finite--temperature theory, with $\s_k \propto 1/k$.

A comment should be made also about the periodic boundary conditions
used for these simulations. This choice guarantees the translation
invariance of the dynamics needed to consider a stable uniform
background. If I had chosen other boundary conditions (Dirichlet or
Neumann, for instance), the translation symmetry would have been broken
and an uniform background would have become non-uniform pretty
soon. Of course, I expect the bulk behavior to be independent of the
particular choice for the boundary conditions in the infinite volume
limit, even if a rigorous proof of this statement is still lacking.

The numerical evidence for the linear dependence of $\tau_L$ on $L$ is
very strong, and the qualitative argument given in the previous
section clearly explains the physics that determines it. Nonetheless a
solid analytic understanding of the detailed (quantitative) mechanism
that produces the inversion of $\dot{\s}_0$ around $\tau_L$ and its
subsequent irregular behavior, is more difficult to obtain. One could
use intuitive and generic arguments like the quantization of momentum
in multiples of $2\pi/L$, but the evolution equations do not have any
simple scaling behavior towards a universal form, when mass dimensions
are expressed in multiples of $2\pi/L$ and time in multiples of
$L$. Moreover, the qualitative form of the evolution depends heavily
on the choice of initial conditions. In fact, before finite volumes
effects show up, the trajectories of the quantum modes are rather
complex but regular enough, having a small-scale quasi-periodic almost
mode-independent motion within a large-scale quasi-periodic
mode-dependent envelope, with a very delicate resonant equilibrium
(cfr. Fig. \ref{fig:m0} and
\ref{fig:m1}). Apparently (cfr. Fig. \ref{fig:usc_mass} and
\ref{fig:usc_zm}), it is a sudden small beat that causes the turn
around of the zero-mode and of the other low-lying modes (with many
thousands of coupled modes, it is very difficult for the delicate
resonant equilibrium to fully come back ever again), but I think that
a deeper comprehension of the non--linear coupled dynamics is needed
in order to venture into a true analytic explanation.

On the other hand it is not difficult to understand why $\tau_L$ does
not depend appreciably on the coupling constant: when finite-volume
effects first come in, that is when the wave propagating at the speed
of light first starts to interfere with itself, the quantum
back-reaction $\lambda\Sigma$ has settled on values of order 1,
because the time $\tau_L \simeq L/2$ is much greater than the spinodal
time $t_{\rm s}$. The slope of the linear envelope of the zero mode does
depend on $\lambda$ because it is fixed by the early exponential
growth. Similarly, it is easy to realize that the numerical
integrations of refs. 
\cite{Boyanovsky:1998ba,Boyanovsky:1998yp,Boyanovsky:1998zg} over continuum
momenta correspond roughly to an effective volume much larger than any
one used here, so that the calculated evolution remained far away from the
onset of finite-volume effects.

%% file: finvHF.tex
Before going into the details of the analysis, let us briefly
summarize the main limitations and some results of the study of a
scalar field out of equilibrium within the gaussian HF scheme
\cite{Boyanovsky:1997mq,Boyanovsky:1994pa,Boyanovsky:1996sv,
Boyanovsky:1995me}. First of all, this scheme
has the advantage of going beyond perturbation theory, in the sense
that the (numerical) solution of the evolution equations will contain
arbitrary powers of the coupling constant, corresponding to a
non--trivial resummation of the perturbative series. For this reason,
the method is able to take into account the quantum back--reaction on
the fluctuations themselves, which shuts off their early exponential
growth. This is achieved by the standard HF factorization of the
quartic interaction, yielding a {\em time dependent}
self--consistently determined mass term, which stabilizes the modes
perturbatively unstable. The detailed numerical solution of the
resulting dynamical equations clearly shows the dissipation associated
with particle production, as a result of either parametric
amplification in case of unbroken symmetry or spinodal instabilities
in case of broken symmetry, as well as the shut off mechanism outlined
above.

However, the standard HF method is really not controllable in the case
of a single self--interacting scalar field, while it becomes exact
only in the $N \to \infty$ limit or in the free case. Moreover,
previous approaches to the dynamics in this approximation scheme had
the unlikely feature of maintaining a weak (logarithmic) cut--off
dependence on the renormalized equations of motion of the order
parameter and the mode functions
\cite{Boyanovsky:1995me}.

A time dependent variational approximation to study the evolution of
quantum fluctuations has been used also in \cite{Braghin:1998sj},
where spatially dependent configurations are considered and even
semi--analytical solutions are found in special cases. A numerical
method is established in this approximation, which is based on
generalized density matrix which obeys a Liouville--von-Neumann type
equation.

Corrections to the usual ``mean field'' (Gaussian Hartree--Bogolubov
approach) approximation at zero and non--zero temperature were
introduced also in \cite{Braghin:1998yw}, where the $n-$point
correlation functions, which are related to important observables,
were computed as response functions of the system to different
external sources, for both the two phases of the potential. Remarkably
enough, the proper renormalization of the coupling constant eliminates
the logarithmic UV divergence in physical quantities.

I have already defined, in section \ref{cft}, all the relevant notations and 
the quantum representation I will be using to study the evolution of the 
system.

I introduce in section \ref{tdhf} our improved time--dependent
Hartree--Fock (tdHF) approximation, which generalizes the standard
gaussian self-consistent approach \cite{Kerman:1976yn} to non--gaussian
wave--functionals; I then derive the mean--field coupled
time--dependent Schroedinger equations for the modes of the scalar
field, under the assumption of a uniform condensate, see eqs
(\ref{Schroedinger}), (\ref{H_k}) and (\ref{omvev}). A significant
difference with respect to previous tdHF approaches \cite{Boyanovsky:1995me}
concerns the renormalization of ultraviolet divergences. In fact, by
means of a single substitution of the bare coupling constant $\lbare$
with the renormalized one $\l$ in the Hartree--Fock hamiltonian, we
obtain cut-off independent equations (apart from corrections in
inverse powers, which are there due to the Landau pole). The
substitution is introduced by hand, but is justified by simple
diagrammatic considerations.

One advantage of not restricting a priori the self-consistent HF
approximation to gaussian wave--functionals, is in the possibility of
a better description of the vacuum structure in case of broken
symmetry. In fact I can show quite explicitly that, in any finite
volume, in the ground state the zero--mode of $\phi$ field is
concentrated around the two vacua of the broken symmetry, driving the
probability distribution for any sufficiently wide smearing of the
field into a two peaks shape. This is indeed what one would
intuitively expect in case of symmetry breaking. On the other hand
none of this appears in a dynamical evolution that starts from a
distribution localized around a single value of the field in the
spinodal region, confirming what already seen in the large $N$
approach \cite{Destri:1999hd}. More precisely, within a further controlled
gaussian approximation of our tdHF approach, one observes that
initially microscopic quantum fluctuations never becomes macroscopic,
suggesting that also non--gaussian fluctuations cannot reach
macroscopic sizes.  As a simple confirmation of this fact, consider
the completely symmetric initial conditions
$\vev{\phi}=\vev{\dot\phi}=0$ for the background: in this case I find
that the dynamical equations for initially gaussian field fluctuations
are identical to those of large $N$ (apart for a rescaling of the
coupling constant by a factor of three; cfr. ref. \cite{Destri:1999hd}), so
that I observe the same asymptotic vanishing of the effective
mass. However, this time no interpretation in terms of Goldstone
theorem is possible, since the broken symmetry is discrete; rather, if
the width of the zero--mode were allowed to evolve into a macroscopic
size, then the effective mass would tend to a positive value, since
the mass in case of discrete symmetry breaking is indeed larger than
zero. Anyway, also in the gaussian HF approach, I do find a whole class of
cases which exhibit the time scale $\tau_L$. At that time, finite
volume effects start to manifest and the size of the low--lying widths
is of order $L$. I then discuss why this undermines the
self--consistency of the gaussian approximation, imposing the need of
further study, both analytical and numerical.

In section \ref{ep} I study the asymptotic evolution in the broken
symmetry phase, in infinite volume, when the expectation value starts
within the region between the two minima of the potential. I am able
to show by precise numerical simulations, that the fixed points of the
background evolution do not cover the static flat region
completely. On the contrary, the spinodal region seems to
be absolutely forbidden for the late time values of the mean
field. Thus, as far as the asymptotic evolution is concerned, our
numerical results lead to the following conclusions. I can
distinguish the points lying between the two minima in a fashion
reminiscent of the static classification: first, the values satisfying
the property $v/\sqrt{3} < \abs{\bar\phi _{\infty}} \leq v$ are
metastable points, in the sense that they are fixed points of the
background evolution, no matter which initial condition comprised in
the interval $\left( -v , v \right)$ I choose for the expectation
value $\bar\phi$; secondly, the points included in the interval $0 <
\abs{\bar\phi _{\infty}} < v/\sqrt{3}$ are unstable points, because if
the mean field starts from one of them, after an early slow rolling
down, it starts to oscillate with decreasing amplitude around a point
inside the classical metastable interval. Obviously, $\bar\phi=v$ is
the point of stable equilibrium, and $\bar\phi=0$ is a point of
unstable equilibrium. Actually, it should be noted that our data do
not allow a precise determination of the border between the dynamical
unstable and metastable regions; thus, the number I give here should
be looked at as an educated guess inspired by the analogous static
classification and based on considerations about the solutions of the
gap equation [see eq. (\ref{newgap1})]

\subsection{The Variational Principle}\label{tdhf}

I consider the $\phi^4$ Hamiltonian (\ref{lN_ham}) and I shall work in
the wavefunction representation (\ref{rep1}), (\ref{rep2}). 

I examine here only states in which the scalar field has a uniform,
albeit possibly time--dependent expectation value. I may then start
from a wavefuction of the factorized form (which would be exact for
free fields)
\begin{equation}\label{wf}
	\Psi(\varphi)=\psi_0(\bds\varphi_0)
		\prod_{k>0} \psi_k(\bds\varphi_k,\bds\varphi_{-k})
\end{equation}
The dependence of $\psi_k$ on its two arguments cannot be assumed to
factorize in general since space translations act as $SO(2)$ rotations
on $\varphi_k^i$ and $\varphi_{-k}^i$ (hence in case of translation
invariance $\psi_k$ depends on $\varphi_k^i$, $\varphi_{-k}^j $only
through $(\varphi_k^i)^2+(\varphi_{-k}^i)^2$.
The approximation consists in assuming this form as valid at all times
and imposing the stationarity condition on the action
\begin{equation}\label{varpri1}
	\delta \int dt\, \vev{i\partial_t-H}=0 \;,\quad
	\vev{\cdot} \equiv \bra{\Psi(t)}\cdot\ket{\Psi(t)}
\end{equation}
with respect to variations of the functions $\psi_k$. To enforce a
uniform expectation value of $\phi$ I should add a Lagrange
multiplier term linear in the single modes expectations
$\vev{\bds\varphi_k}$ for $k\neq 0$. The multiplier is then fixed at the
end to obtain $\vev{\varphi_k}=0$ for all $k\neq 0$. Actually one may
verify that this is equivalent to the simpler approach in which
$\vev{\varphi_k}$ is set to vanish for all $k\neq 0$ before any
variation. Then the only non trivial expectation value in the
Hamiltonian, namely that of the quartic term, assumes the form
\begin{eqnarray}\label{vevphi4}
\lefteqn{ \hspace{-3cm} \vev{V} =  \dfrac\lbare{4 L^D} \left\{
	\left[ \vev{\left( \bds\varphi_0^2 \right)
	^2}-3\vev{\bds\varphi_0^2}^2 \right] +
	\dfrac32 \sum_{k>0} \left[
	\vev{\left( \bds\varphi_k^2+\bds\varphi_{-k}^2 \right)
	^2}-2 \left( \vev{\bds\varphi_k^2}
	+\vev{\bds\varphi_{-k}^2} \right) ^2 \right] \right. {} }
	\nonumber \\ &
	& {}+ 
	3 \left(\sum_k\vev{\bds\varphi_k^2}\right)^2 {} \nonumber \\ & & {}+
	\left. 2 \sum_{k ^{'} \neq \pm k} \vev{ \varphi _k ^{i} \varphi _k ^{j}
	\varphi _{k ^{'}} ^{l} \varphi _{k ^{'}} ^{m}} \left( \delta
	_{il} \delta _{jm} - \delta _{ij} \delta _{lm} \right)
	\right. {} \nonumber \\ & & {} +
	\left. 2 \sum_{k>0} \vev{ \varphi _k ^{i} \varphi _k ^{j}
	\varphi _{-k} ^{l} \varphi _{-k} ^{m}} \left( \delta
	_{il} \delta _{jm} - \delta _{ij} \delta _{lm} \right)
	\right\} {}
\end{eqnarray}

Notice that the terms in the first line would cancel completely out for
gaussian wavefunctions $\psi_k$ with zero mean value. The second line,
where the sum extends to all wavevectors $k$, would correspond instead
to the standard mean field replacement $\vev{\phi^4}\to
3\vev{\phi^2}^2$, in the case $N=1$. The total energy of our trial
state now reads
\begin{equation}\label{energy}
	E=\vev{H} =\dfrac12\sum_k \VEV{-
	\bds\bigtriangledown_{\varphi_k}^2 +
	(k^2+\mbare)\bds\varphi_k^2} + \dfrac\lbare{4}\int d^Dx\,
	\vev{(\bds\phi(x)^2)^2}
\end{equation}
and from the variational principle (\ref{varpri1}) I obtain a set
of simple Schroedinger equations
\begin{equation}\label{Schroedinger}
	i\partial_t\psi_k = H_k \psi_k 
\end{equation}
\vskip -.4truecm
\begin{eqnarray}\label{H_k}
\lefteqn{  H_0 =-\dfrac12\bds\bigtriangledown ^2 _{\varphi_0} +\dfrac12 \left(
	\om_0^2 \right) _{ij} \varphi_0 ^{(i)} \varphi_0 ^{(j)} 
	+\dfrac{\lbare}{4L^D} \left( \bds\varphi_0 ^2 \right) ^2 {} }
	\\ & & {}
  H_k =-\dfrac12 \left( \bds\bigtriangledown ^2 _{\varphi_k} + 
	\bds\bigtriangledown ^2 _{\varphi_-{k}} \right)
	+\dfrac12 \left( \om_k^2 \right) _{ij} \left( \varphi_k ^{(i)} 
	\varphi_k ^{(j)} + \varphi_{-k} ^{(i)} \varphi_{-k} ^{(j)} \right)
	+\dfrac{3\lbare}{8L^D}\left(\bds\varphi _k ^2+ \bds\varphi
	_{-k} ^2\right)^2 {} \nonumber \\  \label{H_k1} & & {} 
	+ \dfrac{\lbare}{2 L ^D} \left[ \left(
	\bds\varphi _k \bds\varphi _{-k} \right) ^2 - \bds\varphi _k
	^2 \bds\varphi _{-k} ^2 \right] 
\end{eqnarray}
which are coupled in a mean--field way only through 
\begin{equation}\label{omvev}
\left( \om_k^2 \right) _{ij} = \left( k ^2 + \mbare + \dfrac{\lbare}{L
^D} \sum _{q\neq k,-k} \vev{\bds\varphi _q ^2} \right) \delta _{ij} +
\dfrac{2 \lbare}{L ^D} \sum _{q\neq k,-k} \vev{\varphi_q ^{(i)}
\varphi_q ^{(j)}} 
\end{equation}
that can be written also in the form
\begin{equation}\label{omvev1}
\left( \om_k^2 \right) _{ij} = \left( k ^2 + \mbare \right) \delta
_{ij} + 3 \lbare \Sigma _k ^{ij}
\end{equation}
having defined
\begin{equation}\label{omvev3}
\Sigma _k ^{ij} = \dfrac1{3 L ^D} \sum _{q\neq k,-k} \left(
\vev{\bds\varphi _q ^2} \delta ^{ij} + 2 \vev{\varphi_q ^{(i)}
\varphi_q ^{(j)}} \right)
\end{equation}
This defines the HF time evolution for the theory. By construction this
evolution conserves the total energy $E$ of eq. (\ref{energy}).

\subsection*{N=1}
I now continue the discussion for the case of a single scalar
field (i.e. $N=1$), postponing to section \ref{N>1} of the appendix
some technical issues for $N>1$. In this case, the third and fourth
rows of eqs. (\ref{vevphi4}) vanish. First of all it should be
stressed that in this particular tdHF approximation, beside the
mean--field back--reaction term $\Sigma_k$ of all other modes on
$\om_k^2$, I keep also the contribution of the {\em diagonal}
scattering through the diagonal quartic terms. In fact this is why
$\Sigma_k$ has no contribution from the $k-$mode itself: in a gaussian
approximation for the trial wavefunctions $\psi_k$ the Hamiltonians
$H_k$ would turn out to be harmonic, the quartic terms being absent in
favor of a complete back--reaction
\begin{equation}\label{Sigma}
	\Sigma = \Sigma_k + \dfrac{\vev{\varphi_k^2}+
	\vev{\varphi_{-k}^2}}{L^D} = \dfrac1{L^D}\sum_k\vev{\varphi_k^2}
\end{equation}
Of course the quartic self--interaction of the modes as well as the
difference between $\Sigma$ and $\Sigma_k$ are suppressed by a volume
effect and could be neglected in the infrared limit, provided all
wavefunctions $\psi_k$ stay concentrated on mode amplitudes
$\varphi_k$ of order smaller than $L^{D/2}$.  This is the
typical situation when all modes remain microscopic and the volume in
the denominators is compensated only through the summation over a
number of modes proportional to the volume itself, so that in the
limit $L\to\infty$ sums are replaced by integrals
\begin{equation}
	\Sigma_k \to \Sigma \to \int_{k^2\le\Lambda^2} 
	\dfrac{d^Dk}{(2\pi)^D} \vev{\varphi_k^2}
\end{equation}
Indeed I shall apply this picture to all modes with $k\neq 0$, while
I do expect exceptions for the zero--mode wavefunction $\psi_0$.

The treatment of ultraviolet divergences requires particular care,
since the HF approximation typically messes things up (see, for
instance, \cite{Lenaghan:1999si}). Following the same approach as in 
the large $N$ approximation 
\cite{Cooper:1994hr,Boyanovsky:1995me,Destri:1999hd}, I could
take as renormalization condition the requirement that the frequencies
$\om_k^2$ are independent of $\Lambda$, assuming that $\mbare$ and
$\lbare$ are functions of $\Lambda$ itself and of renormalized
$\Lambda-$independent parameters $m^2$ and $\l$ such that
\begin{equation}\label{renomvev}
	\om_k^2 = k^2 +m^2 +3\l\left[\Sigma_k\right]_{\rm finite}
\end{equation}
where by $[.]_{\rm finite}$ I mean the (scheme--dependent) finite
part of some possibly ultraviolet divergent quantity.  Unfortunately
this would not be enough to make the spectrum of energy differences
cutoff--independent, because of the bare coupling constant $\lbare$ in
front of the quartic terms in $H_k$ and the difference between
$\Sigma$ and $\Sigma_k$ [such problem does not exist in large $N$
because that is a purely gaussian approximation].  Again this would
not be a problem whenever these terms become negligible as
$L\to\infty$. At any rate, to be ready to handle the cases when this
is not actually true and to define an ultraviolet--finite model also
at finite volume, I shall by hand modify eq. (\ref{vevphi4}) as
follows:
\begin{eqnarray}\label{subst}
	\lbare\int d^Dx\, \vev{\phi(x)^4} &=& \nonumber \\ \l
	L^{-D}\left\{ \vphantom{\dfrac32}\VEV{\varphi_0^4} -
	3\,\vev{\varphi_0^2}^2 \right. &+&\left. \dfrac32 \sum\limits_{k>0}
	\left[
	\vev{(\varphi_k^2+\varphi_{-k}^2)^2}-2\left(\vev{\varphi_k^2}
	+\vev{\varphi_{-k}^2}\right)^2 \right]\right\} \nonumber \\ &
	+ & 3\lbare \,L^D\Sigma^2
\end{eqnarray}
I keep the bare coupling constant in front of the term containing
$\Sigma^2$ because the double sum over the modes couples each one of
then to all the others. This produces a proper renormalization by
means of the usual {\em cactus} resummation \cite{cactus}, which
corresponds to the standard HF approximation. On the other hand,
within the same approximation, it is not possible to renormalize the
part in curly brackets of the equation above, because of the
factorized form (\ref{wf}) that I have assumed for the wavefunction of
the system. In fact, the $4-$legs vertices in the curly brackets are
diagonal in momentum space; at higher order in the loop expansion,
when I contract two or more vertices of this type, no sum over
internal loop momenta is produced, so that all higher order
perturbation terms are suppressed by volume effects. However, we know
that in the complete theory, the wavefunction is not factorized and
loops contain all values of momentum (not only those corresponding to
external legs). This suggests that, in order to get a finite
hamiltonian, I need to introduce in the definition of our model some
extra resummation of Feynmann diagrams, that is not automatically
contained in this self--consistent HF approach. The simplest choice
(maybe the only one consistent with the cactus resummation performed
in the two--point function by the HF scheme) is the resummation of the
complete $1$-loop {\em fish} diagram in the four--point function. This
amounts to the change from $\lbare $ to $\l$ and it is enough to
guarantee the ultraviolet finiteness of the hamiltonian through the
redefinition
\begin{equation}\label{H_k'}
	H_0 \to H_0+\dfrac{\l-\lbare}{4L^D}\varphi_0^4 \;,\quad
	H_k \to H_k +\dfrac{
	3(\l-\lbare)}{8L^D}\left(\varphi_k^2+\varphi_{-k}^2\right)^2
\end{equation}
At the same time the frequencies are now related to the widths
$\vev{\varphi_{-k}^2}$ by
\begin{eqnarray}\label{omvev2}
	\om_k^2 &=& k^2+ M^2 - 3\l\, L^{-D}(\vev{\varphi_k^2}+
	\vev{\varphi_{-k}^2}) \;,\quad k>0 \\ M^2 & \equiv &
	\om_0^2 + 3\l\,L^{-D}\vev{\varphi_0^2}= \mbare +3\lbare \Sigma 
\end{eqnarray}
Apart for $O(L^{-D})$ corrections, $M$ plays the role of
time--dependent mass for modes with $k\neq 0$, in the harmonic
approximation.
  
In this new setup the conserved energy reads
\begin{equation}\label{menergy}
	E =\sum_{k\ge 0}\vev{H_k} -\dfrac34 \lbare\,L^D\,\Sigma^2 +
	\dfrac34\l\,L^{-D}\left[\vev{\varphi_0^2}^2 + \sum\limits_{k>0}
	\left(\vev{\varphi_k^2}+\vev{\varphi_{-k}^2}\right)^2\right]
\end{equation}
Since the gap--like equations (\ref{omvev2}) are state--dependent, we
have to perform the renormalization first for some reference quantum
state, that is for some specific collection of wavefunctions $\psi_k$;
as soon as $\mbare$ and $\lbare$ are determined as functions
$\Lambda$, ultraviolet finiteness will hold for the entire class of
states with the same ultraviolet properties of the reference
state. Then an obvious consistency check for the HF approximation is
that this class is closed under time evolution.

Rather than a single state, I choose as reference the family of
gaussian states parametrized by the uniform expectation value
$\vev{\phi(x)}=L^{-D/2}\vev{\varphi_0}=\bar\phi$ (recall that I have
$\vev{\varphi_k}=0$ when $k\neq0$ by assumption) and such that the HF
energy $E$ is as small as possible for fixed $\bar\phi$.  Then, apart
from a translation by $L^{D/2}\bar\phi$ on $\varphi_0$, these gaussian
$\psi_k$ are ground state eigenfunctions of the harmonic Hamiltonians
obtained from $H_k$ by dropping the quartic terms. Because of the
$k^2$ in the frequencies I expect these gaussian states to dominate
in the ultraviolet limit also at finite volume (as discussed above
they should dominate in the infinite--volume limit for any
$k\neq0$). Moreover, since now
\begin{equation}\label{gauss}
	\vev{\varphi_0^2} = L^D\bar\phi^2+ \dfrac1{2\om_0} \;,\quad 
	\vev{\varphi_{\pm k}^2} = \dfrac1{2\om_k} \;,\quad k\neq 0
\end{equation}
the relation (\ref{omvev2}) between frequencies and widths turns into the
single gap equation
\begin{equation}\label{gap}
	M^2 =\mbare+3\lbare\left(\bar\phi^2
	+\dfrac1{2L^D}\sum_{q^2\le\Lambda^2}\dfrac1{\sqrt{k^2+M^2}}\right)
\end{equation}
fixing the self-consistent value of $M$ as a function of $\bar\phi$.
It should be stressed that (\ref{omvev2}) turns through
eq. (\ref{gauss}) into the gap equation only because of the
requirement of energy minimization. Generic $\psi_k$, regarded as
initial conditions for the Schroedinger equations (\ref{Schroedinger}),
are in principle not subject to any gap equation.

The treatment now follows closely that in the large $N$ approximation
(cfr. section \ref{ln} and ref. \cite{Destri:1999hd}), the only
difference being in the value of the coupling, now three times larger.
In fact, in case of $O(N)$ symmetry, the quantum fluctuations over a
given background $\vev{\bds\phi(x)}=\bar{\bds\phi}$ decompose for each
$k$ into one longitudinal mode, parallel to $\bar{\bds\phi}$, and
$N-1$ transverse modes orthogonal to it; by boson combinatorics the
longitudinal mode couples to $\bar{\bds\phi}$ with strength
$3\lbare/N$ and decouples in the $N\to\infty$ limit, while the
transverse modes couple to $\bar{\bds\phi}$ with strength
$(N-1)\lbare/N\to \lbare$; when $N=1$ only the longitudinal mode is
there.

As $L\to\infty$, $\om^2_k \to k^2 +M^2$ and $M$ is exactly the
physical mass gap. Hence it must be $\Lambda-$independent. At finite
$L$ I cannot use this request to determine $\mbare$ and $\lbare$,
since, unlike $M$, they cannot depend on the size $L$. At infinite volume
I obtain
\begin{equation}\label{M}
	M^2 =\mbare+3\lbare[\bar\phi^2+I_D(M^2,\Lambda)]
\end{equation}
[with the function $I_D(z,\L)$ defined in eq. (\ref{Nm2ren})]. When
$\bar\phi=0$ this equation fixes the bare mass to be
\begin{equation}\label{m2ren}
	\mbare= m^2 -3\lbare I_D(m^2,\Lambda)
\end{equation}
where $m=M(\bar\phi=0)$ may be identified with the equilibrium
physical mass of the scalar particles of the infinite--volume Fock
space without symmetry breaking (see below). Now, the coupling
constant renormalization follows from the equalities
\begin{eqnarray}\label{rengap}
	M^2&=&m^2+3\lbare[\bar\phi^2+I_D(M^2,\Lambda)-I_D(m^2,\Lambda)]
	\nonumber \\
	&=&m^2+3\l\,\bar\phi^2+3\l \left[ I_D(M^2,\Lambda)-
	I_D(m^2,\Lambda) \right]_{\rm finite} 
\end{eqnarray}
and reads when $D=3$
\begin{equation}\label{lren}
	\dfrac\l\lbare = 1-\dfrac{3\l}{8\pi^2} \log\dfrac{2\Lambda}{m\sqrt{e}} 
\end{equation}
that is the standard result of the one--loop renormalization group
\cite{zj}.  When $D=1$, that is a $(1+1)-$dimensional quantum field
theory, $I_D(M^2,\Lambda)-I_D(m^2,\Lambda)$ is already finite and the
dimensionfull coupling constant is not renormalized, $\lbare=\l$.

The Landau pole in $\lbare$ prevents the actual UV limit
$\Lambda\to\infty$. Nonetheless, neglecting all inverse powers of the
UV cutoff when $D=3$, it is possible to rewrite the gap equation
(\ref{rengap}) as
\begin{equation}\label{nice2}
	\dfrac{M^2}{\hat\l(M)} = \dfrac{m^2}{\hat\l(m)} + 3\,\bar\phi^2
\end{equation}
in terms of the one--loop running coupling constant
\begin{equation}
	\hat\l(\mu) = \l \left[ 1 - \dfrac{3\l}{8\pi^2} 
	\log\dfrac{\mu}m \right]^{-1}
\end{equation}
It is quite clear that the HF states for which the renormalization
just defined is sufficient are all those that are gaussian--dominated
in the ultraviolet, so that I have [compare to eq. (\ref{gauss})]
\begin{equation}\label{largek}
	\vev{\varphi_{\pm k}^2} \sim \dfrac1{2\om_k} 
	\;,\quad k^2 \sim \Lambda^2\;,\; \Lambda \to \infty
\end{equation}
If this property holds at a certain time, then it should hold at all
times, since the Schroedinger equations (\ref{Schroedinger}) are
indeed dominated by the quadratic term for large $\om_k$ and
$\om^2_k\sim k^2+{\rm const}+O(k^{-1})$ as evident from
eq. (\ref{renomvev}). Thus this class of states is indeed closed under
time evolution and the parameterizations (\ref{m2ren}) and (\ref{lren})
make the tdHF approximation ultraviolet finite. Notice that the
requirement (\ref{largek}) effectively always imposes a gap equation
similar to eq. (\ref{gap}) in the deep ultraviolet.

Another simple check of the self--consistency of our approach,
including the change in selected places from $\lbare$ to $\l$, as
discussed above, follows from the energy calculation for the gaussian
states with $\vev{\phi(x)}=\bar\phi$ introduced above. Using
eq. (\ref{energy}) and the standard replacement of sums by integrals
in the infinite volume limit, I find
\begin{equation}
	\E(\bar\phi) = \lim_{L\to\infty} \dfrac{E}{L ^D} = 
	\dfrac12\bar\phi^2(M^2-\l\bar\phi^2)+\dfrac12\int_{k^2\le\Lambda^2} 
	\dfrac{d^Dk}{(2\pi)^D} \,\sqrt{k^2+M^2} -\dfrac34 \lbare 
	\left[\bar\phi^2+I_D(M^2,\Lambda)\right]^2
\end{equation}
where $M=M(\bar\phi)$ depends on $\bar\phi$ through the gap equation
(\ref{rengap}). The explicit calculation of the integrals involved
shows that the energy density difference $\E(\bar\phi)-\E(0)$ [which
for unbroken symmetry is nothing but the effective potential
$V_{\rm eff}(\bar\phi)$], is indeed finite in the limit
$\Lambda\to\infty$, as required by a correct renormalization
scheme. Notice that the finiteness of the energy density difference
can be shown also by a simpler and more elegant argument, as
presented below in section \ref{ooed}. This check would fail instead when
$D=3$ if only the bare coupling constant $\lbare$ would appear in the
last formula.

The tdHF approximation derived above represents a huge simplification
with respect to the original problem, but its exact solution still
poses itself as a considerable challenge. As a matter of fact, a
numerical approach is perfectly possible within the capabilities of
modern computers, provided the number of equations
(\ref{Schroedinger}) is kept in the range of few thousands. In this
respect, an interesting comparison can be made: we have, on one hand,
the evolution of relevant observables, as the field condensate and the
quantum widths, in the gaussian approximation, where the quantum
system can be reduced to a classical one; on the other hand,
eqs. (\ref{Schroedinger}) are quantum Schroedinger equations for
general wavefunctions, and once we know their history, the evolution
of the expectation value of any given operator can be computed and
compared with the corresponding one in the gaussian
approximation. This is relevant also for clarifying further an
inconsistency of the tdHF in the gaussian approximation, that we find
by analyzing the evolution in finite volume and we discuss in section
\ref{ooed}. Of course, the numerical algorithm presented in section
\ref{num} of the appendix is not appropriate for this purpose and a
method for evolving the wavefunction numerically must be implemented.

\subsection{On symmetry breaking}

Quite obviously, in a finite volume and with a UV cutoff there cannot
be any symmetry breaking, since the ground state is necessarily unique
and symmetric when the number of degrees of freedom is finite
\cite{gj}. However, I may handily envisage the situation which would
imply symmetry breaking when the volume diverges.

Let us first consider the case that we would call of unbroken
symmetry. In this case the HF ground state is very close to the member
with $\bar\phi=0$ of the family of gaussian states introduced
before. The difference is entirely due to the quartic terms in $H_k$.
This correction vanishes when $L\to\infty$, since all wavefunctions
$\psi_k$ have $L-$independent widths, so that one directly obtains the
symmetric vacuum state with all the right properties of the vacuum
(translation invariance, uniqueness, etc.)  upon which a standard scalar
massive particle Fock space can be based. The HF approximation then
turns out to be equivalent to the resummation of all ``cactus
diagrams'' for the particle self--energy \cite{cactus}. In a finite
volume, the crucial property of this symmetric vacuum is that all
frequencies $\om_k^2$ are strictly positive. The generalization to
non--equilibrium initial states with $\bar\phi\neq0$ is rather
trivial: it amounts to a shift by $L^{D/2}\bar\phi$ on
$\psi_0(\varphi_0)$. In the limit $L\to\infty$ we should express
$\psi_0$ as a function of $\xi=L^{-D/2}\varphi_0$ so that,
$|\psi_0(\xi)|^2\to \delta(\xi-\bar\phi)$, while all other
wavefunctions $\psi_k$ will reconstruct the gaussian wavefunctional
corresponding to the vacuum $\ket{0,M}$ of a free massive scalar
theory whose mass $M=M(\phi)$ solves the gap equation
(\ref{rengap}).  The absence of $\psi_0$ in $\ket{0,M}$ is
irrelevant in the infinite volume limit, since
$\vev{\varphi_0^2}=L^D\bar\phi^2 +$ terms of order $L^0$. The effective
potential $V_{\rm eff}(\bar\phi)= \E(\bar\phi)-\E(0)$, where
$\E(\bar\phi)$ is the lowest energy density at fixed $\bar\phi$ and
infinite volume, is manifestly a convex function with a unique minimum
in $\bar\phi=0$.

Now let us consider a different situation in which one or more of the
$\om_k^2$ are negative. Quite evidently, this might happen only for
$k$ small enough, due to the $k^2$ in the gap equation [thus
eq. (\ref{largek}) remains valid and the ultraviolet renormalization
is the same as for unbroken symmetry]. Actually I assume here that
only $\om_0^2<0$, postponing the general analysis.  Now the quartic
term in $H_0$ cannot be neglected as $L\to\infty$, since in the ground
state $\psi_0$ is symmetrically concentrated around the two minima of
the potential $\frac12 \om_0^2 \varphi_0^2 +\frac\l{4L^D}\varphi_0^4$,
that is $\varphi_0=\pm(-\om_0^2L^D/\l)^{1/2}$.  If I scale
$\varphi_0$ as $\varphi_0=L^{D/2}\xi$ then $H_0$ becomes
\begin{equation}\label{Hscaled}
  H_0 = -\dfrac1{2L^D}\pdif{^2}{\xi^2}+\dfrac{L^D}2 \left(
	\om_0^2\, \xi^2 +\dfrac\l2\xi^4 \right)
\end{equation}
so that the larger $L$ grows the narrower $\psi_0(\xi)$ becomes around
the two minima $\xi=\pm(-\om_0^2/\l)^{1/2}$. In particular
$\vev{\xi^2}\to -\om_0^2/\l$ when $L\to\infty$ and
$\vev{\varphi_0^2}\simeq L^D\vev{\xi^2}$. Moreover, the energy gap
between the ground state of $H_0$ and its first, odd excited state as
well as difference between the relative probability distributions for $\xi$
vanish exponentially fast in the volume $L^D$. 
  
Since by hypothesis all $\om_k^2$ with $k\neq0$ are strictly positive,
the ground state $\psi_k$ with $k\neq0$ are asymptotically gaussian when
$L\to\infty$ and the relations (\ref{omvev2}) tend to the form 
\begin{eqnarray}
	\om_k^2 &=& k^2+ M^2  \equiv  k^2 +m^2 \\
	M^2 &=& -2\om_0^2 = \mbare + 3\lbare (L^{-D}\vev{\varphi_0^2}
	+ \Sigma_0) = \mbare + 3\lbare\om_0^2+ 3\lbare I_D(m^2,\Lambda)] 
\end{eqnarray}  
This implies the identification $\om_0^2=-m^2/2$ and the bare mass
parameterization
\begin{equation}\label{brokenm}
	\mbare = \left(1-\dfrac32 \dfrac\lbare\l\right)m^2 -3\lbare
	I_D(m^2,\Lambda)
\end{equation}
characteristic of a negative $\om_0^2$ [compare to eq. (\ref{m2ren})],
with $m$ the physical equilibrium mass of the scalar particle, as in
the unbroken symmetry case. The coupling constant renormalization is
the same as in eq. (\ref{lren}) as may be verified by generalizing to
the minimum energy states with given field expectation value
$\bar\phi$; this minimum energy is nothing but the HF effective potential
$V_{\rm eff}^{\rm HF}(\bar\phi)$, that is the effective potential
in this non--gaussian HF approximation; of course, since $\psi_0$ is no longer
asymptotically gaussian, I cannot simply shift it by $L^{D/2}\bar\phi$
but, due to the concentration of $\psi_0$ on classical minima as
$L\to\infty$, one readily finds that $V_{\rm eff}(\bar\phi)$ is the
convex envelope of the classical potential, that is its Maxwell
construction. Hence I find
\begin{equation}
	\vev{\varphi_0^2} \underset{L\to\infty}\sim
        \begin{cases}
        -L^D\om_0^2/\l \;,\; & \l\bar\phi^2\le-\om_0^2 \\
        L^D\bar\phi^2  \;,\; & \l\bar\phi^2>-\om_0^2
        \end{cases}
\end{equation}
and the gap equation for the $\bar\phi-$dependent mass $M$ can be
written, in terms of the step function $\Theta$ and the extremal
ground state field expectation value $v=m/\sqrt{2\l}$,
\begin{equation}\label{gapbroken}
	M^2 = m^2 + 3\lbare(\bar\phi^2-v^2) \,
	\Theta(\bar\phi^2-v^2) +3\lbare \left[ 
	I_D(M^2,\Lambda)-I_D(m^2,\Lambda) \right] 
\end{equation}
I see that the specific bare mass parameterization (\ref{brokenm})
guarantees the non--renormalization of the tree--level relation
$v^2=m^2/2\l$ ensuing from the typical symmetry breaking classical
potential $V(\phi)=\frac14\l(\phi^2-v^2)^2$. With the same finite part
prescription as in eq. (\ref{rengap}), the gap equation
(\ref{gapbroken}) leads to the standard coupling constant
renormalization (\ref{lren}) when $D=3$.

In terms of the probability distributions $|\psi_0(\xi)|^2$ for the
scaled amplitude $\xi=L^{-D/2}\varphi_0$, the Maxwell construction
corresponds to the limiting form
\begin{equation}\label{limitform}
        |\psi_0(\xi)|^2 \underset{L\to\infty}\sim
        \begin{cases}
        \tfrac12(1+\bar\phi/v)\delta(\xi-v)+ \tfrac12(1-\bar\phi/v)\,
        \delta(\xi+v) \;,\; &\bar\phi^2 \le v^2 \\
        \delta(\xi-\bar\phi) \;,\;  &\bar\phi^2 > v^2 
        \end{cases}
\end{equation}
On the other hand, if $\om^2_0$ is indeed the
only negative squared frequency, the $k\ne0$ part of this minimum
energy state with arbitrary $\bar\phi=\vev{\phi(x)}$ is better and
better approximated as $\L\to\infty$ by the same gaussian state
$\ket{0,M}$ of the unbroken symmetry state. Only the effective mass
$M$ has a different dependence $M(\bar\phi)$, as given by the
gap equation (\ref{gapbroken}) proper of broken symmetry.

At infinite volume I may write
\begin{equation}
	\vev{\varphi_k^2} = C(\bar\phi)\,\delta^{(D)}(k)+
	\dfrac1{2\sqrt{k^2+M^2}}
\end{equation}
where $C(\bar\phi)=\bar\phi^2$ in case of unbroken symmetry (that is
$\om^2_0>0$), while $C(\bar\phi)={\rm max}(v^2,\bar\phi^2)$
when $\om^2_0<0$.  This corresponds to the field correlation in space
\begin{equation}
	\vev{\phi(x)\phi(y)}=\int\dfrac{d^Dk}{(2\pi)^D}
	\vev{\varphi_k^2} e^{ik\cdot(x-y)}= C(\bar\phi) +\Delta_D(x-y,M)
\end{equation}
where $\Delta_D(x-y,M)$ is the massive free field equal--time two
points function in $D$ space dimensions, with self--consistent mass
$M$. The requirement of clustering
\begin{equation}
	\vev{\phi(x)\phi(y)} \to \vev{\phi(x)}^2 = v^2 
\end{equation}
contradicts the infinite volume limit of 
\begin{equation}
	\vev{\phi(x)}=L^{-D/2}\sum_k \vev{\phi_k}\,e^{ik\cdot x} =
	\vev{\varphi_0} = \bar\phi
\end{equation}
except at the two extremal points $\bar\phi=\pm v$. In fact we know
that the $L\to\infty$ limit of the finite volume states with
$\bar\phi^2<v^2$ violate clustering, because the two peaks of
$\psi_0(\xi)$ have vanishing overlap in the limit and the first
excited state becomes degenerate with the vacuum: this implies that
the relative Hilbert space splits into two orthogonal Fock sectors
each exhibiting symmetry breaking, $\vev{\phi(x)}=\pm v$, and
corresponding to the two independent equal weight linear combinations
of the two degenerate vacuum states. The true vacuum is either one of
these symmetry broken states. Since the two Fock sectors are not only
orthogonal, but also superselected (no local observable interpolates
between them), linear combinations of any pair of vectors from the two
sectors are not distinguishable from mixtures of states and clustering
cannot hold in non--pure phases. It is perhaps worth noticing also
that the Maxwell construction for the effective potential, in the
infinite volume limit, is just a straightforward manifestation of this
fact and holds true, as such, beyond the HF approximation.

To further clarify this point and in view of subsequent
applications, let us consider the probability distribution for the
smeared field $\phi_f=\int d^Dx\,\phi(x)f(x)$, where
\begin{equation}
	f(x)=f(-x)=\dfrac1{L^D}\sum_k f_k \,e^{ik\cdot x} 
	\underset{L\to\infty}\sim \,\int \dfrac{d^Dk}{(2\pi)^D}
	\, \tilde f(k) \,e^{ik\cdot x} 
\end{equation}
is a smooth real function with $\int d^Dx\,f(x)=1$ ({\em
i.e.} $f_0=1$) localized around the origin (which is good as any other
point owing to translation invariance). Neglecting in the infinite
volume limit the quartic corrections for all modes with $k\neq 0$, so
that the corresponding  ground state wavefunctions are
asymptotically gaussian, this probability distribution evaluates to
\begin{equation}
	{\rm Pr}(u\!<\!\phi_f\!<\!u+du)= \dfrac{du}{(2\pi\Sigma_f)^{1/2}} 
	\intf d\xi\, |\psi_0(\xi)|^2
	\exp\left\{\dfrac{-(u-\xi)^2}{2\Sigma_f} \right\}
\end{equation}
where
\begin{equation}
	\Sigma_f = \sum_{k\neq0}\vev{\varphi_k^2}\,f_k^2 
	\;\underset{L\to\infty}\sim \;\int \dfrac{d^Dk}{(2\pi)^D} 
	\,\dfrac{\tilde f(k)^2}{2\sqrt{k^2+m^2}}
\end{equation}
In the unbroken symmetry case we have
$|\psi_0(\xi)|^2\sim\delta(\xi-\bar\phi)$ as $L\to\infty$, while
the limiting form (\ref{limitform}) holds for broken symmetry.
Thus I obtain
\begin{equation}
	{\rm Pr}(u\!<\!\phi_f\!<\!u+du) =p_f(u-\bar\phi)\,du \;,\quad 
	p_f(u) \equiv \left(2\pi\Sigma_f\right)^{-1/2}
	\exp\left(\dfrac{-u^2}{2\Sigma_f} \right)
\end{equation}
for unbroken symmetry and 
\begin{equation}
	{\rm Pr}(u\!<\!\phi_f\!<\!u+du) = \left\{
	\begin{array}{ll}
	\frac12(1+\bar\phi/v)\,p_f(u-v)\,du+\frac12(1-\bar\phi/v)
	\,p_f(u+v)\,du \;, &\bar\phi^2 \le v^2 \\ \\
	p_f(u-\bar\phi)\,du\;, &\bar\phi^2 > v^2 
	\end{array}
	\right.
\end{equation}
for broken symmetry. Notice that the momentum integration in the
expression for $\Sigma_f$ needs no longer an ultraviolet cutoff; of
course in the limit of delta--like test function $f(x)$, $\Sigma_f$
diverges and $p_f(u)$ flattens down to zero. The important observation
is that ${\rm Pr}(u\!<\!\phi_f\!<\!u+du)$ has always a single peak
centered in $u=\bar\phi$ for unbroken symmetry, while for broken symmetry it
shows two peaks for $\bar\phi^2 \le v^2$ and  $\Sigma_f$ small enough.
For instance, if $\bar\phi=0$, then there are two peaks for 
$\Sigma_f<v^2$ [implying that $\tilde f(k)$ has a significant
support only up to wavevector $k$ of order $v$, when $D=3$, or
$m\exp({\rm const}v^2)$ when $D=1$].

To end the discussion on symmetry breaking, I may now verify the
validity of the assumption that only $\om_0^2$ is negative. In fact,
to any squared frequency $\om_k^2$ (with $k\neq0$) that stays strictly
negative as $L\to\infty$ there corresponds a wavefunction $\psi_k$
that concentrates on $\varphi_k^2+\varphi_{-k}^2=-\om_k^2L^D/\l$ ;
then eqs. (\ref{omvev2}) implies $-2\om_k^2 = k^2 + m^2$ for such
frequencies, while $\om_k^2 = k^2 +m^2$ for all frequencies with
positive squares; if there is a macroscopic number of negative
$\om_k^2$ (that is a number of order $L^D$), then the expression for
$\om_0^2$ in eq. (\ref{omvev2}) will contain a positive term of order
$L^D$ in the r.h.s., clearly incompatible with the requirements that
$\om_0^2<0$ and $\mbare$ be independent of $L$; if the number of
negative $\om_k^2$ is not macroscopic, then the largest wavevector
with a negative squared frequency tends to zero as $L\to\infty$ (the
negative $\om_k^2$ clearly pile in the infrared) and the situation is
equivalent, if not identical, to that discussed above with only
$\om_0^2<0$.

\subsection{Out--of--equilibrium dynamics}\label{ooed}

I considered above the lowest energy states with a predefinite
uniform field expectation value, $\vev{\phi(x)}=\bar\phi$, and
established how they drastically simplify in the infinite volume
limit.  For generic $\bar\phi$ these states are not stationary and
will evolve in time. By hypothesis $\psi_k$ is the ground state
eigenfunction of $H_k$ when $k>0$, and therefore $|\psi_k|^2$ would be
stationary for constant $\om_k$, but $\psi_0$ is not an eigenfunction
of $H_0$ unless $\bar\phi=0$. As soon as $|\psi_k|^2$ starts changing,
$\vev{\varphi_0^2}$ changes and so do all frequencies $\om_k$ which
are coupled to it by the eqs. (\ref{omvev2}). Thus the change
propagates to all wavefunctions.  The difficult task of studying this
dynamics can be simplified with the following scheme, that we might
call {\em gaussian approximation}. I first describe it and discuss
its validity later on.

Let us assume the usual gaussian form for the initial state [see
eq. (\ref{gauss}) and the discussion following it]. We know that it is
a good approximation to the lowest energy state with given
$\vev{\varphi_0}$ for unbroken symmetry, while it fails to be so for
broken symmetry, only as far as $\psi_0$ is concerned, unless
$\bar\phi^2 \geq v^2$. At any rate this is an acceptable initial state: the
question is about its time evolution. Suppose I adopt the harmonic
approximation for all $H_k$ with $k>0$ by dropping the quartic term.
This approximation will turn out to be valid only if the
width of $\psi_k$ do not grow up to the order $L^D$ (by symmetry the
center will stay in the origin). In practice I am now dealing with a
collection of harmonic oscillators with time--dependent frequencies and
the treatment is quite elementary: consider the simplest example of one quantum
degree of freedom described by the gaussian wavefunction
\begin{equation}
	\psi(q,t)=\dfrac{e^{-i\a}}{(2\pi\s^2)^{1/4}}
	\exp\left[-\dfrac12\left(\dfrac1{2\s^2}-i\dfrac{s}\s\right)q^2\right]
\end{equation}
where $s$ and $\s$ and the overall phase $\a$ are time--dependent. If
the dynamics is determined by the time--dependent harmonic hamiltonian
$\frac12[-\partial^2_q+\om(t)^2\,q^2]$, then the Schroedinger equation
is solved exactly provided that $s$ and $\s$ satisfy the classical
Hamilton equations
\begin{equation}
	\dot\s = s \;,\quad \dot s = - \om^2\s + \dfrac1{4 \s^3}
\end{equation}
It is not difficult to trace the ``centrifugal'' force $(4\s)^{-3}$
which prevents the vanishing of $\s$ to Heisenberg uncertainty
principle \cite{Habib:1996ee,Cooper:1997ii}.

The extension to our case with many degrees of freedom is
straightforward and I find the following system of equations
\begin{equation}\label{sk}
	i\pdif{}t \psi_0 = H_0\psi_0 \;,\quad
	\der{^2 \s_k}{t^2}= - \om_k^2\,\s_k + \dfrac1{4 \s_k^3} \;,\; k>0
\end{equation}
coupled in a mean--field way by the relations (\ref{omvev2}), which
now read
\begin{eqnarray}\label{momvev2}
	\om_k^2 &=& k^2+ M^2 - 6\l\, L^{-D}\s_k^2 \;,\quad k>0 \\ 
	M^2 &=& \mbare +3\lbare \left(L^{-D}\vev{\varphi_0^2} + \Sigma_0\right)
	\;,\quad \Sigma_0= \dfrac1{L^D} \sum_{k\ne 0}\s_k^2 
\end{eqnarray}
This stage of a truly quantum zero--mode and classical modes with
$k>0$ does not appear fully consistent, since for large volumes some type of
classical or gaussian approximation should be considered for
$\varphi_0$ too. I may proceed in two (soon to be proven equivalent)
ways:
\begin{enumerate}
\item 
I shift $\varphi_0=L^{D/2}\bar\phi+\eta_0$ and then
deal with the quantum mode $\eta_0$ in the gaussian approximation,
taking into account that I must have $\vev{\eta_0}=0$ at all times.
This is most easily accomplished in the Heisenberg picture rather than
in the Schroedinger one adopted above. In any case I find that
the quantum dynamics of $\varphi_0$ is equivalent to the classical
dynamics of $\bar\phi$ and $\s_0\equiv\vev{\eta_0^2}^{1/2}$ described
by the ordinary differential equations
\begin{equation}\label{classical}
	\der{^2 \bar\phi}{t^2} = -\om_0^2\, \bar\phi -\l\, \bar\phi^3
	\;,\quad \der{^2 \s_0}{t^2}= - \om_0^2\,\s_0 + \dfrac1{4 \s_0^3}
\end{equation}
where $\om_0^2=M^2-3\l\,L^{-D}\vev{\varphi_0^2}$ and 
$\vev{\varphi_0^2} = L^D\bar\phi^2+ \s_0^2$.

\item 
I rescale $\varphi_0=L^{D/2}\xi$ right away, so that $H_0$ takes the
form of eq. (\ref{Hscaled}). Then $L\to\infty$ is the classical limit
such that $\psi_0(\xi)$ concentrates on $\xi=\bar\phi$ which evolves
according to the first of the classical equations in
(\ref{classical}). Since now there is no width associated to the
zero--mode, $\bar\phi$ is coupled only to the widths $\s_k$ with $k\neq 0$
by $\om_0^2=M^2-3\l\bar\phi^2$, while $M^2=\mbare
+3\lbare(\bar\phi^2+\Sigma_0)$.
\end{enumerate}

It is quite evident that these two approaches are completely
equivalent in the infinite volume limit, and both are good
approximation to the original tdHF Schroedinger equations, at least
provided that $\s_0^2$ stays  such that
$L^{-D}\s_0^2$ vanishes in the limit for any time. In this case we
have the evolution equations
\begin{equation}\label{Emotion}
	\der{^2 \bar\phi}{t^2} = (2\l\,\bar\phi^2 -M^2)\,\bar\phi \;,\quad
	\der{^2\s_k}{t^2} = - (k^2+M^2)\,\s_k + \dfrac1{4 \s_k^3} 	
\end{equation}
mean--field coupled by the $L\to\infty$ limit of eqs. (\ref{momvev2}),
namely
\begin{equation}\label{unbtdgap}
	M^2=m^2+3\lbare\left[\bar\phi^2 + \Sigma -I_D(m^2,\Lambda)\right]
\end{equation}
for unbroken symmetry [that is $\mbare$ as in eq. (\ref{m2ren})] or
\begin{equation}\label{btdgap}
	M^2=m^2+3\lbare\left[\bar\phi^2 -v^2 + 
	\Sigma -I_D(m^2,\Lambda)\right] \;,\quad m^2=2\l v^2
\end{equation}
for broken symmetry [that is $\mbare$ as in
eq. (\ref{brokenm})]. In any case I define
\begin{equation}
	 \Sigma =\dfrac1{L^D} \sum_k\s_k^2 \;\underset{L\to\infty}\sim\;
	\int_{k^2\le\Lambda^2} \dfrac{d^Dk}{(2\pi)^D}\,\s_k^2
\end{equation}
as the sum, or integral, over all microscopic gaussian widths
[N.B.:this definition differs from that given before in
eq. (\ref{Sigma}) by the classical term $\bar\phi^2$]. Remarkably, the
equations of motion (\ref{Emotion}) are completely independent of the
ultraviolet cut--off and this is a direct consequence of the
substitution (\ref{H_k'}). Had I kept the bare coupling constant
everywhere in the expression (\ref{subst}), I would now have $\lbare$
also in front of the $\bar\phi^3$ in the r.h.s. of the first of the two
equations (\ref{Emotion}) [cfr., for instance, ref. \cite{Boyanovsky:1995me}].

The conserved HF energy (density) corresponding to these equations of
motion reads
\begin{eqnarray}\label{EHF}
	\E &=& \T +\V \;,\quad \T = \dfrac12(\dot{\bar{\phi}})^2 +
	\dfrac1{2L^D}\sum_k \dot\s_k^2  \\ \V &=& \dfrac1{2L^D}\sum_k\left(
	 k^2\,\s_k^2 + \dfrac1{4\s^2_k}\right) +\dfrac12\mbare(\bar\phi^2+
	\Sigma) + \dfrac34\lbare (\bar\phi^2+\Sigma)^2 -\dfrac12\l\bar\phi^4
\end{eqnarray}
Up to additive constants and terms vanishing in the infinite volume
limit, this expression agrees with the general HF energy of
eq. (\ref{menergy}) for gaussian wavefunctions.  It holds both for
unbroken and broken symmetry, the only difference being in the
parameterization of the bare mass in terms of UV cutoff and physical
mass, eqs. (\ref{m2ren}) and (\ref{brokenm}). The similarity to the
energy functional of the large $N$ approach is evident; the only
difference, apart from the obvious fact that $\bar{\phi}$ is a single
scalar rather than a $O(n)$ vector, is in the mean--field coupling
$\s_k$--$\bar\phi$ and $\s_k$--$\Sigma$, due to different coupling
strength of transverse and longitudinal modes
(cfr. ref. \cite{Destri:1999hd}).

This difference between the HF approach for discrete symmetry ({\em
i.e} $N=1$) and the large $N$ method for the continuous
$O(N)$-symmetry is not very relevant if the symmetry is unbroken [it
does imply however a significantly slower dissipation to the modes of
the background energy density].  On the other hand it has a drastic
consequence on the equilibrium properties and on the
out--of--equilibrium dynamics in case of broken symmetry (see below),
since massless Goldstone bosons appear in the large $N$ approach,
while the HF treatment of the discrete symmetry case must exhibits a
mass also in the broken symmetry phase.

The analysis of physically viable initial conditions proceeds exactly
as in the large $N$ approach \cite{Destri:1999hd} and will not be repeated
here, except for an important observation in case of broken
symmetry. The formal energy minimization w.r.t. $\s_k$ at fixed
$\bar\phi$ leads again to eqs. 
\begin{equation}\label{inis1}
	\dot \s_k=0 \;,\quad \s_k^2=\dfrac1{2\sqrt{k^2+M^2}}
\end{equation}
and again these are acceptable initial conditions only if the gap
equation that follows from eq. (\ref{btdgap}) in the $L\to\infty$
limit, namely
\begin{equation}\label{bgap}
	M^2 = m^2 + 3\lbare \left[\bar\phi^2-v^2 +
	I_D(M^2,\Lambda)-I_D(m^2,\Lambda) \right] 
\end{equation}
admits a nonnegative, physical solution for $M^2$. Notice that there
is no step function in eq. (\ref{bgap}), unlike the static case of
eq. (\ref{gapbroken}), because $\s_0^2$ was assumed to be microscopic,
so that the infinite volume $\s_k^2$ has no delta--like singularity in
$k=0$. Hence $M=m$ solves eq. (\ref{bgap}) only at the extremal points
$\bar\phi=\pm v$, while it was the solution of the static gap equation
(\ref{gapbroken}) throughout the Maxwell region $-v\le\bar\phi\le v$.
The important observation is that eq. (\ref{bgap}) admits a positive
solution for $M^2$ also within the Maxwell region. In fact it can be
written, neglecting as usual the inverse--power corrections in the UV
cutoff
\begin{equation}\label{bnice}
	\dfrac{M^2}{\hat\l(M)} = \dfrac{m^2}\l + 3\,(\bar\phi^2 -v^2) =
	3\,\bar\phi^2 -v^2
\end{equation}
and there exists indeed a positive solution $M^2$ smoothly connected
to the ground state, $\bar\phi^2=v^2$ and $M^2=m^2$, whenever
$\bar\phi^2\ge v^2/3$. The two intervals $v^2\ge\bar\phi^2\ge v^2/3$
correspond indeed to the metastability regions, while $\bar\phi^2<
v^2/3$ is the spinodal region, associated to a classical potential
proportional to $(\bar\phi^2 -v^2)^2$. This is another effect of the
different coupling of transverse and longitudinal modes: in the large
$N$ approach there are no metastability regions and the spinodal
region coincides with the Maxwell one. As in the large $N$ approach in
the spinodal interval there is no energy minimization possible, at
fixed background and for microscopic widths, so that a modified form
of the gap equation
\begin{equation}\label{newgap1}
	M^2 = m^2 + 3\lbare \left[\bar{\phi}^2-v^2 + \dfrac1{L^D} 
	\sum_{k^2<|M^2|}\s_k^2 + \dfrac1{L^D}\sum_{k^2>|M^2|}
	\dfrac1{2\sqrt{k^2-|M^2|}} -I_D(0,\Lambda) \right] 
\end{equation}
should be applied to determine ultraviolet--finite initial conditions.

The main question now is: how will the gaussian widths $\s_k$ grow
with time, and in particular how will $\s_0$ grow in case of method 1
above, when I start from initial conditions where all widths are
microscopic? For the gaussian approximation to remain valid through
time, all $\s_k$, and in particular $\s_0$, must at least not become
macroscopic.  In fact I have already positively answered this
question in the large $N$ approach \cite{Destri:1999hd} and the HF equations
(\ref{Emotion}) do not differ so much to expect the contrary now. In
particular, if I consider the special initial condition
$\bar\phi=\dot{\bar\phi}=0$, the dynamics of the widths is identical
to that in the large $N$ approach, apart from the rescaling by a
factor of three of the coupling constant.

\begin{figure} 
\includegraphics[height=8cm,width=15cm]{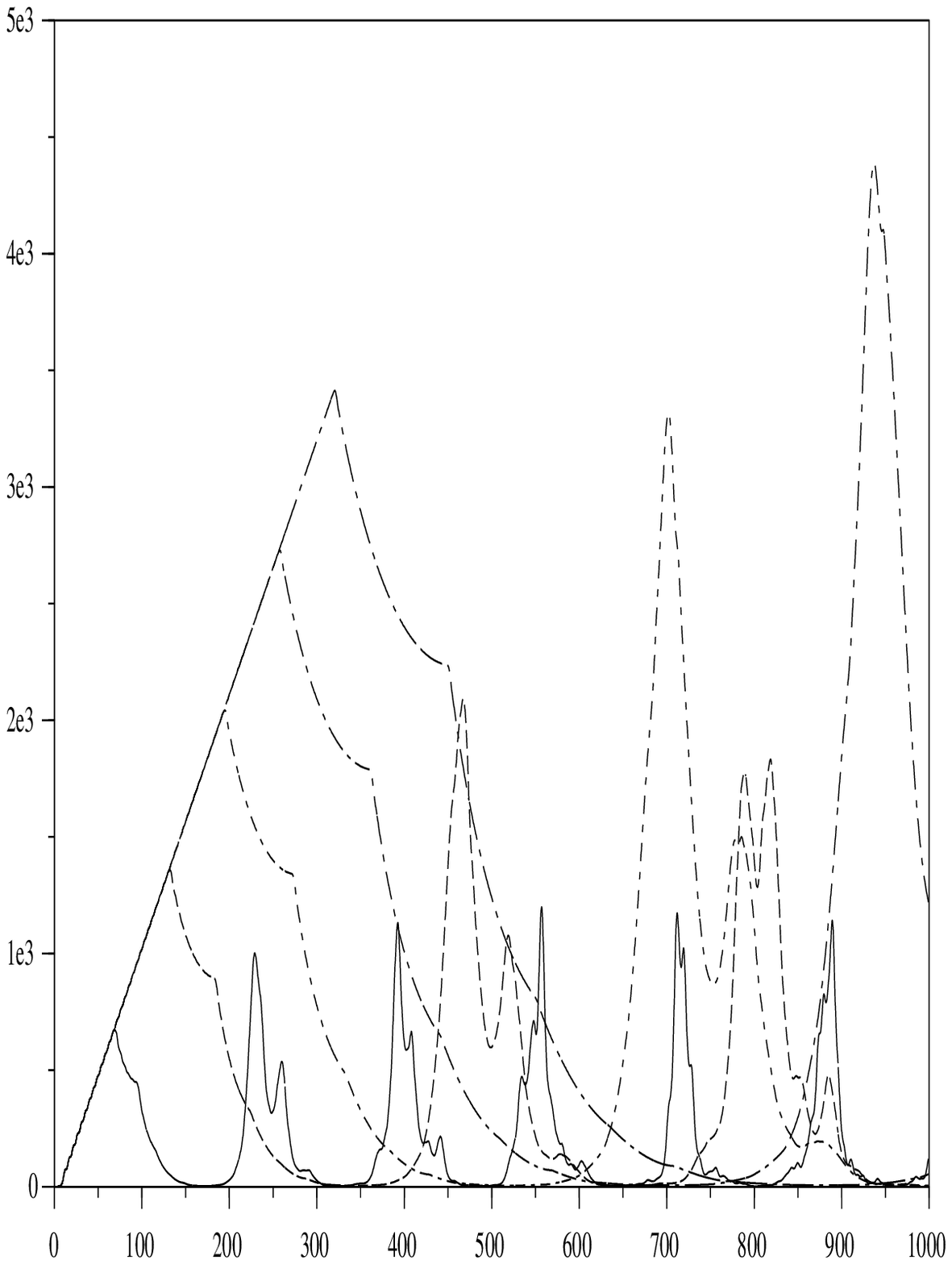}
\caption{\it Zero--mode amplitude evolution for 
different values of the size
$\frac{L}{2\pi}=20,40,60,80,100$, for $\lambda = 0.1$ and broken symmetry,
with $\bar\phi=0$. }\label{1fig:m0}
\end{figure}

In fact, if we look at the time evolution of the zero--mode amplitude
$\s_0$ [see Fig. \ref{1fig:m0}], we can see the presence of the
time--scale $\tau_L$ at which finite volume effects start to
manifest. The time scale $\tau_L$ turns out to be proportional to the
linear size of the box $L$ and its presence prevents $\s_0$ from
growing to macroscopic values. Thus our HF approximation confirms the
large $N$ approach in the following sense: even if one considers in
the variational ansatz the possibility of non--gaussian
wavefunctionals, the time evolution from gaussian and microscopic
initial conditions is effectively restricted for large volumes to
non--macroscopic gaussians. The strong similarity of
Fig. \ref{1fig:m0} with Fig. \ref{fig:m0} is due to the fact that when
$\bar\phi=\dot{\bar\phi}=0$, the evolution equations are the same both
for large $N$ and Hartree--Fock, apart from a rescaling of the
coupling constant by a factor of three (which accounts for the
different slope of the linear growth in the figures); their slight
differences in secondary peaks, instead, may be explained with the
difference in the initial conditions, which are fixed by two different
gap equations, (\ref{newgap}) for large $N$ and (\ref{newgap1}) for
Hartree-Fock.

\begin{figure} 
\includegraphics[height=8cm,width=15cm]{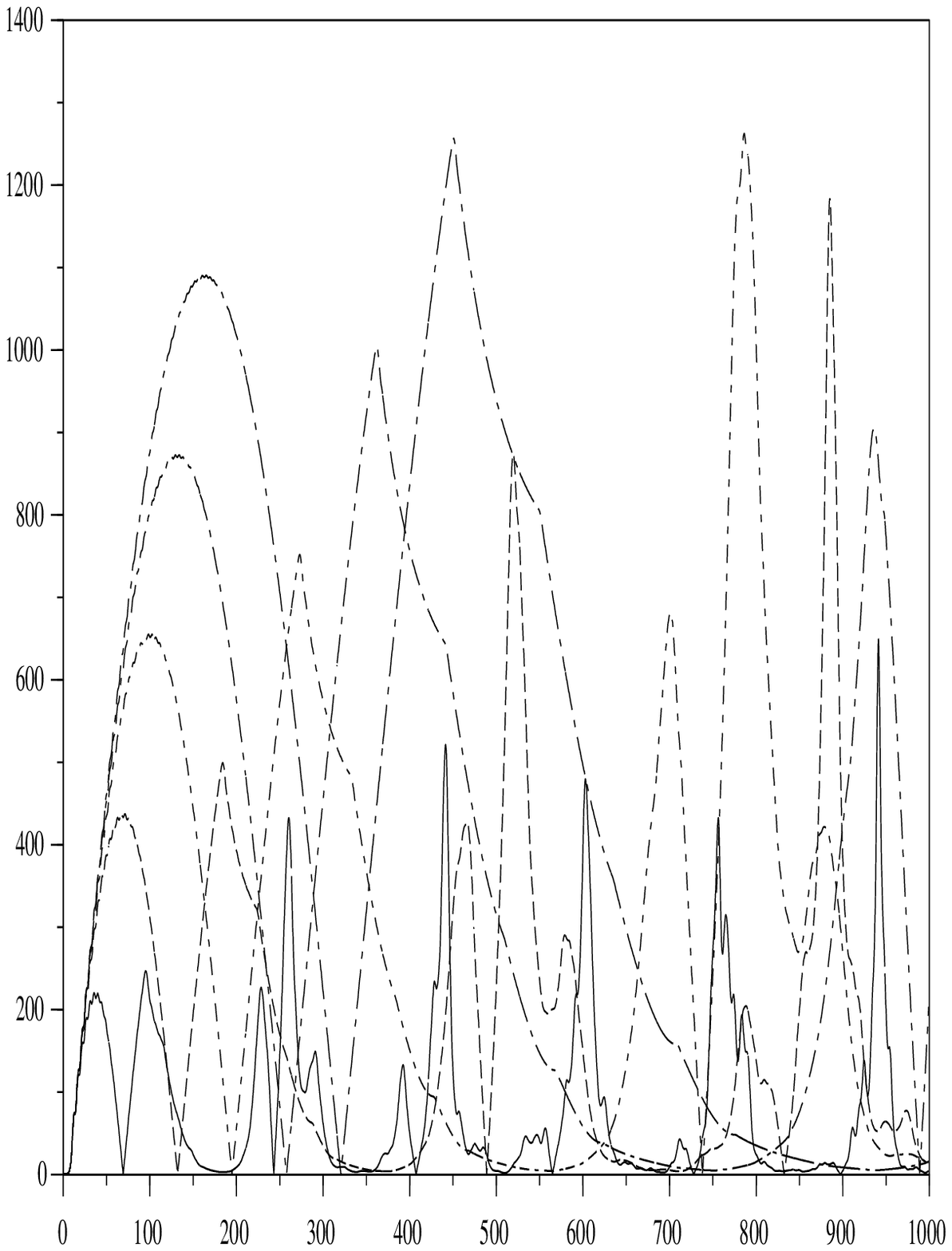}
\caption{\it Next--to--zero mode ($k=2\pi/L$) amplitude evolution for
different values of the size $L=20,40,60,80,100$, for $\lambda =
0.1$ and broken symmetry, with $\bar\phi=0$.}\label{1fig:m1}
\end{figure}

Strictly speaking, however, this might well not be enough, since the
infrared fluctuations do grow beyond the microscopic size to become of
order $L$ [see Fig. \ref{1fig:m1}, where the evolution of the mode with 
momentum $k=2\pi/L$ is plotted]. Then the quartic term in the low$-k$
Hamiltonians $H_k$ is of order $L$ and therefore it is not negligible
by itself in the $L\to\infty$ limit, but only when compared to the
quadratic term, which {\em for a fixed $\om_k^2$ of order $1$} would
be of order $L^2$. But we know that, when $\bar\phi=0$, after the
spinodal time and before the $\tau_L$, the effective squared mass
$M^2$ oscillates around zero with amplitude decreasing as $t^{-1}$ and
a frequency fixed by the largest spinodal wavevector. In practice it
is ``zero on average'' and this reflect itself in the average linear
growth of the zero--mode fluctuations and, more generally, in the
average harmonic motion of the other widths with non--zero
wavevectors. In particular the modes with small wavevectors of order
$L^{-1}$ feel an average harmonic potential with $\om_k^2$ of order
$L^{-2}$. This completely compensate the amplitude of the mode itself,
so that the quadratic term in the low$-k$ Hamiltonians $H_k$ is of
order $L^0$, much smaller than the quartic term that was neglected
beforehand in the gaussian approximation. Clearly the approximation
itself no longer appears fully justified and a more delicate analysis
is required. However, I here restrict myself to the gaussian
approximation.

\subsection{Late--time evolution and dynamical Maxwell construction}\label{ep}

By definition, the gaussian approximation of the effective potential
$V_{\rm eff}(\bar\phi)$ coincides with the infinite--volume limit
of the potential energy $\V(\bar\phi,\{\s_k\})$ of eq. (\ref{EHF})
when the widths are of the $\bar\phi-$dependent, energy--minimizing
form (\ref{inis1}) with the gap equation for $M^2$ admitting a
nonnegative solution. As we have seen, this holds true in the unbroken
symmetry case for any value of the background $\bar\phi$, so that the
gaussian $V_{\rm eff}$ is identical to the HF one, since all
wavefunctions $\psi_k$ are asymptotically gaussians as
$L\to\infty$. In the presence of symmetry breaking instead, this
agreement holds true only for $\bar\phi^2\ge v^2$; for
$v^2/3\le\bar\phi^2<v^2$ the gaussian $V_{\rm eff}$ exists but is
larger than the HF potential $V_{\rm eff}^{\rm HF}$, which is
already flat. In fact, for any $\bar\phi^2\ge v^2/3$, I may write the
gaussian $V_{\rm eff}$ as
\begin{equation}
	V_{\rm eff}(\bar\phi) = V_{\rm eff}(-\bar\phi) =
	V_{\rm eff}(v) + \int_v^{|\bar\phi|}\,du\,u[M(u)^2-2\l\,u^2] 
\end{equation}
where $M(u)^2$ solves the gap equation (\ref{bnice}), namely
$M(u)^2=\hat\l(M(u))(3u^2-v^2)$. In each of the two disjoint regions
of definition this potential is smooth and convex, with unique minima
in $+v$ and $-v$, respectively. These appear therefore as regions of
metastability (states which are only locally stable in the presence of
a suitable uniform external source). The HF effective potential is
identical for $\bar\phi^2\ge v^2$, while it takes the constant value
$V_{\rm eff}(v)$ throughout the internal region
$\bar\phi^2<v^2$. It is based on truly stable (not only metastable)
states. The gaussian $V_{\rm eff}$ cannot be defined in the
spinodal region $\bar\phi^2<v^2/3$, where the gap equation does not
admit a nonnegative solution in the physical region far away from the
Landau pole.

Let us first compare this HF situation with that of large $N$ (see section 
\ref{statprop} and also ref. \cite{Destri:1999hd}). There the different 
coupling of the transverse modes, three time smaller than the HF longitudinal 
coupling, has two main consequences at the static level: the gap equation 
similar to (\ref{bnice}) does not admit nonnegative solutions for
$\bar{\bds\phi}^2<v^2$, so that the spinodal region coincides with the
region in which the effective potential is flat, and the physical mass
vanishes. The out--of--equilibrium counterpart of this is the
dynamical Maxwell construction: when the initial conditions are such
that $\bar{\bds\phi}^2$ has a limit for $t\to\infty$, the set of all
possible asymptotic values exactly covers the flatness region (and the
effective mass vanishes in the limit). In practice this means that
$|\bar{\bds\phi}|$ is not the true dynamical order parameter, whose
large time limit coincides with $v$, the equilibrium field expectation
value in a pure phase. Rather, one should consider as order parameter
the renormalized local (squared) width
\begin{equation}
	\lim_{N \to \infty} \dfrac{\vev{\bds{\phi}(x)
	\cdot\bds{\phi}(x)}_{\rm R}}{N} = \bar{\bds\phi}^2 +
	\Sigma_{\rm R} =  v^2 + \dfrac{M^2}\l
\end{equation}
where the last equality follows from the definition itself of the
effective mass $M$ (see ref. \cite{Destri:1999hd}). Since $M$ vanishes as
$t\to\infty$ when $\bar{\bds\phi}^2$ tends to a limit within the
flatness region, I find the renormalized local width tends to
the correct value $v$ which characterizes the broken symmetry phase,
that is the bottom of the classical potential. I may say that the 
spinodal region, perturbatively unstable, at the non--perturbative
level corresponds to metastable states, all reachable through the 
asymptotic time evolution with a vanishing effective mass.

\begin{figure} 
\includegraphics[height=8.5cm,width=15cm]{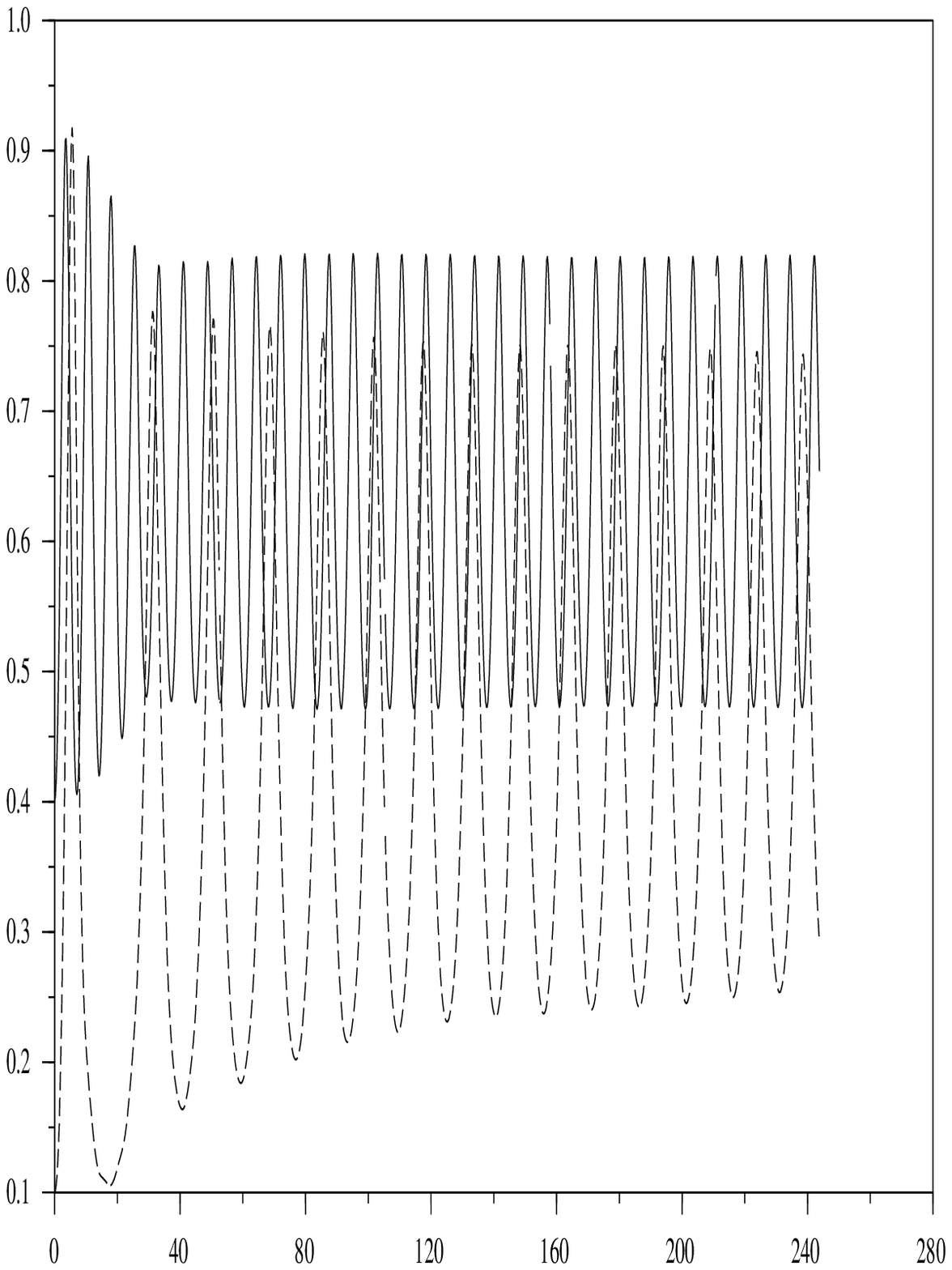}
\caption{\it Evolution of the background for two different initial
conditions within the spinodal interval, in the tdHF
approximation, for $\l=1$: $\bar\phi(t=0)=0.1$ (dotted line) and
$\bar\phi(t=0)=0.4$ (solid line). }\label{fig:max1}
\end{figure}

\begin{figure} 
\includegraphics[height=8.5cm,width=15cm]{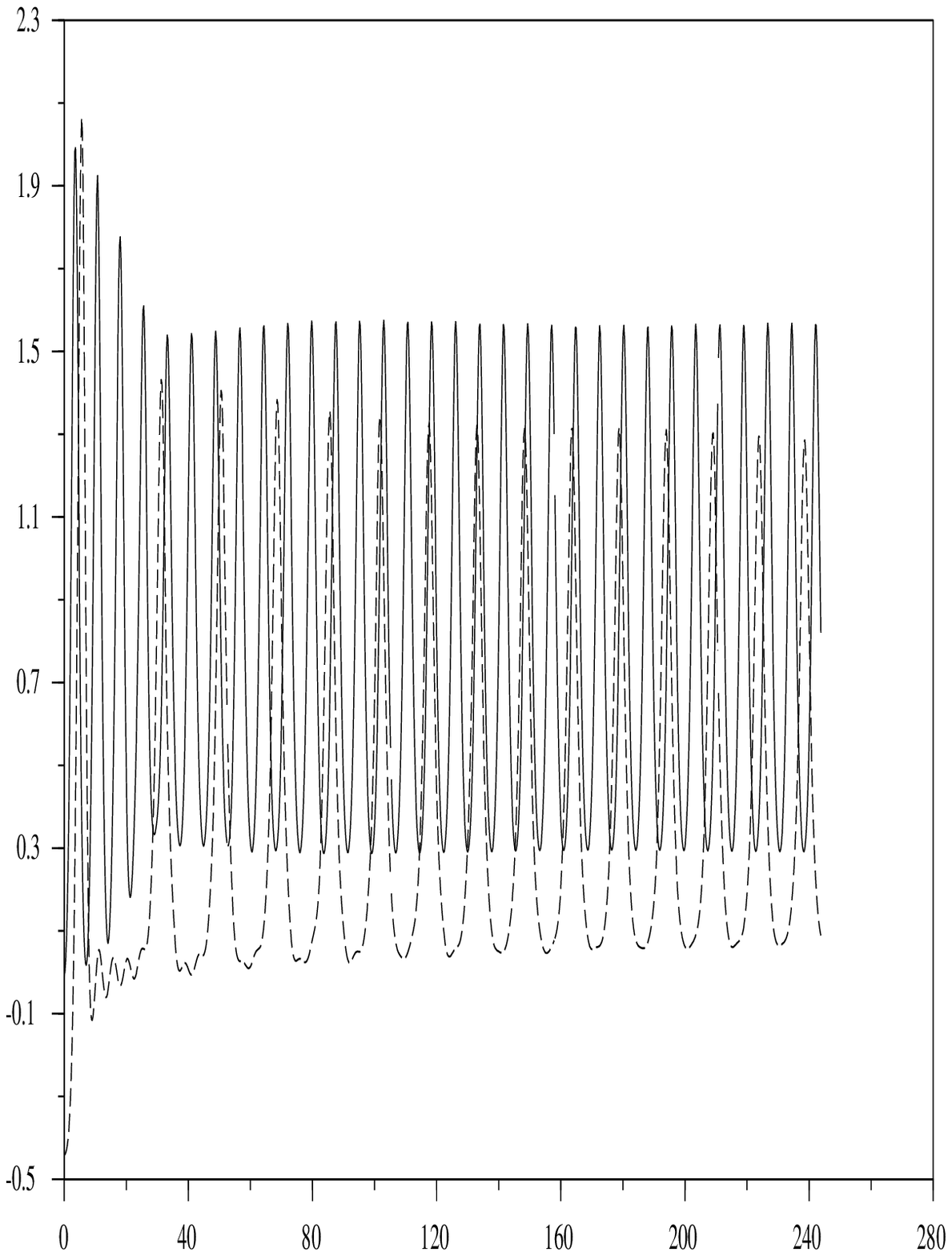}
\caption{\it Evolution of $M^2$ for the two initial
conditions of fig. \ref {fig:max1}. }\label{fig:max2}
\end{figure}

\begin{figure} 
\includegraphics[height=8.5cm,width=15cm]{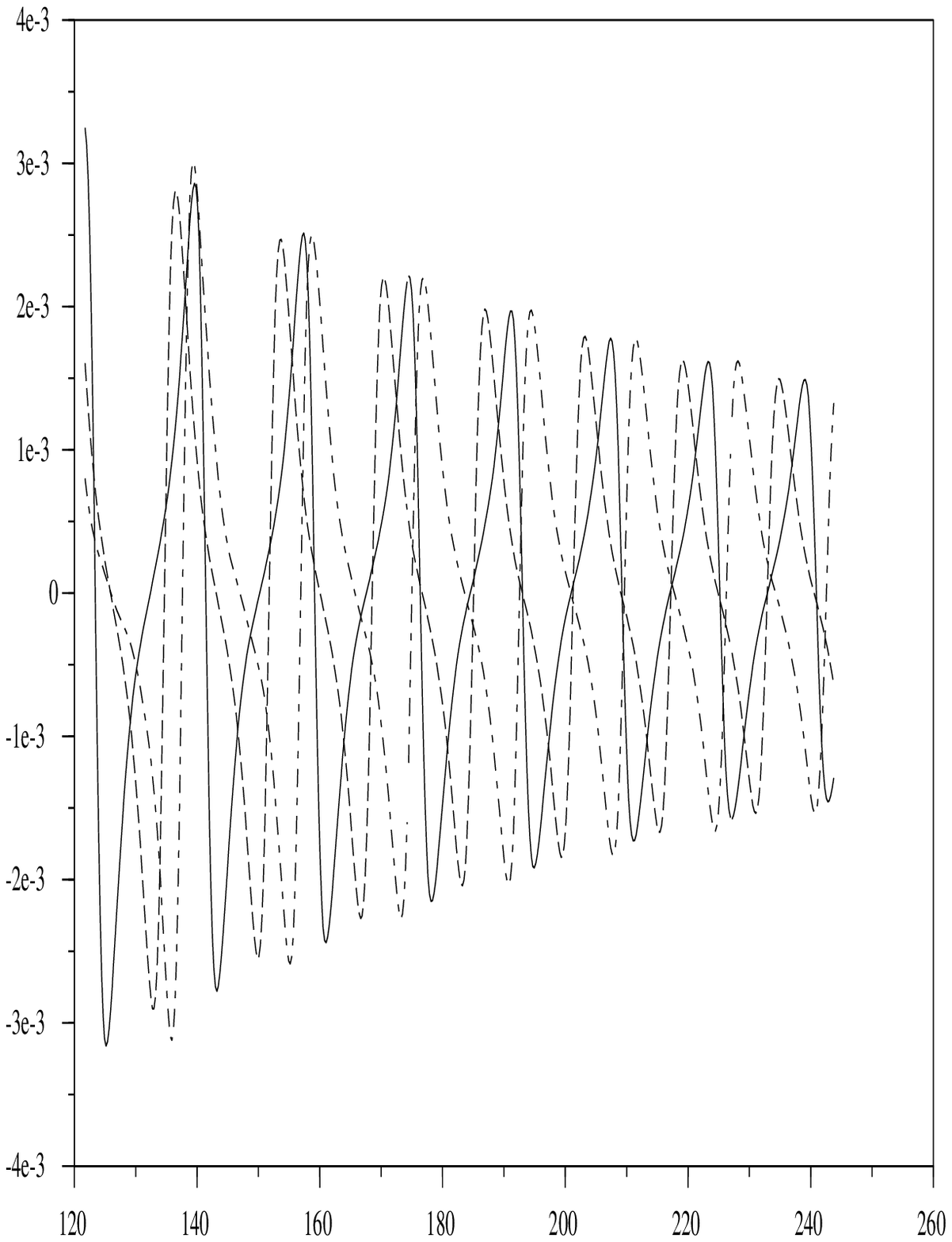}
\caption{\it The average force $f$, defined as
$\bar{f}=\int^Tf(t)dt/T$, plotted vs. $T$, for $\l=0.1$ and
$\bar\phi=10^{-2}$ (solid line), $\bar\phi=10^{-3}$ (dashed line) and
$\bar\phi=10^{-4}$ (dotted-dashed line).}\label{fig:meanforce}
\end{figure}

\begin{figure} 
\includegraphics[height=8.5cm,width=15cm]{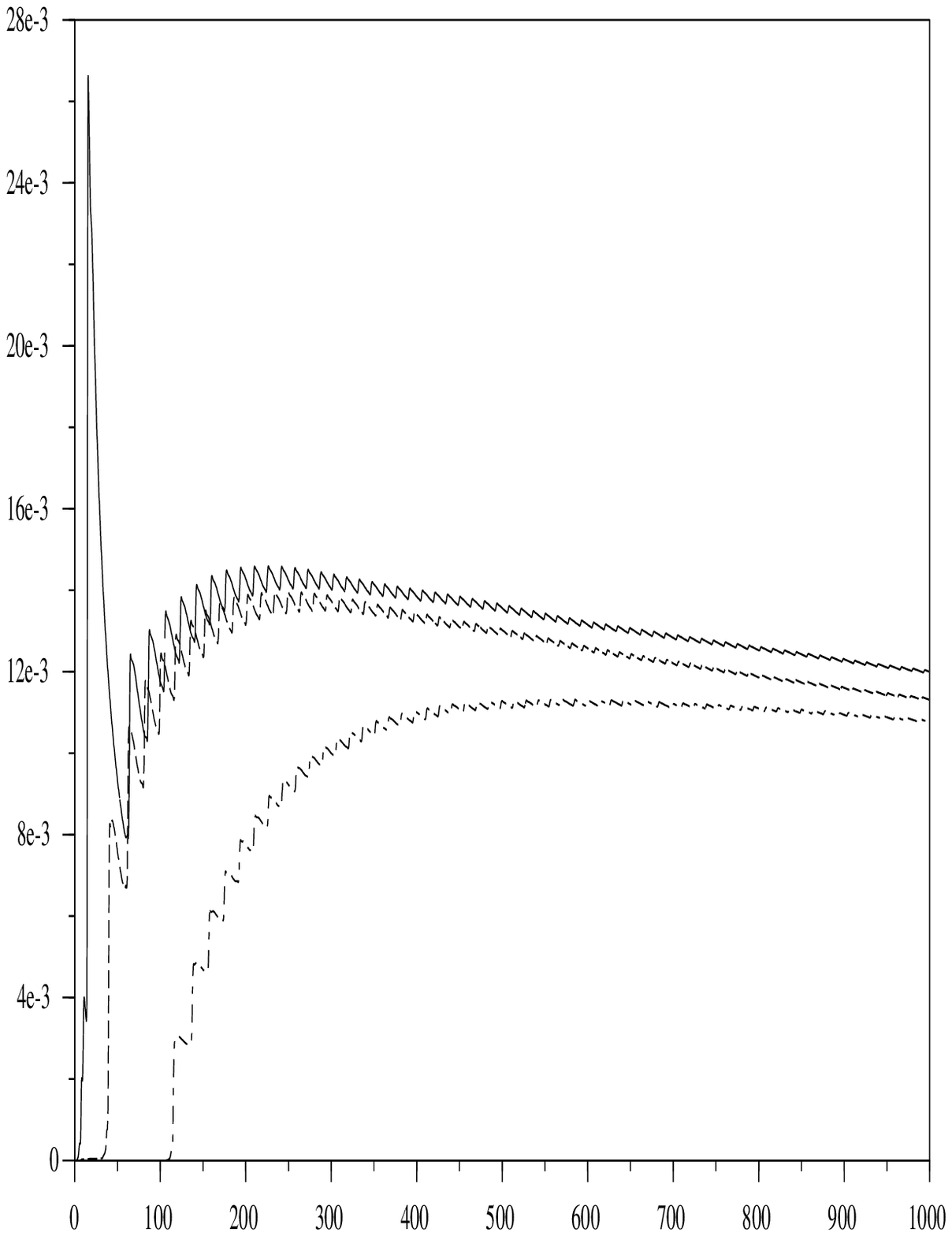}
\caption{\it The mean squared fluctuations of the force $f$, defined
as $\int^T(f(t)-\bar{f})^2dt/T$, plotted vs. $T$, for the three
initial conditions of fig. \ref{fig:meanforce}.}\label{fig:sq_flct}
\end{figure}

In the HF approximation, where at the static level the spinodal region
$\bar\phi^2<v^2/3$ is smaller than the flatness region
$\bar\phi^2<v^2$, the situation is rather different. Our numerical
solution shows that, $\bar\phi$ oscillates around a certain value
$\bar\phi _{\infty}$ with an amplitude that decreases very slowly. As
in large $N$, the asymptotic value $\bar\phi _{\infty}$ depends on the
initial value $\bar\phi (0)$. But, if the background $\bar{\phi}$
starts with zero velocity from a non--zero value inside the spinodal
interval, then it always leaves this region and eventually oscillates
around a point between the spinodal point $v/\sqrt{3}$ and the minimum
of the tree level potential $v$ (see Fig.s \ref{fig:max1} and
\ref{fig:max2}). In other words, if we start with a $\bar\phi$ in the
interval $[-v,v]$, except the origin, we end up with a $\bar\phi
_{\infty}$ in the restricted interval
$[-v,-v/\sqrt{3}]\cup[v/\sqrt{3},v]$. The spinodal region is
completely forbidden for the late time evolution of the mean field, as
is expected for an unstable region. I stress that we are dealing with
true fixed points of the asymptotic evolution since the force term on
the mean field [cfr. eq. (\ref{Emotion}), $f=(2\l\,\bar\phi^2
-M^2)\,\bar\phi$] does vanish in the limit. In fact its time average
$\bar{f}=\int^Tf(t)dt/T$ tends to zero as $T$ grows and its mean
squared fluctuations around $\bar{f}$ decreases towards zero, although
very slowly (see Fig.s \ref{fig:meanforce} and
\ref{fig:sq_flct}). Moreover, for $N=1$ the order parameter reads as
$t \to \infty$
\begin{equation}\label{aop}
	\vev{\phi(x)^2}_{\rm R} = \bar{\phi}^2 +
	\Sigma_{\rm R} = \dfrac{v^2}3 + 
	\dfrac{M^2}{3\l} \;,\quad 
	\Sigma_{\rm R} = \dfrac{v^2 - \bar\phi ^2}{3}
\end{equation}
where the last equality is valid for the asymptotic values and follows
from the vanishing of the force term $f$. From the last formula we see
that when $\bar\phi=0$ at the beginning, and then at all times, the
renormalized back--reaction tends to $v^2/3$, not $v^2$. It ``stops at
the spinodal line''. The same picture applies for a long time, all
during the ``slow rolling down'' (see section \ref{num1}), to
evolutions that start close enough to $\bar\phi=0$. This fact is at the 
basis of the so--called {\em spinodal inflation} \cite{Cormier:1998nt}.

In any case, the dynamical Maxwell construction, either complete or
partial, poses an interesting question by itself. In fact it is not at
all trivial that the effective potential, in any of the approximation
previously discussed, does bear relevance on the asymptotic behavior
of the infinite--volume system whenever a fixed point is
approached. Strictly speaking in fact, even in such a special case it
is not directly related to the dynamics, since it is obtained from a
static minimization of the total energy at fixed mean field, while the
energy is not at its minimum at the initial time and is exactly
conserved in the evolution. On the other hand, if a solution of the
equations of motion (\ref{Emotion}) exists in which the background
$\bar\phi$ tends to a constant $\bar\phi_\infty$ as $t\to\infty$, one
might expect that the effective action (which however is nonlocal in
time) somehow reduces to a (infinite) multiple of the effective
potential, so that $\bar\phi_\infty$ should be an extremal of the
effective potential. This is still an open question that deserves
further analytic studies and numerical confirmation.

\begin{figure} 
\includegraphics[height=8.5cm,width=15cm]{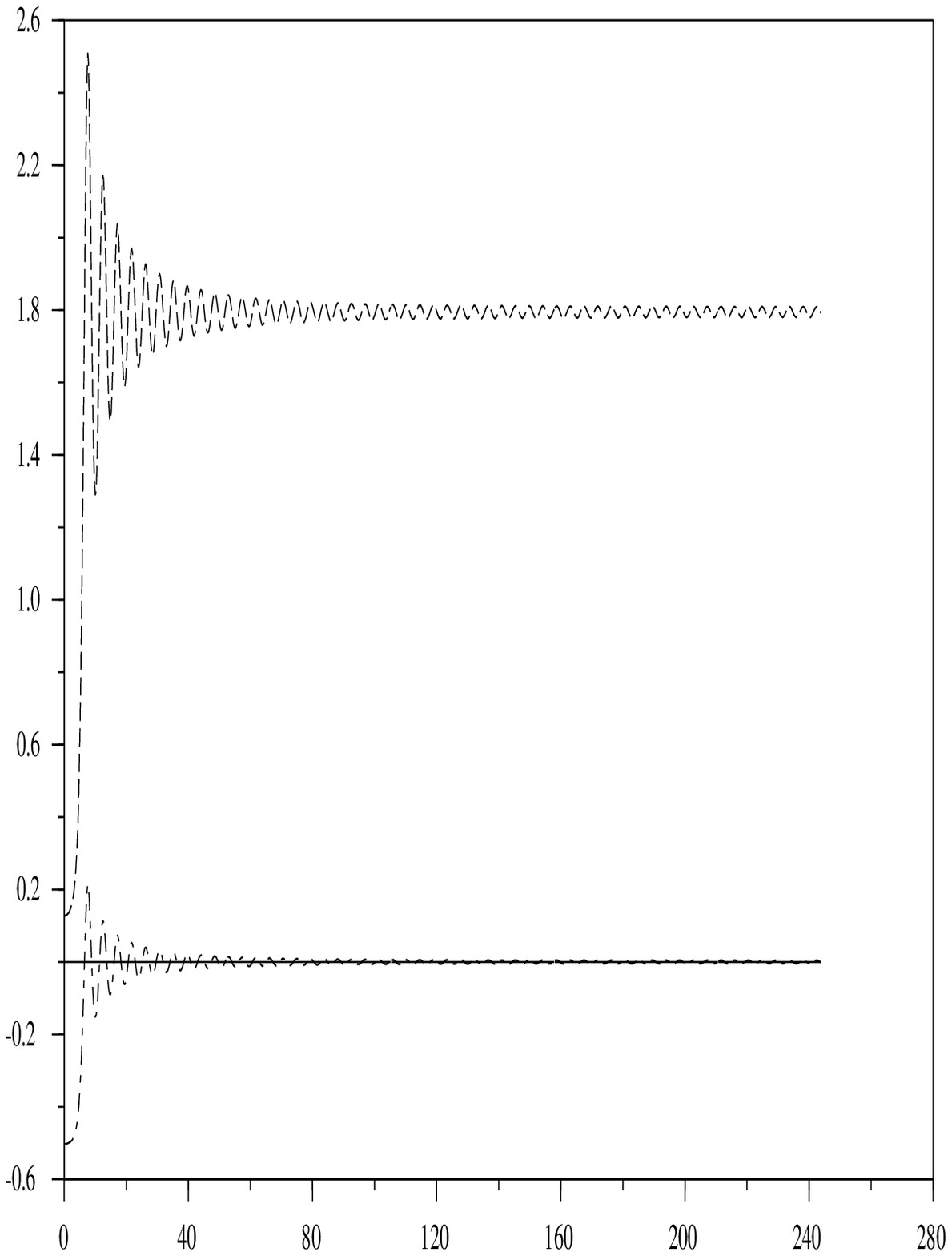}
\caption{\it The evolution of the mean value (solid line), the quantum
back--reaction $\Sigma$ (dashed line) and the squared effective mass
$M^2$ (dotted-dashed line), for $\bar\phi=0$ at $t=0$.}
\label{simm:ev}
\end{figure}

\begin{figure} 
\includegraphics[height=8.5cm,width=15cm]{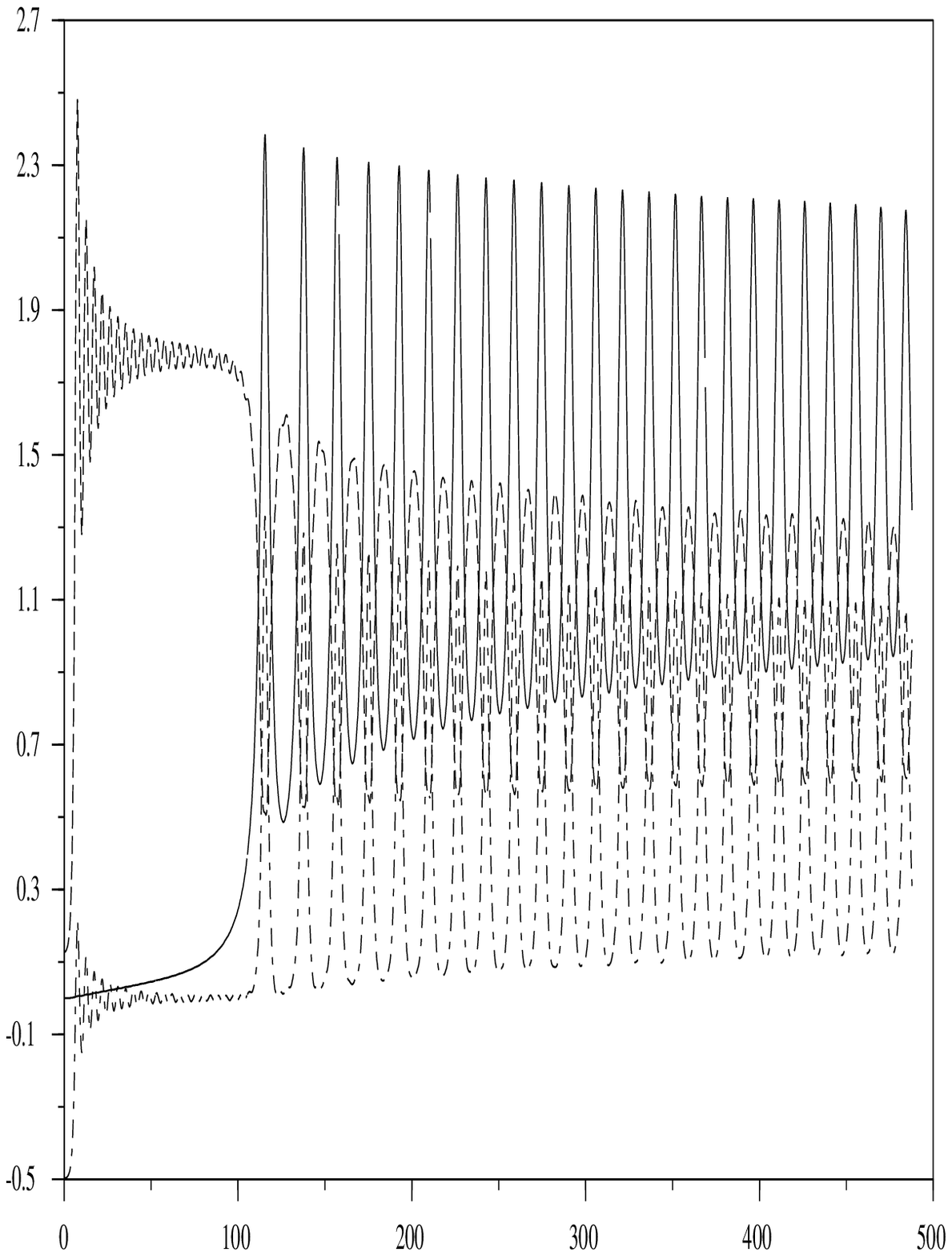}
\caption{\it The evolution of the mean value (solid line), the quantum
back--reaction $\Sigma$ (dashed line) and the squared effective mass
$M^2$ (dotted-dashed line), for $\bar\phi=10^{-4}$ at $t=0$, and
$\l=0.1$. The field rolls down very slowly at the beginning.}
\label{slowrd}
\end{figure}

\begin{figure} 
\includegraphics[height=8.5cm,width=15cm]{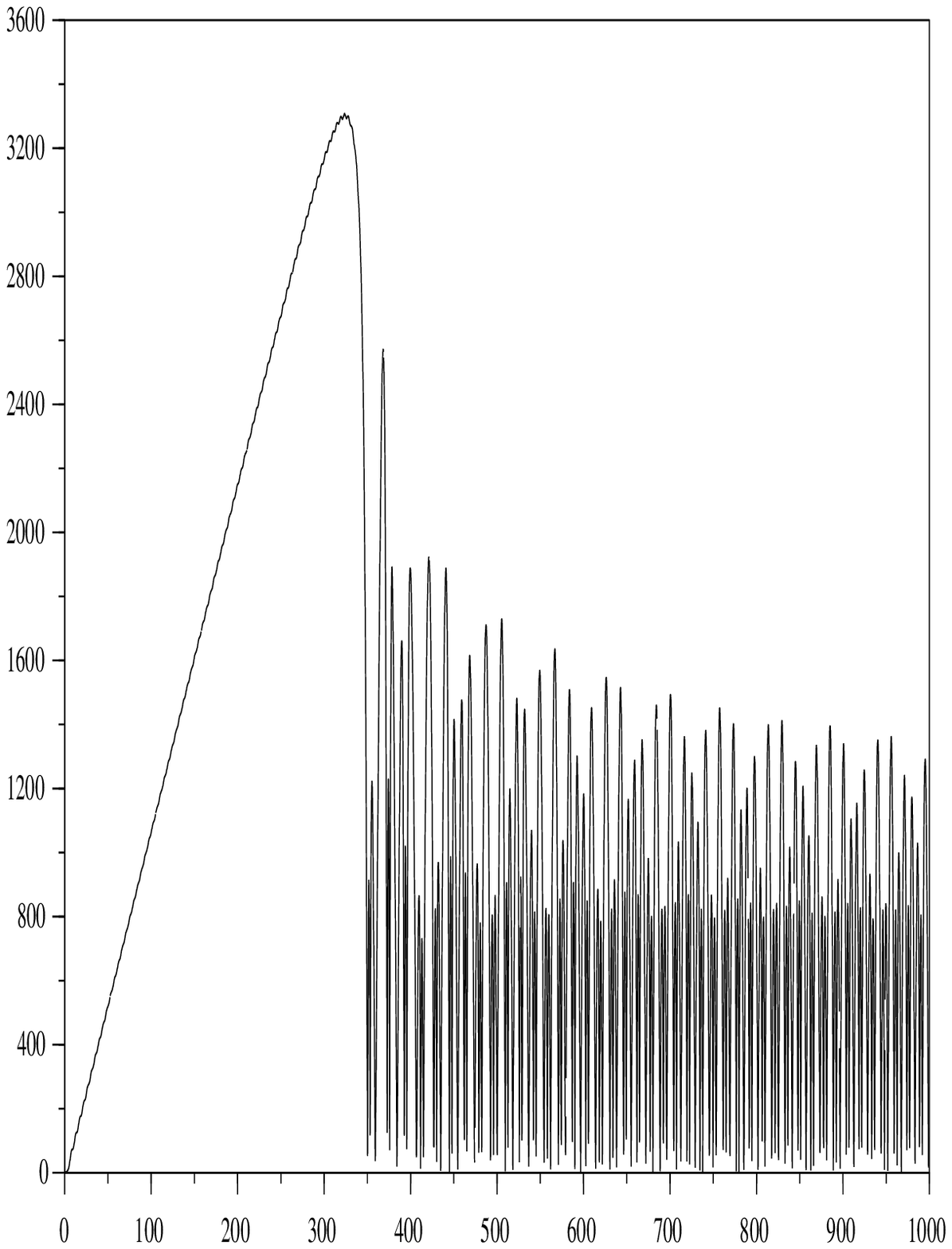}
\caption{\it Evolution of the amplitude of the zero mode for $\l=0.1$
and $\bar\phi=10^{-5}$.}
\label{slowrd:0m}
\end{figure}

It is worth noticing also that when the field starts very close to the
top of the potential hill, it remains there for a very long time and
evolves through a very slow rolling down, before beginning a damped
oscillatory motion around a point in the metastability region. During
the slow roll period, $M^2$ oscillates around zero with decreasing
amplitude and the ``phenomenology'' is very similar to the evolution
from symmetric initial conditions, as can be seen comparing Fig.s
\ref{simm:ev} and \ref{slowrd}. Fig. \ref{slowrd:0m} shows
the evolution of the zero mode amplitude in case of a very slow
rolling down. In such a case, after a very short (compared to the time
scale of the figure) period of exponential growth (the spinodal time),
the quantum fluctuations start an almost linear growth, very similar
to the evolution starting from a completely symmetric initial
state. This, obviously, corresponds to the vanishing of the effective
mass. In the meanwhile, $\bar\phi$ keeps growing and rolling down the
potential hill with increasing speed towards the minimum of the
classical potential, eventually entering the metastable region. At
that time, the effective mass starts to increase again and the zero
mode stops its linear growth, turns down and enters a phase of
``wild'' evolution. This time scale, let us call it $\tau_{\rm srd}$,
depends on the initial value of the condensate: the smaller
$\bar\phi(t=0)$ is, the longer $\tau_{\rm  srd}$ will be. I find
numerically that $\tau_{\rm srd} \propto \left( \bar\phi(t=0)
\right) ^{-1/2}$.

\begin{figure} 
\includegraphics[height=8.cm,width=15cm]{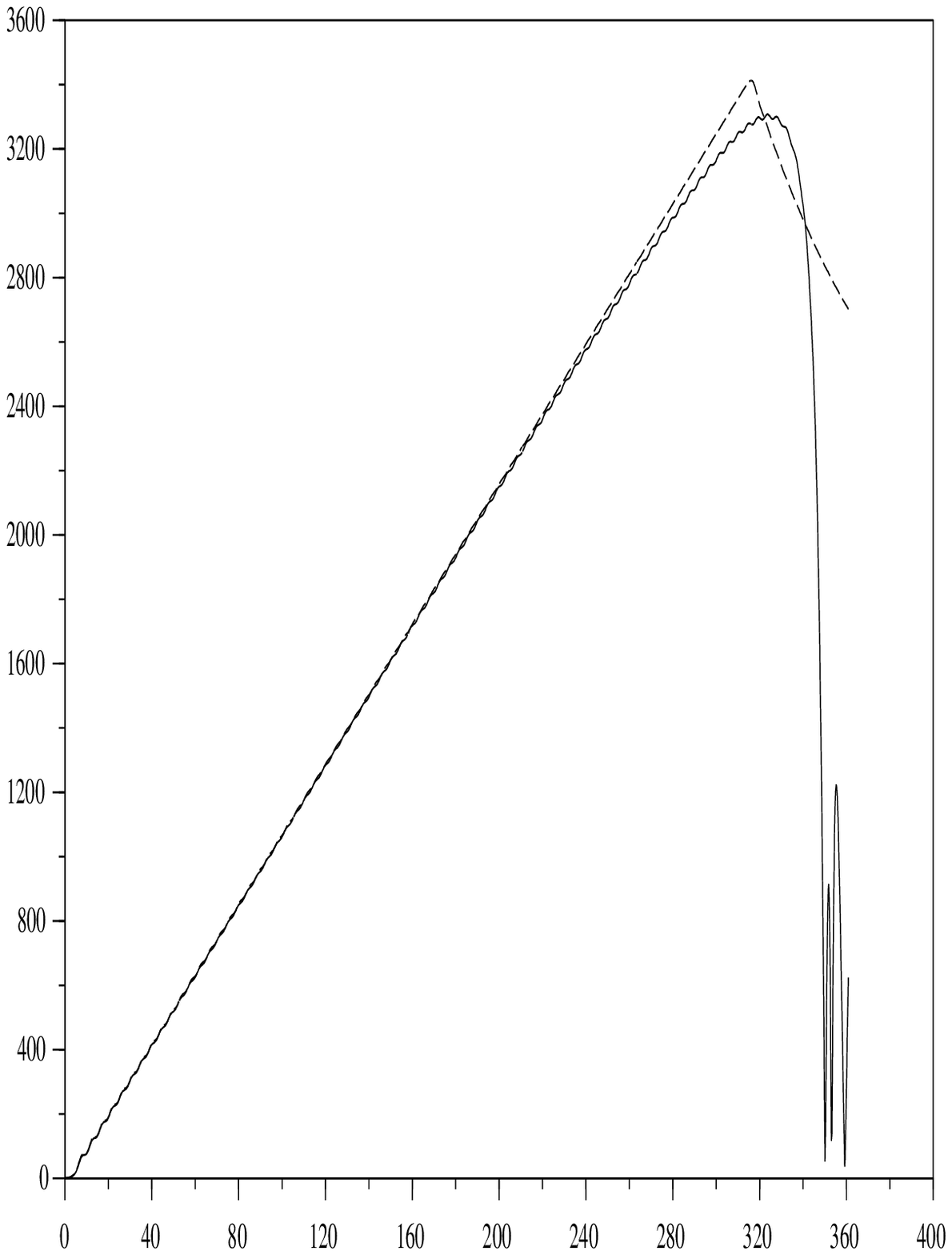}
\caption{\it Comparison between the evolutions of the zero mode
amplitude in the following two situations: the dashed line corresponds
to a finite volume simulation with $L/2\pi = 100$ and $\bar\phi=0$,
while the solid line refers to the infinite volume evolution, with
$\bar\phi=10^{-5}$. Both correspond to $\l=0.1$.}
\label{slowrd_vs_fv}
\end{figure}

If I now study the dynamics in finite volume, starting from
condensates different from zero, I will find a competition between
$\tau_{\rm srd}$ and $\tau _L$, the time scale characteristic of the
finite volume effects, that is proportional to the linear size of the
box I put the system in. Fig. \ref{slowrd_vs_fv} shows clearly
that when $L/2\pi=100$ and $\bar\phi=10^{-5}$, I have $\tau_{\rm
srd} \sim \tau _L$. In any case, either one or the other effect will
prevent the zero mode amplitude from growing to macroscopic values for
any initial condition I may start with. 

It should be noted, also, that the presence of the time scale $\tau
_{\rm srd}$ does not solve the internal inconsistency of the gaussian
approximation described above in section \ref{ooed}. In fact, for any
fixed value $L$ for the linear size of the system, I can find a whole
interval of initial conditions for the mean field, which leave enough
time to the fluctuations for growing to order $L$, much before the
field itself had rolled down towards one of the minima of the
classical potential. For those particular evolutions, I would need to
consider the quartic terms in the hamiltonians that the gaussian
approximation neglects, as already explained. 

In addition, there will be also initial conditions for which $\tau _L
> \tau_{\rm srd}$. In that case, the effective mass soon starts
oscillating around positive values and it is reasonable to think that
it will take a much longer time than $\tau_L$ for the finite volume
effects to manifest. In \cite{Destri:1999hd} I have interpreted the
proportionality between $\tau_L$ and $L$ as an auto interference
effect (due to periodic boundary conditions) suffered by a Goldstone
boson wave, traveling at speed of light, at the moment it reaches the
borders of the cubic box. Here, the massless wave I have in the early
phase of the evolution, rapidly acquires a positive mass, as soon as
the condensate rolls down; this decelerates the wave's propagation and
delays the onset of finite volume effects. The gaussian approximation
appears to be fully consistent when we limit ourselves to the
evolution of these particular configurations.

\subsection{Numerical analysis}\label{num1}

The technical details of the numerical computation are postponed in the 
appendix \ref{num}. Instead, I wuold like to add some comments on the choice 
of initial conditions and the solution of the gap equations, for the case of 
broken symmetry.

In this case the gap equation is a viable mean for
fixing the initial conditions only when $\phi$ lies outside the
spinodal region [cfr. eq (\ref{bnice})]; otherwise, the gap equation
does not admit a positive solution for the squared effective mass and
I cannot minimize the energy of the fluctuations. 
Following the discussion presented in \ref{ooed}, one possible
choice is to set $\s_k^2 = \frac1{2\sqrt{k^2+|M^2|}}$ for $k^2<|M^2|$
and then solve the corresponding gap equation (\ref{newgap1}). I will
call this choice the ``flipped'' initial condition. An other
acceptable choice would be to solve the gap equation, setting a
massless spectrum for all the spinodal modes but the zero mode, which
is started from an arbitrary, albeit microscopic, value. This choice
will be called the ``massless'' initial condition.

\begin{figure} 
\includegraphics[height=8cm,width=15cm]{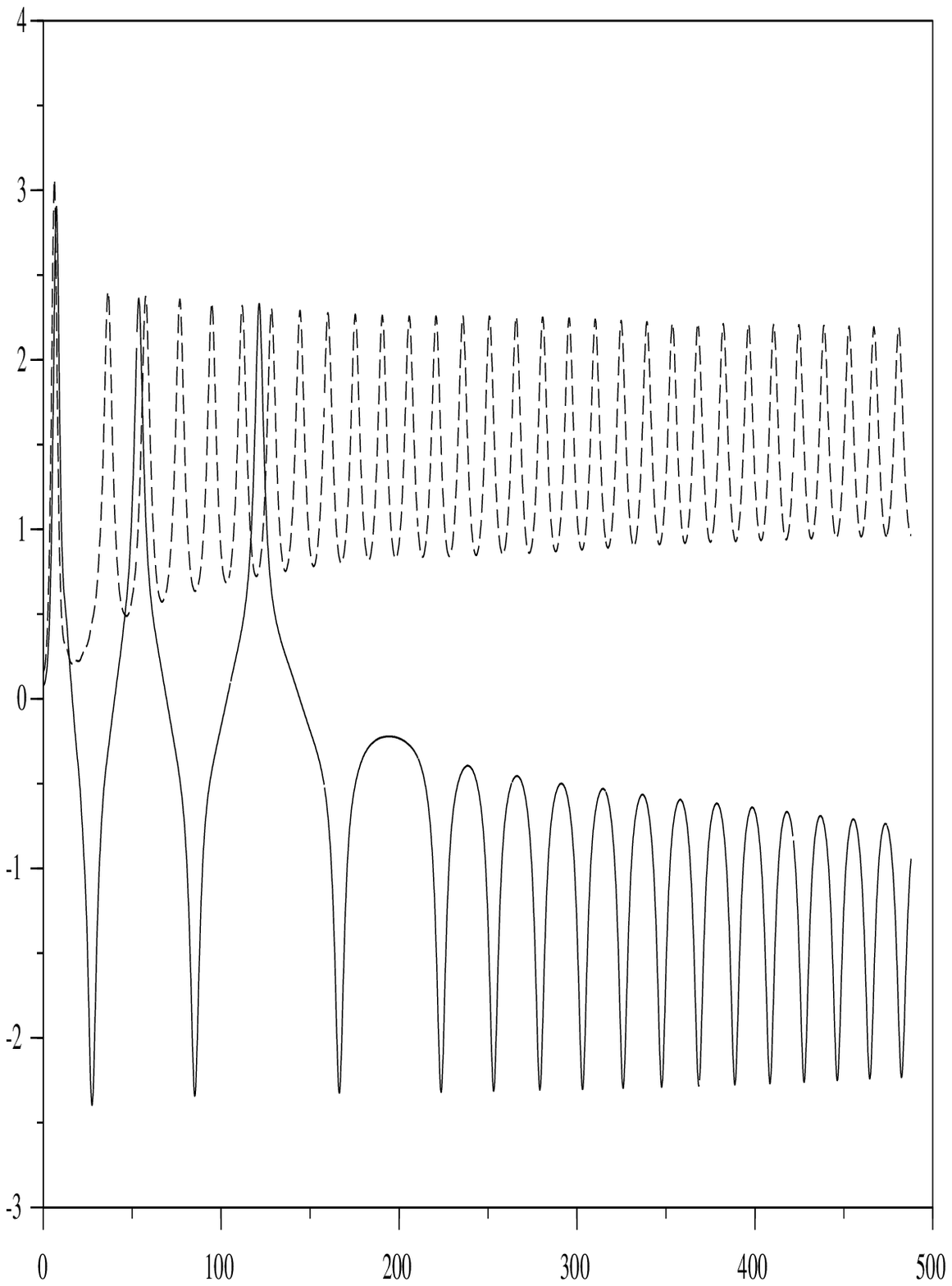}
\caption{\it Evolution of the mean value $\bar\phi$ for $\l=0.1$ and
for two different initial conditions: $\bar\phi=0.08$ (solid line) and
$\bar\phi=0.16$ (dashed line), with the ``flipped'' choice for the
spinodal modes.}
\label{boths:1}
\end{figure}
\vskip 0.5cm
\begin{figure} 
\includegraphics[height=8cm,width=15cm]{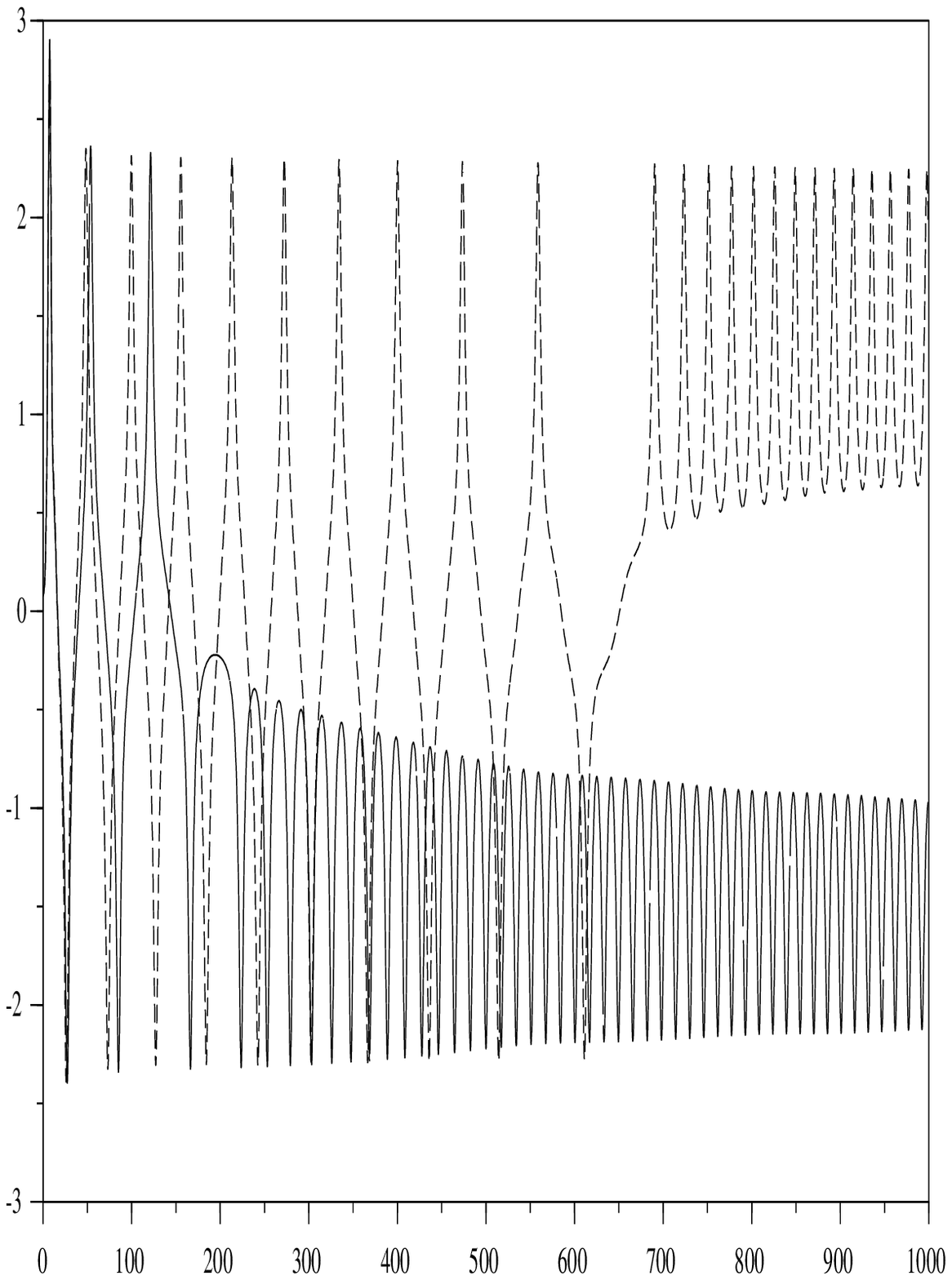}
\caption{\it Evolution of the mean value $\bar\phi$ for $\l=0.1$, with
$\bar\phi(t=0)=0.08$, and two different initial conditions for the
quantum spinodal modes, ``flipped'' (solid line) and massless (dashed line).}
\label{boths:2}
\end{figure}

\begin{figure} 
\includegraphics[height=9cm,width=15cm]{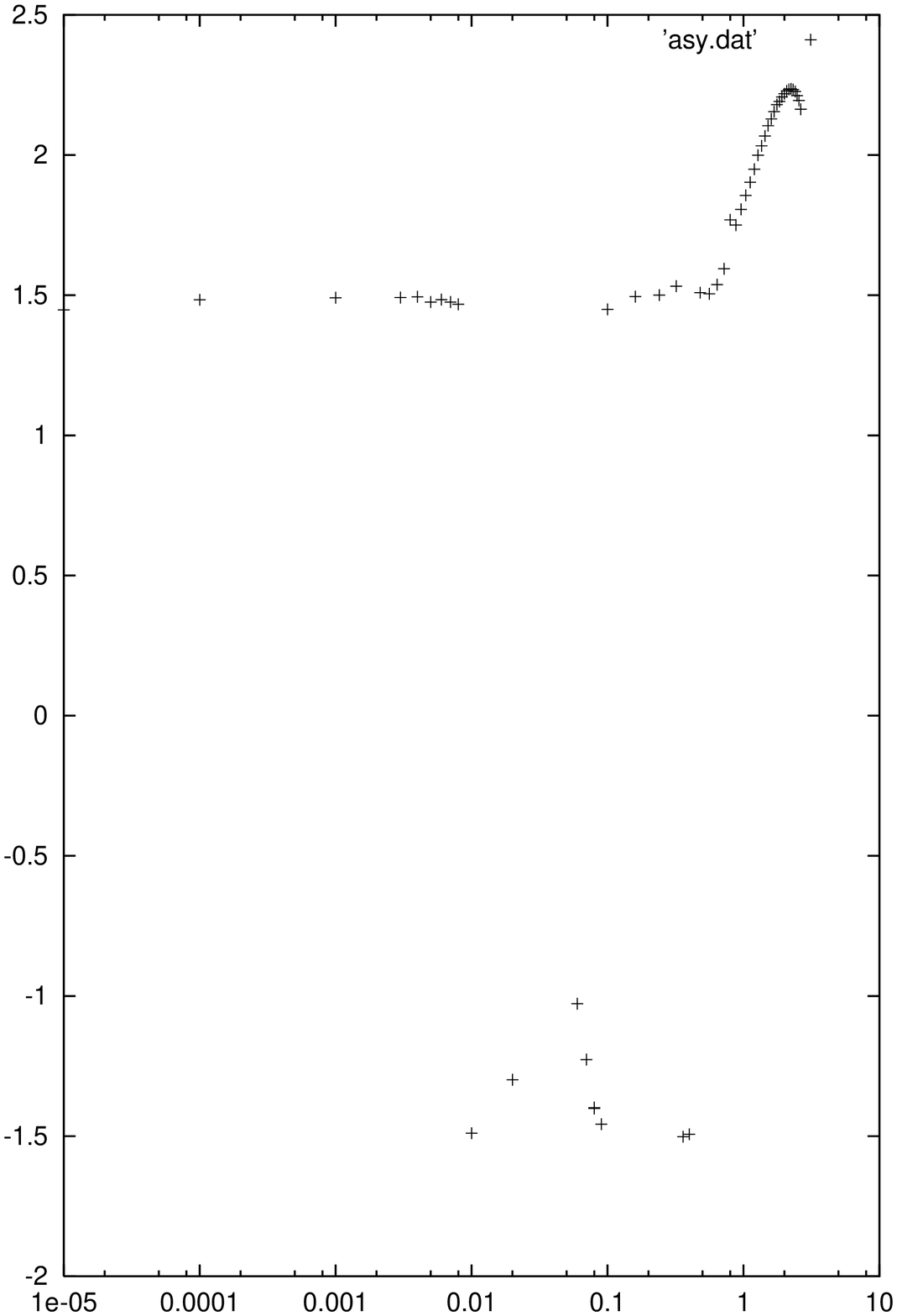}
\caption{\it Asymptotic values of the mean field $\bar\phi$, plotted
vs. initial values $\bar\phi(t=0)$, for $\l=0.1$.}
\label{asy:fig}
\end{figure}

\begin{figure} 
\includegraphics[height=8.5cm,width=15cm]{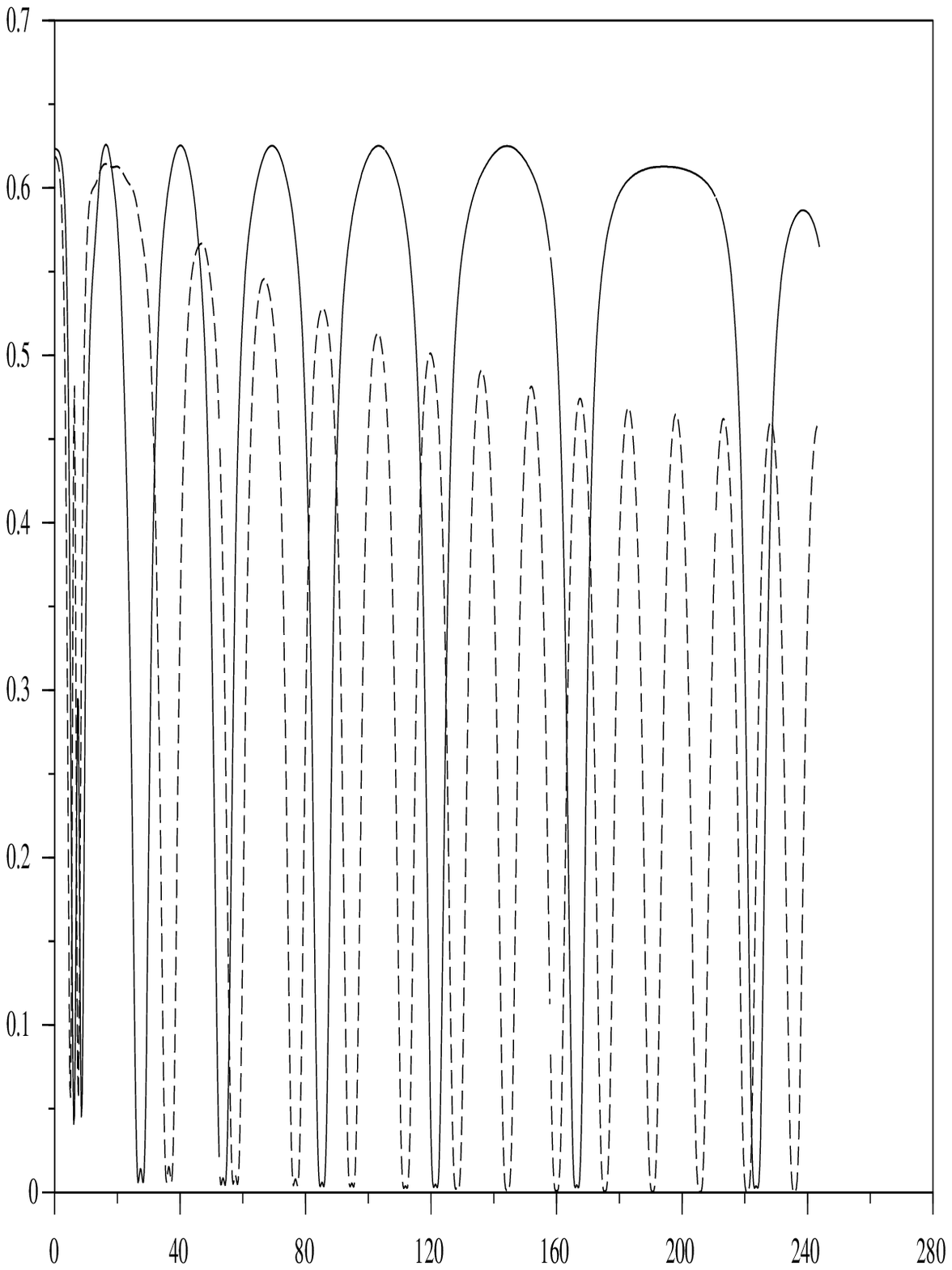}
\caption{\it Comparison between the classical energies for the two
initial conditions of Fig \ref{boths:1}.}
\label{energies}
\end{figure}

\begin{figure} 
\includegraphics[height=8.5cm,width=15cm]{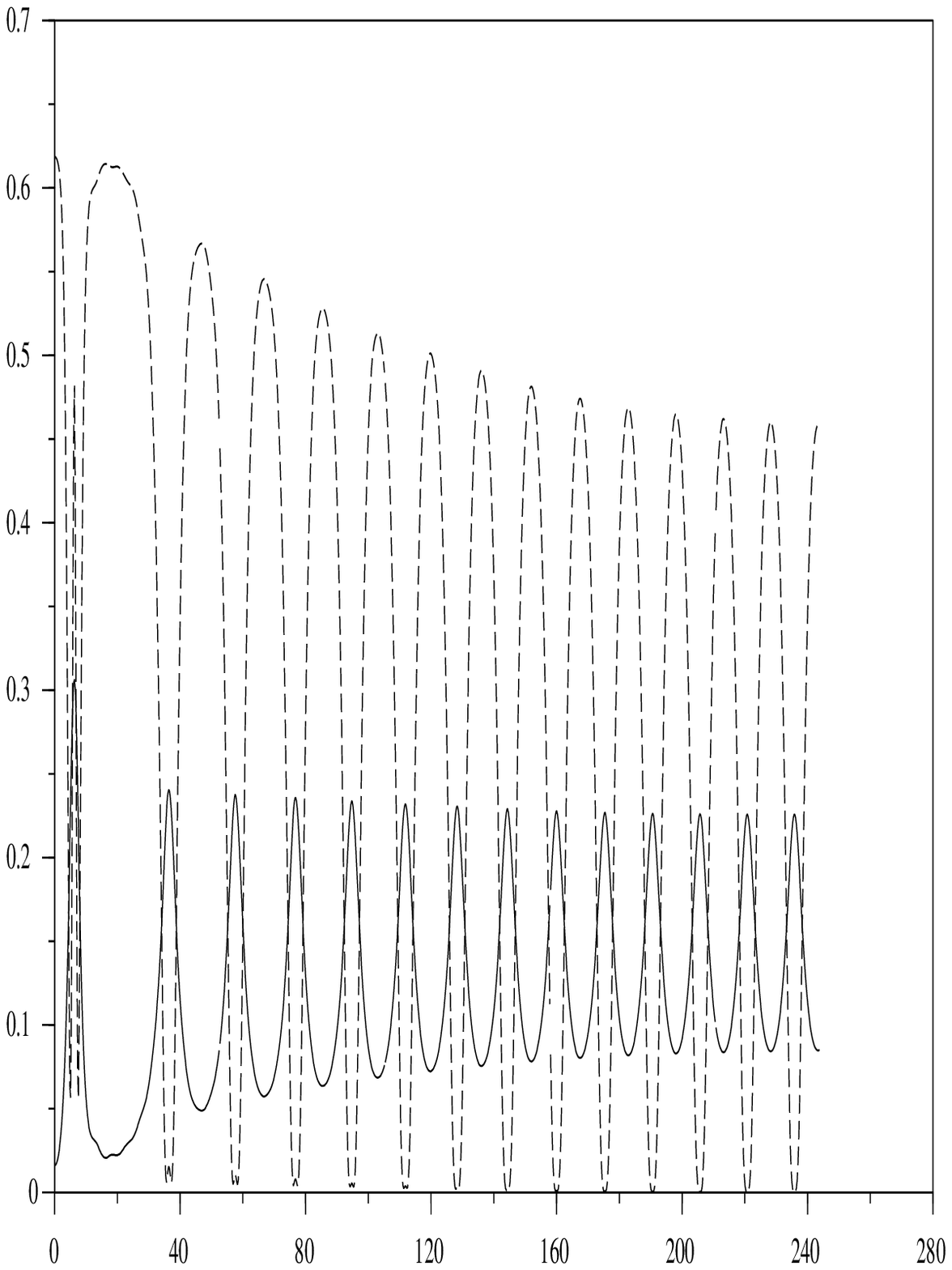}
\caption{\it Evolution of the condensate $\bar\phi$ (solid line) and
of the corresponding classical energy (dashed line), for
$\bar\phi(t=0)=0.16$ and $\l=0.1$ (cfr. Fig.s \ref{energies} and
\ref{boths:1}).}
\label{cond_en1}
\end{figure}

\begin{figure} 
\includegraphics[height=8.5cm,width=15cm]{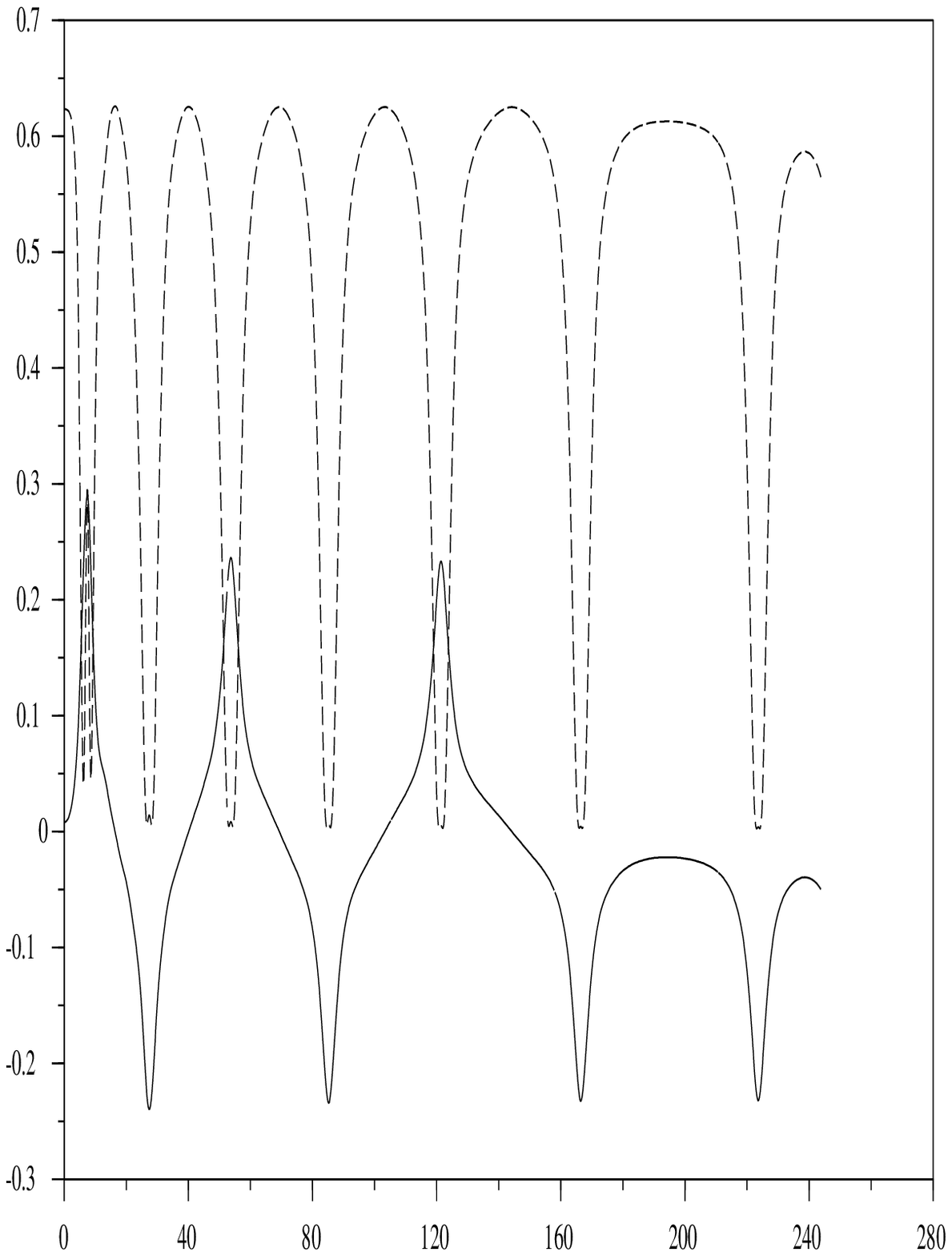}
\caption{\it Evolution of the condensate $\bar\phi$ (solid line) and
of the corresponding classical energy (dashed line), for
$\bar\phi(t=0)=0.08$ and $\l=0.1$ (cfr. Fig.s \ref{energies} and
\ref{boths:1}).}
\label{cond_en2}
\end{figure}

Before passing to discuss the influence of different initial
conditions on the results, let us present the asymptotic behavior I
find when I choose the flipped initial condition. In
Fig. \ref{asy:fig} I have plotted the asymptotic values of the mean
field versus the initial values, for $\l=0.1$. All dimensionfull
quantities are expressed in terms of the suitable power of the
equilibrium mass $m$. For example, the vev of the field is equal to
$\sqrt{5}$ in these units. First of all, consider the initial values
for the condensate far enough from the top of the potential hill, say
between $\bar\phi(t=0)=0.88$ and $\bar\phi(t=0)=2.64$. In that region
the crosses seem to follow a smooth curve, that has its maximum
exactly at $\bar\phi _{\infty} = \sqrt{5}$ (the point of stable
equilibrium). When I start from an initial condition smaller than
$\bar\phi(t=0)=0.88$, the asymptotic value $\bar\phi _{\infty}$ is not
guaranteed to be positive anymore. On the contrary, it is possible to
choose the initial condition in such a way that the condensate will
oscillate between positive and negative values for a while, before
settling around an asymptotic value near either one or the other
minimum, as fig \ref{boths:1} clearly shows. Fig.s \ref{energies},
\ref{cond_en1} and \ref{cond_en2} helps to understand this behavior by
consideration on the energy balance. Both the evolutions are such that
the classical energy, defined as $(\dot{\bar{\phi}})^2/2 + \l
(\bar\phi ^2 - v^2) /4$, is not a monotonically decreasing function of
time. Indeed, energy is continuously exchanged between the classical
degree of freedom and the quantum fluctuations bath, in both
directions. However, the two rates of energy exchange are not exactly
the same and an effective dissipation of classical energy on average
can be seen, at long time at least. Of course, this is not the case
for the initial transient part of the evolution starting from the
initial condition $\bar\phi(t=0)=0.08$; there, the condensate absorbs
energy (on average) from the quantum fluctuations, being able to go
beyond the top of the potential hill, towards the negative
minimum. This happens because in case of broken symmetry, the
minimization of the fluctuation energy, within microscopic gaussian
states, is not possible for initial conditions in the spinodal region
[cfr. the discussion about the gap equation (\ref{bnice}) in section
\ref{ooed}]. After a number of oscillations, the energy starts to flow
from the condensate to the quantum bath again (on the average),
constraining the condensate to oscillate around a value close to one
of the two minima. If we look at fig. \ref{asy:fig} again, we can find
positive asymptotic values as well as negative ones, without a
definite pattern, in the whole interval $[0.01,0.8]$. If we start with
$0<\bar\phi(t=0)<0.01$ we have the slow rolling down, already
described in section \ref{ep} and the mean field oscillates around a
positive value from the beginning, never reaching negative values. A
further note is worth being added here. During the phase of slow
rolling down, the evolution is very similar to a symmetric evolution
starting from $\bar\phi(t=0)=0$; in that case, the dissipation
mechanism works through the emission of (quasi-)massless particles and
it is very efficient because it has not any perturbative threshold. If
the field stays in this slow rolling down phase for a time long
enough, it will not be able to absorb the sufficient energy to pass to
the other side ever again and it will be confined in the positive
valley for ever. Evidently, when $\bar\phi(t=0)>0.01$ this dissipative
process might not be so efficient to prevent the mean field from
sampling also the other valley. Which one of the two valleys will be chosen by 
the condensate is a matter of initial conditions and it is very dependent from 
the energy stored in the initial state, as is shown in fig. \ref{boths:2}, 
where two evolutions are compared, starting from the same value for the
condensate, but with the two initial conditions, ``flipped'' and
massless, for the quantum fluctuations.

%% file: non_hom.tex

I summarize briefly the main definitions and the procedures followed to 
derive the evolution equations (the interested readers may find useful
reading ref. \cite{Fubini:1973mf} for the rigorous quantization of a
field theory on a spherical basis).\\
I define the expectation values of the field and of its conjugated momentum:
\begin{equation}
	\VEV{\hat{\phi} (x)} = \phi (x)
	\;,\qquad \VEV{\hat{\pi} (x)} = \pi (x)
\end{equation}
Then I shift the field operators and I define the fluctuation operators $\eta$ 
and $\sigma$:
\begin{equation}
	\hat{\phi} (x) = \phi (x) + \hat{\eta} (x) \;,\quad 
	\hat{\pi} (x) = \pi (x) + \hat{\sigma} (x)
\end{equation}
The 2-point functions (equal--time Green's functions) of the fluctuation 
operators
\begin{eqnarray}
\lefteqn{	w(x,y) = \VEV{\hat{\eta} (x) \hat{\eta} (y)} {} } \\
	& & {} u(x,y) = \Re\VEV{\hat{\eta} (x) \hat{\sigma} (y)} = 
	\dfrac{1}{2} \left[\VEV{\hat{\eta} (x) \hat{\sigma} (y)}
	 + \VEV{\hat{\sigma} (y) \hat{\eta} (x)} \right] {} \\ & & {}
	s(x,y) = \VEV{\hat{\sigma} (x) \hat{\sigma} (y)} {}
\end{eqnarray}
At $N=\infty$ (cfr. \cite{Yaffe:1982vf}) we get the following classical 
hamiltonian:
\begin{eqnarray}
\lefteqn{	H = \int d^{\rm d}x \left\{ \dfrac12 \pi^2 + 
	\dfrac12 \left|\nabla \phi\right|^2 + 
	V \left(\phi ^2 + {\rm diag}(w) \right) \right\} {} } \\ & & {}
	+ \dfrac{1}{2} \int d^{\rm d}x\, d^{\rm d}x'\, d^{\rm d}x''\,
	 v(x,x') w(x',x'') v(x'',x) {} \\ & & {} + \dfrac12\int d^{\rm d}x 
	\left[ \nabla_{\!\! x} \cdot\nabla_{\!\! x'}
	\,w(x,x')\bigr|_{x=x'} + \dfrac{1}{4}\,{\rm diag}(w ^{-1})
	\right] {}
\end{eqnarray}
The canonical variables enjoy the symplectic structure
(cfr. \cite{Boyanovsky:1998ka,Yaffe:1982vf,Destri:2000sh}):
\begin{eqnarray}
\lefteqn{	\left\{ \phi _j ( x ) ,\pi _k ( x ^{\prime} ) \right\}
_{\rm P.B.} = 
	\delta _{jk} \delta ^d ( x - x ^{\prime} ) {} } \\ & & {}
	\left\{ w ( x , y ) , v ( x ^{\prime} , y ^{\prime} ) \right\}
	_{\rm P.B.} = \delta ^d ( x - x ^{\prime} ) 
	\delta ^d ( y  - y ^{\prime} ) + \delta ^d ( x - y ^{\prime} )
	\delta ^d ( y  - x ^{\prime} ) {}
\end{eqnarray}
The Hamilton equations derived for this system can be set in a linear
form, that is better implemented numerically, by means of the
following conversion from/to canonical variables $w$ and $v$ to/from
{\em linear} variables $w$, $u$ and $s$:
\begin{equation}
s = \dfrac{1}{4} w ^{-1} + v w v + \omega
\end{equation}
where $\omega$ is a constant ($=0$ for pure states). In addition, we have the 
relation $u = w v + \omega t$. Thus, we obtain the following linear equations:
\begin{equation}
\begin{array}{ll}
\dot{\phi} = & \pi \\
\dot{\pi} =  & \nabla^2 \phi - M^2 \phi \\
\dot{w} =    & u + u ^{\rm T} \\
\dot{u} =    & s + \nabla ^2 _{y} w - M_y^2 w \\
\dot{s} =    & \nabla^2_{\!\!x} u + \nabla^2_{\!\!y} u
^{\rm T} - M_x^2 u - M_y^2 u^{\rm T}
\end{array}
\end{equation}
where $M^2 = 2V'\left(\phi^2 +{\rm diag}(w)\right)$ and
$M_x \equiv M(x)$, $w \equiv w(x,y) \equiv w_{xy}$ and so on.\\

When the condensate is spherically symmetric, we may assume the
quantum state to be rotational invariant, and I may define
\begin{equation}
	\phi (x) = \dfrac{1}{\sqrt{4\pi}} \dfrac{\varphi(r)}{r} 
	\;,\quad \pi (x) = \dfrac{1}{\sqrt{4\pi}} \dfrac{p(r)}{r}
\end{equation}
In addition, I can expand the 2-point functions as follows
\begin{equation}
	\Gamma(x,x') = \sum _{l=0} ^{\infty} \dfrac{2l+1}{4\pi\,rr'}\;
	{\rm P}_l(\cos\t)\, \Gamma^{(l)}(r,r')
\end{equation}
having defined $rr'\cos\t = x\cdot x'$, where $\Gamma$ is any of the 2-point 
functions $w$,  $v$, $u$ or $s$. As usual, the boundary conditions at $r=0$ 
are 
\begin{equation}
	\varphi(0) = 0 = p(0) \;,\quad 
	\Gamma^{(l)}(0,r') = \Gamma^{(l)}(r,0) = 0
\end{equation}
In case of a sphere of finite volume, with radius $R$, suitable boundary 
conditions (Dirichlet, von Neumann) can be assumed also at $r=R$. The 
coincidence limit for $w$ reads:
\begin{equation}
{\rm diag}(w)(r) = w(x,x) = 
	\sum _{l=0} ^{\infty} \dfrac{2l+1}{4\pi\,r^2} w^{(l)}(r,r)
\end{equation}

Recalling the useful relation (with $r_i r_j \cos \t_{ij} = x_i \cdot x_j$)
\begin{equation}
\int d\Omega_2 {\rm P}_l(\cos\t_{12}) {\rm P}_{l'}(\cos\t_{23}) =
\dfrac{2l+1}{4\pi} \d_{ll'} {\rm P}_l(\cos\t_{13})
\end{equation}
I can write, for instance
\begin{eqnarray}
\lefteqn{ \int d^3 x_2 d^3 x_3 v(x_1,x_2) w(x_2,x_3) v(x_3,x_4) = {} }
\\ & & {}
 \sum_{l=0}^{\infty} \dfrac{2l+1}{4\pi\, r_1 r_4} {\rm P}_l(\cos\t_{14})
 \int dr_2 dr_3 v ^{(l)} (r_1 ,r_2 ) w ^{(l)} (r_2 ,r_3 ) v ^{(l)}
 (r_3 ,r_4 ) {}
\end{eqnarray}
Thus, it is easy to show that in case of rotational invariance the
Hamiltonian can be written as
\begin{eqnarray}
\lefteqn{	H = \int dr \left\{ \dfrac12 p^2 +\dfrac12 (\partial_r\varphi)^2
	+ 4 \pi r^2 V\left( \phi^2 + {\rm diag}(w)\right) \right\} {} } \\ &
 & {} + \dfrac12 \sum _{l=0}^{\infty}(2l+1) \int dr \left\{
 \left[s^{(l)}(r,r) + \left( -\partial_r^2 +\frac{l(l+1)}{r^2} \right)
 w ^{(l)}(r,r')\bigr|_{r=r'}\right] \right\} {}
\label{invenergy}
\end{eqnarray}
while the evolution equations in case of rotational invariance read:
\begin{equation}\label{rad_eqs}
\begin{array}{ll}
\dot\varphi = & p \\
\dot p = & D^{(0)} \varphi \\
\dot w^{(l)} = & u^{(l)} + u^{(l)T} \\
\dot u^{(l)} = & s^{(l)} + D^{(l)}_{r'} w ^{(l)} \\
\dot s^{(l)} = & D^{(l)} u^{(l)} + \left(D^{(l)} u^{(l)}\right)^T
\end{array}
\end{equation}
where  
\begin{equation}
D^{(l)} = \pdif{^2}{r^2} - \dfrac{l(l+1)}{r^2} - M(r)^2 
\end{equation}

If I want to consider $\Phi^4$ model, I should specify the suitable 
potential, which is $V(z) = 1/2\mbare z + \lbare /4 z^2$.
\vskip 0.5cm
{\bf Discretized equations}\\ 
With an eye to the numerical calculation, that will have to be done on
a computer, I set up a spatial (radial) lattice with spacing $a$ from
$0$ to $R=(N+1)a$, so that the total number of sites is
$N+2$. However, due to Dirichelet's boundary conditions, the two
boundary sites (i.e. $0$ and $N+1$, are not linked to dynamical
variables). Thus, in this case the background field is a vector
$\varphi_j=\varphi(ja)$ and the 2-point functions are standard
matrices $\Gamma_{ij} =
\Gamma(ia,ja)$, with $i,j=0, \cdots, N$. The discretized version for
the second derivative is the standard one: $\pdif{^2}{r^2} \varphi(r)
\bigr|_{r=ja} = (\varphi_{j+1} - 2 \varphi_j +
\varphi_{j-1})/a^2$. The square effective mass becomes:
\begin{equation}
M_j^2 = \mbare + \dfrac{\lbare}{4 \pi j^2 a^2} \left[ \varphi _j^2 
+ \sum_{l=0}^{\infty} (2l+1) w^{(l)}_{jj} \right]
\end{equation}
Now I can try to use the fourth--order Runge-Kutta algorithm (already
used in the homogeneous case) to solve this system of coupled ordinary
differential equations.

\vspace{0.5cm} 
{\bf Initial conditions}\\ 
I can fix the initial conditions in the following way: I start with
an arbitrary profile for $\varphi(r)$ and with $p(r)=0$; then I want
to find suitable initial conditions for the 2-point functions. One
possible choice is to minimize the energy functional with respect to
the fluctuations. This is achieved first by setting $v(x,y)=0$; 
in that case, also $u(x,y)$ is $0$; then I must find some minimal
$w^{(l)}(r,r')$. From the numerical point
of view, I may choose two possible strategies in order to solve this
problem:\\ 1) one might try to solve the non--linear differential
equations for $w^{(l)}$:
\begin{equation}
	s^{(l)} (r,r') + \left[\pdif{^2}{r^2} -
	 \dfrac{l(l+1)}{r^2} - M(r)^2\right]w^{(l)}(r,r') = 0
\label{init}
\end{equation}
where now
\begin{equation}
\int dr s ^{(l)} (r,r') w ^{(l)} (r',r'') = \dfrac14 \delta (r - r'')
\end{equation}
2) otherwise, one may try to minimize directly the energy functional
(\ref{invenergy}), using numerical algorithms like the Simplex,
Conjugated Gradient or Simulated Annealing methods.

\vspace{0.5cm} 
{\bf Free Massive Scalar Field}\\
To clarify matters, especially with respect to the renormalization issue, let
us compute the expansion in partial waves of the Green function
$w(x,y)$ for a free massive scalar field and check that its spherical
components $w ^{(l)}$ satisfy the equations (\ref{init}).
\begin{equation}
w_0(x_1,x_2;m)=w_0(x_1-x_2;m)=\dfrac12 \int \dfrac{d^3k}{(2\pi)^3}
\dfrac{e^{\imath k \cdot (x_1-x_2)}}{\sqrt{k^2+m^2}}
\end{equation}
This integral can be computed in closed form
\begin{equation}
w_0(r_{12};m)=\dfrac{m}{4\pi^2r_{12}} K_1 (mr_{12}) =
\dfrac1{4\pi^2r_{12}^2} +
\dfrac{m^2}{8\pi^2} \ln (mr_{12}) + {\rm finite}
\end{equation}
where the coincidence limit singularities are made explicit.\\ Let us
use a Dirac formalism to indicate the simultaneous eigenstates of the
operators $|\boldsymbol{P} |$, $\bds L ^2$ and $\bds L _z$:
\begin{equation}
\braket{x}{k,l,m}=\sqrt{\dfrac2{\pi}} j_l (kr) Y _{lm} (\hat{x})
\end{equation}
that satisfy the orthogonality and closure rule
\begin{eqnarray}
\lefteqn{ \braket{k,l,m}{k',l',m'} = \dfrac1{k^2} \d (k-k') \d _{ll'}
\d _{mm'} {} }  \\ & & {}
\sum_{l.m} \int dk k^2 \ket{k,l,m} \bra{k,l,m} = {\rm Id} {}
\end{eqnarray}
The following identities hold:
\begin{eqnarray}
\lefteqn { \sum_{l,m} Y _{lm} (\hat{x} _1) Y _{lm} (\hat{x} _2) = \sum _l
\dfrac{2l+1}{4 \pi} P _l (\cos \t _{12})  = \d ^{(2)} (\hat{x} _1 -
\hat{x} _2) {} } \\ & & {}
\int dr\,r ^2 j _l (k_1 r) j _l (k_2 r) = \dfrac{\pi}2 \dfrac1{k_1^2} \d
(k_1 - k_2) {}
\end{eqnarray}
Using the partial wave expansion of the plane wave and after some algebra I 
end up with
\begin{equation}
w_0(x_1,x_2;m) = \sum_l \dfrac{2l+1}{2\pi^3} P _l (\cos \t _{12}) \int dk k^2
\dfrac{j _l (kr) j _l (kr')}{2 \sqrt{k^2+m^2}}
\end{equation}
from where I can read the explicit expression of $w ^{(l)}$:
\begin{equation}
w_0 ^{(l)}(r,r';m) = \dfrac2{\pi^2}rr' \int_0^\Lambda dk k^2 \dfrac{j _l (kr) j _l 
(kr')}{2 \sqrt{k^2+m^2}} \;,\quad \Lambda \to\infty
\label{wl}
\end{equation}
For example, the integral for $l=0$ can be computed exactly in terms
of the Bessel $K_0$:
\begin{equation}
w_0 ^{(0)}(r,r';m) = \dfrac1{2 \pi^2} \left[ K_0 (m|r-r'|) - K_0(m(r+r') \right]
\end{equation}
The spherical components of $w^{-1}$ are given by:
\begin{equation}
w_0 ^{(l)-1}(r,r';m) = \dfrac2{\pi^2}rr' \int dk k^2 2 \sqrt{k^2+m^2} j _l
(kr) j _l (kr')
\end{equation}
One can easily verify that $w ^{(l)}$ is a solution of the
self-consistent equation (\ref{init}); in fact, the spherical Bessel
functions $j _l (kr)$ are eigenfunctions of the Bessel operator
\begin{equation}
\left( - \partial _r ^2 + \dfrac{l(l+1)}{r^2} \right) [rj _l (kr)] =
k^2 [rj _l (kr)]
\end{equation}

Let us now consider the Hamiltonian (\ref{invenergy}). For a free
massive scalar field, the potential reduces to the form:
\begin{equation}
\dfrac{m^2}2 \sum_l (2l+1) \int dr w_0 ^{(l)}(r,r;m)
\end{equation}
that can be written as
\begin{equation}
m^2\int \dfrac{dk}{\pi} \dfrac{k^2}{2 \sqrt{k^2+m^2}}\int dr r^2 \sum_l 
\dfrac{2l+1}{\pi} j _l ^2 (kr)
\end{equation}
The (functional) series in the internal integral is a constant exactly equal to
$1/2$ (cfr. \cite{grad}, 8.536 1., page 980); thus I am left with the
integral on the quantum fluctuations times a volume factor, due to
translation invariance
\begin{equation}
\sum_l (2l+1) \int dr\, w _0 ^{(l)}(r,r;m) = V \int
\dfrac{d^3k}{(2\pi)^3} \dfrac1{2\sqrt{k^2+m^2}} 
\end{equation}
In other words, the volume factor can be written as
\begin{equation}
\d^{(3)}(k)\bigr|_{k=0} = \dfrac{V}{(2\pi)^3} = \sum_l
\dfrac{2l+1}{2 \pi^3} \int dr r^2 j _l ^2 (kr)
\end{equation}

\vspace{0.5cm}
{\bf Renormalization}\\
The coincidence limit of the 2-point function yields ultraviolet
divergences that must be properly subtracted before solving the
evolution equations numerically. I consider the case of {\em unbroken
symmetry} for simplicity. The space-time dependent effective mass must
be written in terms of finite quantities, and this set our 
renormalization conditions. First, I parametrize $\mbare$ using the
equilibrium free field 2-point function for a massive field of
renormalized mass $m$:
\begin{equation}
	\mbare = m^2 - \lbare {\rm diag}(w_0) = m^2 - 
	\lbare\sum_l \dfrac{2l+1}{4\pi r^2} w^{(l)}_0(r,r;m)
\end{equation}
where $w ^{(l)} _0 (r,r')$ is given by eq. (\ref{wl}). When the sum
over $l$ runs from $0$ to $\infty$, the complete free ultraviolet
divergence is correctly rebuilt:
\begin{equation}
	\left({\rm diag}(w_0)\right)(r) = 
	\int_0^\Lambda dk\, \dfrac{k^2}{2\sqrt{k^2+m^2}}
	\sum_l \dfrac{2l+1}{2\pi^3} j_l(kr)^2 =
	\int_0^\Lambda dk\, \dfrac{k^2}{2\sqrt{k^2+m^2}}
\end{equation}
I still have a logarithmic divergence, due to the difference 
$M(r)^2-m^2$, which I absorb in the definition of the 
renormalized coupling $\l$:
\begin{eqnarray}\label{good}
\lefteqn{ M(r)^2 = m^2 + \l\left[\phi(x)^2+ {\rm diag}(w)_{\rm
R}\right] {} } \\ &  
	&= {} m^2 + \dfrac{\lbare}{4\pi r^2} \left\{ \varphi(r)^2 + 
	\sum_{l=0}^{l_{\rm max}} (2l+1) \left[w^{(l)}(r,r) - 
	w^{(l)}_0(r,r;m) \right] \right\} {}
\end{eqnarray}
When I stop the sum over the partial waves at a finite $l _{\rm max}$,
I should subtract the ultraviolet divergences before performing the
sum. The partial waves $w^{(l)}_0 (r,r;m)$ should be computed once and
for all at the beginning, performing the integral (\ref{wl}) for the
values of $r$ corresponding to the lattice chosen and with an upper
momentum cut-off equal to $\pi/a$. It should be recalled that, {\em
for fixed} $l$, each $w^{(l)}_0(r,r;m)$ has only a logarithmic
divergence in the ultraviolet cut-off $\Lambda$, as can be easily
inferred expanding for large arguments the spherical Bessel function
in eq. (\ref{wl}), plus finite parts that do depend on $r$. Thus,
subtracting the divergence for each $l$ before performing the sum
could be quantitatively very different (for given $l_{\rm max}$,
$\Lambda$ and $R$) from subtracting beforehand the entire {\em
constant} $\left({\rm diag}(w_0)\right)$:
\begin{equation}
M ^2 (r) = m^2 + \lbare \left[\phi(x)^2+ \sum_{l=0} ^{l _{\rm max}}
\dfrac{2l+1}{4\pi r^2} w ^{(l)} (r,r) - {\rm diag}(w_0)\right]
\end{equation}
With the subtraction scheme as in (\ref{good}), the functional gap
equation
\begin{eqnarray}\label{gapw}
\lefteqn{ M(r)^2 = m^2 + \dfrac{\lbare}{4\pi r^2} \left\{ \varphi(r)^2 + 
	\sum_{l=0}^{l_{\rm max}} (2l+1) \left[w^{(l)}(r,r) - 
	w^{(l)}_0(r,r;m) \right] \right\} {} } \nonumber \\ & & 
	{} \left[-\pdif{^2}{r^2} + \dfrac{l(l+1)}{r^2} + M(r)^2
	 \right]w^{(l)}(r,r') = \dfrac14 w^{(l)\,-1}(r,r') {}
\end{eqnarray}
that determine the initial conditions, trivially admits the
equilibrium solution $\phi=0$, $M(r)^2 = m^2$, $w^{(l)}=w_0^{(l)}$.
Eq. (\ref{gapw}) is formally solvable via mode expansion: suppose we
have the complete solution of the eigenvalue problem
\begin{equation}
	\left[-\pdif{^2}{r^2} + U_l(r)\right] [r\chi^{(l)}_k(r)]
	=  k^2 [r\chi^{(l)}_k(r)]
\end{equation}
where $U_l(r)=l(l+1)/r^2 + M(r)^2 -m^2$  cannot be negative
since $M(r)^2 >m^2$ for unbroken symmetry and is assumed to vanish
for large $r$ fast enough; then
\begin{equation}
	w^{(l)}(r,r') = \dfrac2{\pi}rr' \int_0^\Lambda dk\, k^2 
	\dfrac{\chi^{(l)}_k(r)\chi^{(l)}_k(r')}{2\sqrt{k^2+m^2}}
\end{equation}
and the gap equation reads
\begin{equation}
	M(r)^2 = m^2 + \dfrac{\lbare}{4\pi} \left\{ \varphi(r)^2 + 
	\sum_{l=0}^{l_{\rm max}}(2l+1) \int_0^\Lambda dk\,k^2 
	\dfrac{\chi^{(l)}_k(r)^2 - j_l(kr)^2}{2\sqrt{k^2+m^2}} \right\}
\end{equation}

%% file: ch3.tex
\section{Motivations}

As we have seen, much work has been done about the quantum evolution
out of equilibrium of the $\phi^4$ model in $3+1$ dimensions
\cite{Cooper:1994hr,Habib:1996ee,Cooper:1997ii,Boyanovsky:1995yk,
Boyanovsky:1995me,Boyanovsky:1998zg}. As is well known \cite{Wilson:1974jj},
the renormalized theory is {\em trivial}. Practically, this means that
we should consider the model as an effective theory, keeping the
ultraviolet cut-off $\Lambda$ much smaller than some Landau scale. The
logarithmic dependence on $\Lambda$ should disappear from the renormalized
quantities, while a weak inverse power dependence remains. 

If we want to push the application of non equilibrium techniques to
more fundamental theories, like QCD, we should consider that the
ultraviolet properties change drastically. In those cases, in fact,
there is no Landau Pole in the ultraviolet and the renormalized
coupling becomes smaller and smaller as the momentum scale
increases. This corresponds to the property of asymptotic freedom,
whose presence justifies self--consistently the perturbative
renormalization procedure and allows in principle to perform the
infinite cut-off limit smoothly.

Motivated by this consideration, I analyze in this chapter the
dynamical properties of the $O(N)$ non linear $\sigma$ model in $1+1$
dimensions and in the large $N$ limit. 

The linear and non linear $\sigma$ models in $3+1$ dimensions were
introduced in elementary particle theory in order to provide a useful
model of the low--energy strong interaction sector, which was able to
realize the $SU(2) \times SU(2)$ current algebra and the Partial
Conservation of Axial Current (PCAC) and satisfy the corresponding low
energy theorems \cite{Gell-Mann:1960np}. Afterwards, the non linear $\sigma$
model has been considered fruitfully in many areas of Quantum Field
Theory and Statistical Mechanics, mainly in the description of $2D$
spin chains and, quite recently, of disordered conductors and of quantum
chaos \cite{simons}. 

For the sake of completeness, I should say that the classical non-linear $\s$ 
model in the large $N$ limit and in $3+1$ dimensions, has been used for the 
study of the phase ordering dynamics in a FRW background, revealing the 
existence of scaling solutions of the classical field equations 
\cite{Turok:1991qq,Spergel:1991ee,Turok:1990gw}. 
However, in that case the main disadvantages are that one looses the
renormalizability and is constrained to stay at the classical level.

As is well known, the theory in $(1+1)$D is renormalizable and
asymptotically free in the ultraviolet
\cite{zj}. Thus, $\Lambda$ can be pushed to infinity rigorously and
there should exist a renormalized out--of--equilibrium dynamics,
completely independent of the ultraviolet cut-off. 

It is very important to study and understand the scaling properties of
the dynamics with respect to the ultarviolet cutoff $\Lambda$. In the
case of the {\em trivial} linear model, in fact, $\Lambda$ should be
kept in any case much smaller than the Landau pole. If it is also
taken bigger than the largest unstable mode, the corresponding
dynamics is quite insensitive to the specific value of $\Lambda$
chosen, because once the whole unstable band has been included, making
$\Lambda$ bigger means adding only oscillating perturbative modes. In
the case of the non linear model, instead, there is not any
instability and the effect of a bigger number of modes should be
checked carefully.

To get the non linear model at leading order in $1/N$ I will have to
perform two limits: the limit $N \to \infty$ and the limit of large
coupling which enforces the constraint. While it is not clear, a priori,
that these two limits commute, I give here a rigorous proof that this
is, in fact the case. Thus, the non linear model at $N = \infty$ is
uniquely defined.

\section{The model in $0+1$ dimensions}
In this section I rigorously prove the commutativity of the two
limits in the simple case of the quantum mechanics constrained on the
sphere.

I first set $N = 1$ and I choose a symmetry breaking potential $V$
of the form:
\begin{equation}\label{pot}
V \left( x \right) = \dfrac{\lambda}{4 !} \left( x ^2 - v ^2 \right) = 
\dfrac12 \mu ^2 x ^2 + \dfrac{\lambda}{4 !}  x ^4 + \dfrac{3 \mu ^4}{2 
\lambda}
\end{equation}
where I identify $v ^2$ with $-6 \mu ^2 / \lambda$. Now, it is not
difficult to see that if I perform the limit $\lambda \rightarrow
\infty$, keeping $v ^2$ fixed, also the second derivative of the
potential at the two minima tends to infinity, meaning that at the
classical level, the variable $x$ is constrained to assume only values
compatible with $x ^2 = v^2$. It is straightforward to treat the vector
model ($N>1$), the only change being transformation of $x$ to a
$N$--component vector.

Now I derive the quantum dynamics in the large $N$ limit, applying a
general technique already used in the analysis of the $\phi^4$
dynamics in finite volume \cite{Destri:1999hd} and based on
well--known work by Yaffe \cite{Yaffe:1982vf}. Consider the quantum
mechanical system described by the potential (\ref{pot}). The
corresponding hamiltonian is $\hat{H}=\hat{p} ^2 /2 +
V(\hat{q})$. Performing first the large $N$ limit, I get the classical
hamiltonian
\begin{equation}
h ^{(u)} = \dfrac{1}{2} p ^2 + \dfrac{1}{2} s ^2 + \dfrac{1}{8 \sigma ^2}
+ V \left( q ^2 + \sigma ^2 \right)
\label{gen_cl_ham_qm}
\end{equation}
Where $q$ is now the average value and $\sigma$ is the width (the
quantum fluctuation) of the position operator, and $p$ and $s$ are
related to the corresponding quantities for the momentum operator. If
I use the potential $V$ to enforce the constraint (i.e. $\lambda \to
\infty$, keeping $v^2$ fixed), I get the {\em quantum} constraint $q
^2 + \sigma ^2 = v^2$. 

This last result deserves a more detailed comment: we see that while
at the classical level the $x$-variable lives on the iper-sphere
exactly, the mean value of the quantum operator $\hat{q}$ can assume
values inside the iper-sphere and not only on its surface, thanks to
the presence of the quantum fluctuations. On the other hand, in
analyzing the quantum model, we should think in terms of wave
functions on the sphere, whose square modulus represents a density
probability for a precise unitary vector. On this basis, the mean
value should be understood as a continuum weighted sum (integral) of
vectors, divided by a suitable normalization and nothing prevents it
to be different from $v$; actually, it may assume any value in the
interval $\left[ 0 , v \right]$, $q = 0$ corresponding to the
spherically symmetric state, where all the directions have the same
weight. In this case the wave function is uniform on the iper-sphere
(it is $Y _{00}$ in the language of the 3-D spherical harmonics).

If I perform the Legendre transform on (\ref{gen_cl_ham_qm}), I obtain 
the corresponding classical Lagrangian, in terms of the constrained 
variables $\left( q , \sigma \right)$ and their time first derivatives. I 
now define the projective coordinate $\tau$:
$$
q = \dfrac{1 - \tau ^2}{1 + \tau ^2} \,\,\, \sigma = \dfrac{2 \tau}{1 + 
\tau ^2}
$$
its conjugated momentum $T$
$$
p = - \tau T \,\,\, s = \dfrac{T}{2} \left( 1 - \tau ^2 \right)
$$
and I obtain the hamiltonian in terms of unconstrained canonically
conjugated variables:
\begin{equation}
h = \dfrac{1}{8} \left( 1 + \tau ^2 \right) ^2 \left( T ^2 + \dfrac{1}{4 
\tau ^2} \right)
\label{ham_proj}
\end{equation}

Let us now see what happens if I take the two limits in the reversed
order. In order to do that I must first define what I mean by
quantum mechanics on the sphere $S ^N$.

Let us consider a sphere of radius $R$ in the vector space $R ^{N+1}$, 
defined by the implicit equation:
$$
\sum _{i = 1} ^{N+1} x _i ^{2} = R ^2
$$
The projective coordinates are defined by:
$$
x _i = \dfrac{2 \alpha _i / R}{1 + \alpha \cdot \alpha / R ^2} \,\,\, x _{N+1} 
= \dfrac{ \alpha \cdot \alpha / R ^2 - 1}{1 + \alpha \cdot \alpha / R ^2}
$$
The metric on the sphere in the projective coordinates is given by:
$$
g _{jk} = \Omega ^2 \delta _{jk} \; \quad \rm{with} \; \Omega = \dfrac2{1
+\alpha \cdot \alpha / R ^2}
$$
I define the hamiltonian for the free motion as the Laplacian
operator on the sphere
\begin{equation}
\bigtriangleup = - \dfrac{1}{\sqrt{g}} \partial _j \sqrt{g} g ^{jk} \partial _k 
= - \Omega ^{-N} \partial _j \Omega ^{N-2} \partial _j
\label{laplacian}
\end{equation}
I neglect, here, a possible term proportional to the scalar curvature, 
because in the case of the sphere it is a constant and it modifies the 
spectrum only additively.

I can choose two different representations of the Hilbert Space of the 
system: either
$$
{\cal L} _2 \left( R ^N ; d \mu = \Omega ^N d ^N \alpha \right)
$$
or
$$
{\cal L} _2 \left( R ^N ; d \mu _0 = d ^N \alpha \right)
$$
The Laplacian in the second representation is given by
$$
\overline{\bigtriangleup} = \Omega ^{N/2} \bigtriangleup \Omega ^{-N/2} = 
-\dfrac{1}{2} \left( \Omega ^{-2} \partial ^2 + \partial ^2 \Omega
^{-2} \right) + \left( \dfrac12 + \dfrac2N \right) \dfrac{\alpha
\cdot \alpha}{N} - \dfrac{N}4 + 1
$$
where I have set $R = \sqrt{N}$ in order to implement the large $N$ limit.
I can express the result in terms of the Yaffe operators \cite{Yaffe:1982vf}:
$$
\dfrac{\alpha \cdot \alpha}{N} = 2 A \,\,\, -\dfrac{\partial ^2}{N} =
2 C
$$
to obtain the classical hamiltonian describing the large $N$ limit:
\begin{equation}
h = \dfrac18 \left( 1 + \tau ^2 \right) ^2 \left( T ^2 + \dfrac1{4 
\tau ^2} \right)
\label{proj_ham}
\end{equation}
Being this the same result obtained previously, it proves the
commutativity of the two limits.

In conclusion, I managed to show, in this very specific case
(large $N$ limit of the $O \left( N \right)$ non-linear $\sigma$
model), that the quantization procedure commutes with the
implementation of the constraint.

It is worth noticing that the formalism can be extended to include also mean 
value vectors pointing in arbitrary directions and having $n$ components.

\section{The $O(\infty)$ non linear $\sigma$ model in $1+1$ dimensions}
\label{eveq1}

\subsection{Definitions}
\label{def}
The classical Lagrangian of the $O(N+1)$ $\s$ model is given by
\begin{equation}\label{clagr}
	L = \dfrac12 \, \partial_{\mu} \bds{\phi} \cdot 
	\partial_{\mu} \bds{\phi}
\end{equation}
Where $\bds{\phi}$ is a multiplet transforming under the fundamental
representation of $O(N+1)$ and constrained to the $N-$dimensional
sphere of radius $\l^{-1/2}$:
\begin{equation}
	\phi^2 \equiv \bds{\phi} \cdot \bds{\phi} = 1/\l
\end{equation}
$\l$ may be regarded as the coupling constant, since the sphere
flattens out in the $\l\to0$ limit. The Hamiltonian corresponding to
(\ref{clagr}) reads
\begin{equation}\label{chami}
	H = \dfrac12 \int dx \left[ J^2 + (\partial_x\phi)^2 
	\right] \;,\quad J^2 = \sum_{i<j} J_{ij}^2
\end{equation}
where $J_{ij}=\phi_i\pi_j-\phi_j\pi_i$ is the angular momentum on the
sphere, $\pi_j$ being the momentum conjugated to $\phi_j$.  This
Hamiltonian can also be obtained as the $g\to\infty$ limit of the
linear model
\begin{equation}
	H_{\rm L} = \int dx \left[ \dfrac12 \pi^2 + \dfrac12 
	(\partial_x\phi)^2 + V(\phi^2)\right]
\end{equation}
where $\bds\phi$ is now unconstrained and the potential $V$ may be taken
of the form
\begin{equation}
	V(u) = \dfrac{g}4 (u - 1/\l)^2
\end{equation}
The quantum version of the linear model defines a textbook
Quantum Field Theory (apart from the nontrivial strong coupling limit
$g\to\infty$). The quantum version of the nonlinear model
(\ref{chami}) may be written instead
\begin{equation}\label{qhami}
	H = \dfrac12 \int dx \left[ -\bigtriangleup + 
	\,\om ^2\,(\partial_x \a)^2 \right]
\end{equation}
where I have used the projective coordinates $(\a_1,\ldots,\a_N)$
on the sphere, namely
\begin{equation}\label{proj}
	\phi_j = \om\,\a_j \;, \quad 
	\lbare^{1/2} \phi_{N+1} = \om - 1 \;,\quad
	\om = \dfrac{2}{1 + \lbare \a^2}
\end{equation}
so that
\begin{equation}
	(\partial_x\phi)^2 = \om ^2\,(\partial_x \a)^2 
\end{equation}
and the $O(N+1)-$symmetric functional Laplacian reads
\begin{equation}
	\bigtriangleup(x) = \om(x)^{-N} \dfrac{\delta}{\delta \a_j(x)}
	\,\om(x)^{N-2}\, \dfrac{\delta}{\delta \a_j(x)}
\end{equation}
I have replaced the coupling constant $\l$ with $\lbare$ (the {\em
bare} coupling constant) to stress the fact that in Quantum Field
Theory it is generally cut-off dependent.

\subsection{The $N\to\infty$ limit}
\label{limit}

If we consider the non linear model as a limit of the $\phi^4$ linear
model (being this true at least at the classical level), we have to
take two limits and we might wonder whether it is legitimate to
interchange their order. To be more specific, if I first perform the
large $N$ limit in the linear model, I get a classical $g-$dependent
unconstrained Hamiltonian $H^\infty_{\rm L}$, that admits a definite
non linear limit $H^\infty$ as $g\to\infty$. I verify here that indeed
the same Hamiltonian $H^\infty$ follows if I start directly form the
nonlinear quantum Hamiltonian (\ref{qhami}) and take the $N\to\infty$
\`a la Yaffe. 

Consider the quantum Hamiltonian of the linear model, with the
couplings suitably rescaled to allow the large $N$ limit
\begin{equation}\label{lqhami}
	\hat{H}_{\rm L} = \int dx\left[ \dfrac12 \hat{\pi}^2 + 
	\dfrac12(\partial_x\phi)^2  + V(\hat{\phi}^2)\right] \;,\quad
	V(u) = \dfrac{g}{4N} (u - N/\lbare)^2 
\end{equation}
According to (a slight extension of) Yaffe's rules, the quantum
dynamics described by the $N \to \infty$ limit of the model is
described by a {\em classical} Hamiltonian, which is the large $N$
limit of the expectation value of the quantum hamiltonian
(\ref{lqhami}) on a set of generalized coherent states, labelled by the
parameters defined in eq. (\ref{Nlim}). We end up with the following
classical hamiltonian
\begin{eqnarray}\label{phi_ham}
	H^\infty_{\rm L} &=& \lim_{N\to\infty}\dfrac{\vev{\hat H}_{\rm L}}N 
	= \int dx \left[ \dfrac12 \pi ^2 + 
	\dfrac12(\partial_x\phi)^2 + V(\phi^2 + w(x,x))
	\right]\\ &+& \dfrac12 \int dx\,dx'\,dx''\,v(x,x')w(x',x'')
	v(x'',x)\\ &+& \int dx \left[ \dfrac18 w ^{-1} (x,x) - \dfrac12 
	\partial_x^2 w(x,x')\bigr|_{x'=x} \right] \;,\quad 
	V(u)=\dfrac{g}{4}(u - 1/\lbare)^2
\end{eqnarray}
where the classical canonical variables are defined as
\begin{equation}\label{Nlim}
	\left\{
	\begin{array}{c}
	   \bds\phi(x) \\ \bds\pi(x) \\ w(x,x') \\ v(x,x')
	\end{array} 
	\right\}
	= \lim _{N\to\infty} \dfrac1{N}
	\left\{
	\begin{array}{c}
	   \sqrt{N} \vev{\hat{\bds\phi}(x)} \\ 
	   \sqrt{N} \vev{\hat{\bds\pi}(x)} \\ 
	   \vev{\hat{\bds\phi}(x)\cdot \hat{\bds\phi}(x')}_{\rm conn}\\ 
	   \vev{\hat{\bds\pi}(x)\cdot \hat{\bds\pi}(x')}_{\rm conn}
	\end{array}
	\right\}
\end{equation}
and the non vanishing Poisson brackets read
\begin{eqnarray}\label{PB}
	\left\{ \phi _j (x) \,,\pi _k (x') \right\}_{\rm P.B.} &=&
	\delta _{jk} \delta( x - x' ) \\
	\left\{ w(x,y)\,, v(x',y') \right\}_{\rm P.B.} &=&
	 \delta(x-x') \delta(y-y') + \delta(x-y') \delta(y-x')
\end{eqnarray}
It is understood that the dimensionality of the vectors $\bds\phi(x)$
and $\bds \pi(x)$ is arbitrary but finite [that is, only a finite
number, say $n+1$, of pairs $(\hat{\bds\phi}(x)\,,\hat{\bds\pi}(x))$
may take a non vanishing expectation value as $N\to\infty$]. Thus, the
index $j$ may run form $1$ to $n+1$, where $n+1$ is the number of
field components with non zero expectation value.

The $g\to\infty$ limit on the classical Hamiltonian $H^\infty_{\rm L}$ is
straightforward and reintroduces the spherical constraint in the new form
\begin{equation}\label{newcon}
	\sum _{j=1} ^{n+1} \phi _j ^2 + {\rm diag}(w) = 1/\lbare
\end{equation}
whose conservation in time implies 
\begin{equation}\label{newconp}
	\bds \phi \cdot \bds\pi + {\rm diag}(wv) = 0
\end{equation}
where I have introduced the condensed notation
\begin{equation}
	(ab)(x,y) \equiv \int dz\, a(x,z)\,b(z,y) \;,\quad
	{\rm diag}(a)(x) \equiv a(x,x)
\end{equation}
Let us now come back to the quantum Hamiltonian (\ref{qhami}) of 
the non-linear model. First of all I perform a similitude
transformation of the Laplacian, to cast it in a form suitable for the
application of Yaffe's method:
\begin{equation}
	\!\!\!-\overline{\bigtriangleup} = -\om^{N/2} \bigtriangleup 
	\om ^{-N/2} = \dfrac{1}{2} \left(\om^{-2} {\hat \b}^2 + 
	{\hat \b}^2 \om^{-2} \right) + \left(\dfrac12 + 
	\dfrac{2}{N} \right) \dfrac{{\hat\a}^2}{N} - \dfrac{N}{4} + 1
\end{equation}
where $\hat\a_j(x)$ is the obvious multiplication operator and
$\hat\b _j(x)=-i\delta/\delta\a_j(x)$ its conjugated momentum. Now, after
the rescaling $\lbare\to\lbare/N$ in eqs. (\ref{proj}), by the
usual rules in the $N\to\infty$ limit I obtain the classical
Hamiltonian
\begin{eqnarray}\label{sigma_ham}
	H^\infty &=& \dfrac12 \int dx \left\{\Om^{-2} \left[\b^2 + 
	{\rm diag}(\chi\eta\chi) + \dfrac14 {\rm diag}(\eta^{-1}) 
	\right] \right.\\ &+& \left. \Om^2 \left[  (\partial_x\a)^2
	+ \partial_x\partial_{x'}\eta(x,x')\bigr|_{x=x'}\right] \right\}
\end{eqnarray}
where
\begin{equation}
	\Om = \dfrac{2}{1 + \lbare[\a^2 + {\rm diag}(\eta)]}
\end{equation}
and, just as in eq. (\ref{Nlim}),
\begin{equation}
	\left\{
	\begin{array}{c}
	   \bds\a(x) \\ \bds\b(x) \\ \eta(x,x') \\ \chi(x,x')
	\end{array} 
	\right\}
	= \lim _{N\to\infty} \dfrac1{N}
	\left\{
	\begin{array}{c}
	   \sqrt{N} \vev{\hat{\bds\a}(x)} \\ 
	   \sqrt{N} \vev{\hat{\bds\b}(x)} \\ 
	   \vev{\hat{\bds\a}(x)\cdot \hat{\bds\a}(x')}_{\rm conn}\\ 
	   \vev{\hat{\bds\b}(x)\cdot \hat{\bds\b}(x')}_{\rm conn}
	\end{array}
	\right\}
\end{equation}
are classical canonically conjugated pairs, with Poisson brackets
identical to those in eq. (\ref{PB}). I take the indices of the
classical fields $\bds\a$ and $\bds\b$ to run form 1 to $n$, having
assumed that only the first $n$ components of their quantum
counterparts may have expectation values of order $\sqrt{N}$.

To show that the classical Hamiltonian $H^\infty$ is equivalent to the
$g\to\infty$ limit of $H^\infty_{\rm L}$, I need only to solve the spherical
constraint (\ref{newcon}) that emerges in that limit. This amounts to
the canonical parameterization of the constrained pairs
$(\bds\phi,\bds\pi)$ and $(w,v)$ in terms of the projective ones
$(\bds\a,\bds\b)$ and $(\eta,\chi)$. It reads
\begin{eqnarray}
	\phi_j = \Om\,\a_j \;,\qquad &
	\lbare^{1/2} \phi_{n+1} = \Om - 1 \\
	\pi_j = \Om^{-1} \b_j + \alpha _j\pi_{n+1} \;,\qquad &
	\pi_{n+1} = - \bds\a\cdot\bds\b -{\rm diag}(\eta\chi)\\
	w(x,x') = \Om(x) \Om(x')\,\eta(x,x') \;,\quad &	
	v(x,x') = \dfrac{\chi(x,x')}{\Om(x)\Om(x')} +
	\dfrac{\delta(x-x')}{\Om(x)} \pi_{n+1}(x)
\end{eqnarray}
and in particular it implies, besides (\ref{newcon}) and
(\ref{newconp}),
\begin{eqnarray}
	\pi^2 + {\rm diag}(vwv) &=& \Om^{-2} \left[ 
	\b^2 + {\rm diag}(\chi\eta\chi) \right] \\
	(\partial_x\phi)^2 + \partial_x\partial_{x'}w(x,x')\bigr|_{x=x'} 
	&=& \Om^2 \left[ (\partial_x\a)^2 + 
	\partial_x\partial_{x'}\eta(x,x')\bigr|_{x=x'}\right]
\end{eqnarray}
This result proves the complete equivalence between the $g\to\infty$
limit on the leading $1/N$ term of the linear model (which imposes the
new spherical constraint) and the $N\to\infty$ limit of the quantum
model directly formulated on the constraint manifold.

Before closing this section, it should be noticed that, even though I
gave the basic definitions and performed the entire computation for a
field theory in $1+1$ dimensions, the results in sections \ref{def} and
\ref{limit} remain valid also for a $(D+1)-$dimensional theory, the only
change being in the dimensionality of the integrals.

For completeness' sake I close this section noticing that previous
studies this subject (the definition of the non linear quantum model
as the large coupling limit of the linear model) were performed in
\cite{Chan:1987gc}, using perturbative techniques and the derivative
expansion for the model in 3+1 dimensions. The conclusion was reached
that the divergent terms are universal, while finite parts do differs
when taking the large coupling limit on the quantum corrections on the
linear model, or calculating the same quantum corrections on the
nonlinear model.

\subsection{Dynamical Evolution}
Let us now derive the evolution equation for this system in the case
the field $\hat{\bds\phi}$ has a non zero, albeit uniform, expectation
value $\bds\phi$ in the initial state. The $2-$point
functions depend only on the difference $x-x'$, and can be
parametrized by time--dependent widths $\s_k$:
\begin{equation}\label{trinv}
	w(x,x') = \intL \, \s_k^2 \,e^{ i  k(x-x')}
	\;,\quad  v(x,x') = \intL \, 
	\dfrac{\dot\s_k}{\s_k} \,e^{ i  k(x-x')}
\end{equation}
where $\Lambda$ is the ultraviolet cut-off. 

In this case of translation invariance, in practice one can always
take $n=1$ owing to the $O(n+1)$ symmetry of $H^\infty_{\rm L}$. Thus,
I choose $\bds\phi$ to have only two non--zero components. In other
words the condensate will move on the plane specified by the initial
conditions for $\bds\phi$ and its velocity. Using eq. (\ref{trinv}),
I may write the Lagrangian density corresponding to the ($g\to\infty$
limit) of the Hamiltonian (\ref{phi_ham}) as
\begin{eqnarray}
	L = \dfrac{1}{2} \left( \dot{\phi} _{1} ^{2} +\dot{\phi}_2^2
	 \right) &+& \dfrac{1}{2} \intL\left(\dot{\s}_k^2 - k^2 \s_k^2 - 
	\dfrac1{4 \s_k^2} \right)\\ &- &\dfrac{m^2}{2} 
	\left(\phi_1^2 + \phi_2^2 + \intL \s_k^2 -\dfrac1{\lbare} \right)
\end{eqnarray}
I have kept into account the constraint by introducing the Lagrange
multiplier $m^2$. The corresponding Euler--Lagrange evolution
equations read, in polar coordinates
\begin{eqnarray}
	& \ddot{\rho} + m ^2 \rho - \dfrac{\ell^2}{\rho ^3} = 0
	\label{eqm1}\\
	& \ddot{\s}_k + \left( k ^2 + m ^2 \right) \s _k - 
	  \dfrac{1}{4 \s _k ^3} = 0 \\ 
	& \rho ^2 + \Sigma - \dfrac{1}{\lbare} = 0 \label{constr}
\end{eqnarray}
with the definitions $\ell = \rho ^2 \dot{\theta}$ (the conserved angular
momentum of the condensate) and 
\begin{equation}\label{sigma}
	\Sigma = {\rm diag}(w) = \intL \s_k^2
\end{equation}
The first thing I can do is to look for the minimum of the
Hamiltonian, that is the ground state of the theory which corresponds to the
vanishing $\ddot{\rho}$, $\dot{\rho}$, $\ddot{\s} _k$, $\dot{\s}_k$
and $\ell$. The equations to solve are:
\begin{eqnarray}
	& m ^2 \rho = 0 \\
	& (k ^2 + m ^2)\,\s_k - \dfrac{1}{4 \s _k ^3} = 0
\end{eqnarray}
The solution $m = 0$ is not acceptable, because it yields a massless
spectrum for the fluctuations and gives an infrared divergence that
violates the constraint. This is nothing else than a different
formulation of the well--known Mermin-Wagner-Coleman theorem stating
the impossibility of the spontaneous symmetry breaking in $1+1$
dimensions \cite{Mermin:1966fe,Coleman:1973ci}. Thus, the unique
solution is: $\rho = 0$ and $\s _k = \frac12 (k ^2 + m ^2)^{-1/2}$.

This result allows for an interpretation of the mechanism of dynamical
generation of mass as the competition between the energy and the
constraint: in order to minimize the ``Heisenberg'' term in the
Hamiltonian, the zero mode width, that is $\s _0$, should be as
large as possible; on the other hand, it cannot be greater than a
certain value, because it must also satisfy the constraint. The
compromise generates a mass term, the same for all modes, which I
call $m_{\rm eq}$

I can take the mass at equilibrium as an independent mass scale
defining the theory, as dictated by dimensional transmutation, and
the relation between this mass scale and the bare coupling constant is
read directly from the constraint (\ref{constr}) 
\begin{equation}\label{lbare}
	\dfrac{1}{\lbare} = \dfrac{1}{2\pi} \log\left(\dfrac{\Lambda}
	{m_{\rm eq}} + \sqrt{1 + \dfrac{\Lambda ^2}{m_{\rm eq}^2}} \right) =
	\dfrac{1}{2 \pi} \log \dfrac{2 \Lambda}{m_{\rm eq}} + O\left(
	\dfrac{m_{\rm eq}^2}{\Lambda ^2} \right)
\end{equation}

When the system is out of equilibrium, the Lagrange multiplier $m$ may
depend on time. Its behavior is determined by the fact that the
dynamical variables must satisfy the constraint. After some algebra,
this parameter can be written as:
\begin{equation}\label{Lagrange}
	m^2 = \lbare \left( \dot{\rho}^2 + \dfrac{\ell^2}{\rho^2} + 
	\Theta 	\right) \;,\quad \Theta = \intL
	\left(\dot{\s}_k^2 - k^2 \s_k^2 + \dfrac{1}{4\s_k^2} \right)
\end{equation}
I can describe the quantum fluctuations also by complex mode functions $z _k$,
which are related to the real function $\s _k$ by:
\begin{equation}\label{zdot}
	z_k  = \s _k e^{ i  \t_k} \;,\quad 
	\s_k^2\,{\dot\t}_k = \ell_k=\dfrac12 \;,\quad  
	\left| \dot{z}_k \right| ^2 = \dot{\s}_k ^2 + 
	\dfrac{\ell_k^2}{\s_k^2} 
\end{equation}
One can recognize in the second term on the r.h.s. of the last
equation in (\ref{zdot}) the centrifugal energy induced by Heisenberg
uncertainty principle.

I choose the following initial conditions for this complex mode
functions:
\begin{equation}\label{init_fl}
	z_k(0) = \dfrac{1}{\sqrt{2 \om _k}} \;,\quad
	{\dot z}_k(0) = - i \sqrt{\dfrac{\om_k}2}
\end{equation}
where $\om_k =\sqrt{k ^2 + \alpha ^2}$ and $\alpha$ is an initial
mass scale. It is worth noticing here
that such a form for the initial spectrum of the quantum fluctuations
does not allow for an initial radial speed for the condensate degrees
of freedom, unless I start from $\rho_0=0$. This is easily seen by
differentiating (\ref{constr}) with respect to time. 

Moreover, I should stress that $\a$ might be different from the
initial value of the Lagrange multiplier. In fact, once the initial
value for $\rho$ is fixed, $\alpha$ can be determined by means of the
constraint equation and it turns out to be
\begin{eqnarray}\label{alfadiro}
	\alpha(\rho_0) &=& m_{\rm eq} \exp \left(2 \pi \rho_0^2
	\right) \left\{\dfrac12 \left[ 1+ \sqrt{1+\dfrac{m_{\rm
	eq}^2}{\L^2}} + \exp (4 \pi \rho_0^2) \left( 1-
	\sqrt{1+\dfrac{m_{\rm eq}^2}{\L^2}} \right) \right]
	\right\}^{-1} \nonumber \\ &=& m_{\rm eq} \exp \left(2 \pi
	\rho_0^2 \right) \left[ 1 + O\left( \dfrac{m_{\rm
	eq}^2}{\Lambda ^2} \right) \right]
\end{eqnarray}
On the other hand, the initial value for the Lagrange multiplier is
given by
\begin{eqnarray}\label{m0dia}
m^2_0 = \alpha^2 + \lbare \left( \dot{\rho} ^2 _0 +
\dfrac{\ell ^2}{\rho ^2 _0} - \a^2 \rho_0^2 \right)
\end{eqnarray}
that is equal to the initial mass scale $\alpha ^2$ only if I push
the ultraviolet cut--off to infinity.

To properly control for any time the ultraviolet behavior of the
integrals in eqs. (\ref{sigma}) and (\ref{Lagrange}), one should
perform a WKB analysis \cite{Boyanovsky:1994xf} of the solution. One finds the
following asymptotics for the mode functions:
\begin{equation}
	z_k(t) = z_k(0) \exp \left\{ikt - 
	\dfrac{ i }{2 k} \int_0^t dt'\, m^2(t') - 
	\dfrac{1}{4k^2} \left[ m^2(t) - m^2(0) \right] \right\} 
	\left[ 1 + O \left( \dfrac{1}{k^3} \right) \right]
\end{equation}
From the above formula it is clear that the logarithmic ultraviolet
divergence in $\Sigma$ is completely determined by the initial
spectrum. For the divergent integral $\Theta$ in eq. (\ref{Lagrange}) the
situation is more involved. Explicitly one finds:
\begin{equation}
	\Sigma(t) \equiv \intL |z_k(t)|^2 = 
	\dfrac{1}{2 \pi} \log \dfrac{\Lambda}{\mu} + \Sigma_{\rm F}(\mu;t) 
\end{equation}
and
\begin{equation}
	\Theta(t)=\intL \left(|{\dot z}_k(t)| ^2 - k^2 
	|z_k(t)|^2 \right) = m(t)^2 \,\Sigma(0) + \Theta_{\rm F}(t)
\end{equation}
where 
\begin{eqnarray}
	\Sigma_{\rm F}(\mu;t) &=& \intL \left[ |z_k(t)|^2 - 
	\dfrac{\t(|k| - \mu)}{2|k|} \right] \\
	\Theta_{\rm F}(t) &=& \intL\left[ |{\dot z}_k(t)|^2 - 
	k^2 |z_k(t)|^2-m(t)^2|z_k(0)|^2 \right]
\end{eqnarray}
have finite limits as $\Lambda\to\infty$. I have introduced in the
above formulae a subtraction point $\mu$. There correspond a
renormalized coupling constant $\l$ running with $\mu$, as the
$\Lambda\to\infty$ limit of the relation
\begin{equation}
	\dfrac{1}{2 \pi} \log \dfrac{\Lambda}{\mu} - 
	\dfrac{1}{\lbare(\Lambda)} + \dfrac{1}{\l(\mu)} = 0
\end{equation}
and a renormalized constraint
\begin{equation}
	\rho(t)^2 + \Sigma_{\rm F}(\mu;t) - \dfrac{1}{\l(\mu)} = 0
\end{equation}
With this definitions, the equilibrium mass scale $m _{\rm eq}$ 
can be written as
\begin{equation}
	m _{\rm eq} = 2 \mu \exp \left[-\dfrac{2\pi}{\l(\mu)}\right]
\end{equation}
which by consistency with eq. (\ref{alfadiro}) implies
\begin{equation}
	\l(\a/2) = 1/\rho_0^2
\end{equation}
In conclusion I can rewrite the constraint and the Lagrange
multiplier as
\begin{equation}\label{ren_lm}
	\rho^2 + \Sigma_{\rm F}(\mu) - \dfrac{1}{\l(\mu)} = 0 \;,\quad
	m^2 = \dfrac{1}{\rho_0^2} \left[{\dot\rho}^2 + 
	\dfrac{\ell^2}{\rho^2} + \Theta_{\rm F}\right]
\end{equation}
For large but finite UV cutoff these expressions retain a inverse
power corrections in $\Lambda$. In the actual numerical calculations 
whose results will be presented in the following section, I used the
``bare'' counterparts of eqs (\ref{ren_lm}) with finite cutoff
and the definition (\ref{lbare}) of the bare coupling constant is used
to reduce to inverse power the cutoff dependence.

Let us conclude this section by summarizing the steps I need to do,
before trying to solve numerically the equations of motion. Once I
have fixed the UV cutoff, the equilibrium mass scale $m_{\rm eq}$ and
the initial value for the condensate $\rho_0$, I can determine the
initial mass scale $\a$ in the fluctuation spectrum from
eq. (\ref{alfadiro}), which in turn gives the initial conditions for
the complex mode functions [cfr. eq. (\ref{init_fl})]. Now, I need to
specify the remaining initial values for the condensate, namely its
velocity ${\dot\rho}_0$ and its angular momentum $\ell$, which must be
consistent with the constraint (\ref{newconp}). Finally,
eq. (\ref{ren_lm}) completely determines the initial value for the
Lagrange multiplier $m_0$, which has exactly the same infinite cutoff
limit as $\a$, but differs significantly from it for finite cutoffs.

I want to stress again the need to consider how the dynamics scales
with $\L$, in a theory like this one, which does not show any
parametric resonant band (as we will see in the following
section). Even though the vanishing of the bare coupling constant
should compensate for the increase in the number of modes, when the UV
cutoff is increased, it is important to check it explicitly both on
the analytic computations and the numerical results.

\section{Numerical results}
\label{num2}
I have studied numerically the following evolution equations
\begin{eqnarray}
	&{\ddot\phi} + m^2 \phi= 0 \\ &{\ddot z}_k + \left( k ^2 + m
	^2 \right) z _k = 0 \\ &\dfrac{m^2}\lbare = |\dot\phi|^2 +
	\displaystyle{\intL} \left(|{\dot z}_k|^2 - k ^2 |z_k| ^2 \right)
\end{eqnarray}
where $\phi = \phi_1+i\phi_2 = \rho\,e^{i\t}$, $\rho^2\,\dot\t = \ell$
and $|\dot\phi|^2 = \dot{\rho}^2 + \ell^2/\rho^2$, while the bare
coupling constant $\lbare$ is given by eq. (\ref{lbare}). The initial
conditions for $\phi$, $\dot\phi$ and $z_k$ [see eq.s (\ref{init_fl})]
must satisfy the constraints (\ref{newcon}) and (\ref{newconp}), that
are then preserved by dynamics.
 
In the classical limit the quantum fluctuations $z_k$ disappear from the
dynamics. In that case the stationary solutions are trivial: 
\begin{equation}
	\rho (t) = \lbare ^{-1/2} \;,\quad  m(t) = \lbare \ell
\end{equation}
with arbitrary value for the angular momentum $\ell$. Thus there are
stationary solutions corresponding to circular motion with constant
angular velocity.

When I include the coupling with quantum fluctuations, I still
obtain stationary solutions, parametrized by $\ell$ which assumes
arbitrary positive values. They have the following form:
\begin{equation}\label{stpt}
	\rho (t) =\sqrt{\dfrac{\ell}{m _{\rm eq} x}} \; \quad \; m (t) = m _{\rm eq} x
\end{equation}
where $x$ depends on $\ell$ through 
\begin{equation}
	\dfrac{2\pi\ell}{m_{\rm eq}x} +\sinh^{-1}\left(\dfrac\L{m_{\rm eq}x}
	\right) = \sinh^{-1}\left(\dfrac\L{m_{\rm eq}} \right) 
\end{equation}
which reduces to $x\log x=2\pi\ell/ m_{\rm eq}$ in the infinite
cut--off limit.

\subsection{Evolution of condensate and Lagrange multiplier}

\begin{figure} 
\includegraphics[height=9cm,width=15cm]{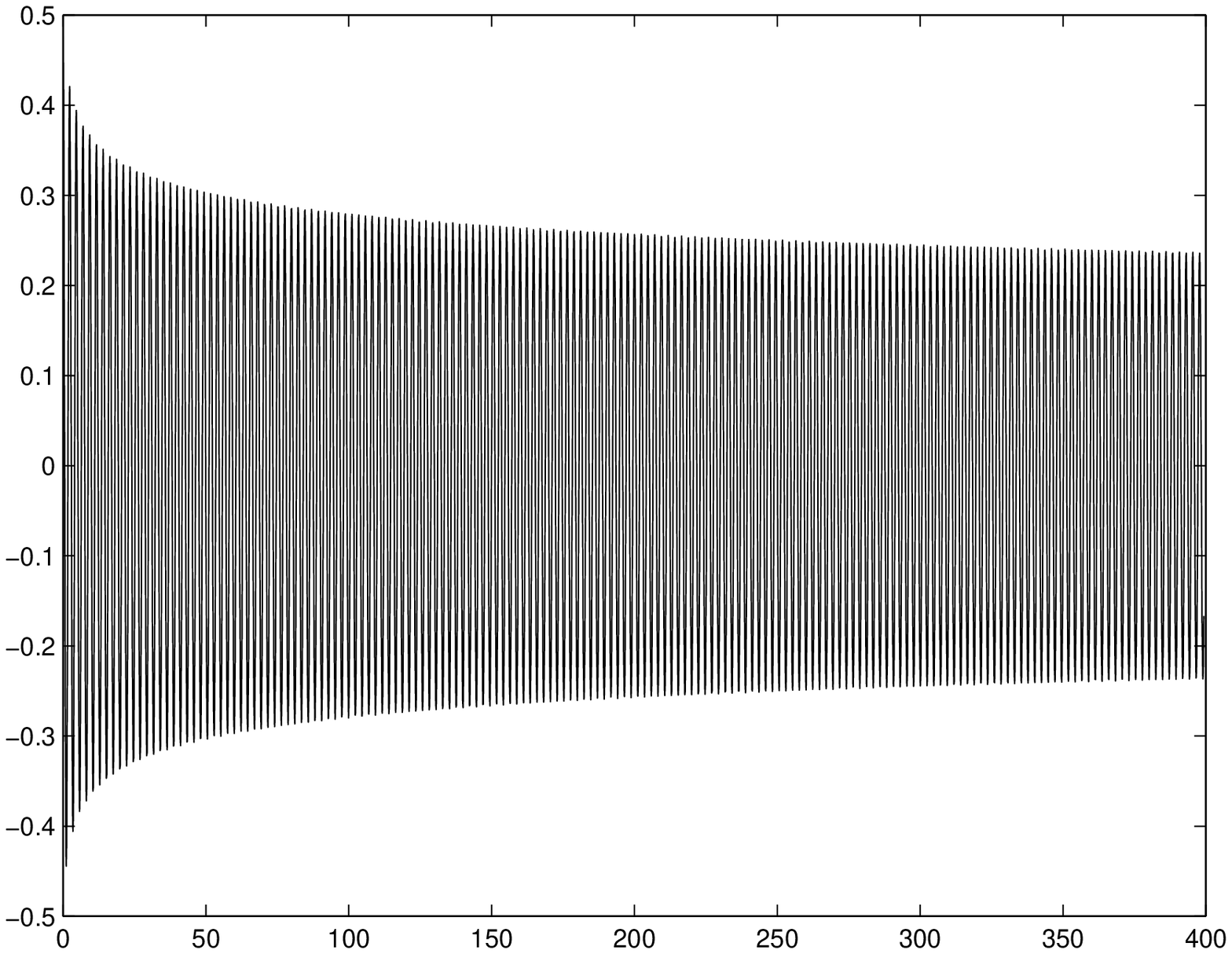}
\caption{\it Evolution of the mean value $\rho(t)$ for $\L/m
_{\rm eq} =10$, $\ell=0$ and $\rho_0=0.5$.}\label{ro}
\end{figure}

\begin{figure} 
\includegraphics[height=9cm,width=15cm]{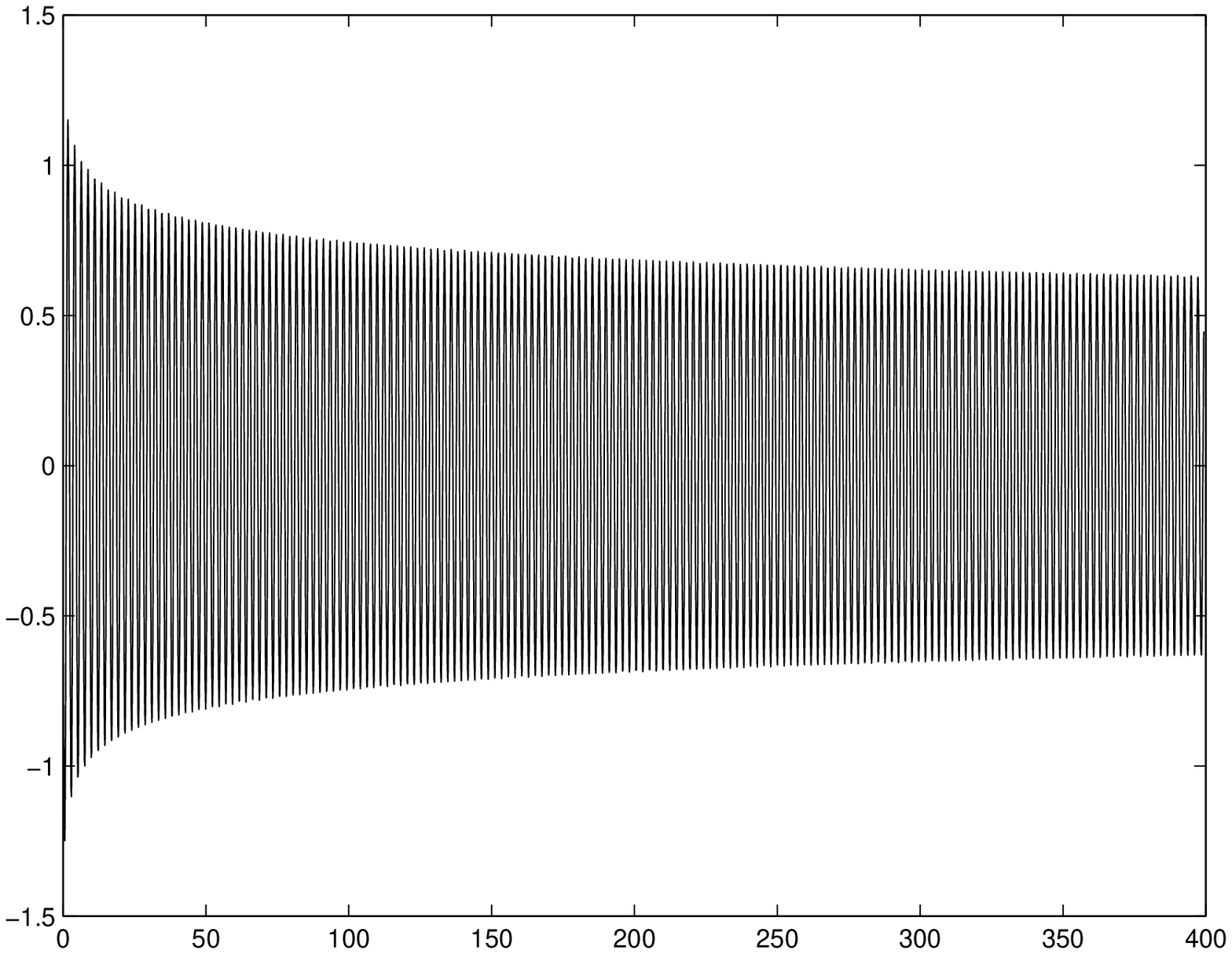}
\caption{\it Evolution of the mean value speed $\dot{\rho} (t)$ for $\L/m
_{\rm eq} =10$, $\ell=0$ and $\rho_0=0.5$.}\label{rop}
\end{figure}

\begin{figure} 
\includegraphics[height=9cm,width=14cm]{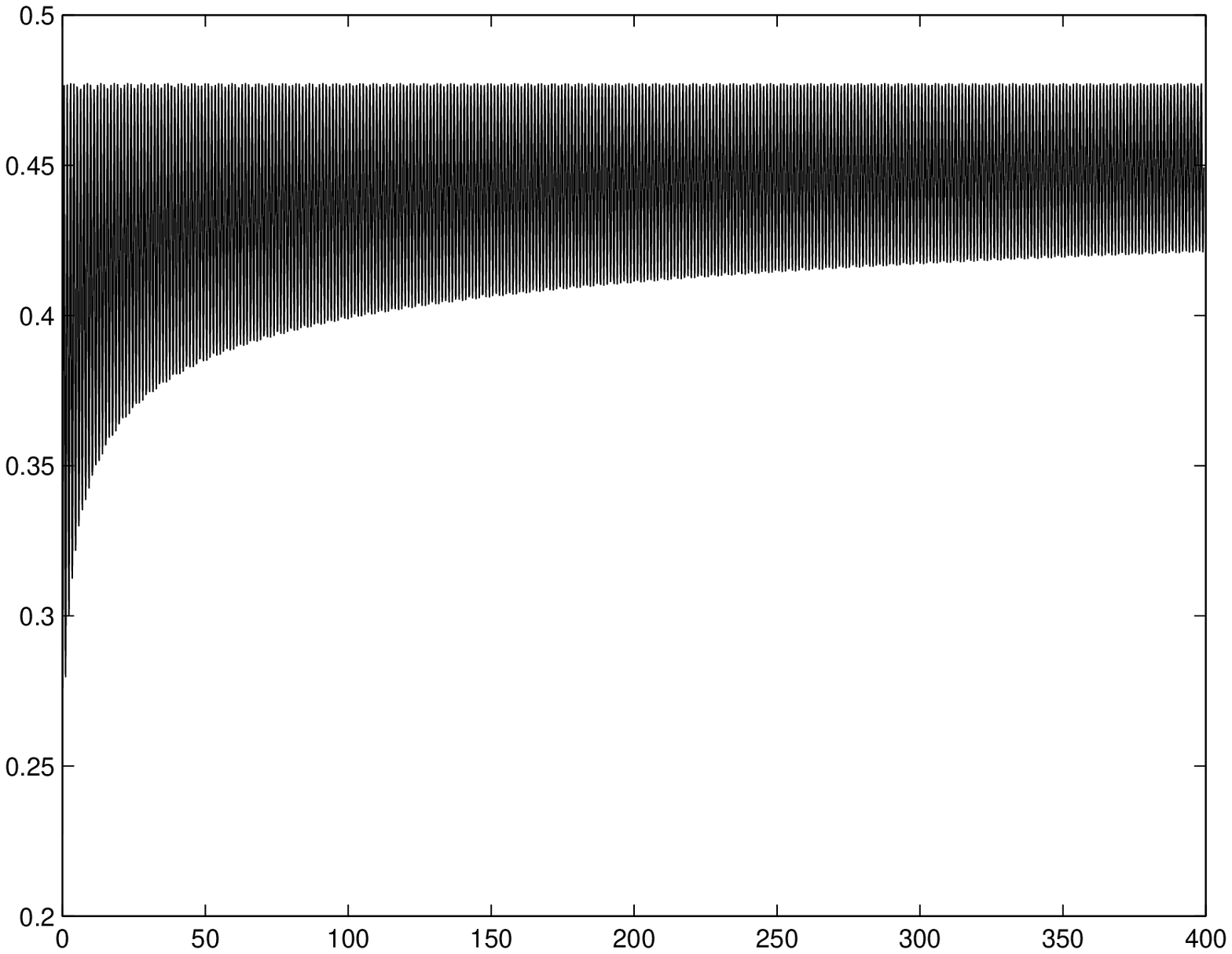}
\caption{\it Evolution of the backreaction $\Sigma (t)$ for $\L/m
_{\rm eq} =10$, $\ell=0$ and $\rho_0=0.5$.}\label{sigma-fig}
\end{figure}

\begin{figure} 
\includegraphics[height=8cm,width=16.5cm]{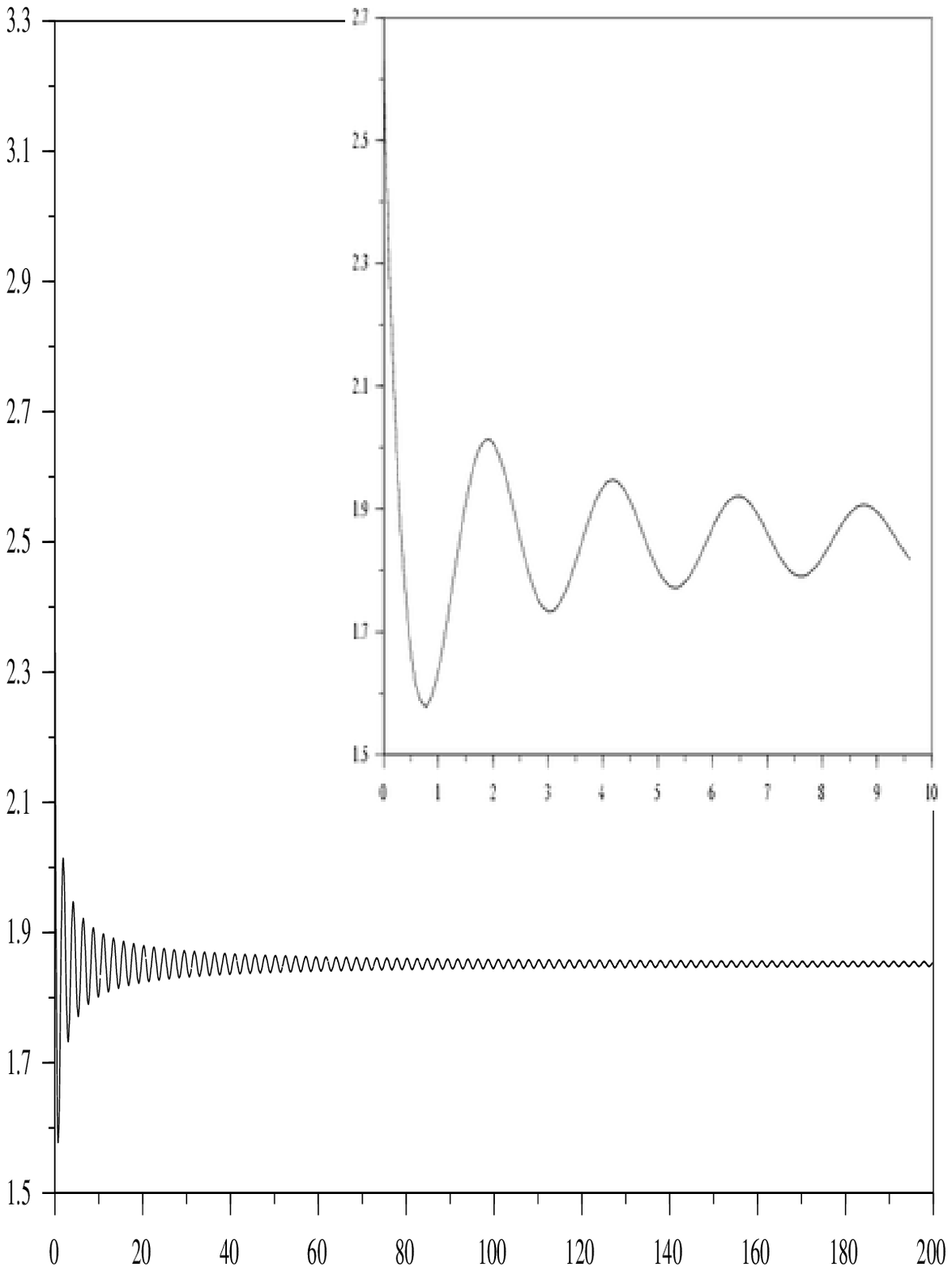}
\caption{\it Evolution of the Lagrange multiplier $m ^2 (t)$ for $\L/m
_{\rm eq} =20$, $\ell=0$ and $\rho_0=0.3$. In the smaller figure
there is zoom of the early times.}\label{lag_mult_ev}
\end{figure}
\vskip 0.5cm
\begin{figure} 
\includegraphics[height=8cm,width=16.5cm]{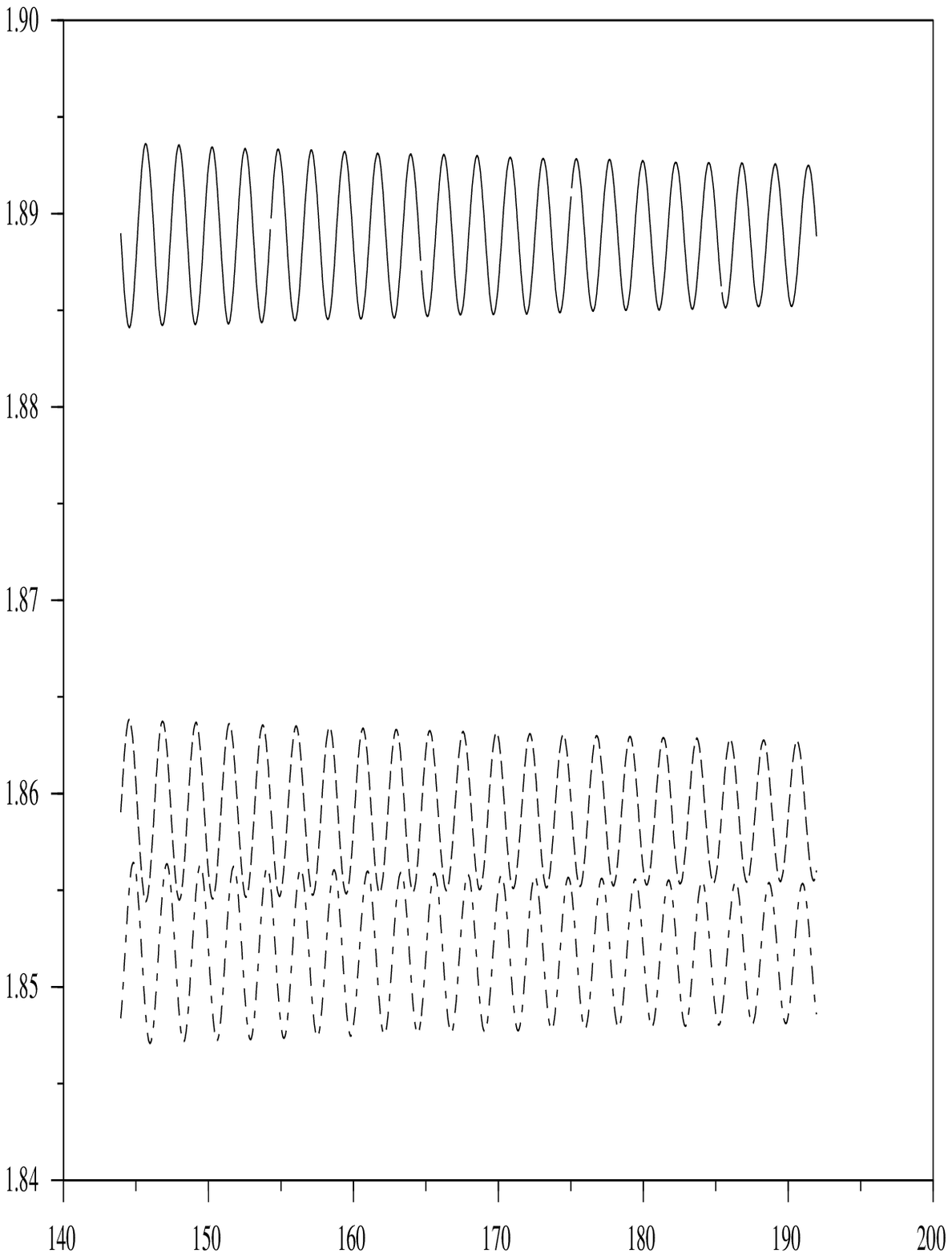}
\caption{\it Asymptotic evolution of $m ^2 (t)$ for three different
values of the ultraviolet cut--off: from top to bottom, $\L/m _{\rm
eq} = 5, 10$ and $20$, $\ell=0$ and $\rho_0=0.3$}\label{asy_sqmass}
\end{figure}

In order to control the dependence of the dynamics on the ultraviolet
cutoff, I solved the equations of motion for values of $\L$ ranging
from $5 m_{\rm eq}$ to $20 m _{\rm eq}$, with an initial condensate
ranging from $\rho_0 = 0.2$ to $\rho_0 = 0.7$. I mainly considered
the case $\ell=0$.  A typical example of the time evolution of the
relevant variables is showed in Fig.s \ref{ro}, \ref{rop} and
\ref{sigma-fig}. Figure \ref{lag_mult_ev} shows the evolution of the
Lagrange multiplier $m(t)^2$ for $\L/m _{\rm eq} = 20$; in this case,
its starting value is $2.630632$. Due to the lack of massless
particles, the damping of the oscillations of $\rho$ and $m^2$ is very
slow, as already noticed in \cite{Cooper:1997ii} for the linear model in $1+1$
dimensions; the dissipation is not as efficient as for the unbroken
symmetry scenario in $3+1$ dimensions, because of the reduced phase
space. A detailed numerical study of the asymptotic behavior and a FFT
analysis of the evolution allows a precise determination of the
asymptotic value and the main frequency of oscillation of the Lagrange
multiplier:
\begin{equation}
m(t)^2 = m _{\infty} ^2 + \dfrac{p(t)}{t}  + O\left(\dfrac{1}{t^2}\right)
\label{asymass}
\end{equation}
where the function $p(t)$ turns out to be
\begin{equation}\label{pert2}
p(t) \simeq A \cos (2 m _{\infty} t + \gamma_1 \log t + \gamma_2) 
\end{equation}
The logarithmic dependence in the phase could be justified by
self--consistent requirements (see below), along the same lines of the
detailed calculations performed in ref. \cite{Boyanovsky:1998zg} in a similar
context. Numerically it is very difficult to extract and I do not
attempt it here. Comparing further our result with that reported in
ref. \cite{Boyanovsky:1998zg}, I should emphasize that I do not find any
oscillatory component of frequency $2 m_0$, as happens instead for the
$\phi^4$ model in $3+1$ dimensions. Moreover, as figure
\ref{asy_sqmass} shows, both the asymptotic mass $m _{\infty}$ and the
amplitude $A$ depend on the ultraviolet cutoff $\Lambda$. This
dependence may be fitted with great accuracy through a low order
polynomial in $1/\L^2$, showing that the standard renormalization
holds at any time, as anticipated by the WKB analysis. Therefore, the
extrapolated parameters $m_{\infty}^2$ and $A$ give us information on
the fully renormalized physical theory (in the large $N$
approximation). The table below collects the values of $m_{\infty}^2$
for different values of $\Lambda$ and of the initial condensate
$\rho_0$.  The last column contains the extrapolation to infinite
cutoff, obtained by the low order polynomial fit. The empty cells in
the last row correspond to a UV cutoff so small that the exact
$\alpha^2$ turns out to be negative; these values are excluded from
the fit.

\vskip .75truecm
\begin{tabular}{|c|c|c|c|c|c|c|c|c|c|}
\hline
$\rho _0$ & $\L=5$ & $\L=6$ & $\L=7$ & $\L=8$ &
$\L=9$ & $\L=10$ & $\L=11$ & $\L=12$ & $\L=13$ \\
\hline
$0.2$ & $1.3073$ & $1.3047$ & $1.3032$ & $1.3022$ & $1.3014$ &
$1.3010$ & $1.3006$ & $1.3004$ & $1.3001$ \\
\hline
$0.3$ & $1.8888$ & $1.8766$ & $1.8693$ & $1.8646$ &
$1.8614$ & $1.8591$  & $1.8574$ & $1.8561$ & $1.8551$ \\ 
\hline
$0.4$ & $3.3869$ & $3.3162$ & $3.2747$ & $3.2482$ &
$3.2303$ & $3.2175$ & $3.2082$ & $3.2011$ & $3.1956$ \\ 
\hline
$0.5$ & $8.7094$ & $7.9915$ & $7.6082$ & $7.3764$ & $7.2246$
& $7.1193$ & $7.0432$ & $6.9861$ & $6.9424$ \\ 
\hline
$0.6$ & $206.03$ & $52.433$ & $35.564$ & $29.2276$ & $25.969$ & $24.016$ 
& $22.732$ & $21.835$ & $21.178$ \\ 
\hline
$0.7$ & & & & & & & & & $238.12$ \\ 
\hline
\end{tabular}
\vskip 1cm
\begin{tabular}{|c|c|c|c|c|c|c|c|c|}
\hline
$\rho _0$ & $\L=14$ & $\L=15$ & $\L=16$ & $\L=17$ & $\L=18$ & 
$\L=19$ & $\L=20$ & $\L=\infty$ \\ 
\hline
$0.2$ & $1.300$ & $1.2998$ & $1.2997$ & $1.2996$ & $1.2995$ & $1.2995$
& $1.2994$ & $1.2989$\\
\hline
$0.3$ & $1.8543$ & $1.8536$ & $1.8531$ & $1.8527$ & $1.8523$ & 
$1.8520$ & $1.8517$ & $1.8493$ \\ 
\hline
$0.4$ & $3.1912$ & $3.1877$ & $3.1848$ & $3.1824$ & $3.1805$ & 
$3.1788$ & $3.1774$ & $3.1643$ \\ 
\hline
$0.5$ & $6.9080$ & $6.8804$ & $6.8580$ & $6.8395$ & $6.8241$ & 
$6.8111$ & $6.8001$ & $6.6990$ \\ 
\hline
$0.6$ & $20.682$ & $20.295$ & $19.988$ & $19.740$ & $19.536$ & 
$19.366$ & $19.223$ & $16.964$ \\ 
\hline
$0.7$ & $177.645$ & $146.523$ & $127.66$ & $115.07$ & $106.11$ &
$99.442$ & $94.302$ & $68.207$ \\
\hline
\end{tabular}
\vskip .75truecm

A similar table can be provided for the amplitude $A$ in the
eq. (\ref{pert2}). The values extrapolated to infinite cutoff in a
similar fashion as before, turn out to be:

\vskip .75truecm
\begin{tabular}{|c|c|c|c|c|c|c|}
\hline
$\rho_0$ & $0.2$ & $0.3$ & $0.4$ & $0.5$ & $0.6$ & $0.7$ \\
\hline
$A(\L=\infty)$ & $0.539$ & $0.701$ & $0.924$ & $1.39$ & $2.34$ &
$4.30$ \\
\hline 
\end{tabular}
\vskip .75truecm \noindent 
However, this fit is not as accurate as that for $m_{\infty}$.

It is interesting to observe that at large UV cutoff
$m_{\infty}$ has an exponential dependence on $\rho_0$ analogous to
that of $m_0$ (which coincides to $\a$ at $\Lambda=\infty$). Most
remarkably the prefactor in the exponent is modified by the time
evolution: I find 
\begin{equation}\label{minfrho}
	m_{\infty}^2 \sim \exp(2\g\, \rho_0^2) \;,\quad 
	3.5 < \g < 4.5
\end{equation}
The determination of $\g$ is rather rough due to the uncertainties
in the values of $m_{\infty}$ extrapolated to $\Lambda=\infty$ at larger
$\rho_0$. Notice in any case that the analog of $\g$ for $m_0$ is
$2\pi=6.28\ldots$. 

\begin{figure} 
\includegraphics[height=9cm,width=15cm]{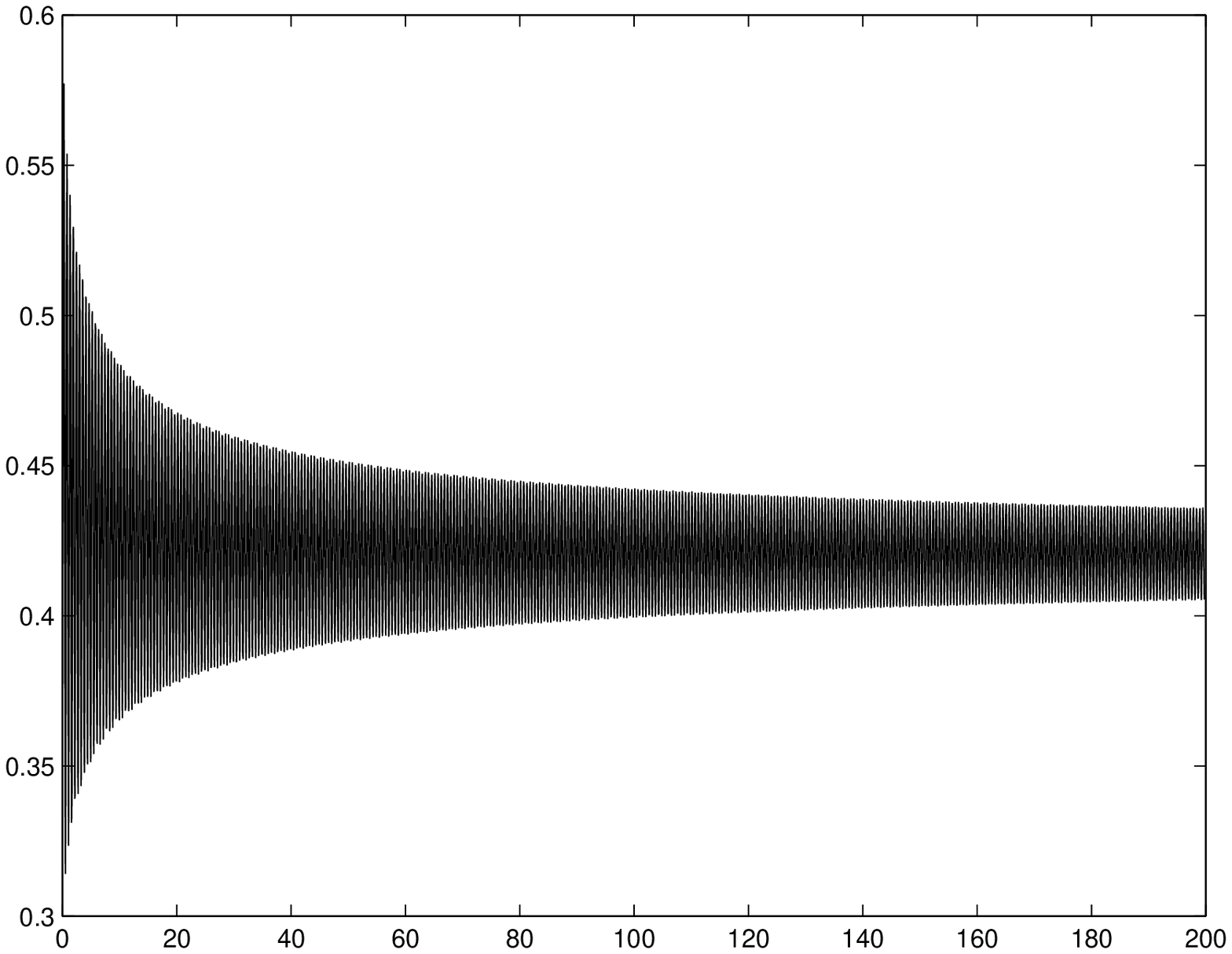}
\caption{\it Evolution of the mean value $\rho$ for $\L/m _{\rm eq} =
10$, $\rho_0=0.3$ and $\ell=1$.}
\label{lneq0_1}
\end{figure}

\begin{figure} 
\includegraphics[height=9cm,width=15cm]{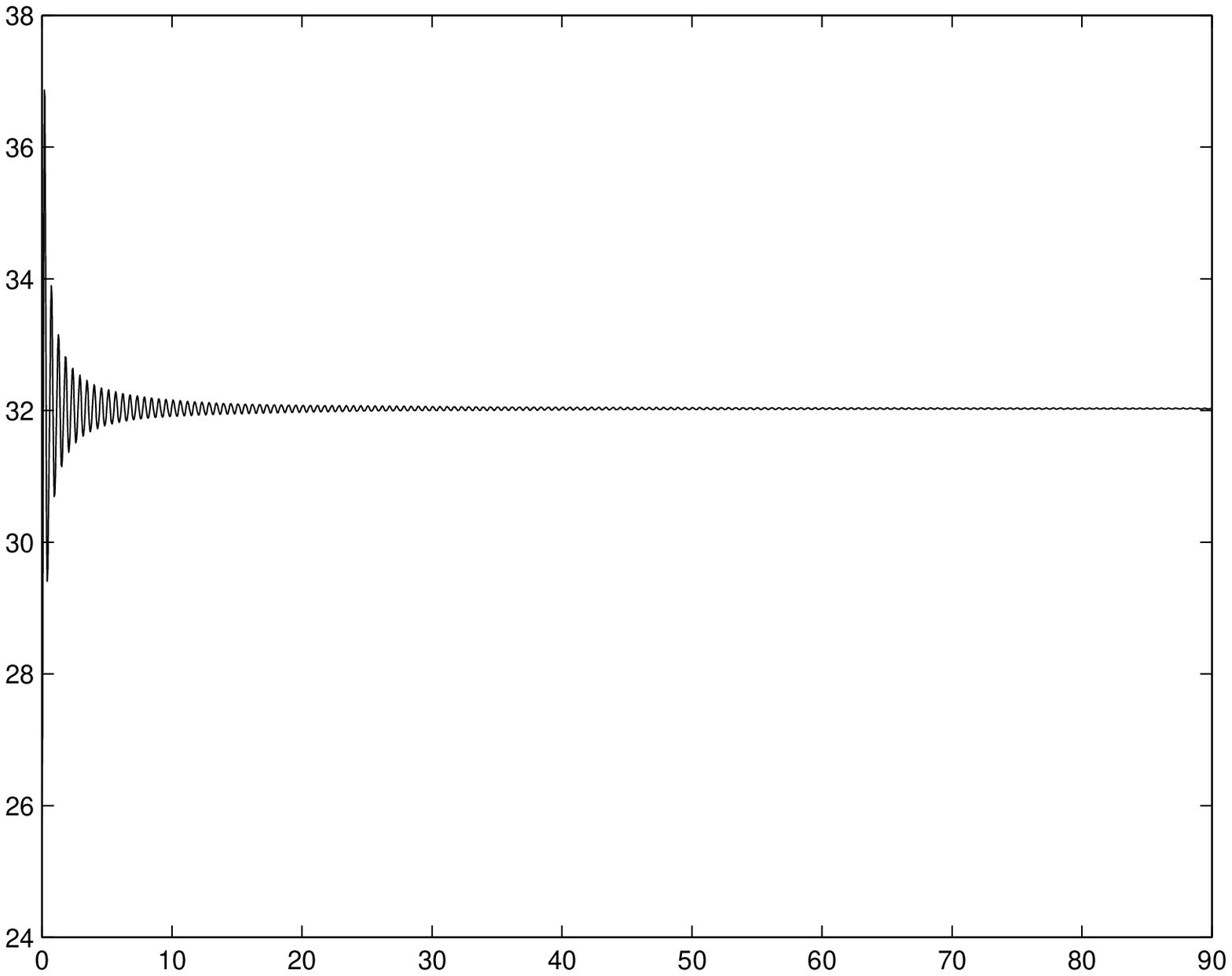}
\caption{\it Evolution of the squared mass $m^2$ for $\L/m _{\rm eq} =
10$, $\rho_0=0.3$ and $\ell=1$.}
\label{lneq0_2}
\end{figure}

I also performed some computations for $\ell>0$, with the following
results: if I start from an out of equilibrium value for $\rho$, it
will relax through emission of particles towards a fixed point,
different from the equilibrium value determined by eq.
(\ref{stpt}). Figures \ref{lneq0_1} and \ref{lneq0_2} show such a
situation for $\ell=1.0$, $\rho(0)=0.3$ and $\L = 10 m_{\rm eq}$. In that
case I have $x=1.000057$, while the mean values of the asymptotic
oscillations are $\rho _{\infty} = 0.4203$ and $m^2_{\infty} = 32.0294$.

\begin{figure} 
\includegraphics[height=8cm,width=15cm]{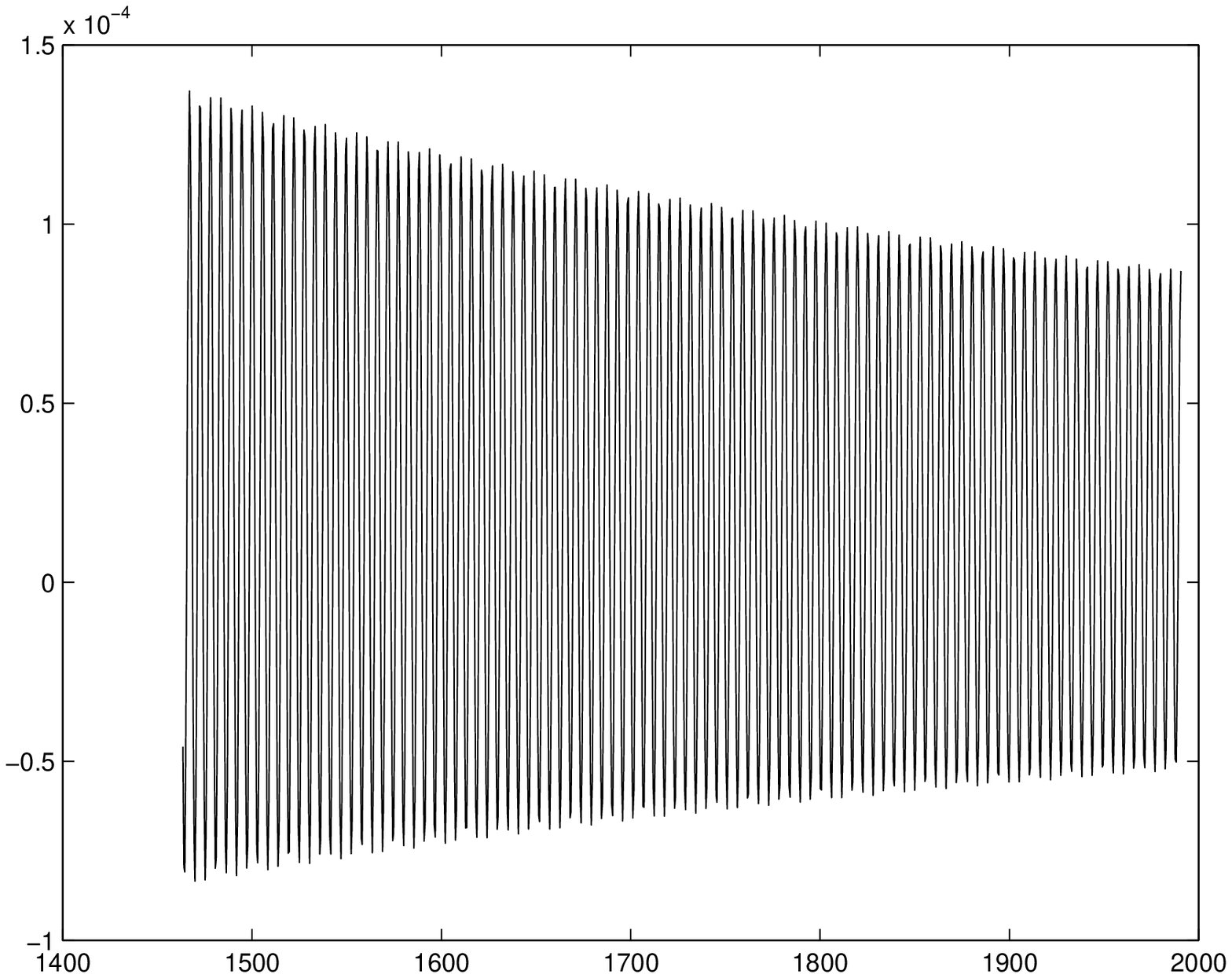}
\caption{\it The average value of the condensate $\rho$, defined as
$\bar{\rho}=\int^T\rho(t)dt/T$, plotted vs. $T$, for $\L/m_{\rm
eq}=20$, $\ell=0$ and $\rho_0=0.2$.}\label{rob}
\end{figure}
\vskip 0.5cm
\begin{figure} 
\includegraphics[height=8cm,width=15cm]{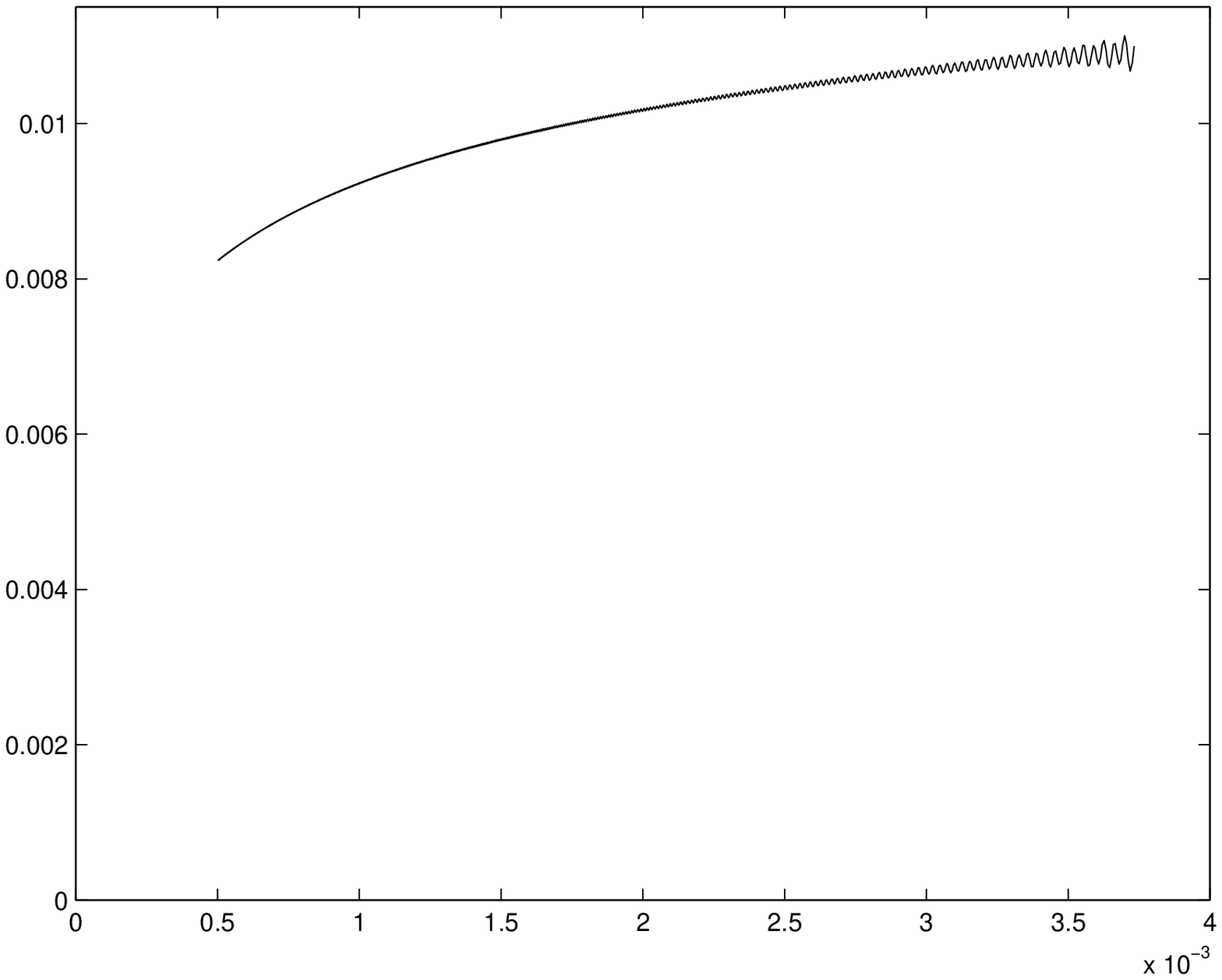}
\caption{\it The mean squared fluctuations of the condensate $\rho$, defined
as $\int^T(\rho(t)-\bar{\rho})^2dt/T$, plotted vs. $1/T$, for the same
values of the parameters as in figure \ref{rob}.}
\label{sq_flct}
\end{figure}

Before closing this section, I should comment a little further on the
evolution of the condensate $\rho$. When $\ell=0$, fig. \ref{rob}
shows that the oscillations are actually around zero. Because of the
reduced momentum phase space, I observe that the damping of the
condensate is not as efficient as in the large$-N$ $\phi^4$ model in
$3+1$ dimension with unbroken symmetry. However, from the available
data, it is not possible to decide whether the amplitude will
eventually vanishes or will tend to a limiting cycle (see
fig. \ref{sq_flct}).

On the other hand, in the case of $\ell\neq 0$, it is already clear
that the condensate does not relax to the state of minimum energy
compatible with the given value of $\ell$, which would correspond to
the circular orbit with radius given by eq. (\ref{stpt}). However, it
may still relax to a circular orbit with a different radius and a
different (larger) energy. More detailed and longer numerical
computations are needed to decide whether the damping reduces the
oscillation amplitude to zero or not.

\subsection{Emission spectrum}
\begin{figure} 
\includegraphics[height=9cm,width=15cm]{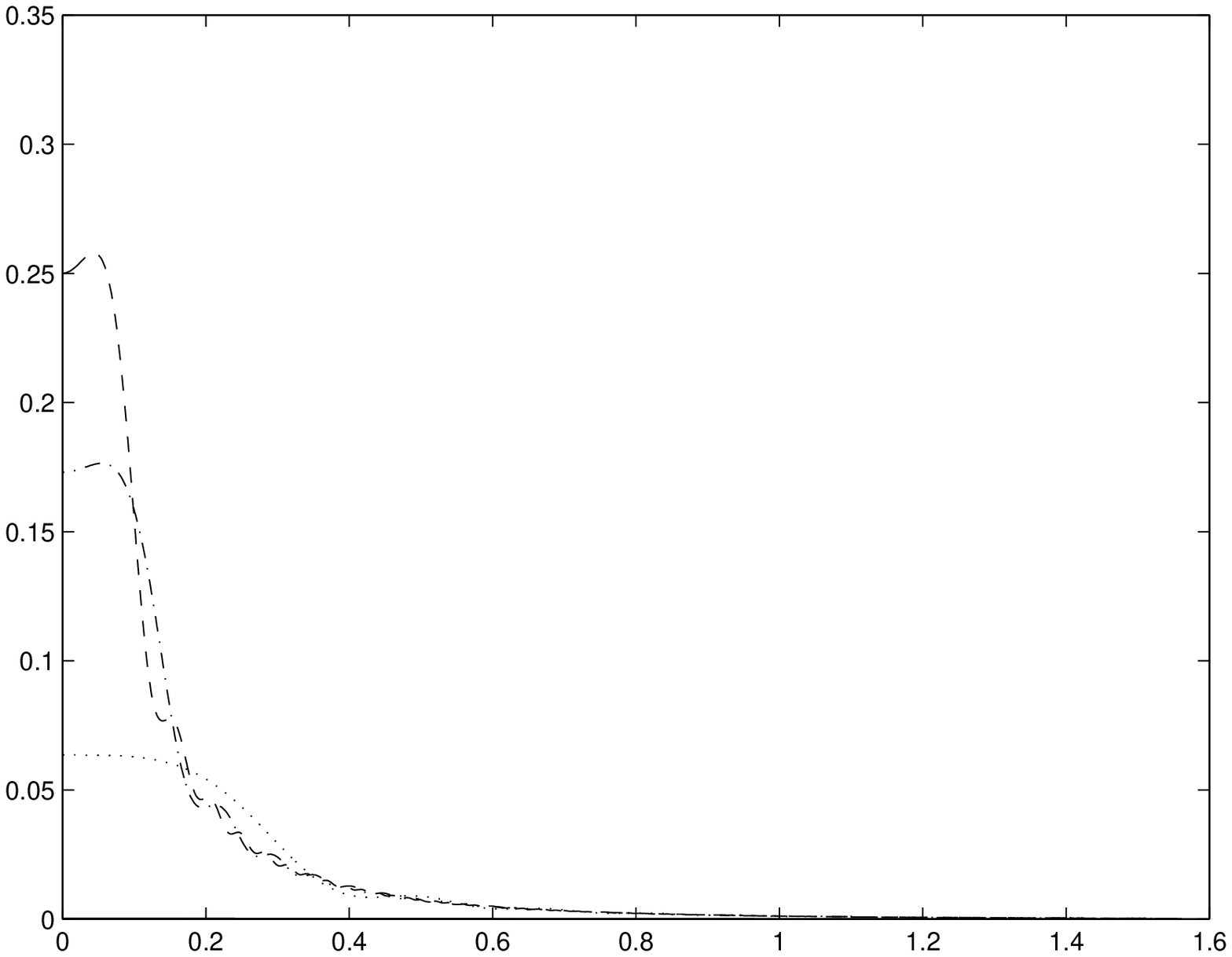}
\caption{\it Adiabatic spectrum for $tm_{\rm eq}=0$ (solid line),
$39.723$ (dotted line), $199.006$ (dashdot line) and $398.11$
(dashed line), for $\L/m _{\rm eq} = 20$, $\ell=0$ and $\rho_0=0.2$.}
\label{spec_1}
\end{figure}

\begin{figure} 
\includegraphics[height=8cm,width=15cm]{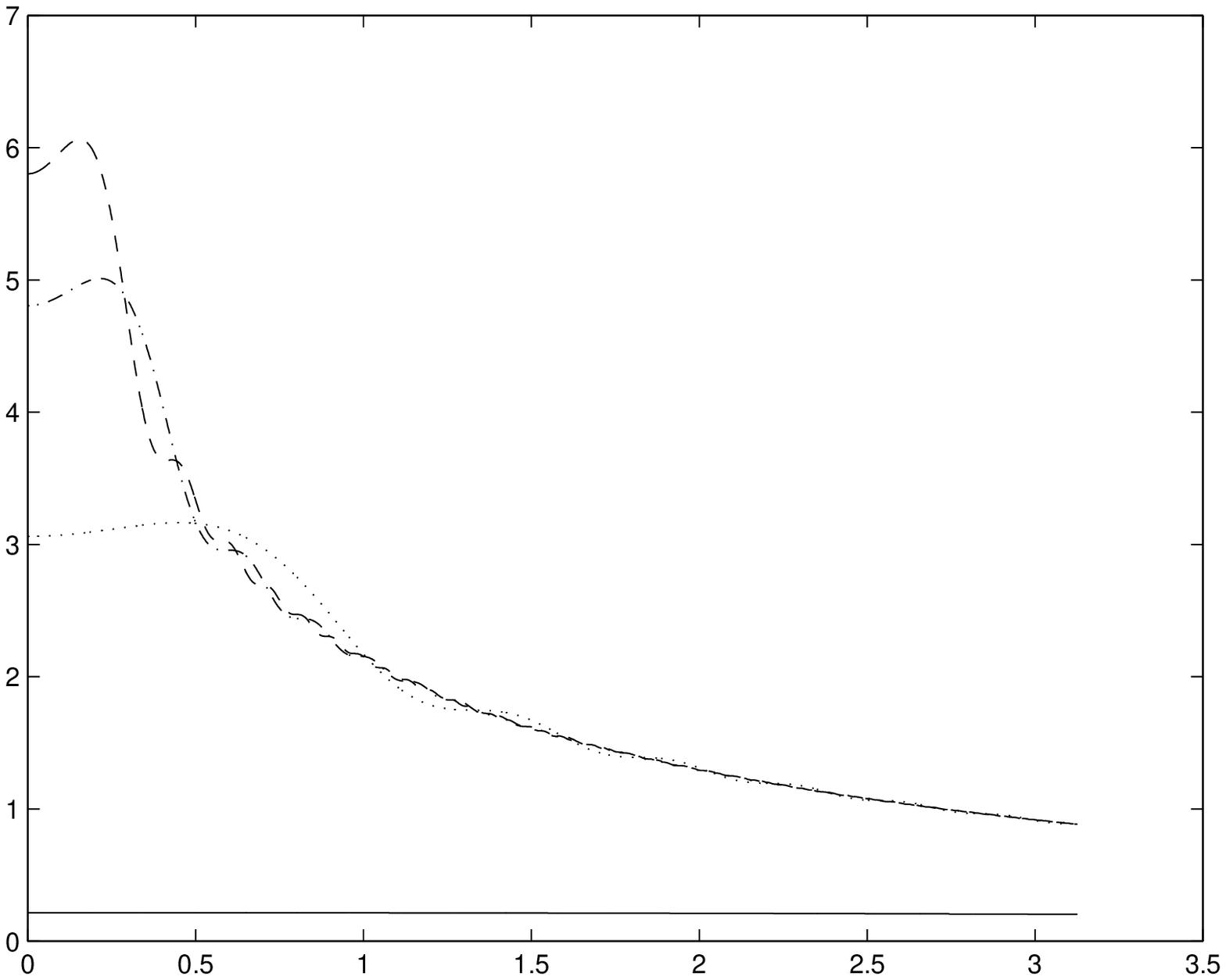}
\caption{\it Adiabatic spectrum for $tm_{\rm eq}=0$ (solid line),
$39.723$ (dotted line), $199.006$ (dashdot line) and $398.11$
(dashed line), for $\L/m _{\rm eq} = 20$, $\ell=0$ and $\rho_0=0.7$.}
\label{adnum0_7}
\end{figure}

\begin{figure} 
\includegraphics[height=8cm,width=15cm]{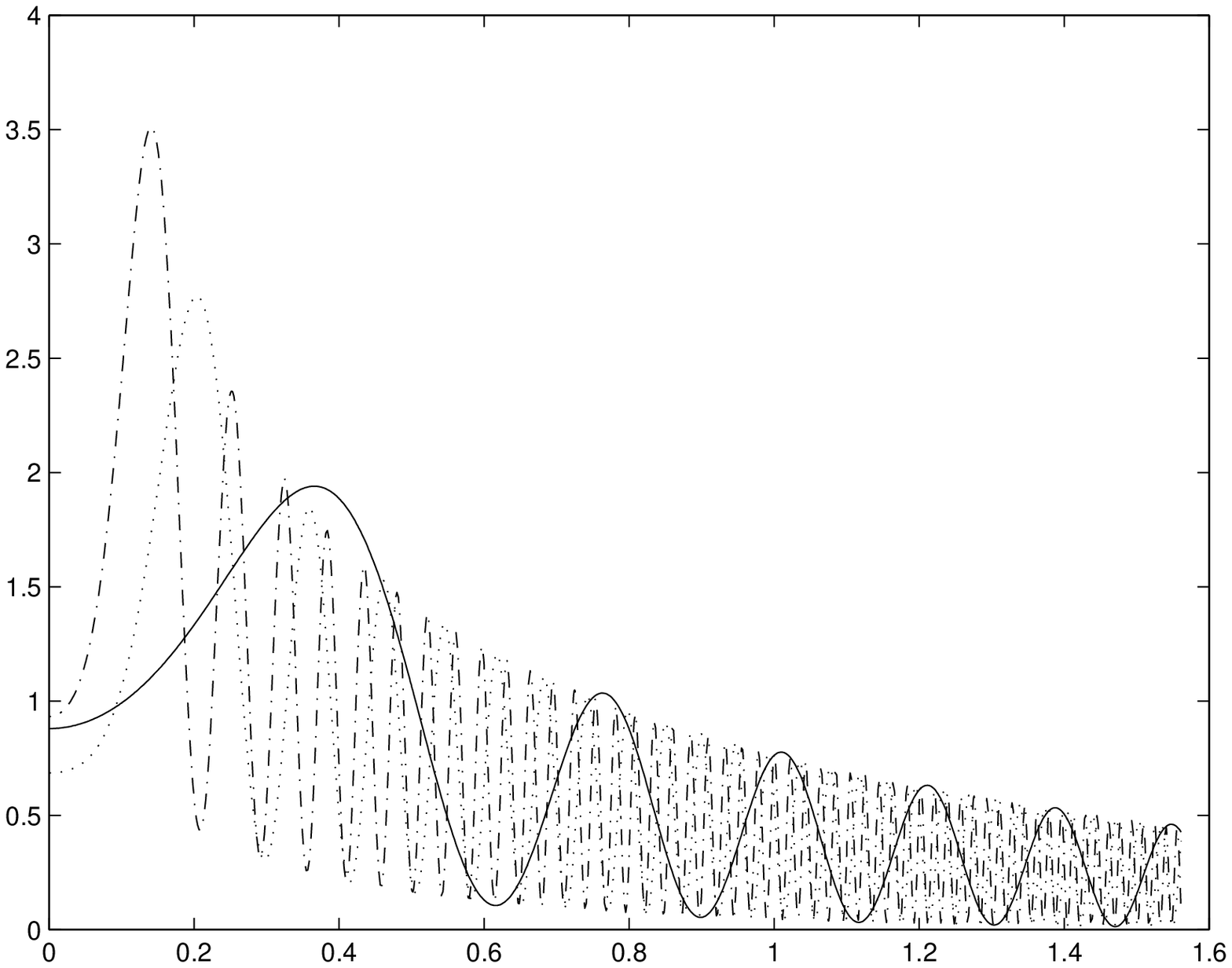}
\caption{\it Spectrum with respect to the initial vacuum, for $tm_{\rm
eq}=39.723$ (solid line), $199.006$ (dotted line) and $398.11$ (dashdot
line), for $\L/m_{\rm eq} = 20$, $\ell=0$ and $\rho_0=0.5$.}
\label{inspec}
\end{figure}

\begin{figure} 
\includegraphics[height=8cm,width=15cm]{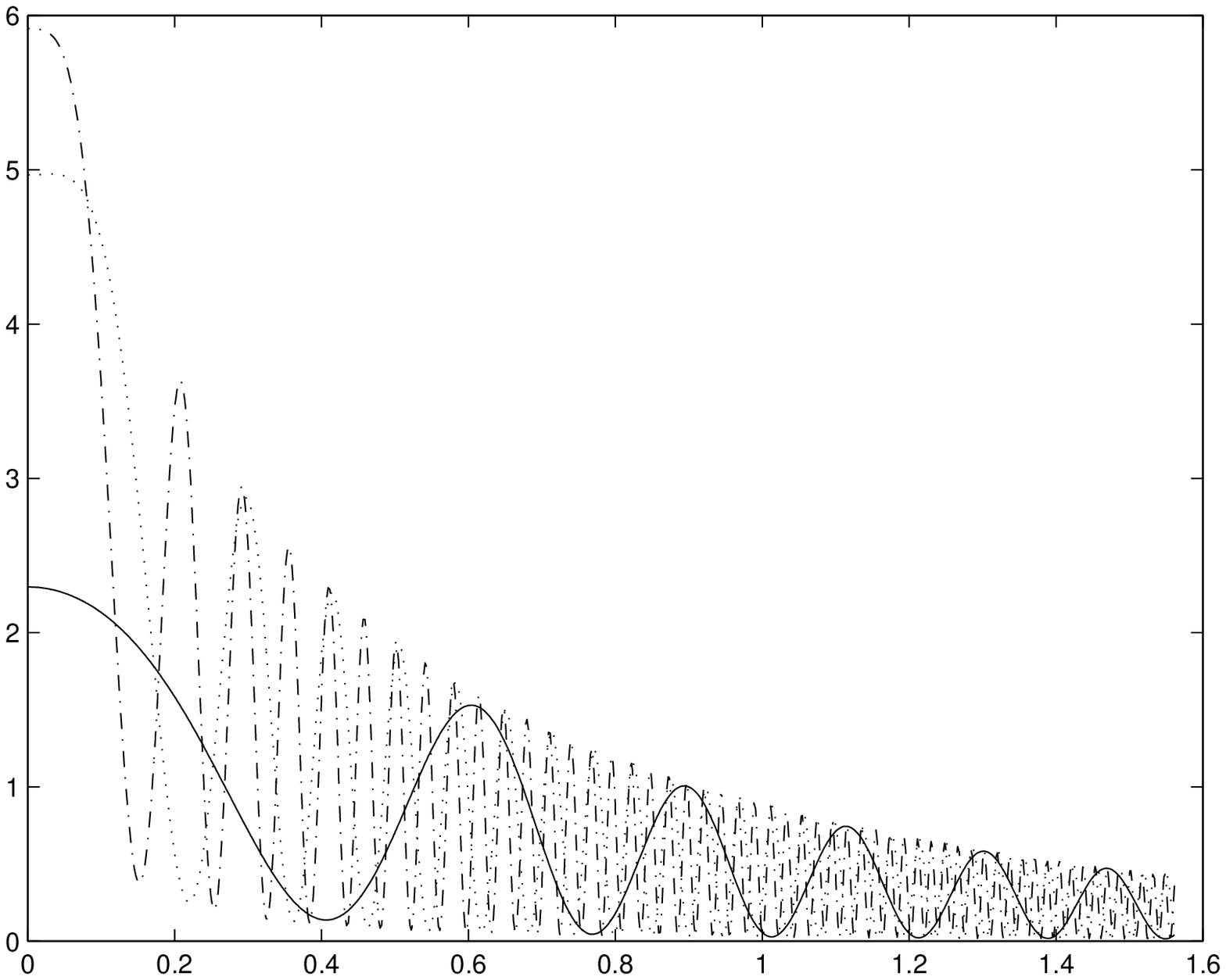}
\caption{\it Spectrum with respect to the true vacuum, for $tm_{\rm
eq}=39.723$ (solid line), $199.006$ (dotted line) and $398.11$ (dashdot
line), for $\L/m_{\rm eq} = 20$, $\ell=0$ and $\rho_0=0.5$.}
\label{eqspec}
\end{figure}

\begin{figure} 
\includegraphics[height=8cm,width=15cm]{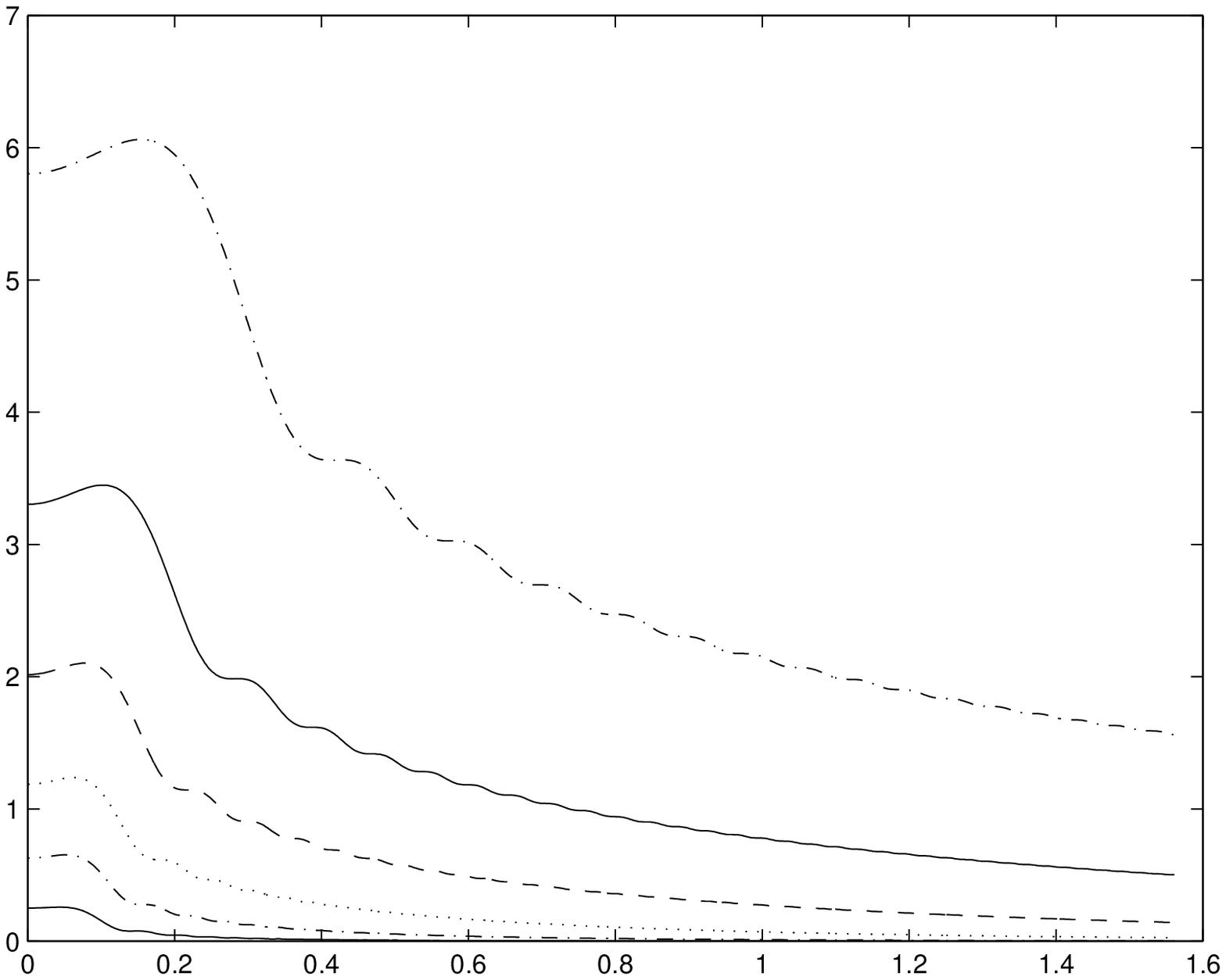}
\caption{\it Adiabatic spectrum for $tm_{\rm eq}=398.11$, $\L/m _{\rm
eq} = 20$ and $\ell=0$. The different curves correspond to different
initial values for the condensate: from top to bottom, $\rho_0=0.7$,
$0.6$, $0.5$, $0.4$, $0.3$ and $0.2$.}
\label{asy_sp}
\end{figure}

Once the evolution equations for the complex mode functions has been
solved, it is possible to compute the spectrum of the produced
particles. First, I should say that the notion of particle number is
ambiguous in a time dependent situation. Nevertheless, I may give a
suitable definition with respect to some particular pointer state. I
choose here two particular definitions, the same already used in the
study of the $\phi^4$ model \cite{Boyanovsky:1998zg}, plus a third one. The 
first choice corresponds to defining particles with respect to the initial
Fock vacuum state, the second with respect to the instantaneous
adiabatic vacuum state, and the third to the equilibrium vacuum (the
true vacuum of the theory). The corresponding expressions in terms of
the complex mode functions are:
\begin{eqnarray}
	&N^{\rm in}_k(t) = \dfrac14 \left[ \om_k |z_k (t)|^2 
		+ \dfrac{|\dot{z}_k(t)|^2}{\om_k} \right] - \dfrac12 \\
	&N^{\rm ad}_k(t) = \dfrac14 \left[ \om^{\rm ad}_k |z_k(t)|^2 + 
		\dfrac{|\dot{z}_k(t)|^2}{\om^{\rm ad}_k} \right] -\dfrac12
		\;,\quad \om^{\rm ad}_k = \sqrt{k^2+m(t)^2} \\
	&N^{\rm eq}_k(t) = \dfrac14 \left[ \om^{\rm eq}_k |z_k(t)|^2 + 
		\dfrac{|\dot{z}_k(t)|^2}{\om^{\rm eq}_k} \right] - \dfrac12
		\;,\quad \om^{\rm eq}_k = \sqrt{k^2+m_{\rm eq}^2} \\
\end{eqnarray}
I report my numerical findings on these quantities in
figs. \ref{spec_1} - \ref{asy_sp}. Since the Lagrange multiplier tends
asymptotically to a constant value $m^2_{\infty}$, the condensate
$\rho(t)$ oscillates with frequency $m_{\infty}$ and the mode
functions $z_q(t)$ with frequency $\omega (q) =
\sqrt{q^2+m^2_{\infty}}$. This implies that particle spectra $N^{\rm
in}_k(t)$ and $N^{\rm eq}_k(t)$ are more and more strongly modulated
as time elapses, as figs. \ref{inspec} and \ref{eqspec} show; on the
contrary, $N^{\rm ad}_k(t)$ is a slowly varying function of the
momentum $k$ (cfr. figs. \ref{spec_1} - \ref{adnum0_7}), because the
oscillations of the mode functions are counterbalanced by the time
dependence of the adiabatic frequencies $\sqrt{k^2+m(t)^2}$. Finally,
fig. \ref{asy_sp} allows for a comparison of the spectra related to
different initial values of the condensate. 
\begin{figure} 
\includegraphics[height=8cm,width=15cm]{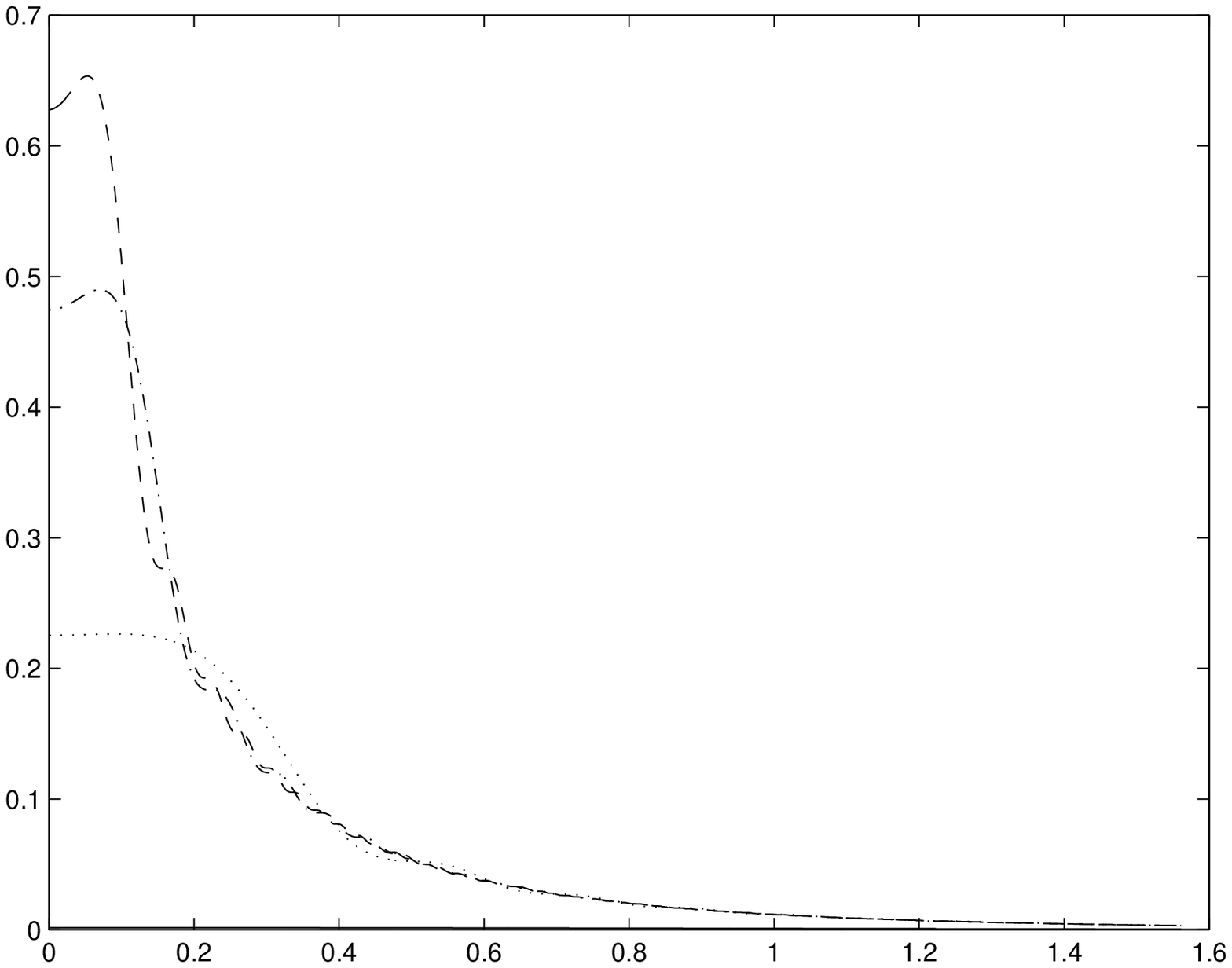}
\caption{\it Adiabatic spectrum for $tm_{\rm eq}=0.0$ (solid line),
$39.723$ (dotted line), $199.006$ (dashdot line) and $398.11$
(dashed line), for $\L/m _{\rm eq} = 20$, $\ell=0$ and $\rho_0=0.3$.}
\end{figure}
\vskip 0.5cm
\begin{figure} 
\includegraphics[height=9cm,width=15cm]{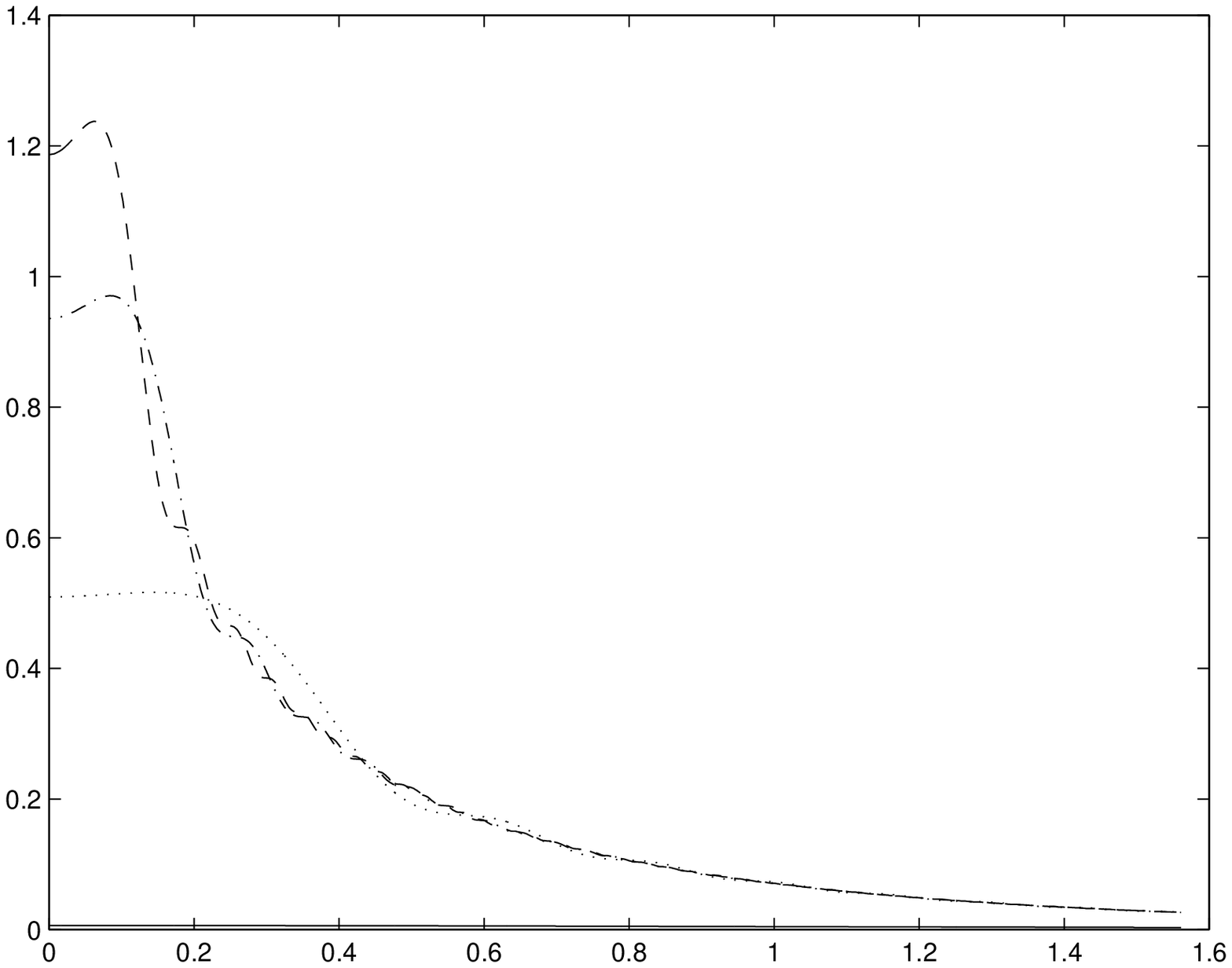}
\caption{\it Adiabatic spectrum for $tm_{\rm eq}=0$ (solid line),
$39.723$ (dotted line), $199.006$ (dashdot line) and $398.11$
(dashed line), for $\L/m _{\rm eq} = 20$, $\ell=0$ and $\rho_0=0.4$.}
\end{figure}
\begin{figure} 
\includegraphics[height=8cm,width=15cm]{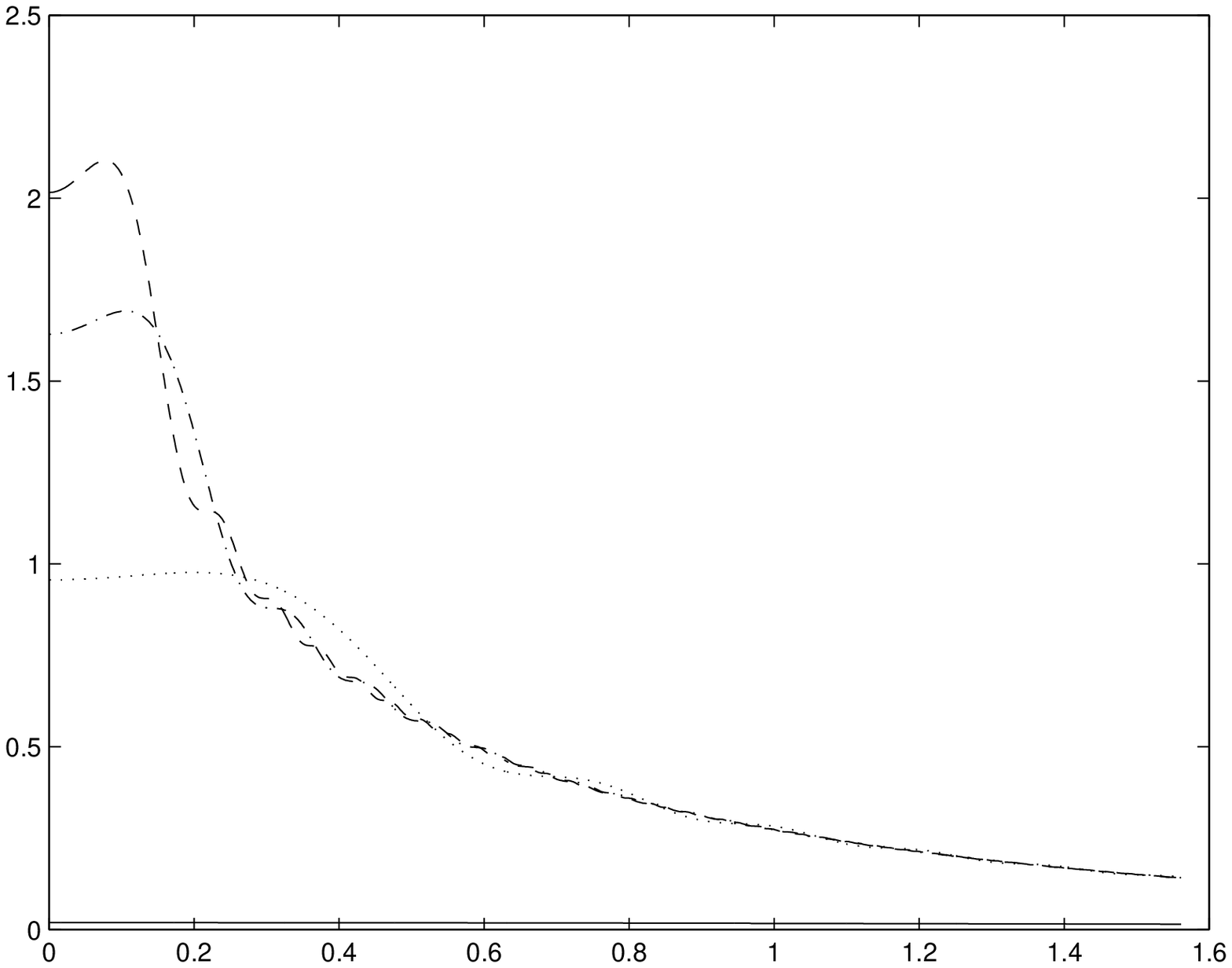}
\caption{\it Adiabatic spectrum for $tm_{\rm eq}=0$ (solid line),
$39.723$ (dotted line), $199.006$ (dashdot line) and $398.11$
(dashed line), for $\L/m _{\rm eq} = 20$, $\ell=0$ and $\rho_0=0.5$.}
\end{figure}
\vskip 0.5cm
\begin{figure} 
\includegraphics[height=9cm,width=15cm]{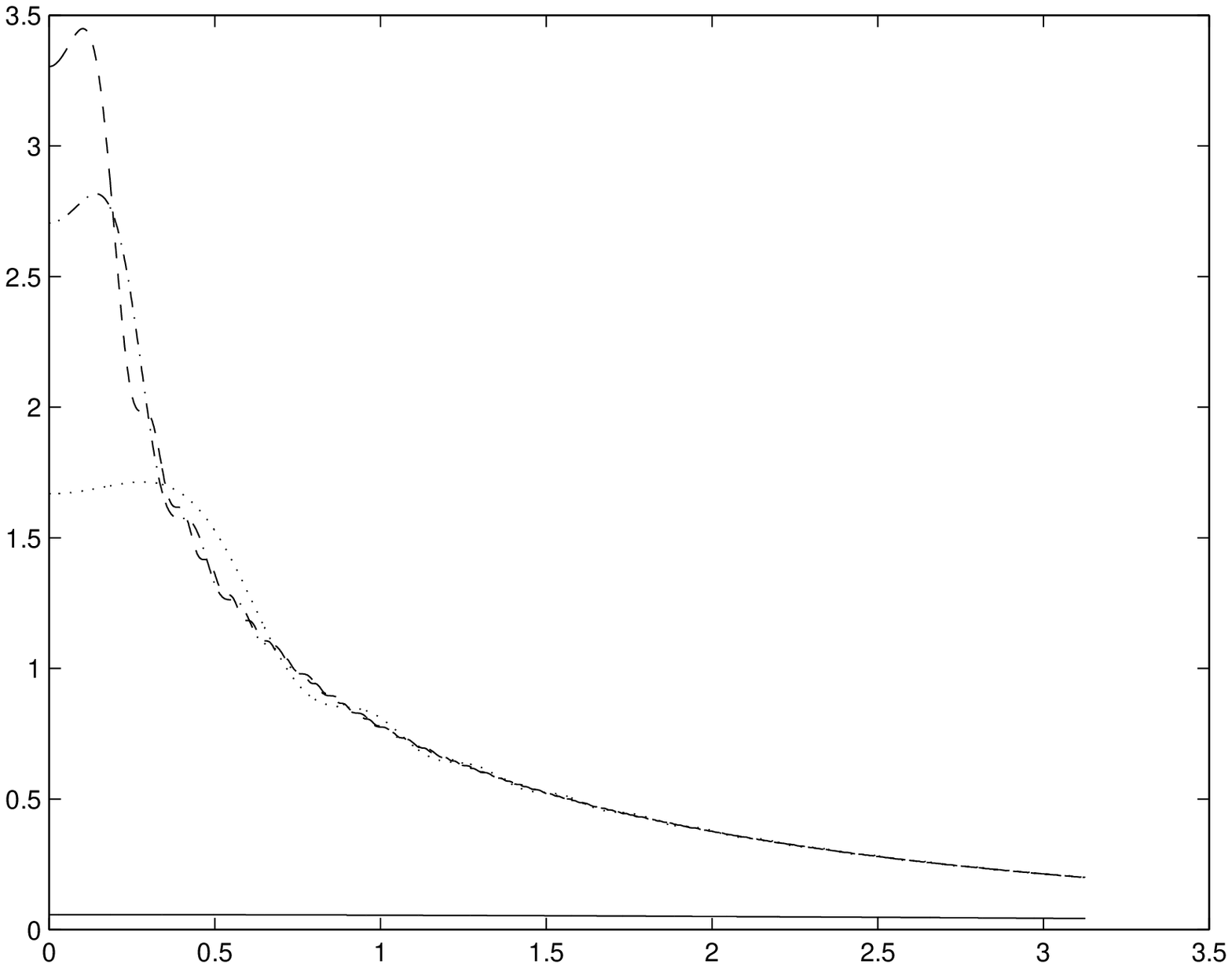}
\caption{\it Adiabatic spectrum for $tm_{\rm eq}=0$ (solid line),
$39.723$ (dotted line), $199.006$ (dashdot line) and $398.11$
(dashed line), for $\L/m _{\rm eq} = 20$, $\ell=0$ and $\rho_0=0.6$.}
\end{figure}
\vskip 0.5cm
\vskip 0.5cm
\vspace{-2cm}
Looking at the momentum distribution of the created particles at
different times, we see the formation of a growing peak corresponding
to soft modes. I can give an analytic, self-consistent description of
this behavior at large times through a perturbative approach, similar
to the one used in ref. \cite{Boyanovsky:1998zg}. I split the time--dependent
Lagrange multiplier in two parts, as in equation (\ref{asymass}) and
I treat the ``potential'' $p(t)/t$ perturbatively, as is done in
\cite{Boyanovsky:1998zg}. I find the following solution:
$$
z_q(t)=A_q e^{i\omega_qt} + B_q e^{-i\omega_qt} - \int_t^{\infty}
\dfrac{\sin \omega_q (t' - t)}{\om_q} \dfrac{p(t')}{t'} z_q(t') dt'
$$
which is equivalent, up to terms of order $O(1/t^2)$, to
\begin{eqnarray}\label{pert_z}
z_q(t) & =&A_q \left[ 1 + \dfrac{A \sin \Psi (t)}{4im_{\infty}t} -
\dfrac{A}{8\omega_qt} \left( \dfrac{e ^{i \Psi (t)}}{\omega_q +
m_{\infty}} + \dfrac{e ^{-i \Psi (t)}}{\omega_q - m_{\infty}} \right)
\right] e ^{i \omega_q t} \\
& +& B_q \left[ 1 - \dfrac{A \sin \Psi (t)}{4im_{\infty}t} -
\dfrac{A}{8\omega_qt} \left( \dfrac{e ^{i \Psi (t)}}{\omega_q -
m_{\infty}} + \dfrac{e ^{-i \Psi (t)}}{\omega_q + m_{\infty}} \right)
\right] e ^{- i \omega_q t} + O\left(\dfrac{1}{t^2}\right)
\end{eqnarray}
with $\Psi (t) = 2 m_{\infty} t + \gamma_1 \log t +
\gamma_2$. The logarithmic dependence is due to the ``Coulomb form''
of the perturbative term $p(t)/t$ in the equations of motion.  The
expression (\ref{pert_z}) displays resonant denominators for
$\omega_q=m_{\infty}$, that is $q=0$.  The perturbative approach is
valid as long as the first order correction is small compared to
zeroth order. Such a condition is satisfied if
\begin{equation}
	\dfrac{{\cal A}}{4t\,\omega_q(\omega_q - m_{\infty})} < 1
\end{equation}
that implies $q^2>{\cal A}/4t$ for non relativistic modes. Thus the
position of the peak found before may be interpreted as the result of
a weak nonlinear resonance. The asymptotic behavior of the condensate
and the mode functions related to soft momenta must be obtained
through non-perturbative techniques, implementing a multitime scales
analysis and a dynamical resummation of sub-leading terms. A
self-consistent justification of the numerical result (\ref{asymass}),
along with the power law relaxation behavior for the expectation value
(with non-universal dynamical anomalous dimensions), are likely to be
obtained following the line of the analysis performed in
\cite{Boyanovsky:1998zg} for the $\phi^4$ model in $3+1$ dimensions.

From the numerical study of the complete spectrum history, I conclude
that no exponentially growing (parametric or spinodal) instabilities
are present in the case at hand, as apparent from Fig.s \ref{spec_1} -
\ref{adnum0_7}, which show the spectrum of produced particles with
respect to the adiabatic vacuum state.

This is due to the quite different nonlinearities of the $\s-$model
and in particular to the nonlinear constraint [see eq. (\ref{constr})]
which sets an upper bound to the quantum infrared fluctuations [see
fig. \ref{sigma-fig}]. In fact, even if the constraint disappears as
the bare coupling constant $\lbare$ vanishes in the infinite UV cutoff
limit (asymptotic freedom), the quantum fluctuations in any given
finite range of momentum remain constrained to finite values, as
implied by the possibility of fully renormalize the model, including
the constraint [see eq. (\ref{ren_lm})].

\section{Summary and outlook}
\label{ol}
The natural continuation of this preliminary analysis is the detailed 
numerical study of the evolution, in order to give a precise picture of the 
process of dissipation via particle production in the framework of this
constrained, asymptotically free model. It should be possible to
determine precisely the power laws that characterize the asymptotic
evolution of relevant variables, like the condensate, the Lagrange
multiplier and the number of created particles. After this, one should
be able to decide whether, at zero angular momentum, the damping leads
to the complete dissipation of the energy stored in the condensate or
the system evolves towards a limit cycle with an asymptotic amplitude
different form $0$. Also a comparison with the linear model in $1+1$
dimension might be useful to understand the peculiarities of the
dynamics in a constrained model.

\begin{figure}
\includegraphics[height=8cm,width=15cm]{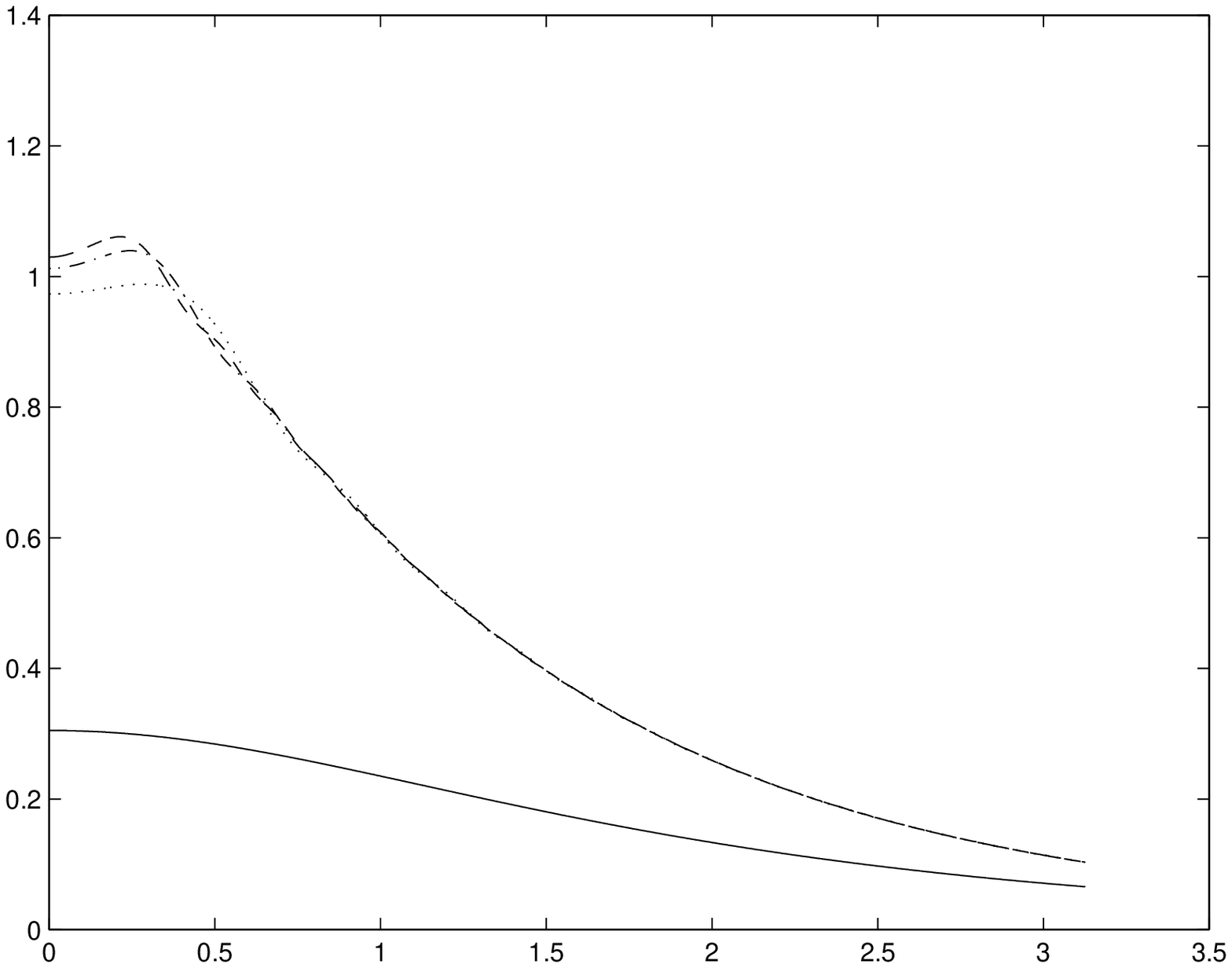}
\caption{\it Adiabatic spectrum for $tm_{\rm eq}=0.0$ (solid line),
$59.895$ (dotted line), $119.835$ (dashdot line) and $179.775$ (dashed
line), for $\L/m _{\rm eq} = 10$, $\ell=1$ and $\rho_0=0.3$.}
\label{lneq0_3}
\end{figure}

Moreover, it would be very interesting to study the dependence of the
evolution on the value of $\ell$, the angular momentum of the field in
the internal space. As the preliminary results presented in this paper
show (see figure \ref{lneq0_3}), the asymptotic state is far from the
state of minimum energy compatible with the given value of
$\ell$. Remarkably, the adiabatic spectrum of produced particles in
case of $\ell \neq 0$ is broader than that one corresponding to
$\ell=0$, suggesting a stronger coupling with hard modes.

%% file: ch4.tex

There are common relevant problems in cosmology, astrophysics and
ultrarelativistic heavy ion collisions, that ask for a deeper
understanding of the physics of the formation of a non--equilibrium
plasma of charged particles beginning from an initial state of large
energy density, its evolution, the onset of electric and magnetic
screening phenomena and the generation of seeds of bulk electric and
magnetic fields \cite{Boyanovsky:1999jh}.

Taking as a first example the world of elementary particles, one of
the main motivations to the use of out of equilibrium techniques comes
from the possibility of studying experimentally the formation and
evolution of a deconfined phase in QCD: the quark gluon plasma
(QGP). It is important, for example, to understand how the quark-gluon
plasma forms and equilibrates from the evolution of the parton
distribution functions, to correctly compute the time scales for
electromagnetic screening that cuts off small angle scattering and how
a hydrodynamic picture of the space-time evolution of the plasma
emerges from first principles. Also the comprehension of the possible
experimental signatures of the quark gluon plasma and chiral phase
transition would be very relevant. Electromagnetic probes (photons and
dileptons) could provide clear signatures for an out of equilibrium
chiral phase transition and the formation of quark phase transition,
because they only interact electromagnetically and their mean free
path is much bigger that the size of the fireball. In addition, even
though these days QCD is generally accepted as the {\em fundamental}
theory of strong interactions, its very rich phenomenology and its
complex phase structure have not been completely unveiled yet, as
recent results at finite temperature and density (color--flavor
locking, color superconductivity and quark--hadron continuity) suggest
\cite{Wilczek:1999fr,Wilczek:1998ur,Alford:1998mk,Schafer:1998ef}.

Regarding cosmology, the fascinating hypothesis of sphaleron induced
B-violating processes inside the Standard Model as the basis for the
baryon asymmetry, calls for a more accurate analysis of the phenomena
involved.

The common ingredient in such different situations is the description
of the transport properties, like the screening masses and the
electrical conductivity, as well as the determination of the
relaxation time scales. To this end, a much deeper understanding of
collective excitations in an ultrarelativistic plasma is required: to
decide if the local thermodynamical equilibrium (LTE) approximation is
justified, we must compute the relaxation time scales, given by the
damping rate or the lifetime of the excitations in the plasma. The
quasi particle description of the collective excitations treats them
as narrow resonances characterized by a Breit-Wigner distribution,
whose width is related to the imaginary part of the self--energy on
the mass shell. In this approximation, the damping rate turns out to
be exponential.

Of course, in order to reach physically sensible results, we have to
improve the models considered so far, by introducing gauge fields: a
gauge invariant description of the real--time dynamics of gauge fields
is required. Recently, new and unusual features of relaxation of soft
degrees of freedom in gauge theories has been discovered
\cite{Boyanovsky:1998pg}. The real time evolution of (abelian) gauge
fields is very important because photoproduction is expected to be a
very clear experimental signature of the chiral phase transition. It
may be also relevant in the description of the creation of primordial
magnetic fields in the early Universe.

In the framework of the non--equilibrium formalism the Hard Thermal
Loop (HTL) resummation scheme of Braaten and Pisarsky can be rederived
\cite{Boyanovsky:1998pg}. This is a useful scheme, which produce gauge 
invariant results and infrared finite transport cross sections which
renders the damping rate of the excitations in the plasma
\cite{Pisarski:1989cs} finite. In fact, the medium effects such as the
Debye screening, the collective plasma modes and Landau damping are
taken into account in the resummed Green's functions. However, soft
transverse (magnetic) photons remain unscreened and induce IR
divergences. This characteristic is common to QED, QCD and scalar
QED. The HTL resummation scheme may be regarded as giving a Wilsonian
effective action (which turns out to be non--local) for soft modes,
after the hard modes have been integrated out. Since the soft modes
have in general a non perturbative dynamics leading to quite large
occupation numbers, it has been hoped to get some information from the
numerical simulation of the classical theory. This approach, although
justified in scalar field theories, should be considered carefully
when applied to gauge theories.

With the aim of understanding relaxational dynamics in the quark gluon
plasma and electroweak plasma, hot scalar QED has attracted much
interest
\cite{Boyanovsky:1998pg,Boyanovsky:1998aa}, because it has the same HTL 
structure as the non abelian case (for the leading term). The real
time evolution of inhomogeneous expectation values with soft length
scales has been considered in the weak field regime, with linearized
equations which admit solutions in closed form through Laplace
transform. In such a regime, the results clearly show the dominance of
off-shell Landau damping processes. It is thus crucial to understand
the time scales of these dissipative processes, their microscopic
description and how the relaxation of soft collective excitations
proceeds. The analytic structure in the $s-$plane gives the time
behavior, which shows a power law tail in the relaxation of transverse
field amplitudes and a logarithmic tail for longitudinal fields; this
is determined by the behavior of the retarded self--energy at the
Landau damping threshold.  This is special of HTL at one loop and may
change, should high orders be included in the computation. In this
linearized regime, it is also possible to compute the Influence
Functional \cite{Feynman:1963fq}, and derive an effective Langevin
equation for the gauge invariant observables. Also the dissipative and
noise correlation functions (kernels) and the fluctuation-dissipation
relation between them, is obtained, proving in detail that a Markovian
approximation fails to describe correctly the dynamics, when the
processes are dominated by Landau damping. 

Similar techniques have been applied to understand the
non--equilibrium dynamics of a thermal plasma in a gravitational field
\cite{Calzetta:1994qe,Campos:1998un}. The plasma is described by a
massless scalar quantum field and the matter--gravity coupling is
treated semiclassically. The quantum matter back reaction on the metric
is considered as a noise term in the Einstein equation, which becomes
a semiclassical stochastic equation. It is equivalent to a Langevin
equation, whose damping and noise kernels satisfy a definite
Fluctuation--Dissipation relation.

Some results have also been obtained recently, for a condensate of
arbitrary amplitude, but uniform in space. In such a situation, a
Bloch-Nordsieck resummation of IR divergences, equivalent to a
dynamical Renormalization Group resummation, has been applied to
scalar QED, in connection with relaxation and damping in real time
\cite{Blaizot:1996hd,Boyanovsky:1998aa}. The resummation of
quasistatic transverse photons gives an anomalous logarithmic
relaxation of the form $\exp(\alpha T t \log t/t_0)$, where $\alpha$
is related to the coupling constant and $T$ is the temperature.

The dynamical RG resummation to obtain the real time dynamics of
relaxation and thermalization is equivalent to the RG resummation of
divergences in Euclidean Field Theories. Here, we have the resummation
of logarithmic secular terms in the perturbative solution of the
evolution equation of expectation values. This analysis implies that a
quasi particle description of the relaxation process is allowed only
when the perturbative solution displays linear secular terms, while
non--linear secular terms in lowest order signal anomalous,
non--exponential relaxation.

An other important issue is the generation of an out of equilibrium
plasma in scalar QED. This problem has been studied within a model
containing N charged scalar fields coupled to a $U(1)$ photon, plus a
neutral scalar field, which acts as an order parameter and
distinguishes between the broken and unbroken symmetry phases. This
model provides a convenient description of the chiral phase transition of QCD 
and allows the study of photon production, the electric and magnetic screening
and the build up of electric conductivity. The approximation consists
in considering the large $N$ limit in the scalar sector, which takes
into account the non perturbative and non linear effects involved in
the dynamics out of equilibrium, and the lowest order in $\alpha$,
which provides a mechanism of photoproduction. 

As far as the photon production is concerned, a consistent kinetic
equation suitable for strongly out of equilibrium situations has been
derived in \cite{Boyanovsky:1999jh}, considering only spontaneous
emission of photons (without the inclusion of stimulated
processes). The exponential growth in the quantum fluctuations of the
matter field, due to parametric amplification or spinodal
instabilities, drives the formation of the non equilibrium quantum
plasma, and a similar explosive production of photons at early times
occurs, while the asymptotic distribution at late times reaches a form
\begin{equation}
N_{ph}(\omega) \simeq \dfrac{\alpha m^2}{\lambda^2 \omega^3}
\end{equation}
In case of spinodal instabilities, the asymptotic distribution
diverges logarithmically with time and a resummation via the dynamical
renormalization group may be in order.

The analysis of the photon production allows for an understanding of
the mechanism of creation of the electromagnetic field, which shows up
in correlated domains of linear size $\xi(t)
\simeq \sqrt{t/m}$. The corresponding power spectrum inside the
unstable band with an amplitude $\simeq \alpha/\lambda^2$

In \cite{Boyanovsky:1999jh} Magnetic and Debye screening masses are
defined out of equilibrium, in order to study the dynamical aspects of
electric and magnetic screening.  While the magnetic mass vanishes out
of equilibrium, it is possible to give a definition of a momentum and
time dependent magnetic mass which has the advantage of displaying
explicitly the different time scales involved in the process.  The
Debye mass is $\alpha m^2 / \lambda$ for broken symmetry and is
proportional to $\sqrt{t}$ for unbroken symmetry, suggesting the
necessity of renormalization group resummation for the secular terms,
similar to that performed in \cite{Boyanovsky:1998aa}.

Such an analysis allows one to follow in detail the onset of the
transverse electric conductivity (in the framework of the linear
response theory \'a la Kubo) during the non equilibrium evolution and
the formation of the quantum plasma. At the end of the early (linear)
evolution, the long wavelength conductivity turns out to be
proportional to $\alpha m / \lambda$.

A comparison with the corresponding equilibrium phenomena is very
useful and shows that many results have the same dependence on the
particle distribution, which is a time dependent quantity in the case
of the non--equilibrium evolution, while is the static Bose--Einstein
distribution in case of thermodynamic equilibrium (cfr. also 
\cite{Niegawa:1997yv}).

Scalar QED is very useful to clarify some phenomena which occur in an
electromagnetic plasma. On the other hand, most of the non equilibrium
phenomena occurring in the ``ordinary'' low--energy matter are
described by fermionic QED. In this context, one of the older
questions is how the vacuum fluctuations of the photon field affect
the effective dynamics of an electron. Moreover, the program of
decoherence proposes that the superselection rules coming from the
conservation of charges could be induced dynamically: superposition of
states with different eigenvalue of the charge operator would rapidly
lose their coherence in time. This issue has been considered in
\cite{Anastopoulos:1998wh}, where the reduced density matrix for the
electron has been computed in perturbation theory by means of the
Feynman--Vernon influence functional technique.

An other approach is based on the extension of the HTL resummation
technique to non equilibrium QFT, by means of the real time formalism
in the Keldish CTP representation \cite{Carrington:1997sq}. This
extension is based on the assumption of quasi stationary distributions,
which occur if the time scale of the microscopic processes in the
medium is much smaller than the time scale of relaxation towards
equilibrium. In this approximation, the HTL photon self energy, the
resummed photon propagator and the damping rate of a hard electron in
a QED plasma have the same form out of equilibrium as in equilibrium.

Regarding the study of time evolution in non abelian gauge theories, it should 
be noticed that some work has been done in the framework of classical 
Yang--Mills theory, solving the evolution equations on a lattice and modeling 
the two ions by means of classical wave--packets of chromoelectromagnetic
field \cite{Krasnitz:1998fy,Krasnitz:1998jb}


%% file: ch5.tex

\section{Summary}\label{conclusion1}

A great effort has been devoted in the last few years in order to
develop a deeper qualitative and quantitative understanding of systems
described by interacting quantum fields out of equilibrium. There is a
class of physical problems that requires the consistent treatment of
time dependent mean--fields in interaction with their own quantum or
thermal fluctuations. I may mention, among others, the problem of
reheating of the universe after the inflationary era of exponential
growth and cooling, and the time evolution of the scalar order
parameter through the chiral phase transition, soon to be probed in
the heavy--ion experiments at BNL--RHIC and CERN--LHC. In these
situations, a detailed description of the time--dependent dynamics is
necessary to calculate the non--equilibrium properties of the
system. Indeed, the development of practical general techniques and
the advent of faster and cheaper computers have made possible the
discovery of novel and unexpected phenomena, ranging from dissipative
processes via particle production to novel aspects of symmetry
breaking
\cite{Cooper:1997ii,Boyanovsky:1995me,Boyanovsky:1998ba,Boyanovsky:1998zg}.

From the technical point of view, it should be pointed out, first of
all, that a perturbative treatment of this dynamical problem is
meaningful only when the early time evolution is considered. The
presence of parametric resonant bands or spinodal instabilities (in
the case, respectively, of unbroken or spontaneously broken
symmetries) rapidly turns the dynamics completely non--linear and
non--perturbative. Thus, the asymptotic evolution at late time can be
consistently studied only if approximate {\em non--perturbative}
methods are applied to the problem \cite{Boyanovsky:1995me}.

Quite recently one of these schemes, namely the large $N$ expansion at
leading order (LN) \cite{largen_exp,Cooper:1997ii}, has been used in
order to clarify some dynamical aspects of the $\phi^4$ theory in $3$
spatial dimensions, reaching the conclusion that the non--perturbative
and non--linear evolution of the system might eventually produce the
onset of a non--equilibrium Bose--Einstein condensation (BEC) of the
long--wavelength Goldstone bosons usually present in the broken
symmetry phase
\cite{Boyanovsky:1998ba,Boyanovsky:1998yp,Boyanovsky:1998zg}. Another
very interesting result in \cite{Boyanovsky:1998yp} concerns the
dynamical Maxwell construction, which reproduces the flat region of
the effective potential in case of broken symmetry as asymptotic fixed
points of the background evolution.

In section \ref{numerical} (see also ref. \cite{Destri:1999hd}) I have
addressed the question of whether a standard BEC could actually take
place as time goes on, by putting the system in a finite volume (a
periodic box of size $L$) and carefully studying the volume dependence
of out--of--equilibrium features in the broken symmetry phase. We
summarize here the main results contained in section \ref{numerical} as
well as in \cite{Destri:1999hd}. The numerical solution shows the
presence of a time scale $\tau_L$, proportional to the linear size $L$
of the system, at which finite volume effects start to manifest, with
the remarkable consequence that the zero-mode quantum fluctuations
cannot grow macroscopically large if they start with microscopic
initial conditions. In fact, the size of low--lying widths at time
$\tau_L$ is of order $L$, to be compared to order $L^{3/2}$ for the
case of standard BEC. In other words we confirmed that the linear
growth of the zero mode width, as found also by the authors of
\cite{Boyanovsky:1998ba,Boyanovsky:1998yp,Boyanovsky:1998zg}, really
signals the onset of a novel form of dynamical BEC, quite different
from the standard one described by equilibrium finite--temperature
field theory. This interpretation is reinforced by the characteristics
of the long--wavelength fluctuations' spectrum.

To go beyond the gaussian approximation and study the interaction
between longitudinal and transverse fluctuations, I have defined in
section \ref{impHF} an extended time dependent Hartree-Fock (tdHF)
approximation for the $\phi^4$ QFT, which includes some non-gaussian
features of the full theory (for an introduction to the standard
tdHF, see ref. \cite{Kerman:1976yn}). I have presented a rather
detailed study of the dynamical evolution out of equilibrium, in
finite volume (a cubic box of size $L$ in $3$D), as well as in
infinite volume. For comparison, I have also analyzed some static
characteristics of the theory both in unbroken and broken symmetry
phases.

By means of a proper substitution of the bare coupling constant with
the renormalized coupling constant (fully justified by diagrammatic
consideration), I have been able to obtain equations of motion
completely independent of the ultraviolet cut-off (apart from a slight
dependence on inverse powers, that is, however, ineluctable because of
the Landau pole). I have described in detail the shape of the ground
state, showing how a broken symmetry scenario can be recovered from
the quantum mechanical model, when the volume diverges. 

Moreover, I have shown that, within this slightly enlarged tdHF
approach that allows for non--gaussian wavefunctions, one might
recover the usual gaussian HF approximation in a more controlled way.
In fact, studying the late time dynamics, I have confirmed the
presence of a time scale $\tau_L$, proportional to the linear size $L$
of the box, at which the evolution ceases to be similar to the
infinite volume one. At the same time, the low--lying modes amplitudes
have grown to order $L$. The same phenomenon has been observed in the
$O(N)$ model \cite{Destri:1999hd}. Looking at this result in the framework of
our extended tdHF approximation, one realizes that the growth of
long--wavelength fluctuations to order $L$ in fact undermines the
self--consistency of the gaussian HF itself. In fact, in our tdHF
approach the initial gaussian wavefunctions are allowed to evolve into
non--gaussian forms, but they simply do not do it in a macroscopic
way, within a further harmonic approximation for the evolution, so
that in the infinite--volume limit they are indistinguishable from
gaussians at all times. But when $M^2$ is on average not or order
$L^0$, but much less, as it happens for suitable initial conditions,
infrared modes of order $L$ will be dominated by the quartic term in
our Schroedinger equations (\ref{Schroedinger}), showing a possible
internal inconsistency of the gaussians approximation.

An other manifestation of the weakness of the HF scheme is the curious
``stopping at the spinodal line'' of the width of the gaussian quantum
fluctuations, when the initial configuration does not break the
symmetry. This does not happen in the large $N$ approach because of
different coupling of transverse modes (the only ones that survive in
the $N\to\infty$ limit) with respect to the longitudinal modes of the
$N=1$ case in the HF approach.

I have also described the non--trivial phenomenology of the
infinite--volume late--time evolution in the gaussian approximation,
showing how the dynamical Maxwell construction differs from the
$N=\infty$ case. In fact, I have observed the presence of an unstable
interval, contained in the static flat region which is forbidden as
attractor of the asymptotic evolution. This region corresponds, more
or less, to the spinodal region of the classical potential, with the
obvious exception of the origin. In particular, I have found that the
energy flux between the classical degree of freedom and the bath of
quantum fluctuations is quite complex and not monotonous. In other
words, since I start from initial conditions where the fluctuation
energy is not minimal, there are special situations where enough
energy is transferred from the bath to the condensate, pushing it
beyond the top of the potential hill. 

As far as the non linear $\sigma$ model in $(1+1)$D is concerned, I
have shown explicitly in chapter \ref{smodel} that the large $N$ limit
and the large coupling limit, which turns the linear model in the non
linear one, commute. I have also derived the evolution equations for
the non linear model at the leading order in the $1/N$ expansion, in
the case of a field condensate different from zero. I implemented the
constraint by the use of a Lagrange multiplier, which I denoted $m^2$,
since it enters the dynamics as a squared mass. I have shown that the
usual renormalization procedure, which makes the bare coupling
constant depend on the UV cutoff, is sufficient to get properly
renormalized, that is UV finite, evolution equations. Moreover, I
characterized the ground state of the model, giving an interpretation
of the dynamical generation of mass (the so--called dimensional
transmutation) in terms of a compromise between energetic requirements
and the constraint. I then solved numerically the evolution equations
and analyzed the history of the condensate and the Lagrange multiplier
as well as the number of particles created during the relaxation of
the condensate (the quantum fluctuations). Remarkably, I do not find
any period of exponential growth for the fluctuations. Actually, no
spinodal instabilities were to be expected, since the symmetry is
always unbroken in $1+1$ dimensions.  But there occurs also no parametric
resonance, as takes place instead in the unbroken symmetry scenario of
the large$-N$ $\phi^4$ model in $3+1$ dimensions.

The estimated dependence of the asymptotic value of the Lagrange
multiplier, $m^2(\infty)$, on the initial condensate $\rho_0$, turns
out to be very well approximated by an exponential, which is the exact
dependence of $m^2(0)$ [at infinite UV cutoff, see
eqs. (\ref{alfadiro}) and (\ref{m0dia})]; remarkably however, the
prefactor in the exponent is changed [see eq. (\ref{minfrho})]. As far
as the emission of particles is concerned, we considered three
different reference states: the initial state, the adiabatic vacuum
state and the equilibrium vacuum state, that is the true ground state
of the theory. The numerical results suggest a weak non linear
resonance, yielding a relaxation of the condensate via particle
production driven by power laws with non universal anomalous
exponents, a result similar to what found in \cite{Boyanovsky:1998zg}
for the asymptotic dynamics of $\phi^4$ in $3+1$ dimensions.  Finally,
since I allow the condensate to have a number $n$ of components larger
than $1$, I was able to study the evolution of configurations with
non--zero angular momentum $\ell$ in the internal space of the field
[see eq. (\ref{eqm1})]. In this case I find numerical evidence for an
adiabatic spectrum broader than in the case $\ell=0$ (see figure
\ref{lneq0_3}), suggesting a stronger coupling with hard modes.

\section{Open issues}

I close this section and this work giving a list of the (in my
opinion) most relevant issues still open, whose solution may result in a
better understanding of both the formal and phenomenological aspects
of QFT out of equilibrium:

\begin{enumerate}
\item The study of the Poincar\'e cycles in the effective hamiltonian
dynamics and their dependence on the parameters
\cite{Cooper:1997ii,Destri:1999hd} may allow for a deeper understanding of the
dynamics in finite volume.
\item The inclusion of $1/N$ corrections, which contain collisional
contributions, is needed to better understand the issue of
thermalization; they, in fact, may lead to a thermal asymptotic
distribution and make the damping more efficient
\cite{Bettencourt:1998xb,Bettencourt:1998nf}. With this regard, it
should be stressed that finite truncations of the large--$N$ expansion
introduce errors which grow with time and make the approximation
unreliable on a time scale proportional to as $N^{1/2}$ \cite{Ryzhov:2000fy}.
\item Further study, both analytical and numerical, is needed in
the generalized tdHF approach of section \ref{impHF}, to better
understand the dynamical evolution of quantum fluctuations in the
broken symmetry phase coupled to the condensate.  An interesting
direction is the investigation of the case of finite $N$, in order to
interpolate smoothly the results for $N=1$ to those of the $1/N$ (at
leading order) approach. It should be noted, in fact, that the theory
with a single scalar field contains only the longitudinal mode (by
definition), while only the transverse modes are relevant in the large
$N$ limit.  Hence a better understanding of the coupling between
longitudinal and transverse modes is necessary. In this direction,
another relevant point is whether the Goldstone theorem is respected
in the HF approximation
\cite{Lenaghan:1999si}. It would be interesting also to study the dynamical
realization of the Goldstone paradigm, namely the asymptotic vanishing
of the effective mass in the broken symmetry phases, in different
models; this issue needs further study in the $2D$ case \cite{Cooper:1997ii},
where it is known that the Goldstone theorem is not valid.
\item The issue of thermalization may be addressed also by deriving a
Quantum Boltzmann (kinetic) equation. The issue is relevant
for the reheating problem and for the dynamics of heavy ion
collisions. It would be useful a systematic quantal generalization of
classical transport theory, in the presence of strong inhomogeneous
mean fields, where quantum self--energy (off--shell virtual processes)
are included on the same footing as collisional or real particle
production processes in the plasma. To this end, the CTP generating
functional may be used in order to derive an effective action
principle which leads to a hierarchy of Schwinger--Dyson equations
\cite{Witte:1994an}. Some preliminary results \cite{Boyanovsky:1998pg}
shows an anomalous relaxation of hard quasiparticles. Considering
higher orders will include collisional processes which will give
exponential decay in time. The competition between collisional and
Landau damping time scales will then depend on the particular model
under study.

In any case, the approach to thermodynamic equilibrium is far from
being a trivial subject. In fact, even though the time evolution of
correlation functions can be described by an exact evolution
(functional) equation for the corresponding generating functionals
\cite{Wetterich:1997ap} (which shows many dynamical features of
quantum mechanics
\cite{Wetterich:1997zt}) and the classical field equations admit
thermodynamic equilibrium as a fixed point, infinitely many conserved
correlation functions prevent the system from approaching the fixed
point, when it starts in its neighborhood. Thus, it seems that
equilibrium can therefore be reached at most for suitably averaged
quantities or for subsystems. A similar analysis in QFT leads to the
same conclusion that a uniform approach to thermal equilibrium is
prevented by the existence of infinitely many other fixed points which
correspond to incoherent mixtures of eigenstates of the quantum
hamiltonian \cite{Wetterich:1997rp}.

Still on this subject, it is worth noticing that an important
understanding both on thermalization and on the quantal generalization
of kinetic equations, has been reached recently by the authors of
ref.s \cite{Berges:2000ur} and \cite{Aarts:2001qa}. Using a three-loop
approximation to th 2PI effective action and solving numerically the
evolution equations for the two point function in $1+1$ dimensions, it
is possible to show that all correlation functions approach the
thermal distribution asymptotically, proving that higher loops contain
indeed the collisional contributions.

\item The study of Disoriented Chiral Condensates has been started
modelling the pion phenomenology with the $O(4)$ linear $\sigma$
model. It would be interesting to look for qualitative and
quantitative differences in the predictions, when they are extracted
from the non-linear model.
\item To better understand out of equilibrium phenomena in many areas
of physics, ranging from condensed matter and statistical mechanics
systems to cosmology and particle physics, it is necessary to follow
the evolution of non-homogeneous condensates in several models. The
derivation of the evolution equations for a spherically symmetric
condensate, coupled to its quantum fluctuations, I gave in section
\ref{dis_homo}, is a promising first step in this direction and deserves a
deeper analytical and numerical analysis, in view also of its
phenomenological implications. This approach may be married with the
the technique developed by Aarts and Smit \cite{Aarts:1998zu}, 
\cite{Aarts:1998gi}, \cite{Aarts:1998td} and \cite{Aarts:1999zn} in
order to include the quantum fluctuations of both fermionic and
bosonic fields, with inhomogeneous condensates.
\item I commented a lot in chapter \ref{gauge} about the results and
the open problems in non equilibrium gauge theories: they are relevant
for our knowledge of phase transitions in Standard Model and
baryogenesis
\cite{Riotto:1998bt,Buchmuller:2000wq}. Domain walls (non homogeneous
field configurations) for the non abelian gauge fields of the
electroweak theory (usually known with the name of {\em sphalerons})
are thought to be responsible of the matter-antimatter asymmetry we
see in our universe.
Thus, in addition to the issue raised in the previous point, we need
to find a dynamical scheme suitable to gauge theories. To the best of
my knowledge, a non perturbative treatment in the gauge coupling is
still lacking. One may try to define a mean field approximation by
means of a time--dependent variational principle. It is still not
clear, however, whether one should introduce the gauge conditions as
constraints in the variational principle or one should enforce these
conditions in the choice of the trial space \cite{Benarous:1997kp}

Finally, detailed computation of transport coefficients, damping rates
and energy loss characteristics of the quark gluon plasma from the
fundamental microtheory, QCD \cite{QFTooE3} is also needed, in view of
the forthcoming experimental results from BNL and CERN.
\end{enumerate}

%% file: ch6.tex
\renewcommand{\thesection}{\Alph{section}}
\setcounter{section}{0}  
\renewcommand{\theequation}{\thesection-\arabic{equation}}
\setcounter{equation}{0}  
\section{Stability analysis in Quantum Mechanics}

In section \ref{warmup} I derived the {\em classical} Hamiltonians
which describe the dynamics of the isotropic quantum harmonic
oscillator plus a $O(N)$ invariant quartic perturbation. I used the
following approximation schemes: first order in $\hbar$ and the time
dependent Hartree-Fock approximation, in the case of a single degrees
of freedom ($N=1$), and the leading order of the expansion in $1/N$.

In this first appendix I also describe in detail the structure of the
different {\em classical} potentials, computing the position of
stationary points for the three cases above. A similar computation is
very useful also for the Field Theoretical models studied in chapters
\ref{sqft} and \ref{smodel}.
 
I will be looking for the configuration of minimum energy with respect
to variation of the parameters characterizing the relevant states in
each approximations. In other words I will find the variational ground
states in the gaussian approximation for {\em one loop}, {\em
Hartree-Fock} and {\em Large N} potentials.
\subsection{$O \left( \hbar \right)$}
Let us consider now the mean energy (\ref{enclrin}) and (\ref{enflrin});
I want to study the stationary points of the potential
\begin{equation}
v \left( \eta , a \right) = s \eta ^{2} + \dfrac{\eta ^{4}}{2} + \dfrac{g}{3} \left[ 
\left( s + 3 \eta ^{2} \right) a ^{2} + \dfrac{1}{16 a ^2} \right]
\end{equation}
as a function of two variables $\eta$ and $a$ (it is clear that the
dimensionless parameter $g$, corresponding to the quantum coupling
constant, must be non negative).

In the classical case, i.e. $g = 0$, the dependence on the quantum
width $a$ disappears, and it is enough to consider the function of the
single variable $\eta$. Of course, we find the classical minima, which
are $\eta=0$ when $s=1$ and $\eta=\pm 1$ when $s=-1$.

When I turn on the quantum fluctuations (i.e., $g \neq 0$), the
degeneracy on the $a$ direction is removed and the positions of the
stationary points is modified as follows. 
The gradient of the potential $v \left( \eta , a \right)$ is:
\begin{equation}
\begin{array}{ll}
\dfrac{\partial v}{\partial \eta} \; = & 2 \eta \left( s + \eta ^2 + g a ^2
\right) \\
\dfrac{\partial v}{\partial a} \; = & \dfrac{2}{3} g \left[ \left( s + 3 \eta ^2 \right)  a - \dfrac{1}{16 a ^3} \right]
\end{array}
\label{grad}
\end{equation}
while its Hessian matrix is:
\begin{equation}
H \left( \eta , a \right) = \left(
\begin{array}{cc}
2 \left( s + 3 \eta ^2 + g a ^2 \right) & 4 g \eta a \\
4 g \eta a & \dfrac{2}{3} g \left[ \left( s + 3 \eta ^2 \right) + \dfrac{3}{16 a^4} 
\right]
\end{array}
\right)
\end{equation}
I start from the case in which the parameter $s$ has the value $+ 1$;
the equations (\ref{grad}) have the unique solution $\eta
= 0$ and $a = 1/2$ and correspondingly the Hessian is diagonal
with positive eigenvalues; in fact, in this case the determinant of
$H$ is $h = \dfrac{g}{3} \left( 4 + g \right)$, the eigenvalues being
$8g/3$ and $2 ( 1 + g/4)$. Thus, the stationary point found is indeed
a minimum, where the potential has the value $v \left( 0 , 1/2 \right) = g/6$

On the contrary, if I consider the broken symmetry case, when $s = -1$, we
see that the gradient of $v$ does not vanish for any point of the
line $\eta = 0$. Moreover, if I define $t = a ^2$, I get the following 
equations:
\begin{equation}
\eta ^2 = 1 - g t
\end{equation}
\begin{equation}
48 g t ^3 - 32 t ^2 + 1 = 0
\label{pol1}
\end{equation}
A close study of the third degree polynomial leads us to the
conclusion that it has two real positive roots ( $t _1 (g)$ and $t _2
(g)$ ) only when the parameter $g$ satisfy the constraint
\begin{equation}
g \leq \dfrac{16}{9} \sqrt{\dfrac{2}{3}}
\end{equation}
As can be easily inferred by computing the value of the polynomial for
$t = 1/g$, the two solutions are such that
\begin{equation}
0 < t _1 (g) \leq \dfrac{4}{9 g} \leq t _2 (g) < \dfrac{1}{g}
\end{equation}
The computation of the determinant of the Hessian matrix yields
\begin{equation}
h _{1,2} = 48 g ^3 \left( \dfrac{1}{g} - t _{1,2} (g) \right) \left(
\dfrac{4}{9 g} - t (g) _{1,2} \right)
\end{equation}
while its trace is
\begin{equation}
{\rm tr} H _{1,2} = 4 \left[1 - g t _{1,2} (g) \right] + \dfrac{8}{3}
g \left[2 - 3 g t _{1,2} (g) \right]
\end{equation}
So, I deduce that the points $\left( \pm \sqrt{1 - g t _1 (g)} ,
\sqrt{t _1 (g)} \right)$ are minima, while the points \\ $\left( \pm
\sqrt{1 - g t _2 (g)} , \sqrt{t _2 (g)} \right)$ are neither maxima
nor minima. The minimum values are
$$
v \left( \pm \sqrt{1 - g t _1 (g)} , \sqrt{t _1 (g)} \right) = -
\dfrac{1}{2} +
\dfrac{4}{3} g t _{1} (g) - \dfrac{3}{2} g ^2 t _{1} (g) ^2 < 0
$$

In conclusion, the potential has two local minima but no global
minimum, because it is not bounded from below. This means that if I
choose an initial value for $a$ that is too high (or too low), I let
the system to go beyond the hill near the local minima and it will not
be able to come back ever again. This conclusion is in agreement with
the analysis made in section \ref{sez_en}, namely with the exponential
growth of quantum fluctuations and the corresponding instability.

\subsection{Hartree-Fock}
If I now consider the formula for the Energy in the Hartree-Fock
approximation, eq. (\ref{ehf}), the potential turns out to be:
\begin{equation}
v \left( \eta , a \right) = s \eta ^{2} + \dfrac{\eta ^{4}}{2} + \dfrac{g}{3} \left[ 
\left( s + 3 \eta ^{2} \right) a ^{2} + \dfrac{1}{16 a ^2} + \dfrac{g}{2} a ^4 
\right]
\end{equation}
This computation is equivalent to the search for the ground state by a 
variational principle, using the gaussian packet (\ref{pacchetto}) as
a trial state and then minimizing with respect to its parameters
$\sigma$ and $X$.

The classical approximation ($g = 0$) is, as before, quite trivial.

When $g \neq 0$, the first partial derivatives (the gradient) of the
function $v \left( \eta , a \right)$ are:
\begin{equation}
\begin{array}{ll}
\dfrac{\partial v}{\partial \eta} \; = & 2 \eta \left( s + \eta ^2 + g a ^2
\right) \\
\dfrac{\partial v}{\partial a} \; = & \dfrac{2}{3} g \left[ \left( s + 3
\eta ^2 \right)  a - \dfrac{1}{16 a ^3} + g a ^3 \right]
\end{array}
\label{grad_hf}
\end{equation}
while the Hessian matrix is 
\begin{equation}
H \left( \eta , a \right) = \left(
\begin{array}{cc}
2 \left( s + 3 \eta ^2 + g a ^2 \right) & 4 g \eta a \\ 4 g \eta a &
\dfrac{2}{3} g \left( s + 3 \eta ^2 + \dfrac{3}{16 a^4} + 3 g a ^2
\right)
\end{array}
\right)
\end{equation}

If the parameter $s$ takes the value $+ 1$ the system (\ref{grad_hf}) has 
only one solution: $(\eta = 0 , a = \sqrt{T(g)} )$, where $T(g)$ is the unique 
positive real root of the polynomial
\begin{equation}
16 g t ^3 + 16 t ^2 - 1
\label{pol2}
\end{equation}
Corresponding to this solution the Hessian matrix is diagonal with
positive eigenvalues; in fact, the determinant of $H$ takes the value
$h = \frac{8}{3} g \left[ 1 + g T(g) \right] \left[ 2 + 3 g T (g)
\right]$, the two eigenvalues being $2 [ 1 + g T (g)]$ and $4 g [ 2 +
3 g T (g)]/3$. The stationary point found is actually a minimum, whose
value is 
$$
v ( 0 , \sqrt{T(g)} ) = \dfrac{2}{3} g T(g) \left( 1  + \dfrac{3}{4} g T(g) 
\right)
$$

The most interesting case is instead when the parameter $s$ takes the
value $s = -1$: a solution is $(\eta = 0 , a = \sqrt{\tau ( g )})$,
where $\tau (g)$ is the only (positive) real root of the polynomial:
\begin{equation}
16 g t ^3 - 16 t ^2 - 1
\label{pol3}
\end{equation}
The Hessian matrix corresponding to this solution is diagonal again
and it has positive eigenvalues; in fact, the determinant is $h =
\frac{8}{3} g \left[ g \tau ( g ) - 1 \right] \left[ 3 g \tau ( g ) -
2 \right]$. Note that $\tau ( g ) > 1/g$. Thus, the stationary point
is a minimum, whose value is
\begin{equation}\label{symin}
v ( 0 , \sqrt{\tau(g)} ) = - \dfrac{2}{3} g \tau (g) \left( 1 -
\dfrac{3}{4} g
\tau (g) \right)
\end{equation}
which is positive for $g > 16/3\sqrt{3}$.
There are also two more solutions, solving the equations
\begin{equation}
\eta^2 = 1 - g t
\end{equation}
\begin{equation}
32 g t ^3 - 32 t ^2 + 1 = 0
\label{pol4}
\end{equation} 
where I define $t = a ^2$. The third degree polynomial has two real 
positive solutions ($t _{1} (g) $ and $t _{2} (g)$) only when 
\begin{equation}
g \leq \dfrac{8}{3} \sqrt{\dfrac{2}{3}}
\end{equation}
($t _{1} (g) $ and $t _{2} (g)$) satisfy the relation
\begin{equation}
0 < t _{1} (g) \leq \dfrac{2}{3 g} \leq t _{2} (g) \leq \dfrac{1}{g} < \tau (g)
\end{equation}
For these values of $g$, the ``symmetric minimum'' (\ref{symin}) is
negative. The computation of the determinant of the Hessian
matrix yields the result:
\begin{equation}
h = 32 g ^3 \left( \dfrac{1}{g} - t _{1,2} \right) \left( \dfrac{2}{3 g} - t 
_{1,2} \right)
\end{equation}
while the trace is:
\begin{equation}
{\rm tr} H = 4 \left[1 - g t _{1,2} (g) \right] + \dfrac{4}{3} g
\left[4 - 3 g t _{1,2} (g) \right]
\end{equation}
from which I realize that the stationary points $\left( \pm \sqrt{1 -
g t _1 ( g)} , \sqrt{t _1 (g)} \right)$ are minima, while the
stationary points $\left( \pm \sqrt{1 - g t _2 (g)} , \sqrt{t _2 (g)} \right)$
are not minima neither maxima. The value of the two minima is:
\begin{equation}\label{bsymin}
v \left( \pm \sqrt{1 - g t _1 (g)} , \sqrt{t _1 (g)} \right) = -
\dfrac{1}{2} +
\dfrac{4}{3} g t _{1} (g) - g ^2 t _{1} (g) ^2
\end{equation}
that is always negative for any value of $g$. Comparing the two
expressions (\ref{symin}) and (\ref{bsymin}), it is possible to decide
which one is the global minimum.

In conclusion, the stability properties of the HF potential are completely
different from those I had in the previous section: thanks to the
Hartree--Fock self--consistent term, the potential has now at least one global 
minimum. This prevents the fluctuations from growing exponentially for later 
times.

\subsection{Large $N$}
\label{gn}
I repeat here the analysis of the potential $v \left( \eta , a \right)$
for the model with $O \left( N \right)$ symmetry; now, the function is
[cfr. (\ref{en_N})]:
\begin{equation}\label{pot_n}
v \left( \eta , a \right) = s \eta ^{2} + \dfrac{\eta ^{4}}{2} + \dfrac{g}{3} \left[ 
\left( s + \eta ^{2} \right) a ^{2} + \dfrac{g}{6} a ^4 + \dfrac{1}{16 a ^2} \right]
\end{equation}
Again, the computation corresponds to the search for the best
approximation to the ground state by means of a variational principle
in which I use the gaussian packet
\begin{equation}
\braket{ \vec{x}}{ \Psi} = \dfrac{1}{(2 \pi \sigma) ^{N/4}}\exp 
\left[ - \sum _{i=1} ^{N-1} \dfrac{x _i ^2}{4 \sigma} \right] \exp
\left[ - \dfrac{(x _N - \sqrt{N} X ) ^2}{4 \sigma} \right]
\label{st_gauss}
\end{equation}
as a trial state and I minimize with respect to $\sigma$ e $X$.

For $g \neq 0$ we get the following structure. The gradient is
\begin{equation}
\begin{array}{ll}
\dfrac{\partial v}{\partial \eta} \; = & 2 \eta \left( s + \eta ^2 + \dfrac{g}{3}
a ^2 \right) \\
\dfrac{\partial v}{\partial a} \; = & \dfrac{2}{3} g \left[ \left( s + \eta
^2 \right) a - \dfrac{1}{16 a ^3} + \dfrac{g}{3} a ^3 \right]
\end{array}
\label{grad_hf_N}
\end{equation}
while the Hessian matrix is now
\begin{equation}
H \left( \eta , a \right) = \left(
\begin{array}{cc}
2 \left( s + 3 \eta ^2 + \dfrac{g}{3} a ^2 \right) & \dfrac{4}{3} g \eta a \\
\dfrac{4}{3} g \eta a & \dfrac{2}{3} g \left(  s + \eta ^2 + \dfrac{3}{16 a^4} + g a 
^2 \right)
\end{array}
\right)
\end{equation}
If the parameter $s$ is $+ 1$, the system made by the equations 
(\ref{grad_hf_N}) has the following unique
solution: $(\eta = 0, a = \sqrt{T(g)} )$, where $T(g)$ is the only real
positive root of the polynomial
\begin{equation}
16 g t ^3 + 48 t ^2 - 3
\label{pol5}
\end{equation}
Corresponding to this solution, the Hessian matrix has positive
eigenvalues; in fact, the determinant of $H$ takes the value $h =
\frac{8}{3} g \left[ 1 + \frac{g}{3} T(g) \right] \left[ 2 + g T (g)
\right]$. Thus, the stationary point found is a minimum, whose value is
$$
v ( 0 , \sqrt{T(g)} ) = \dfrac{2}{3} g T(g) \left[ 1  + \dfrac{g}{4} T(g) \right]
$$
The couple $\left( \eta = 0 , a = \sqrt{\tau (g)} \right)$ is the unique
solution, in the case of $s=-1$, where $\tau(g)$ is the real positive root of 
the polynomial
\begin{equation}
16 g t ^3 - 48 t ^2 - 3
\label{pol6}
\end{equation}
An expansion of the solution $\tau (g)$ in power series of $g$ yields:
$$
\tau (g) = \dfrac{3}{g} + \dfrac{g}{2 ^4 3} + \dfrac{g ^3}{2 ^7 3 ^3} + O(g ^5)
$$
The Hessian matrix is diagonal and its eigenvalues are both positive,
the determinant being $h = \frac{8}{3} g \left[ \frac{g}{3} \tau(g) -
1 \right] \left[ \tau (g) - 2 \right]$, thanks to the fact that $\tau
(g) > 3/g$; then, the stationary point is a minimum again, whose value is
$$
v ( 0 , \sqrt{\tau (g)} ) = - \dfrac{2}{3} g \tau (g) \left[ 1 - \dfrac{g}{4} 
\tau ( g ) \right] \approx -\dfrac{1}{2} + \dfrac{g ^2}{2 ^4 3 ^2} + 
\dfrac{7}{2 ^9 3 ^4} + O(g ^6)
$$
In the classical limit ($g \rightarrow 0$) this minimum is at
infinity and it has the same value as the ones which belong to the
lines $\eta=\pm 1$.

\subsubsection{A definition of the effective potential in Quantum
Mechanics}
Since I use a similar procedure in the QFT model I study in chapter
\ref{sqft}, it is interesting to see how one can define the effective
potential in the simpler setting of quantum mechanics. 

In section \ref{statprop} I define the effective potential as the
minimum of the potential energy $\V$ at fixed field condensate
$\bar{\bds\phi}$. Let us now apply this definition to the expression
(\ref{pot_n}) for the {\em classical} potential. I first of all
specify one value for $\eta$. Then I minimize the energy with respect to
the quantum width $a$, keeping the value of $\eta$ fixed, which is done
by solving the stationary condition
\begin{equation}\label{cl_gap}
48 g t^3 + 48 (s + \eta^2) t^2 -3 = 0
\end{equation}
where $t$ is defined as $a^2$, and $\eta$ appears as a parameter. This
cubic equation may be considered as the counterpart of the
field--theoretical gap equation [cfr. (\ref{massive})], in this
quanto--mechanical case. Substituting the solution $t(\eta)$ of
(\ref{cl_gap}) in the function of two variables $v \left( \eta , a
\right)$ turns it in a different function of the single variable $\eta$,
which is the {\em effective potential} $v_{\rm eff} (\eta)$:
\begin{equation}
v_{\rm eff}(\eta) = \left( s + \dfrac{2g}{3} t(\eta) \right) \eta^2 +
\dfrac{\eta^4}{2} + \dfrac{2g}{3}s t(\eta) + \dfrac{g^2}6 t(\eta)^2
\end{equation}
of the $O(N)$ quartic harmonic oscillator in the large $N$ limit.

\subsection{Numerical Computation for $g = 0.1$}
I present here some numerical result in the case the dimensionless
coupling constant has the value $g = 0.1$.
\begin{itemize}
\item The solutions of the equation (\ref{pol1}) are $t _1 = 0.1792016$ and 
$t _2 = 6.6619726$; the minima correspond to $(\eta = \pm 0.9909994 , 
a = 0.4233221 )$, where the value of the potential is $-0.4765882$.
\item The solution of the equation (\ref{pol2}) is $0.2469689$ and corresponds
to $(\eta = 0 , a = 0.4969597)$, where the potential takes the value
$0.0167696$.
\item In the case (\ref{pol3}), the root of the polynomial is $10.006242$,
corresponding to the stationary point $(\eta = 0 , a = 3.1632645)$ and to
the potential value $v = -0.1664584$.
\item The equation (\ref{pol4}) has the solution $t = 0.1783747$; the two
minima are \linebreak $(\pm 0.9910411,0.4223443)$ and the potential is
$-0.4764288$.
\item The equation (\ref{pol5}) has only one positive real solution, whose
value is $0.2489690$, the corresponding minimum is $( \eta = 0 , a =
0.4989680)$ where the potential takes the value $0.0167012$.
\item Finally, the solution of the equation (\ref{pol6}) is $30.002083$
corresponding to the stationary point $( \eta = 0 , a = 5.4774157)$ and
to the minimum $v = - 0.4999306$.
\end{itemize}
I can summarize the results in the following table:
$$
\begin{array}{|c|c|c|c|} \hline
case        & eq. n.     & solutions                   & v ( \eta , a ) \\ 
\hline
1L / s = +1 &            & (0 , 1/2)                   & 0.0166667   \\ 
\hline
1L / s = -1 & \ref{pol1} & (\pm 0.9909994 , 0.4233221) & -0.4765882  \\
\hline
HF / s = +1 & \ref{pol2} & ( 0 , 0.4969597 )           &  0.0167696  \\
\hline
HF / s = -1 & \ref{pol3} & ( 0 , 3.1632645 )           &  -0.1664584 \\
            & \ref{pol4} & ( \pm 0.9910411, 0.4223443) &  -0.4764288 \\
\hline
N=\infty / s = +1 & \ref{pol5} & ( 0 , 0.4989680 )     &  0.0167012  \\ 
\hline
N=\infty / s = -1 & \ref{pol6} & ( 0 , 5.4774157 )     &  - 0.4999306 \\
\hline
\end{array}
$$

\setcounter{equation}{0}  
\section{Details of the numerical analysis}\label{num}

I present here the precise form of the evolution equations for the
field background and the quantum mode widths, which control the
out--of--equilibrium dynamics of the $\phi ^4$ model in finite volume
at the leading order in the $1/N$ approach, as described in section
\ref{ooedN}, and in the time dependent Hartree--Fock approach of section 
\ref{ooed}. 

Actually, for understanding the structure of the equations, the
quanto--mechanical case of section \ref{warmup} is
enough. Eqs. (\ref{adim_hfeq}) and (\ref{lnadimeq}) may be written as
\begin{equation}
\left[ \dfrac{d ^2}{dt ^2} + \left( \om^2 - 2 \eta^2 \right) \right]
\eta = 0 \;,\quad
\left[ \dfrac{d ^2}{dt ^2}  + \om^2 \right] a - \dfrac{1}{16 a ^3} = 0
\end{equation}
where $\om^2 = s + 3 \eta ^2 + g a^2$, or
\begin{equation}
\left[ \dfrac{d ^2}{dt ^2}  + \om^2 \right] \eta = 0 \;,\quad
\left[ \dfrac{d ^2}{dt ^2} + \om^2 \right] a - \dfrac{1}{16 a ^3} = 0 
\end{equation}
where $\om^2 = s + \eta ^2 + g a^2$.

The field theoretical is quantitatively more complicated, as each mode
corresponds to a different equation and all equations are mean field
coupled; nevertheless, it is conceptually very similar.

I come immediately to it and I restrict here my attention to the
tridimensional case. Let us begin by noticing that each eigenvalue of
the Laplacian operator in a $3D$ finite volume is of the form $k_n ^2
= \left( \frac{2 \pi}{L}
\right) ^2 n$, where $n$ is a non--negative integer obtained as the sum of
three squared integers, $n = n _x ^2 + n _y ^2 + n _z ^2$. Then we
associate a degeneracy factor $g_n$ to
each eigenvalue, representing the
number of different ordered triples $(n _x, n _y, n _z)$ yielding the
same $n$. One may verify  that $g_n$ takes on the {\em continuum} value of
$4 \pi k ^2$ in the infinite volume limit, where $k=\left( \frac{2
\pi}{L} \right) ^2 n$ is kept fixed when $L \to \infty$.

Now, the system of coupled ordinary differential equations is, in case
of the large $N$ approach,
\begin{equation}\label{snumeq}
\left[ \dfrac{d ^2}{dt ^2} + M ^2\right] \phi
= 0 \;, \quad 
\left[ \dfrac{d ^2}{dt ^2} + \left( \dfrac{2 \pi}{L}
\right) ^2 n + M ^2 \right] \s_n - \dfrac1{4\s_n^3} =0
\end{equation}
while I have
\begin{equation}
\left[ \dfrac{d ^2}{dt ^2} + \left(M ^2 - 2 \l \phi^2 \right) \right]
\phi = 0 \;, \quad \left[ \dfrac{d ^2}{dt ^2} + k_n^2+ M ^2
\right] \s_n - \dfrac1{4\s_n^3} =0
\end{equation}
for the time dependent Hartree-Fock approach [cfr eq. (\ref{Emotion})]. 
Here the index $n$ ranges from $0$ to ${\cal N}^2$, ${\cal N}=\Lambda
L/2\pi$ and $M^2(t)$ is defined by the eq. (\ref{Nunbgap}) in case of
unbroken symmetry and by eq. (\ref{Nbgap}) in case of broken symmetry for the 
large $N$ approach, while the suitable definitions are eqs. (\ref{unbtdgap}) 
and (\ref{btdgap}). The back--reaction $\Sigma$ reads, in the notations of 
this appendix
\begin{equation}
\Sigma = \dfrac{1}{L^D} \sum_{n=0}^{{\cal N}^2} g_n \s_n ^2 
\end{equation}
Technically it is simpler to treat an equivalent set of
equations, which are formally linear and do not contain the singular
Heisenberg term $\propto \s_n^{-3}$. This is done by introducing the
complex mode amplitudes $z_n=\s_n\exp(i\t_n)$, where the phases $\t_n$
satisfy $\s_n^2\dot\t_n=1$. Then I find a discrete version of the
equations studied for instance in ref. \cite{Boyanovsky:1995me},
\begin{equation}\label{numeq}
\left[ \dfrac{d ^2}{dt ^2} + \left( \dfrac{2 \pi}{L}
\right) ^2 n + M ^2 \right] z_n=0 \;,\quad 
\Sigma = \dfrac{1}{L^D} \sum_{n=0}^{{\cal N}^2} g_n |z_n| ^2
\end{equation}
subject to the Wronskian condition
\begin{equation}
	z_n\,\dot{\bar{z_n}} - \bar{z_n}\,\dot z_n = -i
\end{equation}
One realizes that the Heisenberg term in $\s_n$ corresponds to the 
centrifugal potential for the motion in the complex plane of $z_n$.
Looking at the figs. \ref{1fig:m1} or \ref{slowrd:0m}, we can see that
the motions of the quantum modes correspond qualitatively to orbits
with very large eccentricities. In fact, there are instants in which
$\s_n$ is very little and the angular velocity $\dot\t_n$ is very
large. This is the technical reason for preferring the equations in
the form (\ref{numeq}).

Let us now come back to the equations (\ref{numeq}). To solve these
evolution equations, I have to choose suitable initial conditions
respecting the Wronskian condition.  In case of unbroken symmetry,
once I have fixed the value of $\phi$ and its first time
derivative at initial time, the most natural way of fixing the initial
conditions for the $z_n$ is to require that they minimize the energy
at $t=0$. I can obviously fix the arbitrary phase in such a way to
have a real initial value for the complex mode functions
\begin{equation}
z _n ( 0 ) = \dfrac{1}{\sqrt{2 \Omega _n}} \hspace{1 cm} \dfrac{d z _n}{dt} ( 0
) = \imath \sqrt{\dfrac{\Omega _n}{2}}
\end{equation}
where $\Omega _n = \sqrt{k ^2 _n + M ^2 ( 0 )}$. The initial squared
effective mass $M ^2 (t= 0)$, has to be determined self-consistently,
by means of its definition (\ref{Nunbgap}) or (\ref{unbtdgap}).

In case of broken symmetry, the gap equation, (\ref{Nbgap}) or
(\ref{btdgap}), is a viable mean for fixing the initial conditions
only when $\phi$ lies outside the spinodal region [cfr. eq
(\ref{NiceN}) or (\ref{bnice})]; otherwise, the gap equation does not
admit a positive solution for the squared effective mass. In that
case, I have to resort to other methods, in order to choose the
initial conditions. Following the discussion presented in \ref{ooedN},
one possible choice is to set $\s_k^2 = \frac1{2\sqrt{k^2+|M^2|}}$ for
$k^2<|M^2|$ and then solve the corresponding gap equation
(\ref{newgap}). An other acceptable choice would be to solve the gap
equation (\ref{newgap}), once I have set a massless spectrum for all
the spinodal modes but the zero mode, which is started from an
arbitrary, albeit microscopic, value.

There is actually a third possibility, that is in some sense half a
way between the unbroken and broken symmetry case. I could allow for
a time dependent bare mass, in such a way to simulate a sort of {\em
cooling down} of the system. In order to do that, I could start with
a unbroken symmetry bare potential (which fixes initial conditions
naturally via the gap equation) and then turn to a broken symmetry one
after a short interval of time. This evolution is achieved by a proper
interpolation in time of the two inequivalent parameterizations of the
bare mass, eqs. (\ref{Nm2ren}) and (\ref{minimum}) for large $N$ or 
(\ref{m2ren}) and (\ref{brokenm}) for Hartree--Fock.

In case of large $N$, I looked for the influence this different
choices could produce in the results and indeed they depend very
little and only quantitatively from the choice of initial condition I
make. As far as the Hartree--Fock approximation is concerned, I
commented extensively in section \ref{num1} the consequences of
choosing different gap equations to fix the initial conditions for the
quantum fluctuations.

Finally, I used a $4$th order Runge-Kutta algorithm to solve the
coupled differential equations (\ref{numeq}), performing the
computations in boxes of linear size ranging from $L = 20\pi$ to $L =
400\pi$ and verifying the conservation of the Wronskian to order
$10^{-5}$. Typically, I have chosen values of $\cal N$ corresponding
to the UV cutoff $\Lambda$ equal to small multiples of $m$ for
unbroken symmetry and of $v\sqrt{\l}$ for broken symmetry. In fact,
the dynamics is very weakly sensitive to the presence of the
ultraviolet modes, once the proper subtractions are performed. This is
because only the modes inside the unstable (forbidden or spinodal)
band grow exponentially fast, reaching soon non perturbative
amplitudes (i.e. $ \approx \lambda ^{-1/2}$), while the modes lying
outside the unstable band remains perturbative, contributing very
little to the quantum back--reaction \cite{Boyanovsky:1998zg} and
weakly affecting the overall dynamics. The unique precaution to take
is that the initial conditions be such that the unstable band lay well
within the cutoff. As a final comment, I want to stress that also in
the case of the non linear $\s$ model in $(1+1)$D, the dynamics is
insensitive to the addition of ultraviolet modes, once the proper
renormalization is performed. Most important, this happens even in the
absence of parametric instabilities because the theory is
asymptotically free.

\setcounter{equation}{0}  
\section{Gap Equations from tdHF for $N>1$}
\label{N>1}
As we have seen, it is generally accepted that, in order to study the
in medium properties of systems at finite density and temperature, the
trivial perturbative approach is not well suited. When dealing with
field theory at finite temperature, some sort of resummation is needed
and a mean field approach has been much used in the past.

As an example, the $\sigma$ meson and pions' masses have been studied
in the framework of the $O(N)$ symmetric linear $\sigma$ model
\cite{Lenaghan:1999si}, using the Cornwall-Jackiw-Tomboulis (CJT)
formalism and in the Hartree--Fock approximation. The CJT effective
action for composite operators is a functional of the expectation
value of the field and of the quantum propagator
\begin{equation}
\Gamma ( \phi _c (x), G(x,y) )
\end{equation}
and is the generating functional of the two particle irreducible
vacuum graphs. 

The stationary condition for the effective potential yields the usual
Schwinger--Dyson equations which reduce, in the HF scheme, to gap
equations for the in--medium dressed masses. In
ref. \cite{Lenaghan:1999si} the gap equations for finite $N$ and in
the large $N$ limit are derived and a renormalization attempted both
in the cutoff and counterterm schemes. The renormalization procedures
consistent with the chiral limit and a finite value for the UV cutoff
turns out to be possible only in large $N$. In other words, the
Nambu--Goldstone's theorem would not be satisfied at finite
temperature. This is because if one performs the renormalization after
a non--perturbative partial resummation, renormalization constants may
acquire some dependence on the medium properties. Also the in--medium
modifications of meson properties, like their mass and decay width,
are studied, computing the contribution to meson self--energy of the
{\em setting sun} diagram with internal dressed (HF) propagators.

It is interesting to compare this approach with others based on a
different definition of masses, or implementing different
resummation. For example, recently
\cite{Nemoto:1999qf} it has been shown that the Nambu--Goldstone's
theorem is safe, provided one defines the masses as the curvature of
the effective potential in the CJT formalism and in the Hartree--Fock
approximation. In this case, the renormalization is performed by an
extension of the auxiliary field method and some conclusions about how
the order of the chiral phase transition depends on the inclusion of
the quantum and thermal fluctuations are reached. Other resummation
schemes have been proposed, which guarantee that the Goldstone's
theorem is satisfied. One of them is the large--$N$ limit. Another
one, the 2PPI expansion, was introduced to study the the $O(N)$ linear
$\sigma$ model at finite temperature \cite{Verschelde:2000ta}, and has
the merit of summing the seagull and bubble graphs to all orders,
going beyond the gaussian Hartree approximation.

In this appendix, I am interested in showing how to derive the gap
equations in the gaussian approximation for the longitudinal ($\sigma$
meson) and transverse (pions) modes, starting from eqs. (\ref{omvev}),
(\ref{omvev1}) and (\ref{omvev3}). I will also make some comments one
the difficulties one finds when a renormalization is attempted.

Eq. (\ref{omvev1}) may be written also as
\begin{eqnarray}
(\omega_k^2)_{ij} = (k^2 + m_b^2) \delta_{ij} + \lambda_b \left[ \mathrm{Tr}
 (\Sigma) \delta_{ij} + 2 \Sigma_{ij} \right]
\end{eqnarray}
with $\Sigma$ defined as
\begin{equation}
\Sigma^{ij} = \dfrac{1}{L^D} \sum_{k ' \neq \pm k} \vev{ \varphi^i_{k '} \varphi^j_{k '}}
\end{equation}
The total energy, divided in the kinetic and potential parts, reads
\begin{eqnarray}
\mathcal E = \mathcal T( \dot{\bds\phi} , (\dot{\sigma}_k^2)_{ij} ) +
\mathcal V ( \bds\phi , (\sigma_k^2)_{ij} )
\end{eqnarray}
where the $(\sigma_k)_{ij}$ are the widths of the gaussian
wavefunctional and $\bds\phi$ is the field condensate. The potential
$\mathcal V$ [cfr. eq. (\ref{ElargeN}) for large $N$] may be written in
compact form
\begin{eqnarray}
\mathcal V = \dfrac{1}{2 L^D} \mathrm{Tr} \sum_k \left[(k^2 + m_b^2) 
(\sigma_k^2) + \dfrac{1}4 (\sigma_k^2)^{-1} \right] - \dfrac{\lambda_b}{2}
(\bds\phi^2)^2 + \dfrac{1}{2} m_b^2 \bds\phi^2 \\ \nonumber +
\dfrac{\lambda_b}{4} (\mathrm{Tr} \Sigma)^2 + \dfrac{\lambda_b}{2}
\mathrm{Tr} ( \Sigma^2)
\end{eqnarray}
I now distinguish the longitudinal width $\sigma_{kL}^2$ from the
transverse one $\sigma_{kT}^2$ and write
\begin{eqnarray}
(\mathrm{Tr} \Sigma)^2 & = & (\bds\phi^2 +\Sigma_L + (N-1)\Sigma_T)^2 \\
\mathrm{Tr} (\Sigma^2)   & = & (\bds\phi^2 +\Sigma_L)^2 + (N-1)\Sigma_T^2
\end{eqnarray}
where 
\begin{equation}
\Sigma_L = \dfrac{1}{L^D} \sum_k \sigma_{kL}^2 \qquad
\Sigma_T = \dfrac{1}{L^D} \sum_k \sigma_{kT}^2
\end{equation}
If I now fix the value of the condensate and try to minimize w.r. to
the quantum fluctuations, I get the two coupled bare gap equations:
\begin{eqnarray}
M_L^2 = m_b^2 + \lambda_b \left[ 3 \bds \phi^2 + \dfrac{1}{L^D} \left(
3 \sum_k \dfrac{1}{2 \sqrt{k^2 +M_{kL}^2}} + (N-1) \sum_k \dfrac{1}{2
\sqrt{k^2 + M_{kT}^2}} \right) \right]
\end{eqnarray}
\begin{eqnarray}
M_T^2 = m_b^2 + \lambda_b \left[\bds \phi^2 + \dfrac{1}{L^D} \left(
\sum_k \dfrac{1}{2 \sqrt{k^2 +M_{kL}^2}} + (N+1) \sum_k \dfrac{1}{2
\sqrt{k^2 + M_{kT}^2}} \right) \right]
\end{eqnarray}
In the infinite volume limit, these two gap equations can be written in
terms of the divergent momentum integral defined in (\ref{Nm2ren}):
\begin{eqnarray}
\label{gapeqnoriginarie}
M_L^2 & = & m_b^2 + 3 \lambda_b \bds \phi^2 + 3 \lambda_b I_D(M_L^2,\L)+
\lambda_b (N-1) I_D(M_T^2,\L) \\ \nonumber M_T^2 & = & m_b^2 + \lambda_b
\bds \phi^2 + \lambda_b I_D(M_L^2,\L) + \lambda_b (N+1) I_D(M_T^2,\L)
\end{eqnarray}

If we consider a quantum state, which is spherically symmetric in the
internal state, $\bds\phi = 0$ and $M_L^2 = M_T^2$. This allows for a
definition of the equilibrium, renormalized mass $m$ and a
parametrization of the bare mass, which reads
\begin{equation}
m_b^2 = m^2 - (N+2) \lambda_b I_D(m^2,\L)
\end{equation}
The gap equations become
\begin{eqnarray}
M_L^2 = m^2 + 3 \lambda_b \bds \phi^2 &+& 3 \lambda_b [I_D(M_L^2,\L) -
I_D(m^2,\L)] \nonumber \\ &+& \lambda_b (N-1) [I_D(M_T^2,\L) - I_D(m^2,\L)]
\end{eqnarray}
\begin{eqnarray}
M_T^2 = m^2 + \lambda_b \bds \phi^2 &+& \lambda_b [I_D(M_L^2,\L) -
I_D(m^2,\L)] \nonumber \\ &+&
\lambda_b (N+1) [I_D(M_T^2,\L) - I_D(m^2,\L)]
\end{eqnarray}

The renormalization of the coupling constant is still to be
performed. If I trivially generalize to $N>1$ the $N=1$ one--loop
resummed perturbative relation between the bare and running coupling
constant, I get
\begin{eqnarray}\label{l1l}
\lambda(\mu) = \dfrac{\lbare(\L)}{1-(N+2) \dfrac{\lbare(\L)}{8 \pi^2} \ln
\dfrac{\L}{\mu}}
\end{eqnarray} 
Now, the gap equations can be solved, but the two squared masses retain a
logarithmic dependence on the UV cutoff (plus the usual inverse power
corrections, always present due to the the Landau pole). Notice that,
for $N>1$, this is not the solution of the renormalization group flow,
which is given replacing the $N+2$ coefficient in the denominator with
$(N+8)/3$ \cite{amit}. In any case, it is not possible to make the
logarithmic dependence disappear. In addition, if I consider guassian
wavefunctional with mean value different from zero, their energy
w.r. to the ground state is not UV finite, as happened for the $N=1$
case of section \ref{tdhf}; thus a substitution similar to
(\ref{subst}) is necessary. 

These unpleasant features signal that the tdHF defined in section
\ref{impHF} is not completely renormalizable, (at least in the
gaussian approximation), as, on the contrary, is the full
theory. Something crucial for the renormalizability is lost when one
reduces himself to this approximation scheme. In order to have finite
renormalized masses and energy differences, I need to implement a more
sophisticated resummation than the simple one--loop relation
(\ref{l1l}). An enlightening view on the procedure to follow is
contained in ref. \cite{Nemoto:1999qf}, where a counterterm
renormalization scheme is used. Along this line, finite renormalized
gap equations can be obtained \cite{prep}. In any case, a diagrammatic
explanation of the resummation needed is far from being clear.

\setcounter{equation}{0}  
\section{Evolution of a spherically symmetric condensate in $2+1$ dimensions}

\input{duepiuno.tex}

\setcounter{equation}{0}  
\section{Technical Issues}

I would like to close with a last appendix, briefly describing the
hardware used to perform the numerical computations. I obtained the
results presented in chapters \ref{sqft} and \ref{smodel}, developing
a C code for the specific purpose of solving the coupled differential
equations of appendix \ref{num}. I run the code on the PC cluster of
Physic Department of Milan University, which is composed by 23 PC's,
equipped with Pentium processors with 166MHz of clock frequency.

As we have seen in section \ref{dis_homo}, the simulation of
microscopic models requires accurate and efficient methods for solving
the integro--differential equations that describe the evolution of
inhomogeneous mean fields self--consistently interacting with internal
and external fluctuations and driving forces. In general, at the
mesoscale level, the system will be described by nonlinear stochastic
PDEs that replace microscopic physics with terms represented by
colored, spatially--correlated and/or multiplicative noise
\cite{QFTooE3,Boyanovsky:1998pg,Calzetta:1994qe}.

Further development of the numerical methodology for complex nonlinear
stochastic PDEs will be an important base for this research project.

Following this direction, I am going to use the Beowulf cluster of the
Physic Department ``G. Occhialini'' of University of Milan-Bicocca,
which is composed by 32 Athlon processors with 500 MHz of frquency clock,
linked by means of ethernet boards. The equations to be solved
(\ref{rad_eqs}) are mean field coupled and they can be {\em weakly}
parallelized. They, in fact, require an intensive work on each node,
to perform the integration step on the whole spatial lattice; on the
other hand, the nodes need to communicate each other the information
to build the mean field only immediately before starting the
integration step. The task may be accomplished by the following
scheme:
\vspace{1.5cm}

\centerline{Parallelization on $l$}

\centerline{define a spatial lattice on the radial coordinate $r$}
\centerline{and cycle on these actions:}

\vspace{1cm}
\centerline{each node perform the temporal evolution step from $t$}
\centerline{to $t+dt$, for the entire lattice and for a subset of $l$'s}
\centerline{$\Downarrow$}
\centerline{sum over the nodes to compute the {\em mean field}}
\centerline{${\rm diag}(w)(r)$ at time $t+dt$}
\centerline{$\Downarrow$}
\centerline{tell each node the value of ${\rm diag}(w)(r)$ at time $t+dt$}

%% file: duepiuno.tex
I consider here a derivation similar to that of section
\ref{dis_homo} in $(2+1)$D. Recalling that the laplacian in spherical
coordinates reads in this case,
\begin{equation}
\nabla ^2 F = \dfrac1r \der{}{r}\left(r\der{F}{r}\right) +
\dfrac1{r^2}\dfrac{d^2F}{d \t^2}
\end{equation}
and, defining the radial [$u(r)$] and angular [$\T(\t)$] functions as
$F=u/\sqrt{r}\T$, I obtain the following equations for the functional basis:
\begin{equation}
\begin{array}{ll}
& \dfrac{d^2\T_\nu}{d \t^2} + \nu^2 \T_\nu = 0 \\
\\
& \dfrac{d^2u_\nu}{d r^2} - \dfrac{\nu^2-1/4}{r^2}\,u_\nu = 0
\end{array}
\end{equation}
which have the following solutions:
\begin{equation}
F_\nu(r,\t)= (A_\nu\,r^\nu + B_\nu\,r^{-\nu})\,\exp(i\,\nu\,\t)
\end{equation}

\vspace{0.5cm}
{\bf Spherical symmetry}\\
in case of rotational invariance, we have 
\begin{equation}
	\phi (x) = \dfrac{\varphi(r)}{\sqrt{2\pi\,r}}
	\;,\quad \pi (x) = \dfrac{p(r)}{\sqrt{2\pi\,r}}
\end{equation}
and we can expand the 2-point functions as follows\\
($rr'\cos\t = x\cdot x'$):
\begin{equation}
	\Gamma(x,x') = \sum _{\nu=-\infty} ^{\infty}
	\dfrac{\exp(i\,\nu\,\t)}{2\pi\,\sqrt{rr'}}\;\Gamma^{(\nu)}(r,r')
\end{equation}
where $\Gamma$ is any of the 2-point functions $w$, $v$, $u$ or $s$.
As usual, the boundary conditions at $r=0$ are 
\begin{equation}
	\varphi(0) = 0 = p(0) \;,\quad 
	\Gamma^{(\nu)}(0,r') = \Gamma^{(\nu)}(r,0) = 0
\end{equation}
In case of a finite volume with radius $R$, suitable b.c. (Dirichlet,
Neumann) can be assumed also at $r=R$.
The coincidence limit for $w$ reads:
\begin{equation}
{\rm diag}(w)(r) = w(x,x) = 
	\sum _{\nu=-\infty} ^{\infty} \dfrac1{2\pi\,r} w^{(\nu)}(r,r)
\end{equation}

\vspace{0.5cm}
{\bf A useful relation}\\
$r_i r_j \cos \t_{ij} = x_i \cdot x_j$
\begin{equation}
\int d\t_2 \exp(i\,\mu\,\t_{12}) \exp(i\,\nu\,\t_{23}) =
2\pi \d_{\mu \nu} \exp(i\,\nu\,\t_{13})
\end{equation}
This formula let us write, for instance
\begin{equation}
\begin{array}{ll}
\displaystyle{\int} d^2 x_2 d^2 x_3 v(x_1,x_2) w(x_2,x_3) v(x_3,x_4) &= \\ 
\\
& \hspace{-3cm}
 \sum_{\nu=-\infty}^{\infty} \dfrac{\exp(i\,\nu\,\t_{14})}{2\pi\, \sqrt{r_1 r_4}}
 \int dr_2 dr_3 v ^{(\nu)} (r_1 ,r_2 ) w ^{(\nu)} (r_2 ,r_3 ) v ^{(\nu)}
 (r_3 ,r_4 )
\end{array}
\end{equation}

\vspace{0.5cm} 
{\bf Hamiltonian}\\ 
It is easy to show that in case of rotational invariance in $2$
spatial dimensions, the Hamiltonian can be written as
\begin{equation}
\begin{array}{ll}
	H &= \displaystyle{\int} dr \left\{ \dfrac12 p^2 +\dfrac12
	(\partial_r\varphi)^2 + 2\pi r V\left( \varphi^2/ 2\pi r +
	{\rm diag}(w)\right) \right\} \\ \\ &+ \frac12 \sum
	_{\nu=-\infty}^{\infty} \displaystyle{\int} dr \left\{
	\left[s^{(\nu)}(r,r) + \left( -\partial_r^2
	+\dfrac{\nu^2-1/4}{r^2} \right) w
	^{(\nu)}(r,r')\bigr|_{r=r'}\right] \right\}
\end{array}
\label{invenergy1}
\end{equation}
Equations in case of rotational invariance:
\begin{equation}
\begin{array}{ll}
\dot\varphi = & p \\
\dot p = & D^{(0)} \varphi \\
\dot w^{(\nu)} = & u^{(\nu)} + u^{(\nu)T} \\
\dot u^{(\nu)} = & s^{(\nu)} + D^{(\nu)}_{r'} w ^{(\nu)} \\
\dot s^{(\nu)} = & D^{(\nu)} u^{(\nu)} + \left(D^{(\nu)} u^{(\nu)}\right)^T
\end{array}
\end{equation}
where  
\begin{equation}
D^{(\nu)} = \pdif{^2}{r^2} - \dfrac{\nu^2-1/4}{r^2} - M(r)^2 
\end{equation}

\vspace{0.5cm}
{\bf $\Phi^4$ potential}\\
I now specify the potential suitable for $\Phi^4$ model. In that case
I have\\ $V(z) = 1/2\mbare z + \l /4 z^2$.

\vspace{0.5cm} {\bf Discretized equations}\\
I set up a spatial (radial) lattice with spacing $a$ from $0$ to
$R=Na$, so that the total number of sites is $N+1$. In this case the
background field is a vector $\varphi_j=\varphi(ja)$ and the
2-point functions are standard  matrices $\Gamma_{ij} =
\Gamma(ia,ja)$, with $i,j=0, \cdots, N$. The discretized version for
the second derivative is the standard one: $\pdif{^2}{r^2} \varphi(r)
\bigr|_{r=ja} = (\varphi_{j+1} - 2 \varphi_j +
\varphi_{j-1})/a^2$. The square effective mass becomes:
\begin{equation}
M_j^2 = \mbare + \dfrac{\l}{2\pi\,j a} \left[ \varphi _j^2 
+ \sum_{\nu=-\infty}^{\infty} w^{(\nu)}_{jj} \right]
\end{equation}
Now I can try to use the fourth--order Runge-Kutta algorithm (already
used in the homogeneous case) to solve this system of coupled ordinary
differential equations.

\vspace{0.5cm} 
{\bf Initial conditions}\\ 
I can fix the initial conditions in the following way: I start with
an arbitrary profile for $\varphi(r)$ and with $p(r)=0$; then I want
to find suitable initial conditions for the 2-point functions. One
possible choice is to minimize the energy functional with respect to
the fluctuations. This is achieved first by setting $v(x,y)=0$; 
in that case, also $u(x,y)$ is $0$; then I must find some minimal
$w^{(\nu)}(r,r')$. From the numerical point
of view, I may choose two possible strategies in order to solve this
problem:\\ 1) one might try to solve the non--linear differential
equations for $w^{(\nu)}$:
\begin{equation}
	s^{(\nu)} (r,r') + \left[\pdif{^2}{r^2} -
	 \dfrac{\nu^2-1/4}{r^2} - M(r)^2\right]w^{(\nu)}(r,r') = 0
\label{init1}
\end{equation}
where now
\begin{equation}
\int dr s ^{(\nu)} (r,r') w ^{(\nu)} (r',r'') = \frac14 \delta (r - r'')
\end{equation}
2) otherwise, one may try to minimize directly the energy functional
(\ref{invenergy1}), using numerical algorithms like the Simplex,
Conjugated Gradient or Simulated Annealing methods.

\vspace{0.5cm} 
{\bf Free Massive Scalar Field}\\
To clarify matters, expecially w.r.t. the renormalization issue, let
us compute the expansion in partial waves of the Green function
$w(x,y)$ for a free massive scalar field and check that its spherical
components $w ^{(\nu)}$ satisfy the equations (\ref{init1}).
\begin{equation}
w_0(x_1,x_2;m)=w_0(x_1-x_2;m)=\dfrac12 \displaystyle{\int}
\dfrac{d^2k}{(2\pi)^2}
\dfrac{e^{\imath k \cdot (x_1-x_2)}}{\sqrt{k^2+m^2}}
\end{equation}
This integral can be computed in close form (cfr. \cite{grad}, 3.338
2., page 309 and 6.554 1., page 682) 
\begin{equation}
\begin{split}
w_0(r_{12};m)&=\dfrac1{8\pi^2}\int dk\, \dfrac{k}{\sqrt{k^2+m^2}}
\int d\t\,\exp (ikr_{12}\cos\t) \\ &= \dfrac1{4\pi} \int dk\,
\dfrac{k}{\sqrt{k^2+m^2}}\,J_0(kr_{12})\\
\end{split}
\end{equation}
\begin{equation}
w_0(r_{12};m)=\dfrac{1}{4\pi r_{12}} \exp(-mr_{12}) = \dfrac1{4\pi r_{12}}
+ {\rm finite}
\end{equation}
where the coincidence limit sigularity is made explicit.\\
The following standard properties hold:
\begin{equation}
\begin{array}{ll}
\sum_{\nu=-\infty}^{\nu=\infty} \exp[i\nu(\t_1-\t_2)] &= \d (\t_1-\t_2)\\
\\
\int dr\,r J _\nu (k_1 r) J _\nu (k_2 r) &= \dfrac1{k_1} \d (k_1 -
k_2) \quad \forall \nu
\end{array}
\end{equation}
Using the partial wave expansion of the plane wave (in $2$ spatial
dimensions) 
\begin{equation}
\exp(ikr\cos\t) = \sum_{\nu=-\infty}^{\nu=\infty} i^\nu \exp(i\nu\t) J_\nu(kr)
\end{equation}
and after some algebra I end up with
\begin{equation}
w_0(x_1,x_2;m) = \sum_{\nu=-\infty}^{\nu=\infty}
\dfrac{\exp(i\nu(\t_{12})}{2\pi} \int dk k
\dfrac{J _\nu (kr_1) J _\nu (kr_2)}{2 \sqrt{k^2+m^2}}
\end{equation}
from where I can read the explicit expression of $w ^{(\nu)}$:
\begin{equation}
w_0 ^{(\nu)}(r_1,r_2;m) = \sqrt{r_1r_2}\int_0^\Lambda dk k
\dfrac{J _\nu (kr_1) J _\nu (kr_2)}{2 \sqrt{k^2+m^2}} \;,\quad \Lambda \to\infty
\label{wl1}
\end{equation}
The spherical components of $w^{-1}$ are given by:
\begin{equation}
w_0 ^{(\nu)-1}(r_1,r_2;m) = \sqrt{r_1r_2} \int dk k 2
\sqrt{k^2+m^2} J _\nu (kr_1) J _\nu (kr_2)
\end{equation}
One can easily verify that $w ^{(\nu)}$ is a solution of the
self-consistent equation (\ref{init1}); in fact, the Bessel
functions $J _\nu (kr)$ are eigenfuctions of the Bessel operator
\begin{equation}
\left( - \partial _r ^2 + \dfrac{\nu^2-1/4}{r^2} \right) [\sqrt{r} J_\nu
(kr)] = k^2 [\sqrt{r} J_\nu (kr)]
\end{equation}
Let us now consider the Hamiltonian (\ref{invenergy1}). For a free
massive scalar field, the potential reduces to the form:
\begin{equation}
\dfrac{m^2}2 \sum_\nu \int dr w_0 ^{(\nu)}(r,r;m)
\end{equation}
that can be written as
\begin{equation}
\dfrac{m^2}2 \int dk \dfrac{k}{2 \sqrt{k^2+m^2}}\int dr r \sum_\nu J _\nu ^2 (kr)
\end{equation}
The (functional) series in the internal integral is a constant exactly
equal to $1$ (cfr. \cite{grad}, 8.536 3., page 980); thus I am left
with the integral on the quantum fluctuations times a surface factor,
due to translation invariance
\begin{equation}
\sum_\nu \int dr\, w _0 ^{(\nu)}(r,r;m) = \dfrac{S}2 \int
\dfrac{d^2k}{(2\pi)^2} \dfrac1{\sqrt{k^2+m^2}} 
\end{equation}
In other words, the surface factor can be written as
\begin{equation}
\d^{(2)}(k)\bigr|_{k=0} = \dfrac{S}{(2\pi)^2} = \sum_\nu
\dfrac1{2 \pi} \int dr r J _\nu ^2 (kr)
\end{equation}

\vspace{0.5cm}
{\bf Renormalization}\\
The coincidence limit of the 2-point function yields ultraviolet
divergences that must be properly subtracted before solving the
evolution equations numerically. I consider the case of {\em unbroken
symmetry} for simplicity. The space-time dependent effective mass must
be written in terms of finite quantities, and this sets our 
renormalization conditions. First, I parametrize $\mbare$ using the
equilibrium free field 2-point function for a massive field of
renormalized mass $m$:
\begin{equation}
	\mbare = m^2 - \l \,{\rm diag}(w_0) = m^2 - 
	\l\sum_\nu \dfrac{w^{(\nu)}_0(r,r;m)}{2\pi\,r}
\end{equation}
where $w ^{(\nu)} _0 (r,r')$ is given by eq. (\ref{wl1}). When the sum
over $\nu$ runs from $-\infty$ to $\infty$, the complete free ultraviolet
divergence is correctly rebuilt:
\begin{equation}
	\left({\rm diag}(w_0)\right)(r) = 
	\int_0^\Lambda dk\, \dfrac{k^2}{2\sqrt{k^2+m^2}}
	\sum_l \dfrac1{2\pi} J_\nu(kr)^2 =
	\int_0^\Lambda \dfrac{d^2k}{(2\pi)^2}\, \dfrac1{2\sqrt{k^2+m^2}}
\end{equation}
We do not have a logarithmic divergence, because the theory is 
superrenormalizable. Thus, I do not need to distiguish between a bare
and renormalized coupling constant. The effective squared mass is
given by
\begin{equation}\label{good1}
\begin{array}{ll}
	M(r)^2 &= m^2 + \l\left[\phi(r)^2+ {\rm diag}(w)_{\rm
	R}\right]\\ \\ &= m^2 + \dfrac{\l}{2\pi r} \left\{
	\varphi(r)^2 + \displaystyle{\sum_{\nu=-m_{\rm max}}^{m_{\rm
	max}}} \left[w^{(\nu)}(r,r) - w^{(\nu)}_0(r,r;m) \right]
	\right\}
\end{array}
\end{equation}
When I stop the sum over the partial waves at a finite $m _{\rm
max}$, I should subtract the ultraviolet divergences before
performing the sum. The partial waves $w^{(\nu)}_0 (r,r;m)$ should be
computed once and for all at the beginning, performing the integral
(\ref{wl1}) for the values of $r$ corresponding to the lattice chosen
and with an upper momentum cut-off equal to $\pi/a$. Recalling that,
{\em for fixed} $\nu$, each $w^{(\nu)}_0(r,r;m)$ has only a logarithmic
divergence in the ultraviolet cut-off $\Lambda$, as can be easily
inferred expanding for large arguments the Bessel function in
eq. (\ref{wl1}), plus finite parts that do depend on $r$. Thus,
subtracting the divergence for each $m$ before performing the sum
could be quantitatively very different (for given $m_{\rm max}$,
$\Lambda$ and $R$) from subtracting beforehand the entire {\em
constant} $\left({\rm diag}(w_0)\right)$:
\begin{equation}
M ^2 (r) = m^2 + \l \left[\phi(r)^2+ \sum_{\nu=-m_{\rm max}} ^{m _{\rm max}}
\dfrac1{2\pi r} w ^{(\nu)} (r,r) - {\rm diag}(w_0)\right]
\end{equation}
With the subtraction scheme as in (\ref{good1}), the functional gap
equation
\begin{equation}\label{gapw1}
\begin{array}{ll}
	& M(r)^2 = m^2 + \dfrac{\l}{2\pi r} \left\{ \varphi(r)^2 +
	\displaystyle{\sum_{\nu=-m_{\rm max}}^{m_{\rm max}}}
	\left[w^{(\nu)}(r,r) - w^{(\nu)}_0(r,r;m) \right] \right\} \\
	\\ &\left[-\pdif{^2}{r^2} + \dfrac{\nu^2-1/4}{r^2} + M(r)^2
	\right]w^{(\nu)}(r,r') = \dfrac14 w^{(\nu)\,-1}(r,r')
\end{array}
\end{equation}
that determine the initial conditions, trivially admits the
equilibrium solution $\phi=0$, $M(r)^2 = m^2$, $w^{(\nu)}=w_0^{(\nu)}$.
Eq. (\ref{gapw1}) is formally solvable via mode expansion: suppose we
have the complete solution of the eigenvalue problem
\begin{equation}
	\left[-\pdif{^2}{r^2} + U_\nu(r)\right] [\sqrt{r}\chi^{(\nu)}_k(r)]
	=  k^2 [\sqrt{r} \chi^{(\nu)}_k(r)]
\end{equation}
where $U_\nu(r)=(\nu^2-1/4)/r^2 + M(r)^2 -m^2$  cannot be negative
since $M(r)^2 >m^2$ for unbroken symmetry and is assumed to vanish
for large $r$ fast enough; then
\begin{equation}
	w^{(\nu)}(r,r') = \sqrt{rr'} \int_0^\Lambda dk\, k 
	\dfrac{\chi^{(\nu)}_k(r)\chi^{(\nu)}_k(r')}{2\sqrt{k^2+m^2}}
\end{equation}
and the gap equation reads
\begin{equation}
	M(r)^2 = m^2 + \dfrac{\l}{2\pi r} \left\{ \varphi(r)^2 + 
	\sum_{\nu=-m_{\rm max}}^{m_{\rm max}} \int_0^\Lambda dk\,k
	\dfrac{\chi^{(\nu)}_k(r)^2 - J_\nu(kr)^2}{2\sqrt{k^2+m^2}} \right\}
\end{equation}

%% file: dedica.tex

Voglio sinceramente ringraziare chi nel corso di questi anni ha voluto
(o spesso dovuto) condividere con me i dubbi e le incertezze che mi
hanno accompagnato nel tentativo di svolgere al meglio questo duro ma
affascinante mestiere: (in rigoroso ordine alfabetico) Bartomeu Alles
Salom, Vito Antonelli, Lorenzo Belardinelli, Paolo Bertona, Marco
Bianchetti, Francesco Bigazzi, Paolo Bonini, Daniel Boyanovsky, Andrea
Brognara, Paolo Butera, Francisco J. Cao, Filippo Castiglia, Paolo
Ceccherini, Gennaro Corcella, Hector J. de Vega, Ruggero Ferrari,
Giulio Gianbrone, Adolfo Gianbastiani, Nicola Giovanardi, Richard
Holman, Petteri Ker\"anen, Cristina Lencioni, Andrea Lucenti, Giuseppe
Marchesini, Stefano Micheli, Fabio Monforti, Carlo Oleari, Andrea
Pasquinucci, Michele Pepe, Michela Petrini, Marco Picariello, Dirk
H. Rischke, Michele Simionato, Fabio Simonetti, Michele Sturlese,
Valentina Torti, Marco Vanzini, Filippo Vernizzi, ma soprattutto mia
moglie Tania.

Infine, un ringraziamento particolare va a chi ha saputo trasmettermi
in ogni occasione, con incontenibile entusiasmo, la sua passione per la
Fisica, e con pazienza ha tentato di condurmi sulla strada della
ricerca scientifica. Grazie, Claudio !
\\
\\
14 novembre 2000
\begin{flushright}
Emanuele Manfredini
\end{flushright}

%% file: canzone.tex

{\em Canzone delle domande consuete}\\

[Dall'album "quello che non ..."]\\

Ancora qui a domandarsi e a far finta di niente\\
come se il tempo per noi non costasse l'uguale,\\
come se il tempo passato ed il tempo presente\\
non avessero stessa amarezza di sale.\\
Tu non sai le domande, ma non risponderei\\
per non strascinare parole in linguaggio d'azzardo;\\
eri bella, lo so, e che bella che sei; dicon tanto un silenzio e uno sguardo.\\
Se ci sono non so cosa sono e se vuoi\\
quel che sono o sarei, quel che sar\`o domani ...\\
non parlare non dire pi\`u niente se puoi,\\
lascia fare ai tuoi occhi e alle mani.\\
Non andare ... vai. Non restare ... stai.\\
Non parlare ... parlami di te.

Tu lo sai, io lo so, quanto vanno disperse,\\
trascinate dai giorni come piena di fiume\\
tante cose sembrate e credute diverse\\
come un prato coperto a bitume.\\
Rimanere cos\`\i, annaspare nel niente,\\
custodire i ricordi, carezzare le et\`a;\\
\`e uno stallo o un rifiuto crudele e incosciente\\
del diritto alla felicit\`a ?\\
Se ci sei, cosa sei ? Cosa pensi e perch\'e ?\\
Non lo so, non lo sai, siamo qui o lontani ?\\
Esser tutto, un momento, ma dentro di te.\\
Aver tutto, ma non il domani.\\
Non andare ... vai. Non restare ... stai.\\
Non parlare ... parlami di te.

E siamo qui, spogli, in questa stagione che unisce\\
tutto ci\`o che sta fermo, tutto ci\`o che si muove;\\
non so dire se nasce un periodo o finisce,\\
se dal cielo ora piove o non piove,\\
pronto a dire "buongiorno", a rispondere "bene"\\
a sorridere a salve, dire anch'io "come va ?"\\
Non c'\`e vento stasera. Siamo o non siamo assieme ?\\
Fuori c'\`e ancora una citt\`a ?\\
Se c'\`e ancora balliamoci dentro stasera,\\
con gli amici cantiamo una nuova canzone ...\\
... tanti anni, e sono qui ad aspettar primavera\\
tanti anni, ed ancora in pallone.

Non andare ... vai. Non restare ... stai.\\
Non parlare ... parlami di te.\\
Non andare ... vai. Non restare ... stai.\\
Non parlare ... parlami di noi.

\vskip 0.5cm
Parole e musica di F.Guccini.\\

\begin{flushright}
19 dicembre 1999
\end{flushright}

%% file: pub-articolo.tex
\begin{center}
{\em ...tratto da un articolo di Marco Lodoli}

               Il grande fascino delle rose mai colte
\end{center}                                                                  
        C'\`e una pagina del       
        racconto {\em La mela d'oro} di 
        Hugo von Hoffmansthal che 
        mi capita di rileggere    
        spesso, per capirla a     
        fondo. Una donna si       
        ridesta da un breve sonno 
        con l'impressione di aver 
        sognato o pensato nel     
        dormiveglia qualcosa di   
        strano. Quelle immagini   
        vaghe le stanno addosso   
        anche ora che \`e ben     
        sveglia. E Hoffmansthal   
        scrive: {\em Non era tanto    
        un'insoddisfazione della  
        propria vita, quanto una  
        rappresentazione          
        lusinghevole di come      
        avrebbe potuto essere     
        diversa, una febbre       
        silenziosa in cui si      
        svolgevano con eccessivo  
        diletto vicende non       
        vissute... Tornava a lei  
        il senso della sua natura 
        di fanciulla, anima e     
        corpo, e avvolta in un    
        destino che non era       
        divenuto il suo,          
        conduceva con amici e     
        nemici sconosciuti, con   
        larve indistinte,         
        colloqui in cui si        
        effondeva tutto ci\'o che 
        giaceva in lei di         
        inespresso e infruttuoso, 
        una tale profusione di    
        inesauribili possibilit\'a,
        di gioco e d'abbandono,    
        che trascinava con s\`e e  
        sommergeva la realt\'a...  
        Non sapeva che era         
        appunto la moltitudine di  
        queste possibilit\'a       
        interiori che la           
        preservava da desideri     
        volgari}. Credo che mai    
        sia stata descritta con    
        tanta finezza quella       
        sensazione, che ognuno di  
        noi talvolta avverte, di   
        essere circondati dalle    
        infinite vite che          
        potevano essere e non      
        sono state. Quelle         
        occasioni non colte, quei  
        baci non dati, quegli      
        arditi viaggi che, una     
        volta intrapresi,          
        avrebbero modificato       
        completamente la nostra    
        esistenza, rimangono a     
        danzare come farfalle      
        attorno al fiore           
        difficile di quest'unica   
        vita, all'unico gambo che 
        ha saputo spuntare con    
        forza nella terra.        
        Tuttavia ci\'o che \`e    
        rimasto inaccaduto non    
        deve diventare un         
        rimpianto doloroso,       
        un'infelicit\'a che ci rode   
        e ci assottiglia, ma anzi   
        - come suggerisce           
        Hoffmansthal - \`e un bene  
        invisibile che ci           
        allontana da una adesione
        troppo letterale e
        meschina alla nostra
        esistenza. Ognuno di noi
        \`e il risultato pi\'u
        prezioso tra le
        innumerevoli possibilit\`a
        offerte, \`e ci\'o che via
        via
  faticosamente prende 
forma nel mezzo di un
gassoso universo di
 ipotesi, le quali per\'o
rimangono a vagare nella
mente e a dare ai nostri  
   giorni uno sfondo         
   infinito, come fanno le   
        stelle sopra ai campi
     lavorati o gli sterminati
    volumi della biblioteca
      di Babele dietro alle
        poche precise parole che
      abbiamo saputo scrivere.  

...

 Cosa ci sarebbe accaduto
 se quel giorno, invece di
      entrare in quel bar,
      invece di rispondere a
      quella telefonata... Ogni
 secondo contiene almeno
       due secondi, quello che
          abbiamo vissuto e
     quell'altro, che \`e andato
 a posarsi sull'infinito
   Orizzonte degli Scarti e
     che non troveremo pi\`u, o
   solo in sogno, come una
    luce che allarga questa
    vita stretta. 

...

\begin{center}
          L'articolo completo \`e stato pubblicato a pag.57 del {\em
Diario della Settimana},
                            n.39 anno 3
\end{center}